\documentclass[a4paper,fleqn,usenatbib,useAMS]{mnras}
\usepackage[T1]{fontenc}
\usepackage{ae,aecompl}
\usepackage{graphicx}   % Including figure files
\usepackage{amsmath}    % Advanced maths commands
\usepackage{amssymb}     % Extra maths symbols
\usepackage{multicol}   % Multi-column entries in tables
\usepackage{bm}
\usepackage{pdflscape}

\title[Variability of TW Hya in the years 2013-2017] 
{Photometric variability of TW~Hya from seconds to years as seen from space and the ground in 2013-2017}
\author[M. Siwak et al.]
{Michal Siwak$^{1}$\thanks{E-mail: siwak@astro.as.up.krakow.pl},
Waldemar Ogloza$^{1}$,
Anthony F.\ J.\ Moffat$^2$,
Jaymie M.\ Matthews$^3$,\newauthor  
Slavek M.\ Rucinski$^4$,
Thomas Kallinger$^5$,
Rainer Kuschnig$^5$,
Chris Cameron$^{6,7}$,\newauthor
Werner W.\ Weiss$^5$,
Jason F.\ Rowe$^2$,
David B.\ Guenther$^8$,	
Dimitar Sasselov$^9$\\
$^1$Mount Suhora Astronomical Observatory, Cracow Pedagogical University,
ul.\ Podchorazych 2, 30-084 Krakow, Poland\\
$^2$D\'{e}partment de Physique, Universit\'{e} 
de Montr\'{e}al, C.P.6128, Succursale: Centre-Ville,
Montr\'{e}al, QC, H3C~3J7, Canada\\
$^3$Department of Physics \& Astronomy, University of
British Columbia, 6224 Agricultural Road, Vancouver, B.C., V6T~1Z1, Canada\\
$^4$Department of Astronomy and Astrophysics,
University of Toronto, 50 St.\ George St., Toronto,
Ontario, M5S~3H4, Canada\\
$^5$Universit\"{a}t Wien, Institut f\"{u}r Astrophysik, 
T\"{u}rkenschanzstrasse 17, A-1180 Wien, Austria\\
$^6$Department of Mathematics, Physics \& Geology, Cape Breton University,
1250 Grand Lake Road, Sydney,NS, B1P 6L2, Canada\\
$^7$Canadian Coast Guard College, Dept. of Arts, Sciences, and Languages,
Sydney, Nova Scotia, B1R 2J6, Canada\\
$^8$Institute for Computational Astrophysics,
Department of Astronomy and Physics,
Saint Mary's University, Halifax, N.S., B3H~3C3,\\ Canada\\
$^9$Harvard-Smithsonian Center for Astrophysics,
60 Garden Street, Cambridge, MA 02138, USA
}
\date{Accepted ;  Received ; in original form }

\pubyear{2018}

\begin{document}
\label{firstpage}
\maketitle

\begin{abstract}
This is the final photometric study of TW~Hya based on new {\it MOST\/} satellite observations. 
During 2014 and 2017 the light curves showed stable 3.75 and 3.69~d quasi-periodic oscillations, 
respectively. 
Both values appear to be closely related with the stellar rotation period, as they might be created 
by changing visibility of a hot-spot formed near the magnetic pole directed 
towards the observer. 
These major light variations were superimposed on a chaotic, flaring-type activity caused by hot-spots 
resulting from unstable accretion -- a situation reminiscent of that in 2011, when TW~Hya showed signs 
of a moderately stable accretion state.
In 2015 only drifting quasi-periods were observed, similar to those present in 2008-2009 data and typical 
for magnetised stars accreting in a strongly unstable regime.\newline 
A rich set of multi-colour data was obtained during 2013-2017 with the primary aim 
to characterize the basic spectral properties of the mysterious occultations in TW~Hya. 
Although several possible occultation-like events were identified, they are not as well defined as 
in the 2011 {\it MOST} data.
The new ground-based and {\it MOST} data show a dozen previously unnoticed 
flares, as well as small-amplitude, 11~min -- 3~hr brightness variations, 
associated with 'accretion bursts'.
It is not excluded that the shortest 11-15~min variations could 
also be caused by thermal instability oscillations in an accretion shock.
\end{abstract}


\section{Introduction}
\label{intro}

TW~Hya was shown to be a genuine Classical T~Tauri-type Star (CTTS) \citep{ruc83} 
in a young (about 7-10 Myr) association called TWA \citep{kastner97,barrado}. 
It is one of the last two stars in the association which still show vigorous accretion 
(see the results of \citealt{tofflemire17} for the young binary TWA~3A) and is also the closest 
(59.5~pc, \citealt{gaia16}) T~Tauri-type star to us.
The pole-on geometry of its transitional protoplanetary disc visibility (inclination of 5-15~deg 
between stellar rotational axis and observer was derived for the star -- see \citealt{ruc08} 
for a review and also in \citealt{donati11}) favours 
TW~Hya for detailed studies with modern imaging instrumentation: in addition to the previously known
disc gap at 80~AU found in {\it HST} images \citep{debes13}, \cite{akiyama15} found a gap localized 
in the disc at a Uranus distance of 20~AU using {\it Subaru-HiCIAO}. 
A most detailed image of TW~Hya provided by {\it ALMA} \citep{andrews16} revealed numerous rings -- 
the closest localized at 1~AU from the star has been popularly interpreted as caused by an Earth-like 
planet sweeping disc matter in its orbit. 
Sophisticated, deep spectral differential imaging in the Pa$\beta$ line looking for accretion signatures 
of possible planets was sensitive to 1.45-2.3 Jupiter-mass planets but did not reveal 
significant signals \citep{uyama17}.
Similar upper limits at 1-2.3 Jupiter masses for planets expected to form in the four well-known 
gaps in the disc were placed from long-exposure $L'$-band coronographic imaging 
observations by \citet{ruane17}.
In addition to the above, \citet{debes17} analyzed {\it HST-STIS} images obtained over 18 years. 
These images reveal that the inner disc is likely inclined and precessing, blocking central star-light 
and casting a shadow on the more external disc parts. 
The shadow is moving coherently with a period of 15.9~yr, 

In spite of superior modern instrumentation, investigation of  innermost disc dynamics 
is still possible only through photometry and spectroscopy. 
For instance, these old but well-established techniques enable insight into the properties 
of hot-spots, which are the disc-plasma fingerprints on a star, produced during magnetospheric 
accretion \citep{konigl91}. 
Photometric observations of TW~Hya were started almost four decades ago and were primarily aimed 
at determining the rotational period of the star -- the basic clock in the star-disc system. 
The first photometric results of \cite{ruc83} and \cite{ruc88} were inconclusive: 
these authors were unable to find any stable periodicity. 
Although \cite{mekkaden98} found a periodic value at 2.196~d, \cite{herbst88} found two totally 
different but significant peaks at 1.28~d and 4.5~d. 
\cite{alencar02} analyzed the power spectrum of veiling variations and obtained three major periodicities 
at $1.4\pm0.1$~d, $2.85\pm0.25$~d and $3.75\pm0.45$~d. 
A similar study by \citet{batalha02} pointed to $4.4\pm0.4$~d as the stellar rotation period. 
The period found by \citet{lawson05} at 2.80~d confirmed that no single value, coherent over 
many years, can be established from ground-based observations of this active star.  

Explanation of this ambiguity was possible after the launch of Microvariability \& Oscillations 
of STars ({\it MOST}) -- the first space telescope dedicated to long-term, high-precision 
photometry of $\sim0-12$~mag stars \citep{WM2003, M2004}. 
\cite{ruc08} observed TW~Hya continuously for 11.4 and 46.7 days in 2007 and 2008, respectively. 
During the longer run, the star showed irregular 'shot-noise' or flaring-type 
activity, with typical amplitudes of about 0.5~mag, typical for Type II, irregular and 
accretion-burst classes, as defined by \citet{herbst94}, \citet{alencar10} 
and \citet{cody14} \& \citet{stauffer14}, respectively. 
\cite{ruc08} found this flicker-noise character of the Fourier spectrum without any dominant 
periodicity -- only during the 2007 observations did the star show a  possibly stable 3.7~d
quasi-period, but the run was too short to definitely confirm this conclusion.
In the longer 2008 light curve analyzed with the wavelet technique the authors discovered 
a few quasi-periodic oscillations (QPOs) showing period shortening over the run. 
These findings were interpreted as possibly caused by condensed hot plasma clumps localized 
between 15-2~R$_{\sun}$ in the disc, gradually moving on spiral orbits toward the star. 
Similar quasi-periodic oscillatory variations, which appeared in the accessible range 
of about 10-1.3~d and shortened their periods by typically a factor of two within 
a few weeks, were confirmed during the third, 40.3-day long run by \cite{siwak11}. 

Yet the results of the fourth set of 2011 {\it MOST} observations were exceptional 
when compared with those from the previous seasons \citep{siwak14}: while the overall light variations 
retained the general characteristics of flicker noise, the 2011 season variations did not show any 
obvious period shortening of oscillation features. 
This time, the Fourier and the wavelet spectra were dominated by a single, almost stable oscillation 
with a period of $4.18\pm0.25$~d. 
We interpreted these findings within the framework proposed by \citet{romanova04, romanova08, romanova09}, 
in which the regime of accretion is responsible for a variety of light variations 
observed in magnetised stars, including CTTSs.
According to these authors, the accretion regime can be stable, moderately-stable or unstable at any particular time. 
Moreover, it can also alternate among different states depending primarily on the temporary mass-transfer rate. 
The resulting light curves of a CTTS may change from (more or less) regular 
to (very) irregular, which also affects the Fourier spectra. 
The Fourier spectrum will be dominated by a single or two well-defined stable peak(s) if the accretion regime is stable. 
In this state, two antipodal high-latitude hot-spots are produced by accretion 
funnels arising at the disc co-rotation radius, and the peaks are expected to reveal 
the rotational period of the star. 
Once the mass accretion rate increases, instabilities start to appear at the {\it inner disc--magnetosphere} 
interface: the accretion regime becomes moderately stable and the Fourier spectrum shows additional 
frequencies arising from hot-spots created by stochastic equatorial tongues of accreting plasma. 
The spectra will eventually become very noisy, without any dominant peaks during 
an unstable state of accretion. 
In this state several QPOs may be produced simultaneously by hot-spots 
revolving on a star with Keplerian frequencies of an inner disc. 
Their periods, being a fraction of a true rotational period of the star, may show considerable 
variations if the mass transfer rate inside the particular narrow tongue changes during its short 
life-time \citep{kulkarni09}.

According to the above scheme, due to the absence of any stable periodicity in the 2008 and 2009 {\it MOST\/} 
light curves, we inferred that a strongly unstable regime of accretion, solely through fast moving equatorial 
tongues, operated in TW~Hya at the time. 
For the same reason, we proposed that the dominant oscillation in 2011 was caused 
by rotational modulation produced by a single, large hot-spot formed 
by a stable accretion funnel, 
and the fairly stable 4.18~d signal could represent the true rotational period of the star. 
However, it was not clear how the 4.18~d periodicity relates to the previously observed 3.57~d 
period in spectroscopic data by \cite{donati11}. 
Inspired by this result we continued monitoring of TW~Hya with {\it MOST} during the 2014, 2015 and 2017 seasons.
\newline 
Except for the fairly regular light variations, the 2011 {\it MOST} light curve showed a previously 
unnoticed feature: 
a series of semi-periodic, short (10-20~min) and shallow (2-3 per cent) well-defined dips, initially 
interpreted as caused by occultations of hot-spots by hypothetical 'dusty clumps'. 
As the dips were observed during one season only and through the {\it MOST} filter that integrates 
almost the entire visual spectrum, we decided to characterise their spectral properties using Johnson and Sloan filters. 
We describe multi-colour observations obtained during 2013--2017 at the South African Astronomical 
Observatory ({\it SAAO}) and Cerro Tololo Inter-American Observatory ({\it CTIO}) in Section~\ref{observations}. 
In the same section we describe observations of two new instances of the moderately stable 
accretion state in TW~Hya, which apparently took place in 2014 and in 2017 and were observed 
by {\it MOST\/} during the fifth and seventh series of observations. 
The sixth 2015 run revealed only drifting QPOs, typical for an unstable accretion state. 
We analyse these new data and immediately discuss the obtained results in Section~\ref{results}. 
A summary of the major results is given in Section~\ref{summary}.

%--------- Table 1 - the log of all observations ------------
\begin{table*}
\caption{The extended log of {\it MOST, SAAO} and {\it CTIO} observations. 
Special features noticed during particular nights, i.e. flares (and their major parameters with errors 
in brackets) and possible occultations are listed.}
%\small{
\begin{tabular}{c c c c}
\hline
Date        & Filters& Flare                  & 'Occultation' \\ 
            &        &$HJD_{max}$; $T_{0.5}$~[min]; $U_{max}$ [norm~flux] &   \\ \hline
2013-04-11  & g'r'i' & --      & --\\
2013-04-12  & g'r'i' & --      & possibly\\
2013-04-13  & g'r'i' & --      &   -- \\
2013-04-14  & g'r'i' & --      & possibly\\  
2013-04-15  & g'r'i' &   --    &   -- \\
2013-04-18  & g'r'i' &   -- \\      
2013-04-19  & g'r'i' &   --    &   -- \\  
2013-04-20  & g'r'i' &   --    &   -- \\  
2013-04-21  & g'r'i' &   --    &   -- \\  
2013-04-22  & g'r'i' &   --    &   --\\  
2013-04-23  & g'r'i' &   --    &   -- \\  
2013-04-24  & g'r'i' &   --    &   -- \\  
2013-04-25  & g'r'i' &   --    &   -- \\  
2013-04-26  & g'r'i' &   --    &   -- \\  
2013-04-27  & g'r'i' &   --    &   -- \\  
2013-04-28  & g'r'i' &   --    &   -- \\  
2013-04-29  & g'r'i' &   --    &   -- \\
2013-04-30  & g'r'i' &   --    & possibly\\    
2013-05-01  & g'r'i' &   --    &   -- \\    
2013-05-02  & g'r'i' &   --    &   -- \\    
2013-05-04  & g'r'i' &   --    &   -- \\  
2013-05-05  & g'r'i' &   --    &   -- \\    
2013-05-06  & g'r'i' &   --    &   -- \\    
2013-05-07  & g'r'i' &   --    &   -- \\ \hline
2014-02-24 : 04-11 & 'white'&   --    &   -- \\  \hline 
%--2014-04-11& 'white'&         &      \\ \hline
2014-03-06  & u'g'z' &   --    &   -- \\
2014-03-07  & u'g'z' &   --    &   -- \\ 
2014-03-08  & u'g'z' &   --    &   -- \\
2014-03-09  & u'g'z' & flare or an accretion burst at 2456726.8340 &   -- \\
2014-03-10  & u'g'z' &   --    & possibly \\
2014-03-11  & u'g'z' & 2456728.8864(3); 10.7(2.5); 1.17(3) &   -- \\
2014-03-12  & u'g'z' &   --    & possibly \\
2014-03-13  & u'g'z' &   --    &   -- \\
2014-03-14  & u'g'z' &   --    &   -- \\
2014-03-15  & u'g'z' &   --    &   -- \\
2014-03-16  & u'g'z' & flare or an accretion burst at 2456733.7075 &  -- \\
2014-03-17  & u'g'z' &   --    &   -- \\
2014-03-18  & u'g'z' &   --    &   -- \\
2014-03-19  & u'g'z' &   --    &   -- \\
2014-03-20  & u'g'z' &   --    & possibly\\ \hline 
2015-03-04  & UBVu'g'& flare or an accretion burst at 2457086.8121 & -- \\
2015-03-05  & UBu'g' & flare or an accretion burst at 2457087.6698 & -- \\
2015-03-06  & UBVu'g'&   --    &  -- \\
2015-03-07  & UBVu'g'&   --    &  -- \\
2015-03-08  & UBVu'g'& no.1: 2457090.5959(3); 2.5(1.5); 0.66(1) & -- \\
            &        & no.2: flare or an accretion burst at 2457090.6696  &  \\
2015-03-09  & UBVu'g'& 2457091.5754(2); 1.5(1.5); 0.15(1) & -- \\
2015-03-10  & UBVu'g'&   --               &   -- \\
2015-03-11  & UBVu'g'& 2457093.5483(3); 2.9(1.5); 0.06(1) & possibly \\
2015-03-12  & UBVu'g'& flare or an accretion burst at 2457094.8085 & -- \\
2015-03-13  & UBVu'g'&   --    & -- \\
2015-03-14  & UBVu'g'&   --    & -- \\
2015-03-15  & UBVu'g'&   --    & -- \\
2015-03-16  & UBVu'g'&   --    & -- \\
2015-03-17  & UBVu'g'& 2457099.7902(2); 7.2(1.5); 0.065(5) & -- \\  \hline
2015-03-25 : 04-15 & 'white'& no.1: 2457117.245(2); 50(10) ; $>1.24$  & -- \\ 
                                 &             & no.2: 2457117.4384(3); 4.0(5); 0.12(1)  &    \\ \hline
\end{tabular}
%}
\label{Tab.log1}
\end{table*} 
%----------------------------------------------

%--------- Table 1a - the log of observations ------------
\begin{table*}
\contcaption{}
%\small{
\begin{tabular}{c c c c}
\hline
Date        & Filters & Flare                  & 'Occultation' \\ 
            &         &$HJD_{max}$; $T_{0.5}$~[min]; $U_{max}$ [norm~flux] &  \\ \hline
2016-02-23  &  UBV    &   --                    &   -- \\
2016-02-24  &  UBV    &   --                    &   -- \\
2016-02-25  &  UBV    &   --                    &   -- \\
2016-02-26  &   B     &   --                    &   -- \\
2016-02-27  &  UBV    &   --                    &   -- \\
2016-02-28  &  UBV    &  two flares or accretion bursts at 2457447.555 &   -- \\
2016-02-29  &  UBV    &   --                    &   -- \\
2016-03-01  &  UBV    &   --                    &   -- \\ 
2016-03-02  &  UBV    &   --                    &   -- \\
2016-03-03  &  UBV    & no.1: 2457451.5333(5); 5.2(1.5); 0.13(1) &  -- \\
            &         & no.2: 2457451.5822(5); 3.3(1.5); 0.13(1) &  \\
2016-03-04  &  UBV    &   --                    &   -- \\
2016-03-06  &  UBV    &   --                    &   -- \\
2016-03-07  &  UBV    &   --                    &   -- \\
2016-03-08  &  UBV    &   --                    &   -- \\
2016-03-09  &  UBV    &   --                    &   -- \\
2016-03-10  &  UBV    &   --                    &   -- \\
2016-03-12  &  UBV    &   --                    & possibly \\
2016-03-13  &  UBV    &  2457461.3981(7); 7.5(1.5); 0.29(1)  &   -- \\
2016-03-14  &  UBV    &   --                    &  -- \\
2016-03-15  &  UBV    &   --                    &   -- \\ \hline 
2017-03-03:28  & 'white'   &   --          &   -- \\ \hline
2017-03-08  &  u'g'r' &                         &   -- \\     
2017-03-09  &  u'g'r' &   --             &   -- \\
2017-03-10  &  u'g'r' &   --                 &   -- \\
2017-03-11  &  u'g'r' &   --                 &   -- \\
2017-03-12  &  u'g'r' &   --                 &   -- \\
2017-03-13  &  u'g'r' &   --                 &   -- \\
2017-03-14  &  u'g'r' &   --           &   -- \\
2017-03-15  &  u'g'r' &   --                 &   -- \\
2017-03-16  &  u'g'r' &   --                 &   -- \\
2017-03-17  &  u'g'r' &   2457830.361(1); 3.5(1.5); 0.42(3) &   -- \\ 
2017-03-18  &  u'g'r' &   --                 &   -- \\
2017-03-19  &  u'g'r' &   --                 &   -- \\
2017-03-20  &  u'g'r' &   --                 &   -- \\
2017-03-21  &  u'g'r' &   --                 &   -- \\ \hline
\end{tabular}
%}
%\label{Tab.log2}
\end{table*} 
%----------------------------------------------

\section{Observations and data reductions}
\label{observations}

\subsection{{\it MOST\/} observations}

The optical system of the {\it MOST\/} satellite consists of a 
Rumak-Maksutov f/6, 15~cm reflecting telescope. 
The custom broad-band 'white' filter covers the spectral range of 
370 -- 750~nm with effective wavelength located close to the Johnson $V$ band.
The pre-launch characteristics of the mission are described by 
\citet{WM2003} and the initial post-launch performance by \citet{M2004}.

The three new runs of nearly contiguous TW~Hya observations,
utilizing the {\it direct-imaging} data-acquisition mode, took place over:
\begin{enumerate}
\item 41.52 days between 25 February and 7 April, 2014, during 514 satellite orbits, 
\item 21.88 days between 25 March and 15 April, 2015, during 299 satellite orbits, 
\item 28.18 days between 3-28 March, 2017, during 295 satellite orbits.
\end{enumerate}
Because the star is not in the Continuous Visibility Zone of the satellite 
and some short time-critical observations of other targets were done in parallel, 
the effective total time coverage was 8.59, 7.59 and 5.87~d, i.e. 20.7, 34.7 
and 20.8 per cent of each run's total length, respectively.

The individual exposures were 60~s long during the first 5.51 days of the 2014 run 
and 120~s for the rest, including the 2015 and 2017 runs. 
Aperture photometry of the stars was obtained using RAW images by means 
of the {\small \sc DAOPHOT~II} package \citep{stet}. 
As in our previous investigations, a weak correlation between the stellar flux and 
the sky background level within each {\it MOST\/} orbit was noted and removed; 
this was most probably caused by a small photometric non-linearity in the electronic system.
Due to the lack in availability of {\it flat-field} and {\it dark} calibration images, we also removed 
the small correlation between the star brightness and its point-spread function 
centroid position in the sub-rasters.  

% ----------------------- Fig.1 the MOST lc ---------------------
\begin{figure*}
\includegraphics[width=\linewidth]{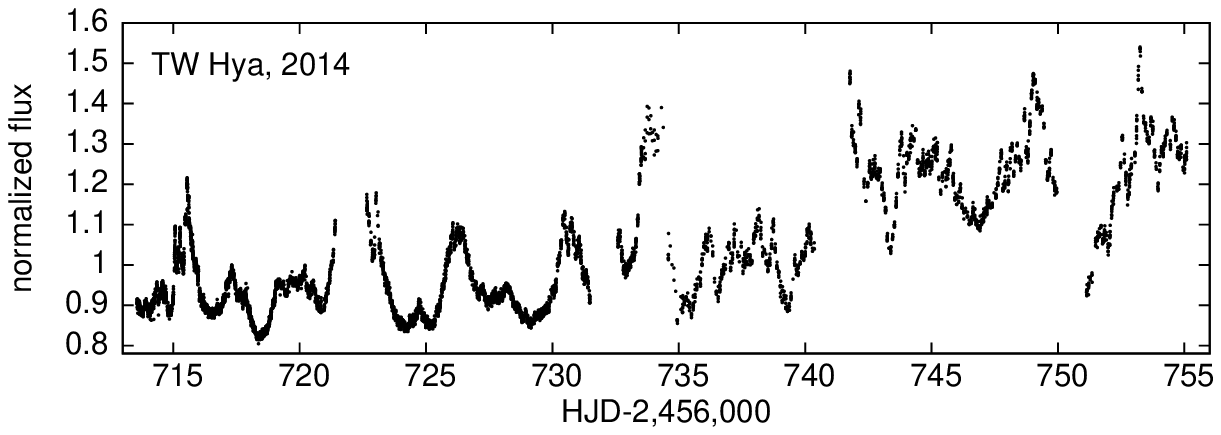}
\includegraphics[width=\linewidth]{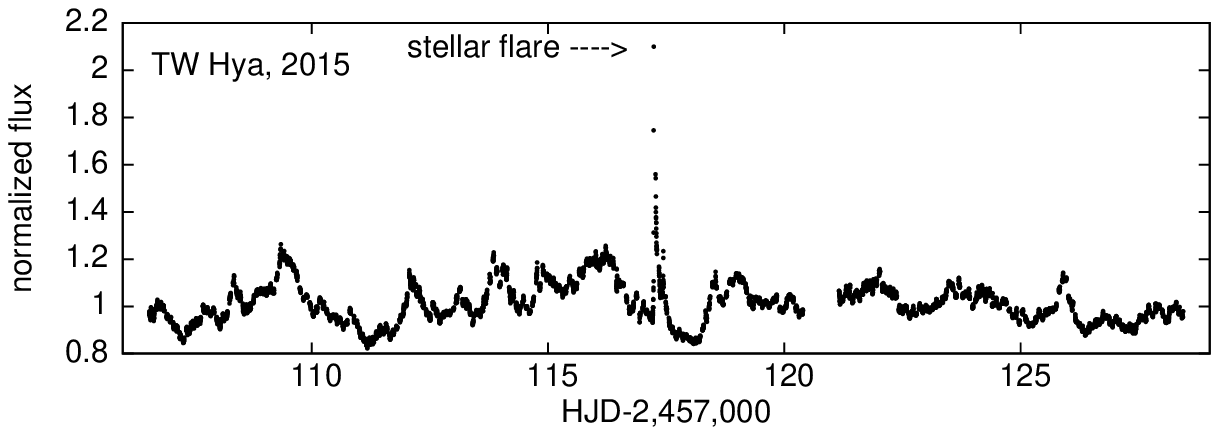}
\includegraphics[width=\linewidth]{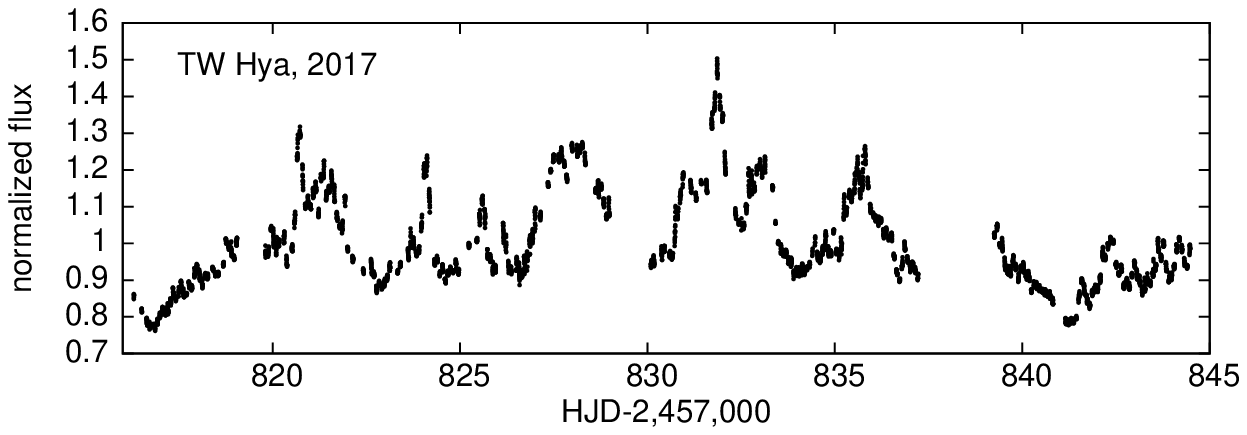}
\caption{
The 2014, 2015 and 2017 {\it MOST\/} light curves of TW~Hya in flux units, 
scaled to unity at the mean brightness level, which was different in each year.}
\label{Fig.most}
\end{figure*}
%----------------------------------------------------------------------

We obtained three well defined light curves for almost the whole duration of 
the observations (Fig.~\ref{Fig.most}). 
The typical error of a single data point is about 0.011~mag. 
The median value of the scatter ($\sigma$) of 514 averaged 2014 data points 
(formed for each satellite orbit of 101~min) is 0.0082, with the full range 
between 0.0002--0.0329 in units of the mean normalised flux for the star. 
The median value of the scatter of the 2015, 299 averaged data points was 0.0087, with full 
range of 0.0031-0.0503, while in the case of the 2017, 295 averaged data points 
the respective values are equal to 0.0064 and 0.0021-0.0335 in units of the mean normalised 
flux for the star -- the values of scatter include the variability intrinsic to the accretion, 
occurring on time-scales shorter than the length of a single {\it MOST\/} orbit.\newline
Unfortunately, the 2014 data were interrupted five times to conduct observations of other 
time-critical targets. 
In addition, the effective coverage time within single orbits became significantly lower 
after heliocentric julian date $HJD\approx2\,456\,731$. 
The 2015 data were interrupted only once, while three major interruptions affected the 2017 
data to a certain extent, especially the longest one at $HJD\approx2\,457\,837$. 
These facts have a negative influence on the investigation of the shortest periods using the wavelet technique. 
We also note that the long on-board stacking intervals of 120~s used for the majority 
of the new {\it MOST} data seriously limited new searches for short lasting 'occultations'.

%--------- Table comparisons - Informations about TW Hya's and comp. stars-------------
\begin{table*}
\caption{The comparison (cmp) and the control (chck) stars used for differential photometry 
of TW~Hya over 2013-2017. The magnitudes of the first star are from the RAVE project \citep{munari14}. 
TYC~7208-1422-1 and USNO-A2~0525-13705888 were usually used to compute a 'mean comparison star'. 
RAVE J110143.8-343930 played the role of the comparison star only during the 2015 {\it SAAO} run.}
\tiny
\begin{tabular}{c c c c c c c c c c c}
\hline
           star       &  B        & V         & g'       & r'        & i'        &2013 &2014 &2015&2016 &2017 \\ \hline\hline
TYC~7208-1422-1 & 12.781(16)& 11.964(16)& 12.333(8)& 11.720(14)& 11.461(23)& -- &cmp  &cmp &cmp  &cmp  \\ \hline  
RAVE J110143.8-343930 & -- & --  & --  & -- & --                                 & -- &chck &chck,cmp  &chck&chck\\ \hline 
USNO-A2~0525-13705888 & -- & --  & --  & -- & --                           & -- &cmp  &cmp &cmp  &cmp\\  \hline
USNO-A2~0525-13705192 & -- & --  & --  & -- & --                                 & cmp & -- & --  & -- & -- \\ \hline 
USNO-A2~0525-13705637 & -- & --  & --  & -- & --                                & cmp & -- & -- & -- & -- \\ \hline 
\end{tabular}
\label{Tab.comp}
\end{table*}
%--------------------------------------------------------------------------

\subsection{Multi-colour observations}
\label{multi-saao1m}

% ----------------------- Fig.2 the g-b lc, fourier and wavelet spectra ---------------------
\begin{figure*}
\includegraphics[width=85mm]{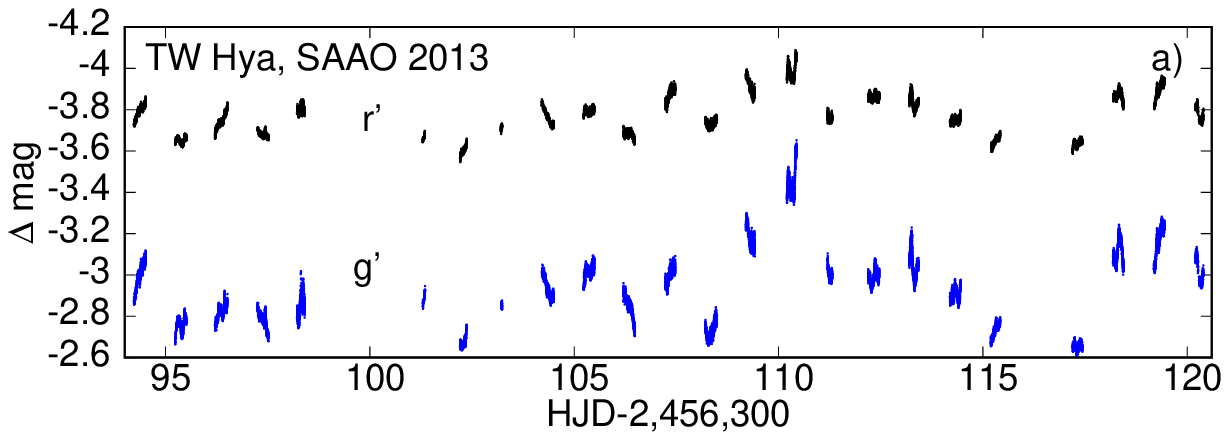}
\includegraphics[width=85mm]{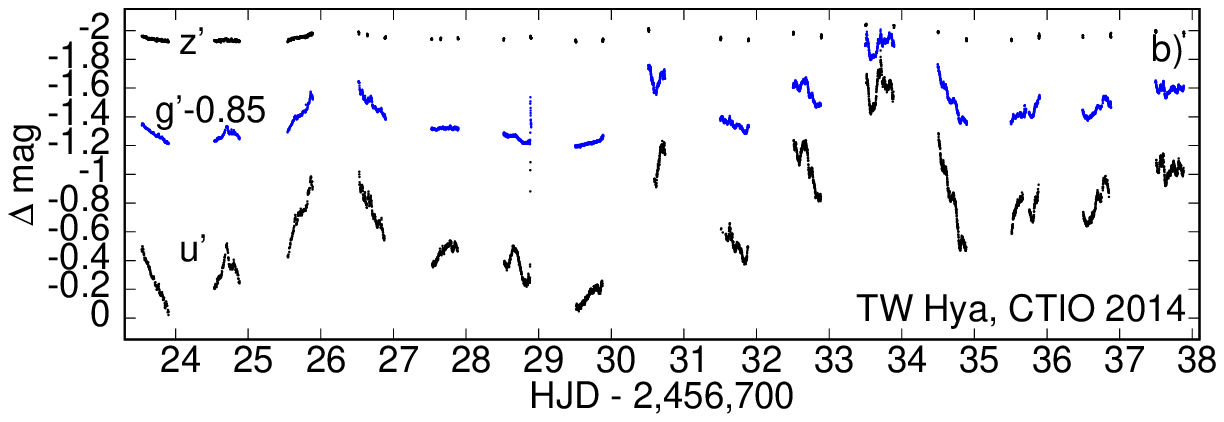}
\includegraphics[width=85mm]{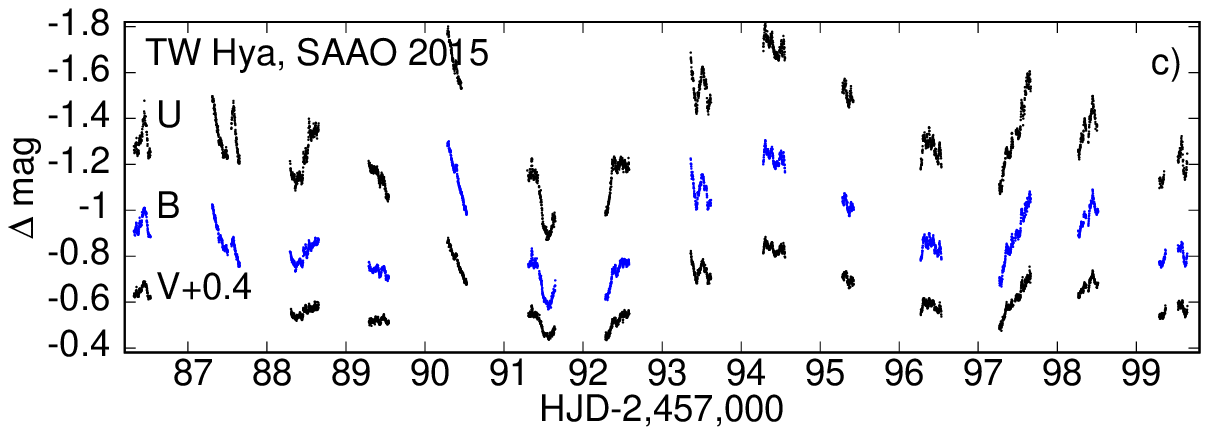}
\includegraphics[width=85mm]{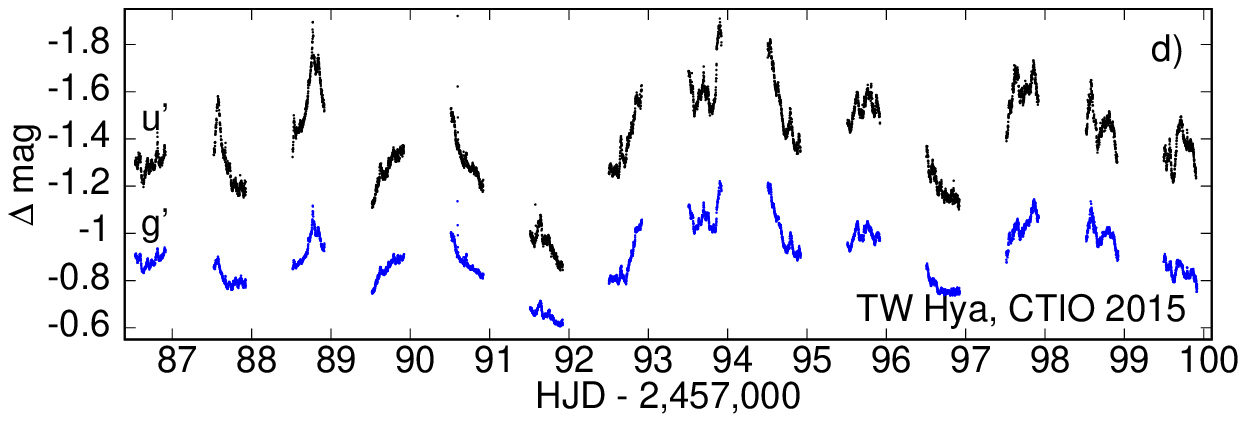}
\includegraphics[width=175mm]{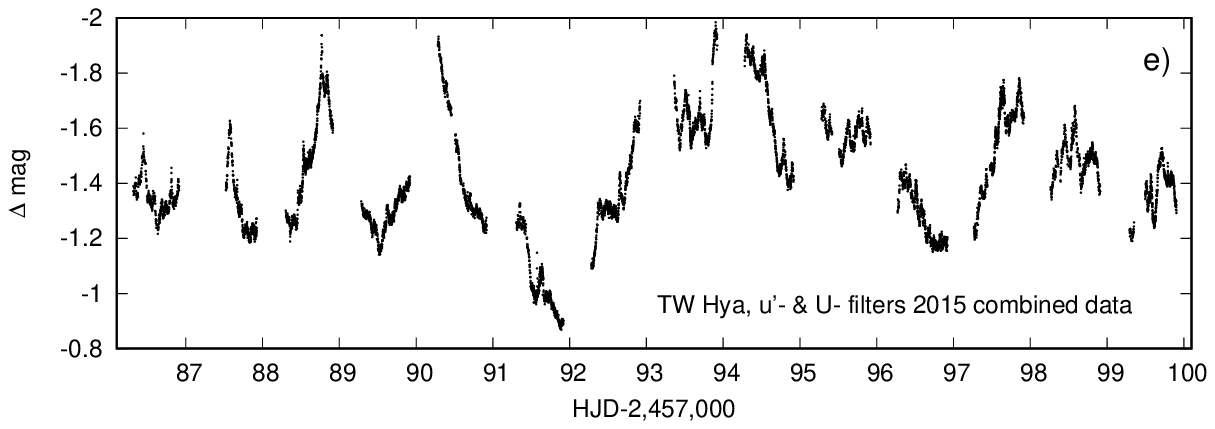}
\includegraphics[width=85mm]{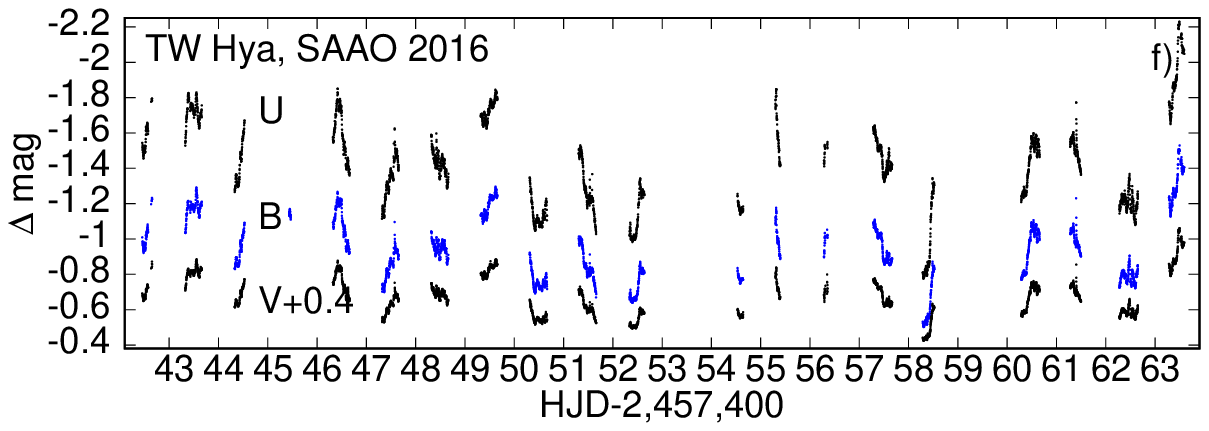}
\includegraphics[width=85mm]{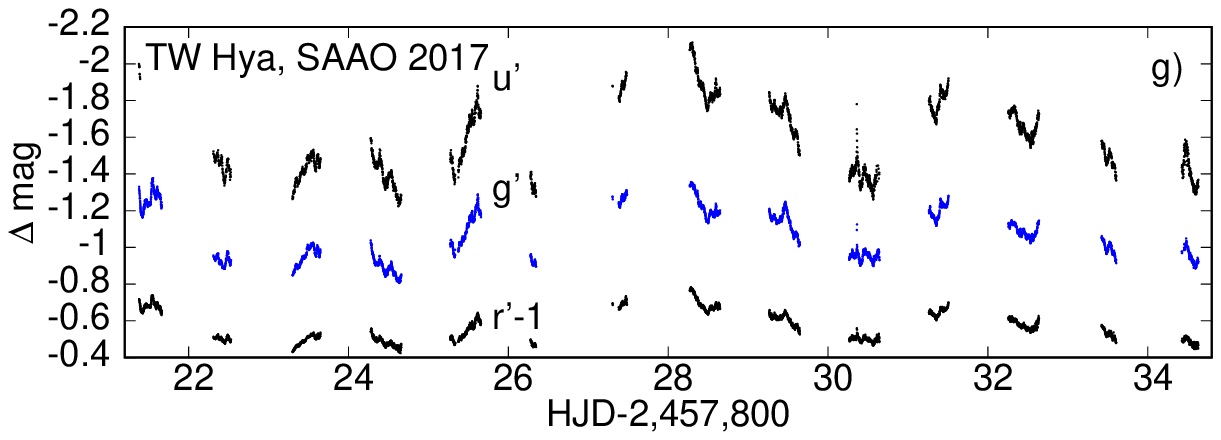}
\includegraphics[width=85mm]{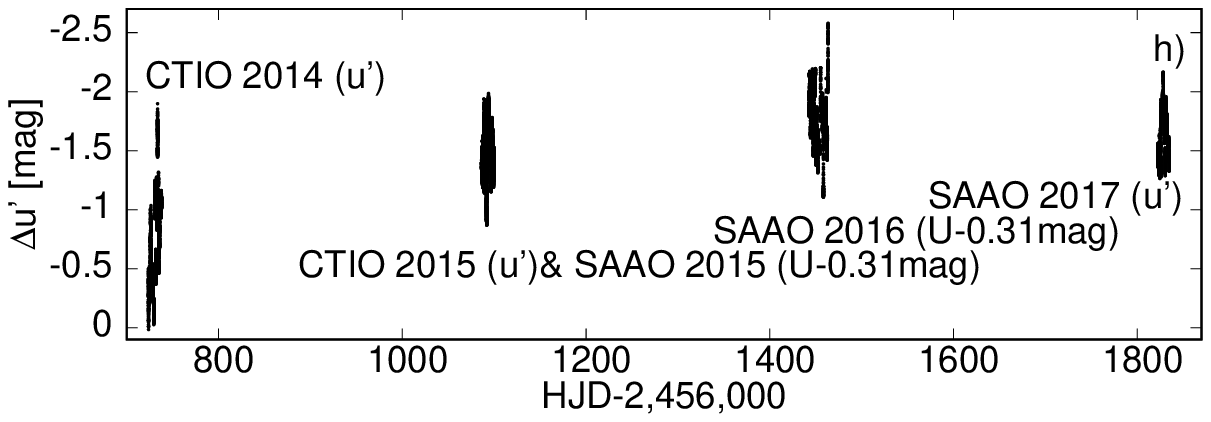}
\includegraphics[width=85mm]{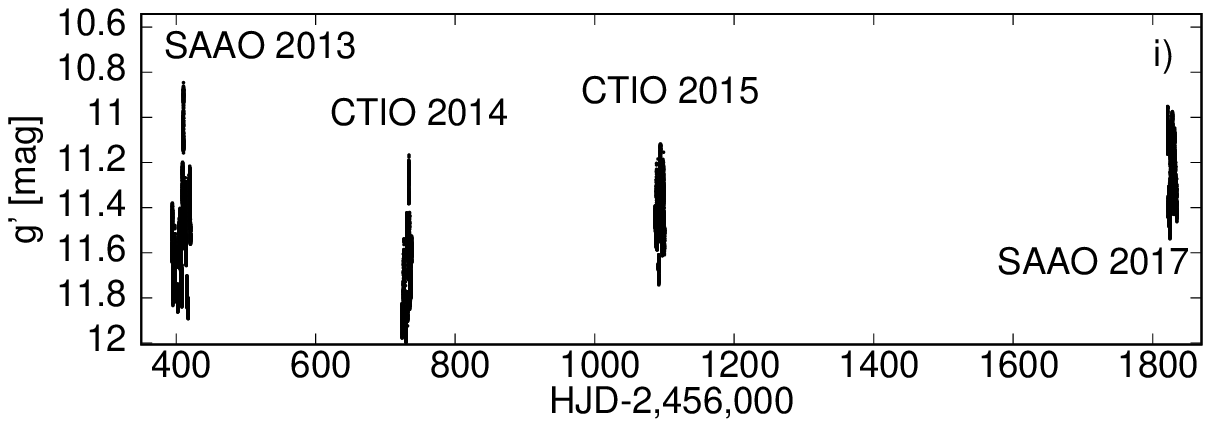}
\caption{The ground-based light curves of TW~Hya obtained in 2013-2017. 
The years, sites and filters are labeled in consecutive panels. 
The large middle panel shows the 2015 u-band light curve, composed of the {\it CTIO u'}-filter 
and the {\it SAAO U}-filter data.
The two bottom panels show the long-term brightness evolution in the $u'$-filter 
and the $g'$-filter -- the last light curve was roughly calibrated to the standard 
Sloan system.}
\label{Fig.gb}
\end{figure*}
%----------------------------------------------------------------------

%************************************************TM

\subsubsection{{\it SAAO} 2013 I: the Sloan $g'r'i'$-filter data from the 0.75-m telescope}
\label{tripol}

We started ground-based monitoring of TW~Hya at the South African Astronomical Observatory 
({\it SAAO}). 
The first data were obtained on 11/12 April and the last on 7/8 May, 2013. 
The total run length was 26.17 days. 
Observations were gathered during 22 nights of different length and photometric quality. 
The observing log is provided in Table~\ref{Tab.log1}.

We used the 0.75-meter telescope equipped with {\it TRIPOL} which is the visitor instrument 
of Nagoya University donated to the {\it SAAO}. 
With the 11.75~m telescope focal length the field of view is $3\times3$~arcmin.
The light is split by two dichroics, then directed to three {\it SBIG ST-9XEI} CCD cameras through 
the Sloan $g'r'i'$ filters, resulting in three simultaneous images for each exposure. 
%In addition to three times higher observing cadence, the crucial advantage of this technique is a possibility 
%to track colour index variations even during non-photometric conditions. 
The cadence of our observations was very high and ranged between 8-23~s (including 3~s read-out time), 
depending on the actual seeing. 
All frames were {\it dark} and {\it flat-field} calibrated in a standard way using our scripts written 
in the {\small \sc MIDAS} package \citep{W1991}. 
For photometric reduction we used the {\small \sc C-Munipack} software \citep{motl11} which utilizes 
the {\small \sc DAOPHOT~II} package.

In Figure~\ref{Fig.gb}a we present $g'$- and $r'$-filter light curves obtained relative 
to the 'mean comparison star' formed from the fourth and fifth stars listed in Table~\ref{Tab.comp}. 
We do not show here the $i'$-filter data, affected by incomprehensible instrumental trends occurring each night.  
We stress that the only two available comparison stars are significantly fainter that TW~Hya, which strongly 
affects the quality of the final light curves. 
Therefore, for studies of subtle light changes in TW~Hya we used just the pure photometric measurements 
of the star, obtained during 13 fully or partly photometric nights. 
These data were corrected for nightly trends using mean extinction coefficients 
for the corresponding filters determined for {\it SAAO}, as described in Appendix~\ref{app1}.
We show the 2013 data obtained during photometric nights in details in Figure~\ref{Fig.saao13} 
of the Appendix~\ref{app2}.

\subsubsection{{\it SAAO} 2013 II: the Johnson $BV$-filter observations from the 0.5-m telescope}

We also observed TW~Hya from December 4, 2013, until January 14, 2014 using the single-channel 
{\it Modular} photometer on the 0.5-m {\it Boller} \& {\it Chivens} telescope. 
The run was shared with multi-colour observations of FU~Ori, which were performed simultaneously with {\it MOST} 
during first halves of the nights, until TW~Hya rose high enough above the horizon for accurate measurements. 
However, during most nights we seriously struggled with bursts of high humidity, eventually leading to the formation 
of ridge clouds over the observatory just after midnight. 
We were able to observe TW~Hya in $BV$-filters continuously for a few hours only during literally a few nights. 
We do not present these light curves, as the way they were collected (without frequent sky-level 
and comparison-star measurements) was focused on the detection of short-duration 10-20~min occultations. 
Unfortunately, the results were negative. 

\subsubsection{{\it CTIO} 2014 and 2015: the Sloan $u'g'z'$-filter data from the 0.9-m {\it SMARTS} 
telescope} 

In 2014 we joined {\it SMARTS} and observed TW~Hya on two occasions: 
for the first time during 15 nights between 6-20 March, 2014, 
and for the second time during 14 nights between 4-17 March, 2015 -- this run was coordinated with the
{\it SAAO} run on the 1-m telescope (see below in Sec.~\ref{saao-mc}). 
The 2014 run was performed simultaneously with the {\it MOST} observations. 
The total run durations were 14.35 and 13.38 days, respectively.\newline
We used the 0.9-meter telescope which is equipped with a Tek2K $2048\times2046$ pixel CCD camera and a set 
of Sloan $u'g'r'i'z'$ filters. 
With the scale of 0.401~arcsec~pix$^{-1}$, the field of view is $13.6\times13.6$~arcmin.
To decrease the readout time, we have reduced the field of view to $6.8\times6.8$~arcmin
and used two amplifiers.
As the camera is cooled down to -120~C with liquid nitrogen, {\it dark} frames were not necessary.
All further calibrations of {\it bias} and {\it flat-field} were made in the same way 
as for the {\it SAAO} data.

A 'mean comparison star', calculated from two very stable (to 0.01~mag in all filters 
over all five years) stars in the field (the first and third stars listed in Table~\ref{Tab.comp}) 
were used for differential photometry. 
The light curves constructed from all-night observations are shown in Figure~\ref{Fig.gb}b\&d. \newline
In 2014 the $u'g'z'$-filters were used.
The observations in the $z'$-filter were obtained with the hope of detection of cold-spot -induced 
modulations in the light curve, but the result was negative -- accretion effects which dominate 
in the $u'g'$ filters data are still pronounced near 9000~\AA, though with amplitudes considerably 
scaled down.
Therefore, after the first three nights in 2014 we considerably limited observations in this band 
to increase the sampling rate in the most detail-rich $u'g'$-filters. 
We totally skipped the $z'$-filter during the 2015 run.

We show the 2014 and 2015 observations in detail in Figure~\ref{Fig.ctio14} and in Figure~\ref{Fig.ctio15} 
in the Appendix~\ref{app2}.
The data shown in these figures were left uncorrected for atmospheric extinction. 
This is due to the fact that proper treatment of the dominant colour extinction term requires actual 
values of the $u'-g'$ color index, whilst during a few nights the $u'$-filter was skipped for several hours 
during poor seeing conditions and/or passing clouds. 
No transformation to the standard system was made and the data were left in the instrumental system.

\subsubsection{The 2015 and 2016 $UBV$ and the 2017 $u'g'r'$ {\it SAAO} light curves from the 1-m telescope}
\label{saao-mc}
Encouraged by the 2014 {\it CTIO} results, in 2015 we decided to increase the effective coverage 
using the {\it SAAO} 1-m {\it Elizabeth} telescope. 
This telescope is equipped with a {\it STE4} $1024\times1024$ pixel CCD camera 
and a set of Johnson-Bessel {\it UBVRI} filters. 
{\it UBV}-filters have been used to cover the spectral range similar to that covered by the Sloan {\it u'g'}-filters 
at {\it CTIO}. 
The 2015 {\it SAAO} run was strictly coordinated with the {\it CTIO} observations, which allowed us to reduce 
daily breaks to 10-11 hours. 
%The total duration of the first {\it SAAO} run was 13.33 days.

To reduce the read-out time to 17 seconds per frame, we binned the chip to $512\times512$ pixels; 
at the same time, the stellar point spread function was still well sampled with 0.31~arcsec~pix$^{-1}$ resolution. 
All reduction steps were made in the same way as for the previous CCD observations. 
The first and second comparison stars were used for differential photometry in 2015 (see Table~\ref{Tab.comp}). 
The data were left in the instrumental system, but were later corrected to colour extinction 
terms (see in Appendix~\ref{app1}) to prepare colour-magnitude diagrams.
The {\it UBV} light curves constructed from all-night observations are shown in Figure~\ref{Fig.gb}c. 
The data obtained during all photometric nights are shown in Figure~\ref{Fig.saao15} in the Appendix~\ref{app2}.\newline 
The major limitation of the combined run was the use of two different photometric systems. 
However, we decided to keep the use of the Sloan filters at {\it CTIO}, which have the huge advantage 
of higher and nearly flat peak transmission curves with better defined profiles, making each band 
practically independent from the other -- this advantage was very important 
to characterize the short-term occultations. 
The only simple combination of light curves obtained from both sites was possible for 
the $u'$- and $U$-filters, as their transmission functions are very similar. 
We found that variability amplitudes in the two data sets overlapping for 1-3 hours are almost 
identical (to 0.005~mag) and the combined light curves can be treated as an (almost) homogeneous set of data. 
The combined u-band 2015 light-curve (corrected for colour extinction effects) 
is shown in Figure~\ref{Fig.gb}e.

The second run on the 1-m {\it Elizabeth} telescope utilizing Johnson filters was performed during 22 nights 
between 23 February -- 15 March, 2016; the results are shown in Figure~\ref{Fig.gb}f and in Figure~\ref{Fig.saao16} 
in the Appendix~\ref{app2}. 
The total run duration was 21.13 days. 
During this run we included into the field of view the most stable comparison star, 
listed as the third one in Table~\ref{Tab.comp}. 
This enabled us to align the 2015 and 2016 {\it SAAO} light curves for display purposes and presentation 
of uniform colour-magnitude diagrams (Sec.~\ref{ci-diagrams}). 
%as if they were received with the use of the first and the third comparison star during both years.
Unfortunately a simultaneous run at {\it CTIO} was impossible at that time. 

The third and last run on the {\it Elisabeth} telescope took place between 8-21 March, 2017. 
It was 13.25 days long and was performed simultaneously with the third {\it MOST} satellite 
run and with high-cadence low-resolution spectroscopic monitoring of TW~Hya by means of 
the 1.9-m {\it SAAO} telescope. 
The combined spectroscopic and photometric results will be the subject of a separate publication.
In 2017 we used the first opportunity to observe the star in $u'g'r'$ Sloan filters. 
The same detector, binning and reduction processes were used: the final light curve is shown 
in Figure~\ref{Fig.gb}g while data obtained during nights with stable weather conditions are shown 
in Figure~\ref{Fig.saao17} in the Appendix~\ref{app2}. 
The most reliable first and third comparison stars (Table~\ref{Tab.comp}) 
formed the 'mean comparison star'.

\section{Analysis of all data and discussion of the results}
\label{results}

\subsection{Fourier analysis of {\it MOST}, {\it SAAO} and {\it CTIO} light curves}
\label{fourierMOST}

We performed analysis of the new light curves in the same way
as in \citet{ruc08} and \citet{siwak11,siwak14,siwak16}. 
The bootstrap sampling technique permitted evaluation of the mean standard
errors of the amplitude $a(f)$, where $f$ is the frequency.
Due to the different sampling rates per orbit over the 2014 run, 
we used mean-orbital data points during the analysis. 
The 2015 and 2017 frequency power spectra were calculated in the same way for uniformity.

\begin{figure}
\includegraphics[width=1.0\linewidth]{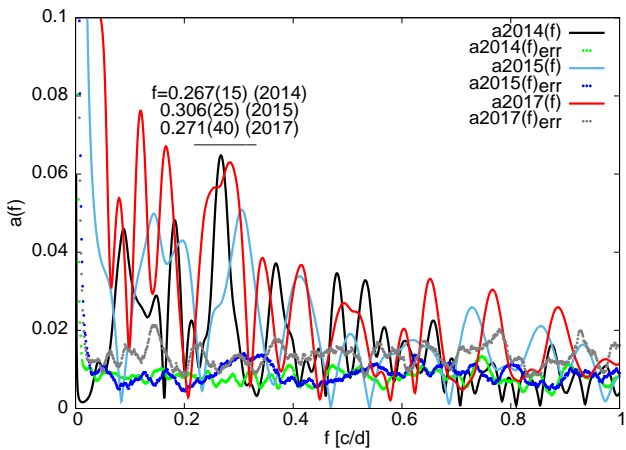}
\caption{
The Fourier spectra of three {\it MOST\/} light curves. 
The peaks indicating the approximate value of the stellar rotational period are underlined and their 
respective central frequencies are shown.}
\label{Fig.fourier}
\end{figure}

The 2014 Fourier spectrum (Fig.~\ref{Fig.fourier}) shows one stable peak centered at $0.267\pm0.015$~cd$^{-1}$ (3.75~d). 
A similar peak appears in the 2015 Fourier spectrum ($0.306\pm0.025$~cd$^{-1}$, i.e. 3.27~d), but its stability will 
be later questioned 
on the basis of the wavelet-analysis result.
The 2017 data show a stable peak at $0.271\pm0.040$~cd$^{-1}$ (3.69~d). 
Note that the error in frequency of the peak is large as the peak likely consists of two close overlapping features. 
We also note that lower-frequency peaks, which indicate 5-10~d oscillations, 
are present in the Fourier spectra but they do not represent significant periodic variations 
based on the light curves themselves.

Taking into account our previous {\it MOST} achievements for TW~Hya, 
attempts to search for a few days' long periodic signals in light curves affected 
by daily breaks and non-uniform weather patterns may appear to be irrational. 
In spite of these concerns, we found some evidences of quasi-periodicities centered 
at:\newline
1)~3.15~d in the 4-week long 2013 {\it SAAO g'r'} data-set,\newline
2)~3.64~d in the 2-week long 2014 {\it CTIO u'g'z'} data-set,\newline 
3)~4.44~d in the combined {\it SAAO} $U$-  and {\it CTIO} $u'$-filter data sets,
although this value lies at the limit of what can be considered as a real quasi-periodic 
signal from the 13.588~d run,\newline
4)~3.45~d in the 3-week long 2016 {\it SAAO UBV} data-set, and\newline
5)~3.46~d in the 2-week long 2017 {\it SAAO u'g'r'} data-set.\newline 
The peaks in the frequency amplitude-spectra obtained from ground-based data are usually wide 
and not symmetric. 
This can be caused by the period drift of the dominant QPO over a run, 
as clearly observed in many {\it MOST} runs (see below), and mostly due to limited temporal coverage.
Although we show the results with precision to the second decimal place, the real uncertainties 
of these values are $\sim0.2-0.3$~d. 
Nevertheless they are in accordance with our previous and current {\it MOST} observations, 
which firmly indicate that the stellar rotational period must be close to the spectroscopic 
value 3.57~d of \citet{donati11}. 

% ----------------------- Fig.3 MOST fourier spectra ---------------------
%----------------------------------------------------------------------

% ----------------------- Fig.4 MOST wavelet spectra ---------------------
\begin{figure*}
\includegraphics[width=0.33\linewidth]{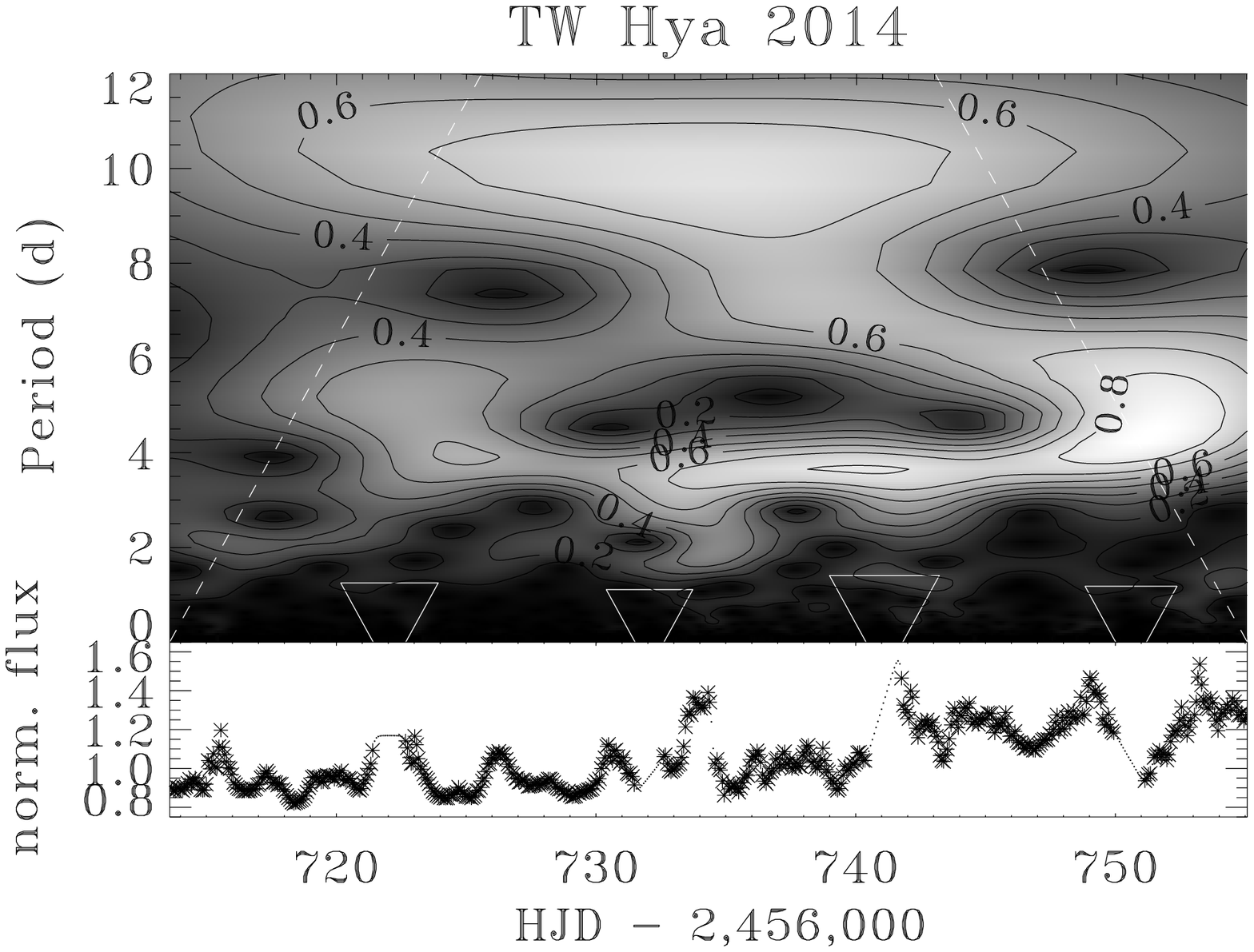}
\includegraphics[width=0.33\linewidth]{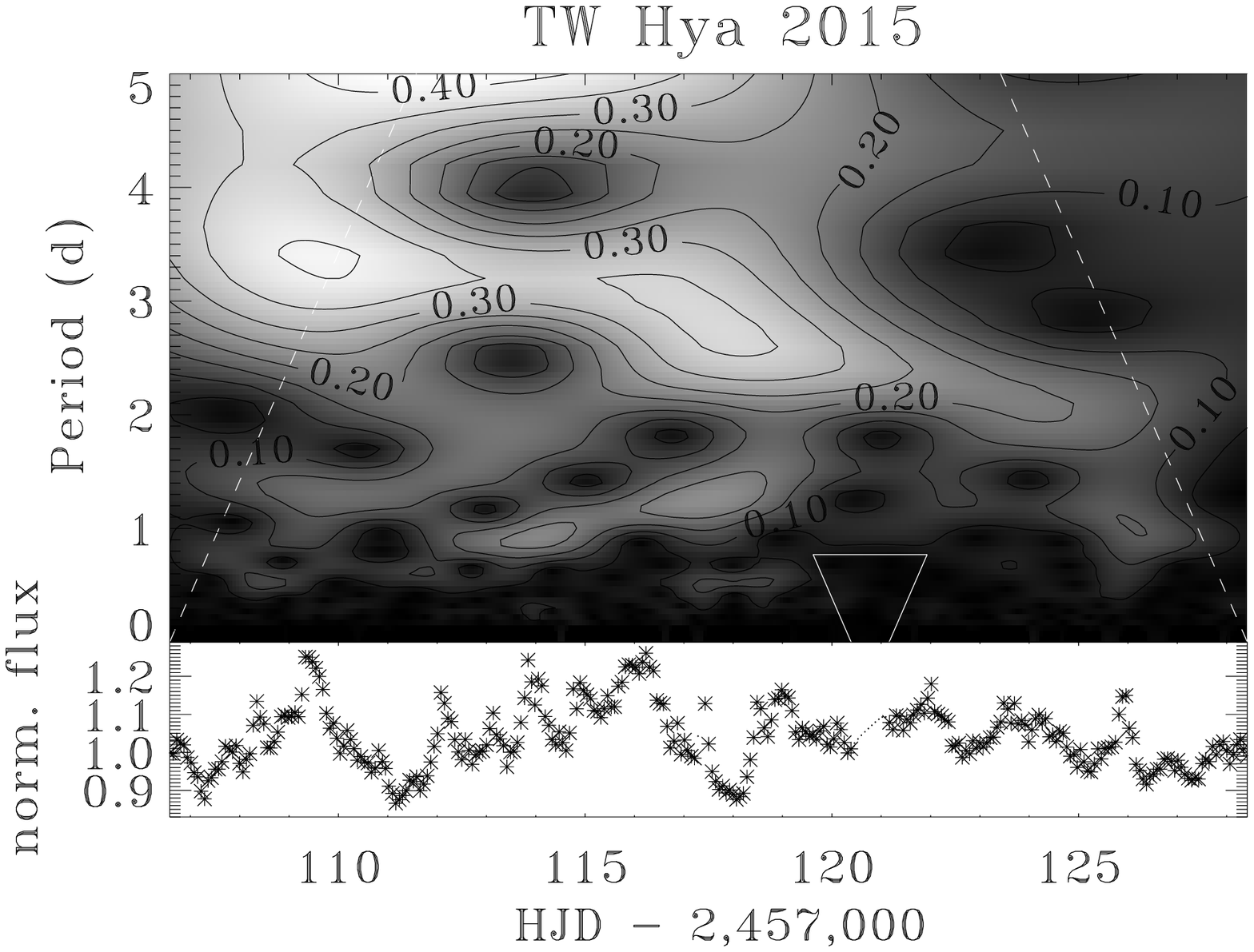}
\includegraphics[width=0.33\linewidth]{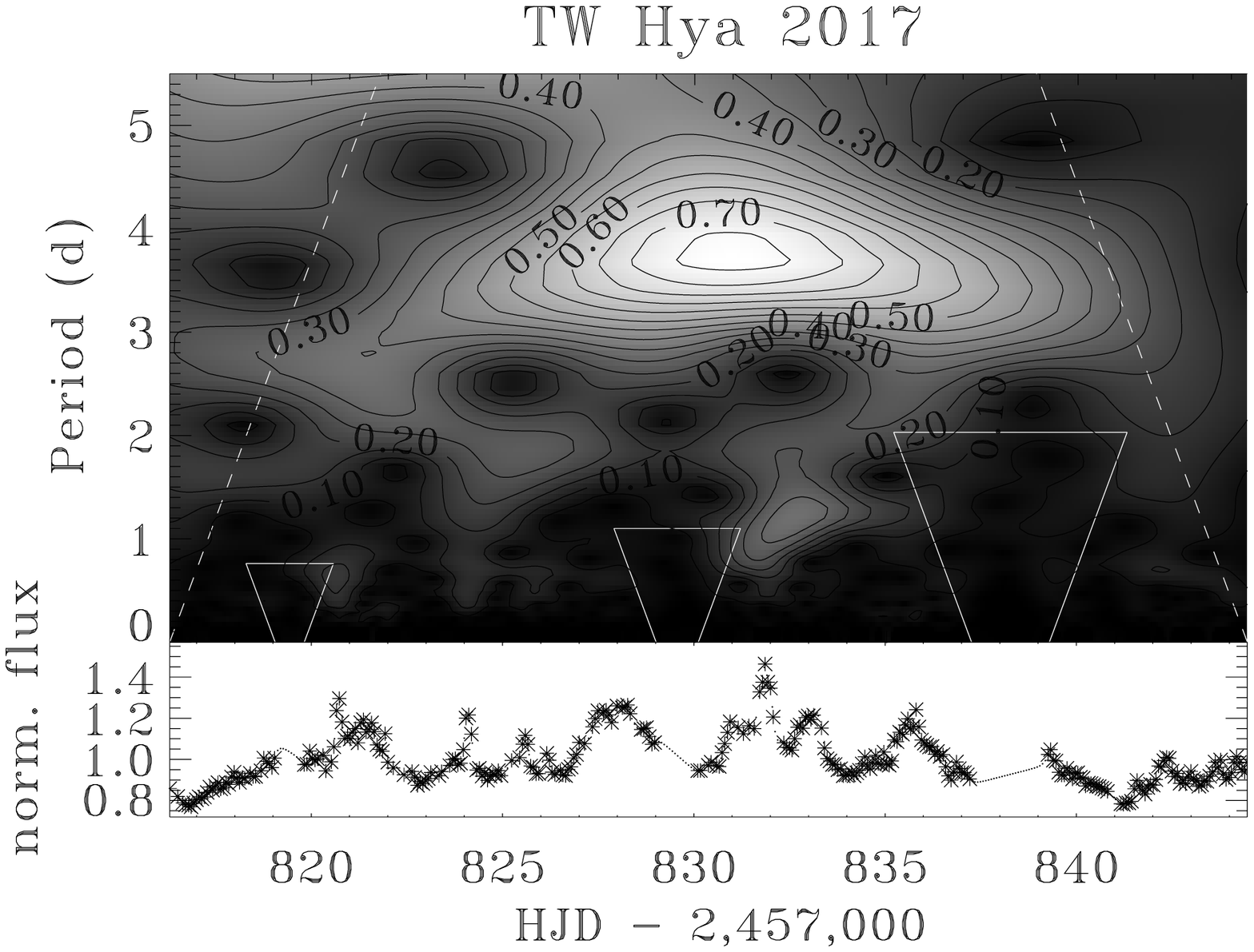}
\caption{
Wavelet spectra (upper panels) computed from a uniformly distributed time-grid with 0.07047~d spacing. 
The {\it MOST} data (in normalised flux units) are shown on bottom panels as asterisks for quick 
cross-identification of the variability. 
The small dots indicate the interpolated gap-filling fragments of these light curves.}
\label{Fig.wav}
\end{figure*}
%----------------------------------------------------------------------

\subsection{Wavelet analysis of the {\it MOST} data}
\label{wavelet}

To obtain uniform data sampling required for the wavelet analysis and to remove a few 
interruptions in the data acquisition (see Sec.~\ref{observations}), 
we interpolated with splines the 514 (2014), the 299 (2015) and the 295 (2017) mean satellite-orbit flux points 
into a grid of 589, 311 and 399 equally-spaced points at 0.07046 day intervals, respectively. 
As a result, the gaps in the data acquisition were filled up with artificial points 
with uniform sampling like the rest of the data. 
As we found previously (\citealt{ruc08, siwak11,siwak14,siwak16}), the Morlet-6 wavelet provided 
the best match between the time-integrated power spectrum and the original frequency 
spectrum of the star. 
The reliable two-dimensional oscillation wavelet spectrum  (Fig.~\ref{Fig.wav}) is contained between 
the two white broken lines. 
Another limitation in our wavelet results is due to the four (2014), the single (2015) 
and the three (2017) interruptions in the data acquisition: the regions obviously affected by these 
breaks lie inside the small trapezium areas defined by white continuous lines. 
Unfortunately, the small trapezium areas may in fact propagate to longer periods and affect 
the wavelet spectra to an unknown degree. 
Therefore detailed visual inspection of the respective light curves is always necessary in order to verify 
whether a particular signal can be treated as real or not.\newline
The new results comply with those obtained during the 2007, 2008, 2009 and 2011 
{\it MOST\/} observations:
\begin{enumerate}
\item The 2014 and 2017 data show stable signals at 3.75 and 3.69 days, both in the wavelet and 
in the Fourier spectrum. 
This result is similar to that obtained recently from the 2011 data, where a stable 4.18~d quasi-periodic 
modulation of large amplitude was observed for most of the run \citep{siwak14}.
\item In contrast to the previous results, the major peak at 0.306~c~d$^{-1}$ in the 2015 frequency 
spectrum does not appear as a stable feature in the wavelet spectrum -- it shows period shortening 
between $\sim3.4-2.2$~d, similar to that observed in this star during the 2008 and 2009 {\it MOST} 
runs \citep{ruc08,siwak11}.
%The new finding in the 2015 wavelet spectrum is the presence of  period lengthening from about 0.9 to 1.5~d 
%visible between $HJD-2\,457\,000\approx113-120$ -- this is a feature never before observed so clearly 
%in any wavelet spectra of the star.
\end{enumerate}
We briefly conclude that the new results are in line with the picture outlined from our last
{\it MOST} observations \citep{siwak14} -- the star apparently shows episodes of unstable (in 2015)  
and moderately stable accretion (in 2014, 2017), which determine the appearance 
of the light curves \citep{kurosawa13}. 
The period values observed this time (3.75 and 3.69~d) are closer to the spectroscopic 
value of rotational period at 3.57~d by \citet{donati11}. 
Thus, the periods observed by {\it MOST} do not appear to be 'false' periods caused by rotational 
modulation of hot-spots created by an {\it ordered unstable regime}, discovered by \citet{blinova16}. 
Instead, the almost-stable quasi-periodicities seen in 2014 and 2017 could be produced by rotational 
modulation in visibility of a large hot-spot, created at the footprint of a steady 
accretion funnel striking the star close to its magnetic pole. 
The smaller peaks in the light curves could then be caused by hot-spots formed at moderate stellar latitudes by equatorial 
tongues.

In addition to the above framework, the 'accretion burst' stochastic 
appearance and triangular shape of many peaks, already noticed in TW~Hya by \citet{ruc08}, 
could also be the result of inhomogeneous accretion, characterised in detail
by \citet{gullbring96}. 
Some of the 'bursts' could also be explained by the development of shocks in the accretion columns 
due to smooth density variations in an inner disc \citep{robinson17}. 
The authors proposed that the shocks amplified in an accretion column can propagate 
along it and ultimately hit the star, leading to rapid, large-amplitude changes in the accretion 
rate. 
They obtained 'sawtooth' burst-profiles, with rapid increase in brightness followed 
by a slower decline. 
Examination of the new {\it MOST} light curves reveals a few possible instances of such behaviour 
during 2015, although generally symmetrical 'triangular' profiles dominate in all {\it MOST} data. 

\subsection{Long-term evolution in ground-based data}
\label{long-term}

To illustrate the long-term brightness evolution of TW Hya we utilized the favoured $g'$- and 
u-band light curves (Figure~\ref{Fig.gb}, bottom panels). 
For homogeneous calibration all $g'$-filter data were re-calculated with respect 
to the well-calibrated comparison star TYC~7208-1422-1 \citep{munari14}. 
The 2013 $g'$-filter light curve had to be aligned to those from 2014-2017 
using shifts determined from the {\it CTIO} observations.

Long term evolution in the u-band can only be presented with respect to comparison stars. 
The two Johnson-$U$ light curves obtained at {\it SAAO} were shifted by 0.31~mag 
(as obtained from the 2015 data) to match the Sloan-$u'$ magnitude level, as indicated on the panel. 
No shift was applied to the 2017 {\it SAAO} $u'$-filter data.
Corrections for colour extinction effects (see Appendix~\ref{app1}) were applied to all these light curves. 
We stress that the data were obtained by means of three photometric systems, with unknown 
transformation equations to the standard system. 
We roughly estimated that this deficiency may lead up to 0.05~mag year-to-year deviations. 
This is fortunately a small effect, given the large amplitude of these long-term light 
changes -- during four years the average brightness level in the u-band changed by about 
1~mag and was followed by similar changes in the $g'$-filter but of smaller, 0.3~mag amplitude. 
Similar changes were noted in the 2001-2008 All Sky Automated Survey ({\it ASAS}) $V$-filter 
data for TW~Hya by \citet{ruc08}. 
%The variations in RU~Lupi \citep{siwak16} were even more interesting as they showed 
%well defined QPO at 258~d in the second half of the {\it ASAS} light curve.

It is possible that the long-term changes could be related to the variation of the mean mass-accretion rate. 
When it increases, the inner disc becomes a source of many instabilities. 
According to \citet{romanova08} numerous unstable tongues of plasma are then formed, which eventually 
hit the star, leading to increased hot-spot number along with associated veiling and mean-seasonal stellar 
brightness increases. 
Thus, just like the daily and weekly variations, the yearly variations also show the largest amplitudes 
in the u-band. 
This view may be partially confirmed in the 2014 data, when the averaged brightness in the $g'$-band 
was the lowest in 5 years and the {\it MOST} data analysed with the wavelet technique indicated 
a moderately stable regime.
However, the same state was found in 2017 when unstable accretion operated in the star, while 
the average $g'$-band star brightness was even higher than during 2015.
Longer continuous monitoring sessions, supported by independent mass accretion rate estimations 
should be performed to examine the legitimacy of the aforementioned numerical scenario for explaining the long-term changes, common for many actively accreting CTTSs.  
We note that other effects like variable circumstellar extinction may also come into play \citep{grankin07}. 
However, this last possibility seems to be unlikely for our target as the dust extinction 
toward TW~Hya has been estimated to be negligible by \citet{herczeg04}.

%----------------------- Fig - flares ---------------------
\begin{figure*}
\includegraphics[width=0.32\linewidth]{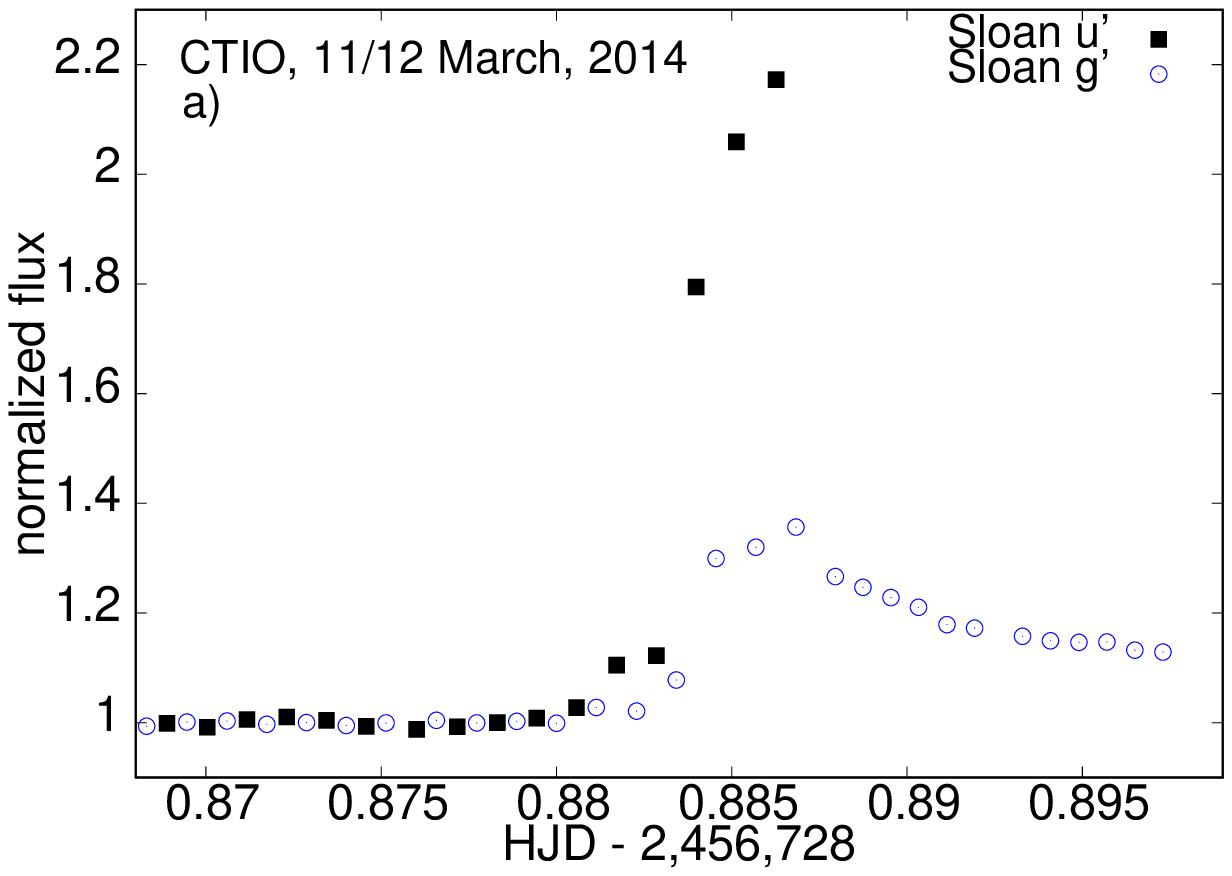}
\includegraphics[width=0.32\linewidth]{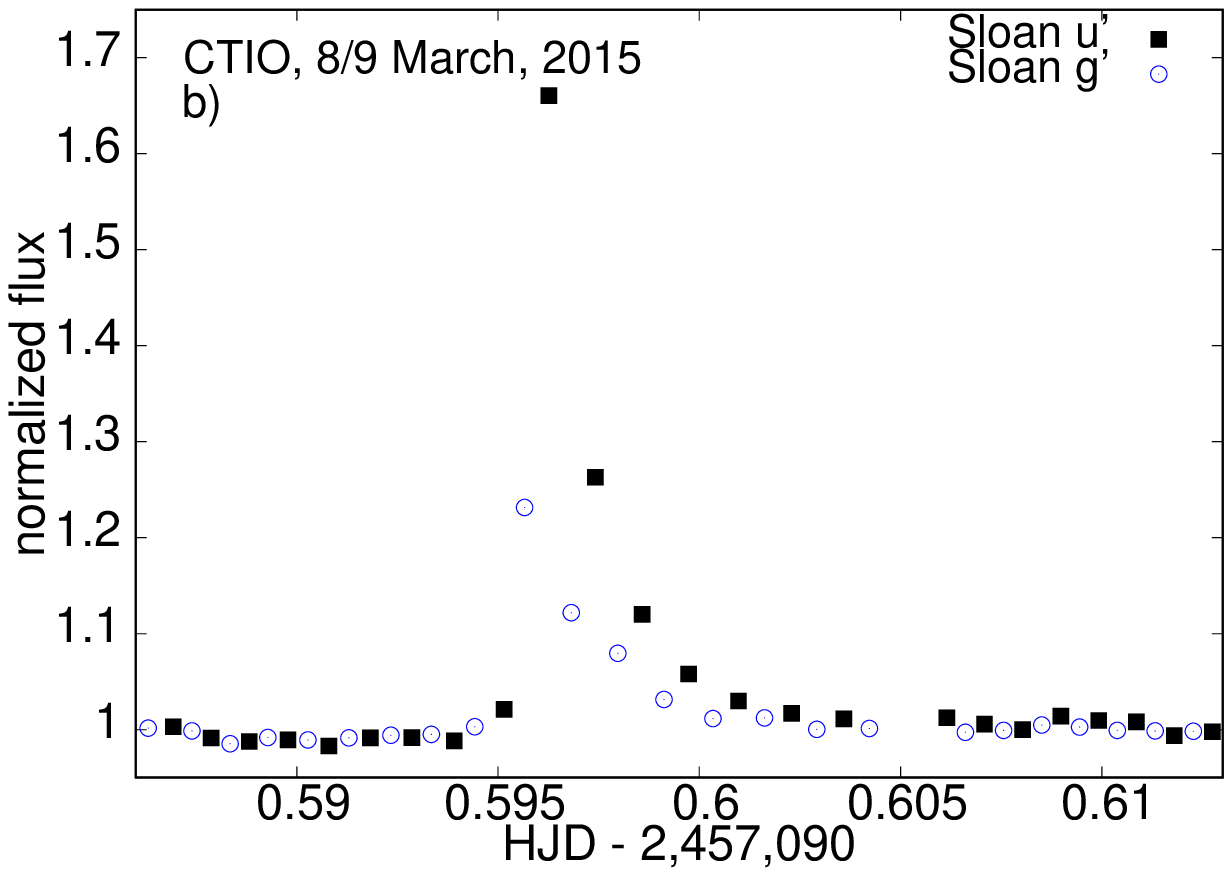}
\includegraphics[width=0.32\linewidth]{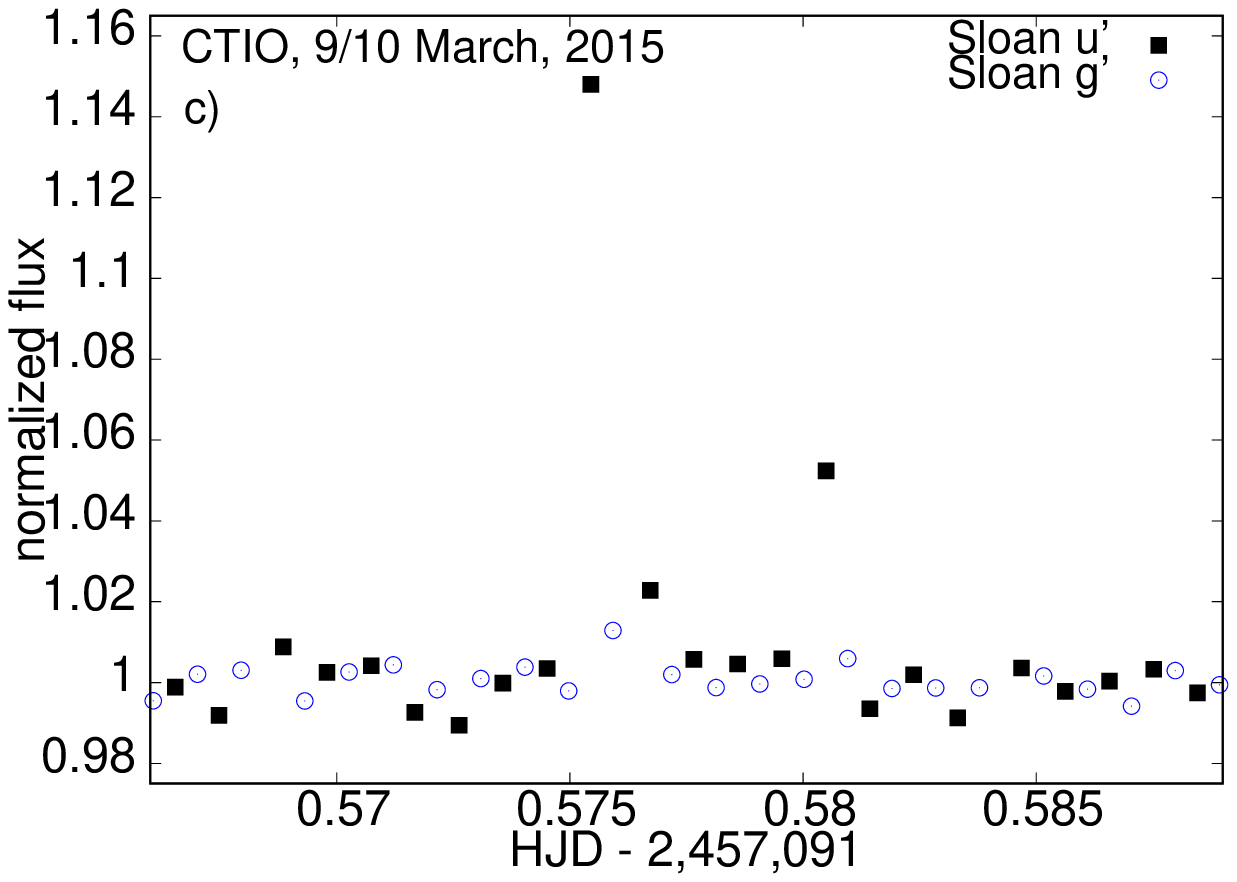}
\includegraphics[width=0.32\linewidth]{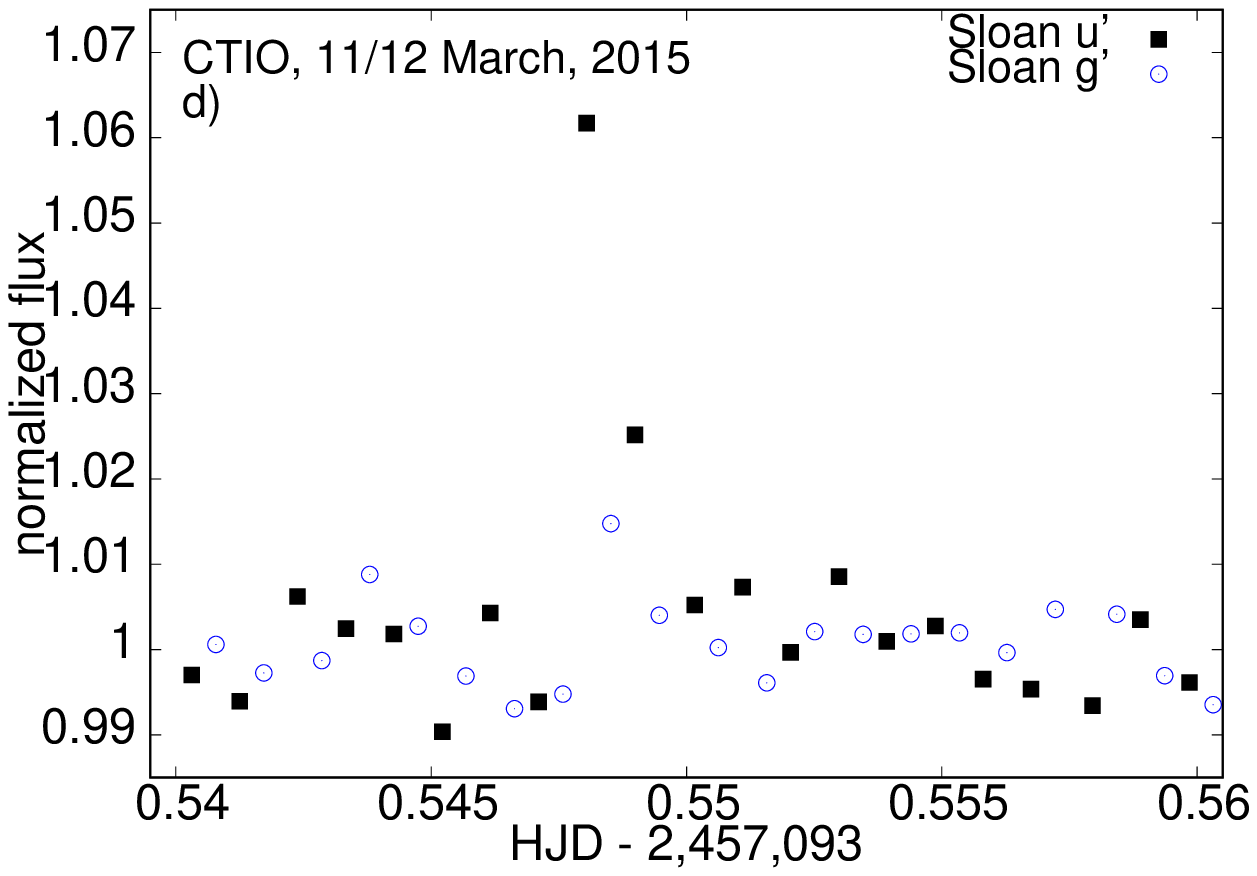}
\includegraphics[width=0.32\linewidth]{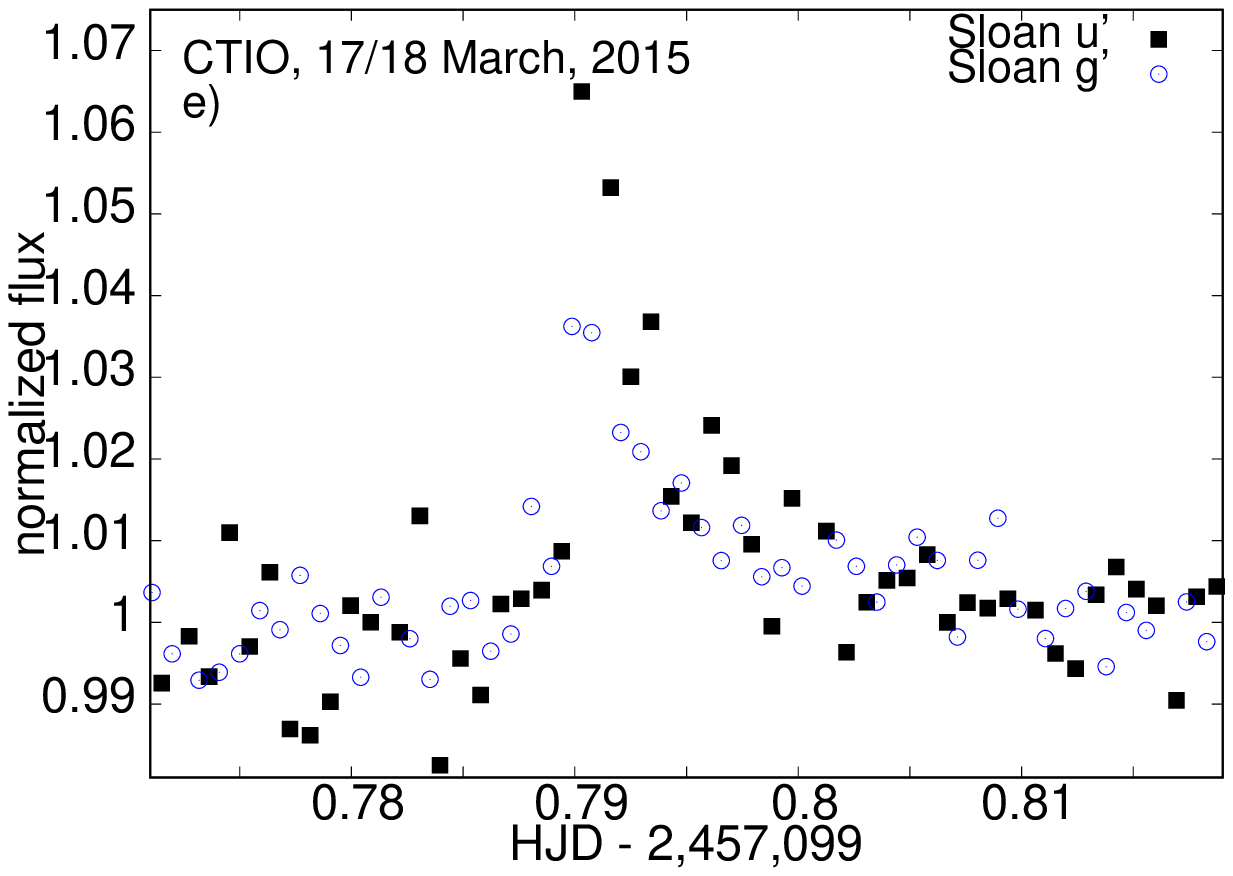}
\includegraphics[width=0.32\linewidth]{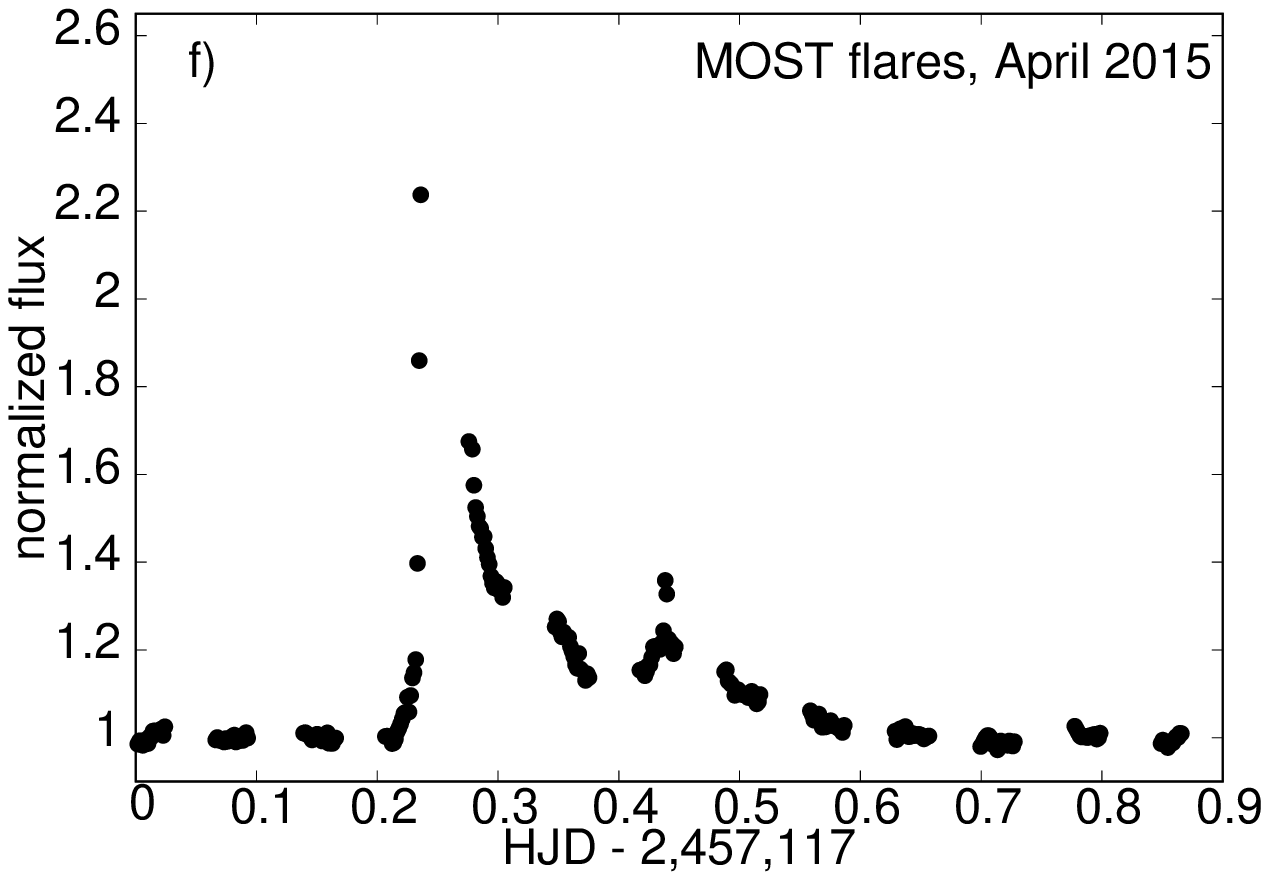}
\includegraphics[width=0.32\linewidth]{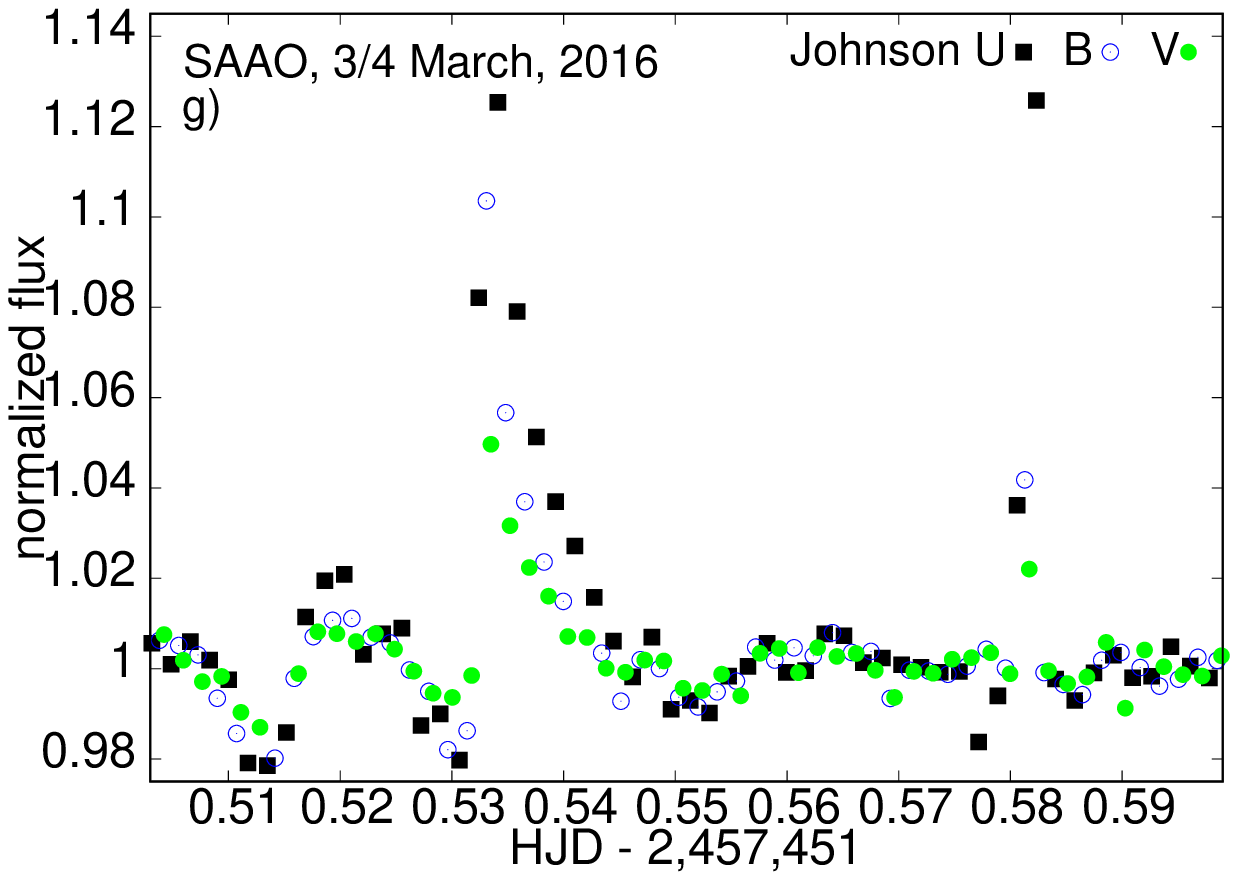}
\includegraphics[width=0.32\linewidth]{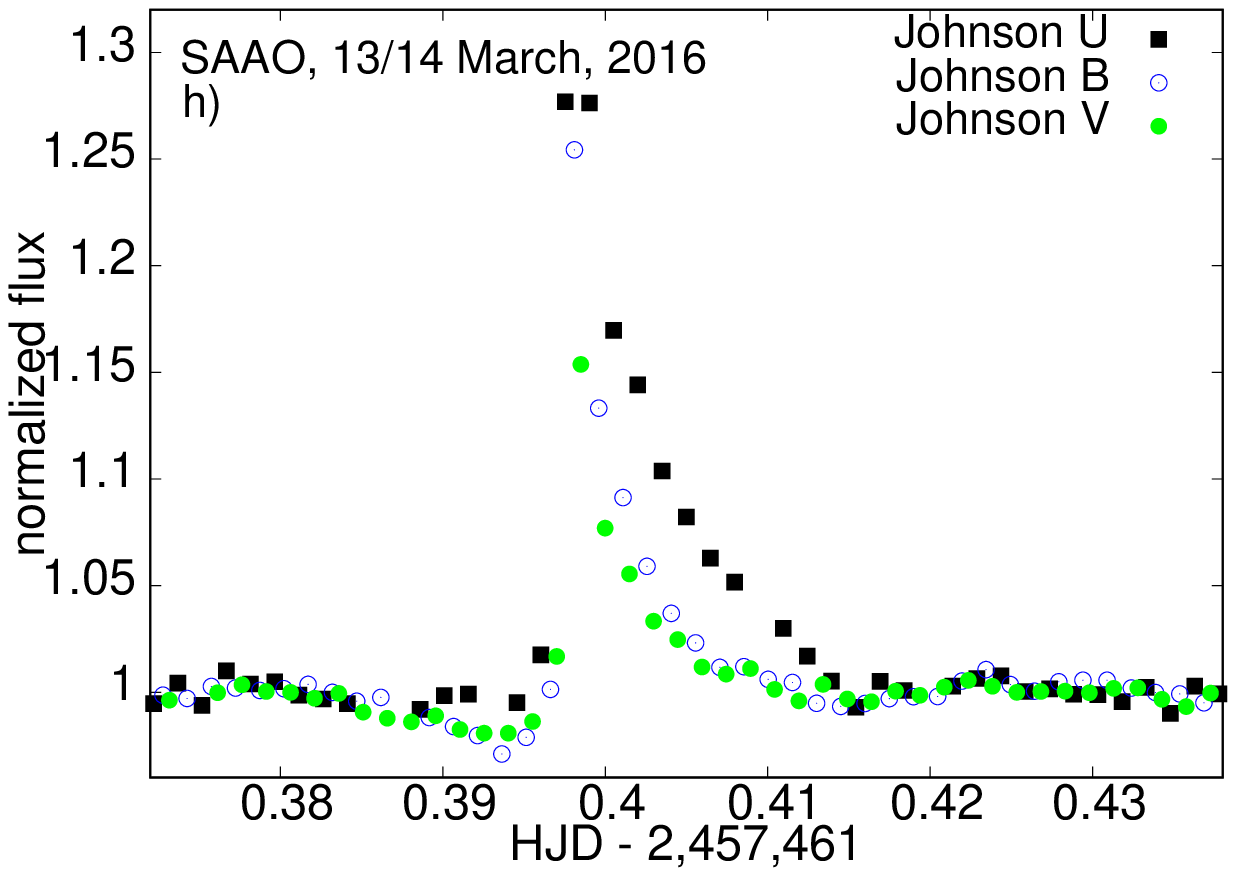}
\includegraphics[width=0.32\linewidth]{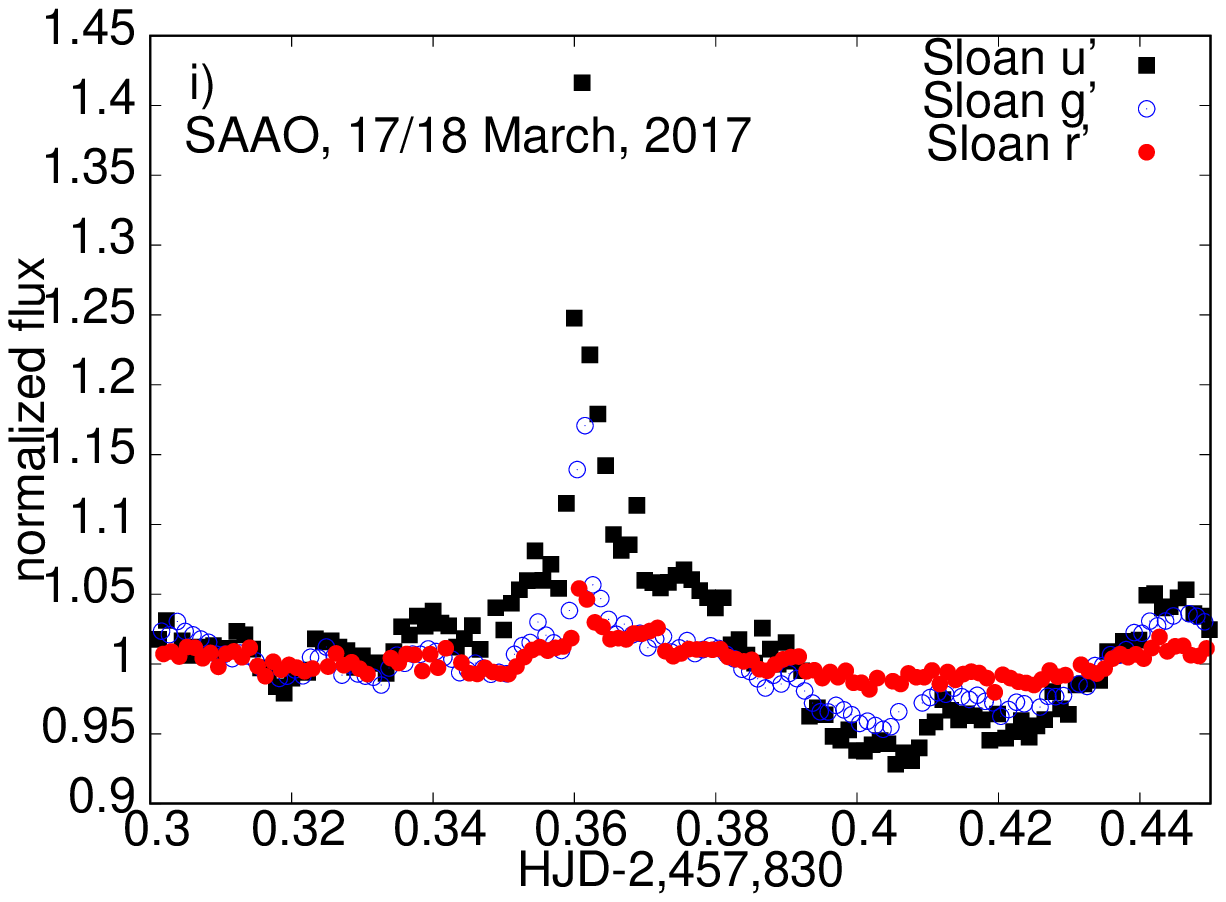}
\caption{Enlargements of nine light curve segments with 12 flares of TW~Hya. 
Their intensities are expressed in underlying continuum flux units, which were estimated 
and removed by means of polynomial fits. 
The 2014 flare (first panel) appeared at twilight and was not observed completely.
In three cases a secondary flare (or a flash) appeared soon after the primary one. 
}
\label{Fig.flares}
\end{figure*}
%----------------------------------------------------------------------

\subsection{Flares -- the rate, amplitudes and durations}
\label{flares}
Contrary to our previous observations, the 2015 {\it MOST} and the ground-based light curves 
revealed at least twelve flares of different amplitudes and shape. 
In order to characterise their properties in a uniform way, we removed the underlying accretion-driven light 
variations using low-order polynomials. 
We then expressed intensities of the flares in flux units normalised to unity at the underlying continuum, 
as shown in Figure~\ref{Fig.flares}. 
Based on the plots we measured their major properties i.e. the start, maximum, end, half-duration ($T_{0.5}$) 
times and the u-band amplitudes $U_{max}$ (if possible) -- the most crucial parameters are shown in Table~\ref{Tab.log1}. 
Based on the parameters one can divide observed flares of TW~Hya into three (or possibly four) families:\newline
\begin{itemize}
\item The first, single-member family consists of the 'super-flare' of 1~mag amplitude in the 'white' filter, 
lasting about 8 hours and observed by {\it MOST} in 2015 (Fig.~\ref{Fig.flares}f). 
Unfortunately, we do not have u-band observations of the flare (where the amplitude could reach 5~mag).
We note that TW~Hya is not in the Continuous Viewing Zone of the satellite and the data acquisition must 
be interrupted for a dozen minutes during each satellite orbit. 
Therefore the peak of the flare was likely missed and the $T_{0.5}$ parameter is determined with low precision. 
\item The second family consists of six flares with fast 1-2~min rise times and purely exponential 
decays, as observed during 2014-03-11/12 (Fig.~\ref{Fig.flares}a), 2015-03-08/09 (Fig.~\ref{Fig.flares}b), 
2015-03-17/18 (Fig.~\ref{Fig.flares}e), 2016-03-03/04 (Fig.~\ref{Fig.flares}g -- the first event during the night), 
2016-03-13/14 (Fig.~\ref{Fig.flares}h), and 2017-03-17/18 (Fig.~\ref{Fig.flares}i). 
Their u-band half-duration times are 2.5-10 minutes, with amplitudes of 0.065-1.17 in normalised 
flux units (0.07-0.85~mag). 
The second 2015 {\it MOST} flare likely also belongs to this family due to its short 
half-duration time and low 0.12 (in normalized flux units) amplitude in the 'white' filter 
of the satellite.
\item The third family consists of very short flares called 'flashes', most pronounced in $U$- and $u'$-filters. 
They are barely distinguishable from the photon-noise in the $g'$- and $B$-filters. 
The 'flashes' were observed on 2015-03-09/10 (Fig.~\ref{Fig.flares}c), 2015-03-11/12 (Fig.~\ref{Fig.flares}d) 
and 2016-03-03/04 (Fig.~\ref{Fig.flares}g -- the second event during the night). 
One of them was observed simultaneously from two sites in 2015, which ruled out the originally 
considered 'cosmic ray' scenario. 
Their u-band amplitudes are 0.04-0.15 (in normalized flux units) and their half-duration times 
are 1.5-3.3 minutes. 
We stress that these estimates are uncertain due to the cadence of our observations only twice higher 
than the lifetimes of the 'flashes'. 
\item We also observed several events which cannot be classified as flares with any certainty. 
Although they seem to have similar amplitudes and duration times as well as similar rise and decay times as flares assigned to the second family, 
they do show smaller differences between amplitudes seen in the $u'g'$- 
and $UBV$-filters.
In many respects they resemble 'accretion-bursts' - common for CTTSs  (see below). 
We also indicate these phenomena in Table~\ref{Tab.log1}.
\end{itemize}

As far as we know, this is probably the first study of flares in TW~Hya at visual wavelengths. 
Prior to the start of this investigation of their occurrence rate in 2007-2017, one must keep in mind 
that different photometric systems were used over these years. 
Additionally, the detection efficiency of flares by {\it MOST} (2007, 2008, 2009, 2011, 2014, 2015, 2017) 
was always reduced due to interruptions in the data acquisition during each orbit. 
Another limitation is imposed by the {\it MOST}-filter, integrating almost whole visual light, which 
means that the small-amplitude flares (maximum 10 per cent in u-band) could have been overlooked.\newline
Because of the above, more general conclusions can be only drawn from analysis of ground-based data. 
We did not detect any flares nor flashes during the 2013 observations. 
In 2014 we observed only a single flare during an effective monitoring time of 5.358 days, but in 2015 
we recorded five flares at two observatories during 8.524 days -- the two {\it MOST} flares 
were excluded from this statistic as they were observed by an instrument of different sensitivity. 
Three flares were observed during the 2016 {\it SAAO} run, which effectively lasted for 5.308 days, 
while only one was observed in the 2017 run, which lasted for 4.026 days.
When weighted by effective monitoring times, the occurrence rate during 2013, 2014, 2015, 
2016 and 2017 was 0, 0.19, 0.59, 0.57, and 0.25 flare per day, respectively. 
If events marked in Table~\ref{Tab.log1} as 'flares or accretion bursts' are included into this statistic, 
the respective rates are 0, 0.56, 0.94, 0.94 and 0.25 flare per day. 
Despite the flare activity being maybe higher during the 2015 and 2016 runs, 
%[<-- I have the impression that you are dealing with small-number statistics, so be careful and don't exaggerate any 
%claims of variable rates! - e.g. "obviously higher" may not be so obvious! TM]
this result may be not significant, as the star was usually observed for a fraction of a month only. 
%On the other hand, our finding [What finding?? TM] is supported by the 'super-flare' (and the second flare of a smaller 
%amplitude) exceptionally observed by {\it MOST} only in 2015. 
%Such a 'super-flare' could not have been overlooked in other {\it MOST} light curves.
We conclude that further observations must be performed to be able to reach any viable conclusion about 
the flaring rate in TW~Hya. 

% -----------------------  Fig.6 the single-night periodograms ---------------------
\begin{figure*}
\includegraphics[width=81mm]{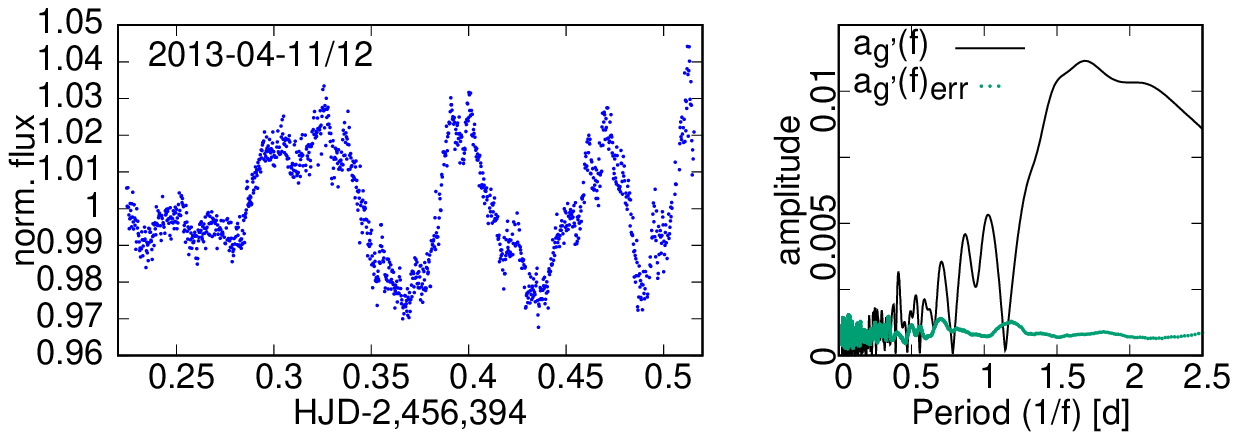}
\includegraphics[width=81mm]{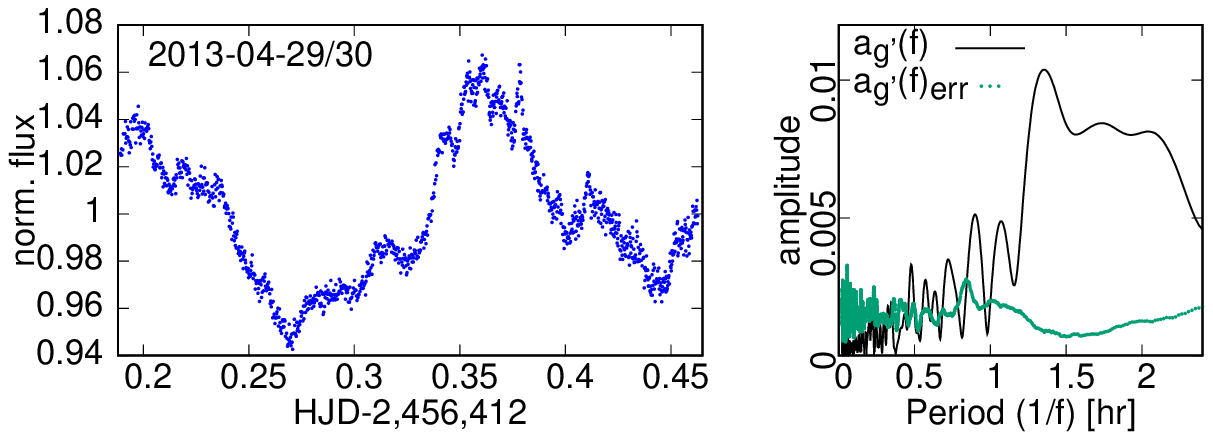}
\includegraphics[width=81mm]{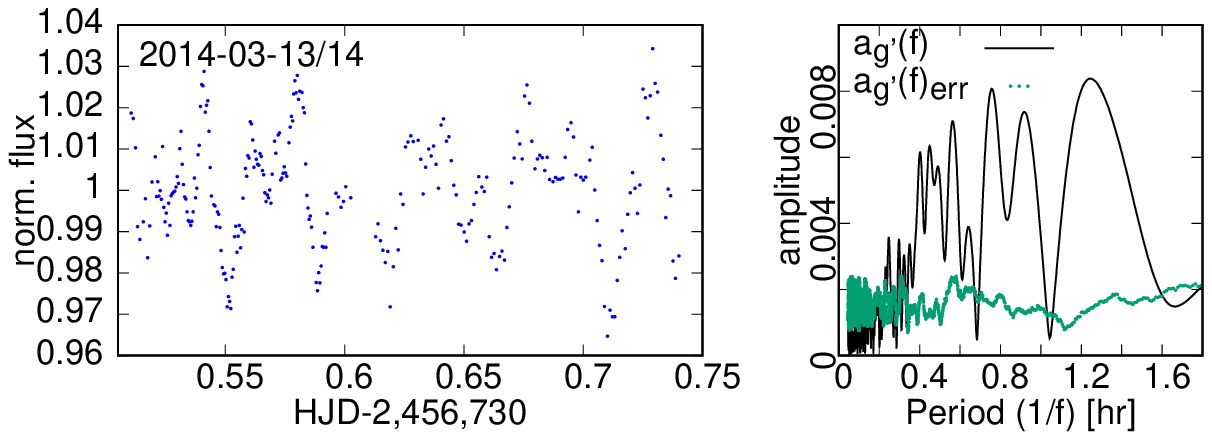}
\includegraphics[width=81mm]{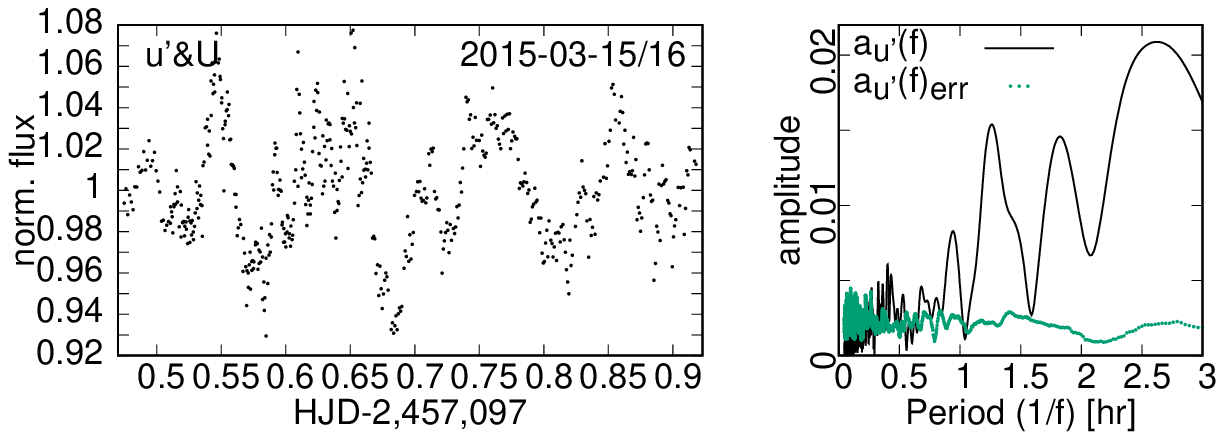}
\includegraphics[width=81mm]{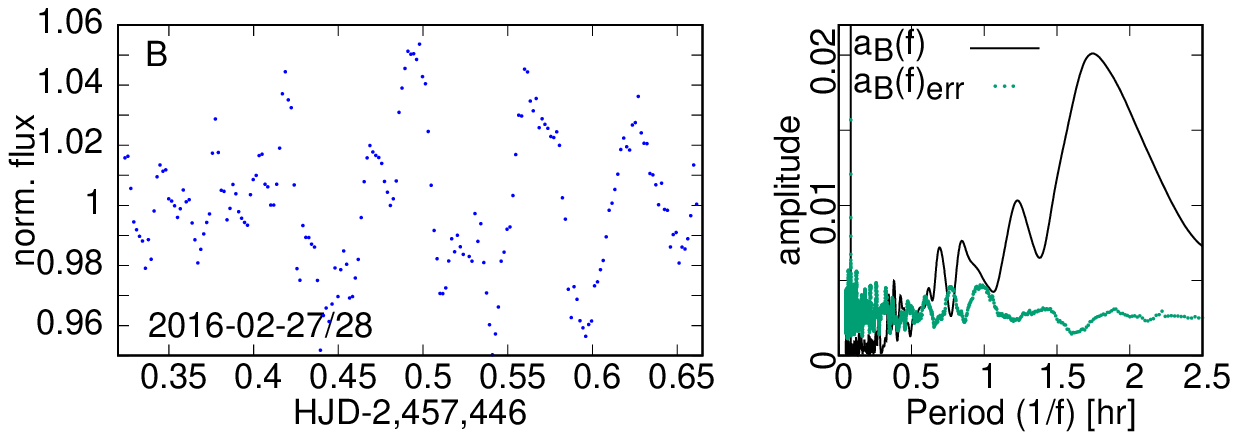}
\includegraphics[width=81mm]{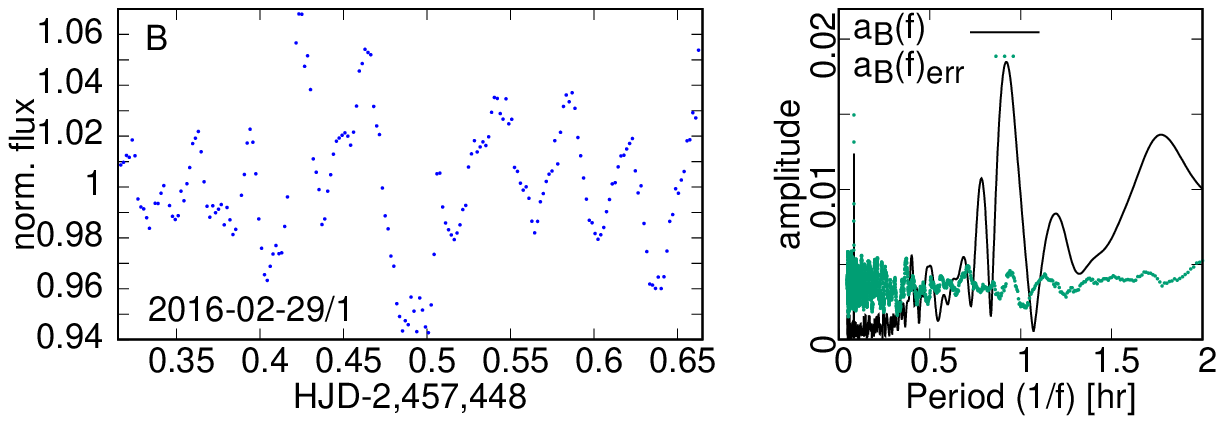}
\includegraphics[width=81mm]{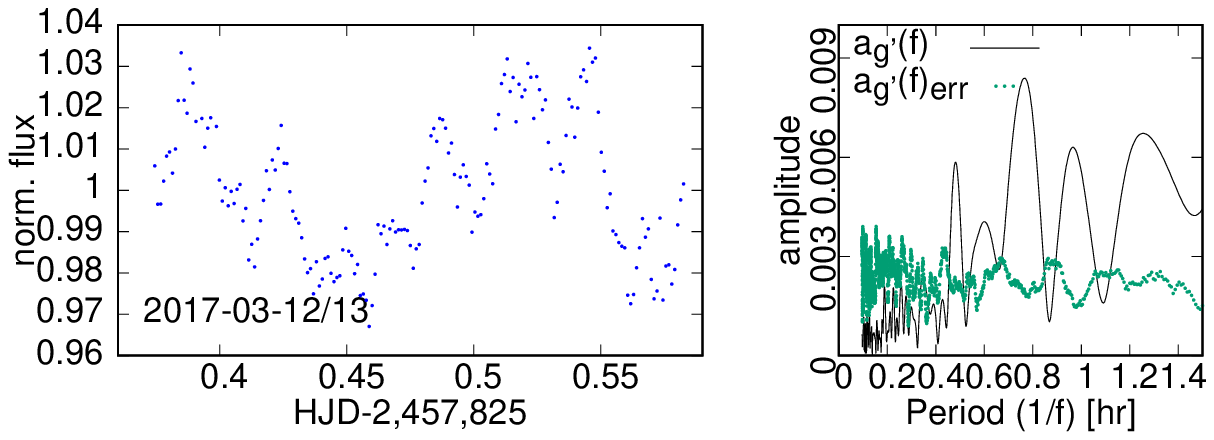}
\includegraphics[width=81mm]{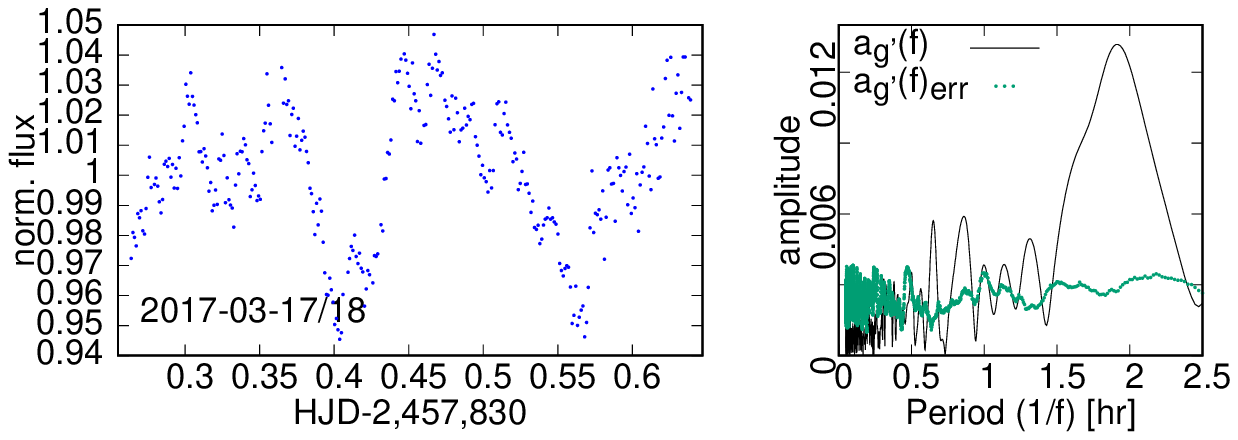}
\caption{The $g'$-filter detrended light curves (left panels) obtained during nights when 
short-periodic QPOs were directly observed. They are expressed in flux units normalised to unity 
at the mean nightly levels. 
The periodograms are shown on the small right panels; the amplitudes of particular filter-data, 
e.g. $a_{g'}(f)$, are expressed by continuous lines, while their errors are indicated by small dots.
For the 2015 oscillations, combined u-band data and their oscillation spectra are also shown. 
$B$-filter light curves were used to investigate variability observed in 2016. 
}
\label{Fig.short}
\end{figure*}
%----------------------------------------------------------------------

% -----------------------  Fig.5 the flicker-noise character ---------------------
\begin{figure}
\includegraphics[width=81mm]{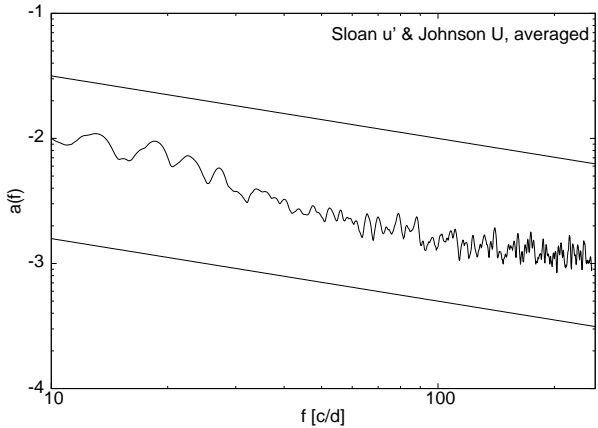}
\caption{Averaged Fourier spectrum (computed from spectra obtained during 16 different nights) 
in a logarithmic scale. 
No distinct periodicities can be identified; instead the spectrum shows a pure flicker-noise 
character indicated by the two straight lines.}
\label{Fig.flicker}
\end{figure}
% -----------------------  Fig.5 appendix the CTIO 2014 periodograms ---------------------

% -----------------------  Fig.5 appendix the CTIO 2014 periodograms ---------------------
\begin{figure*}
\includegraphics[width=80mm]{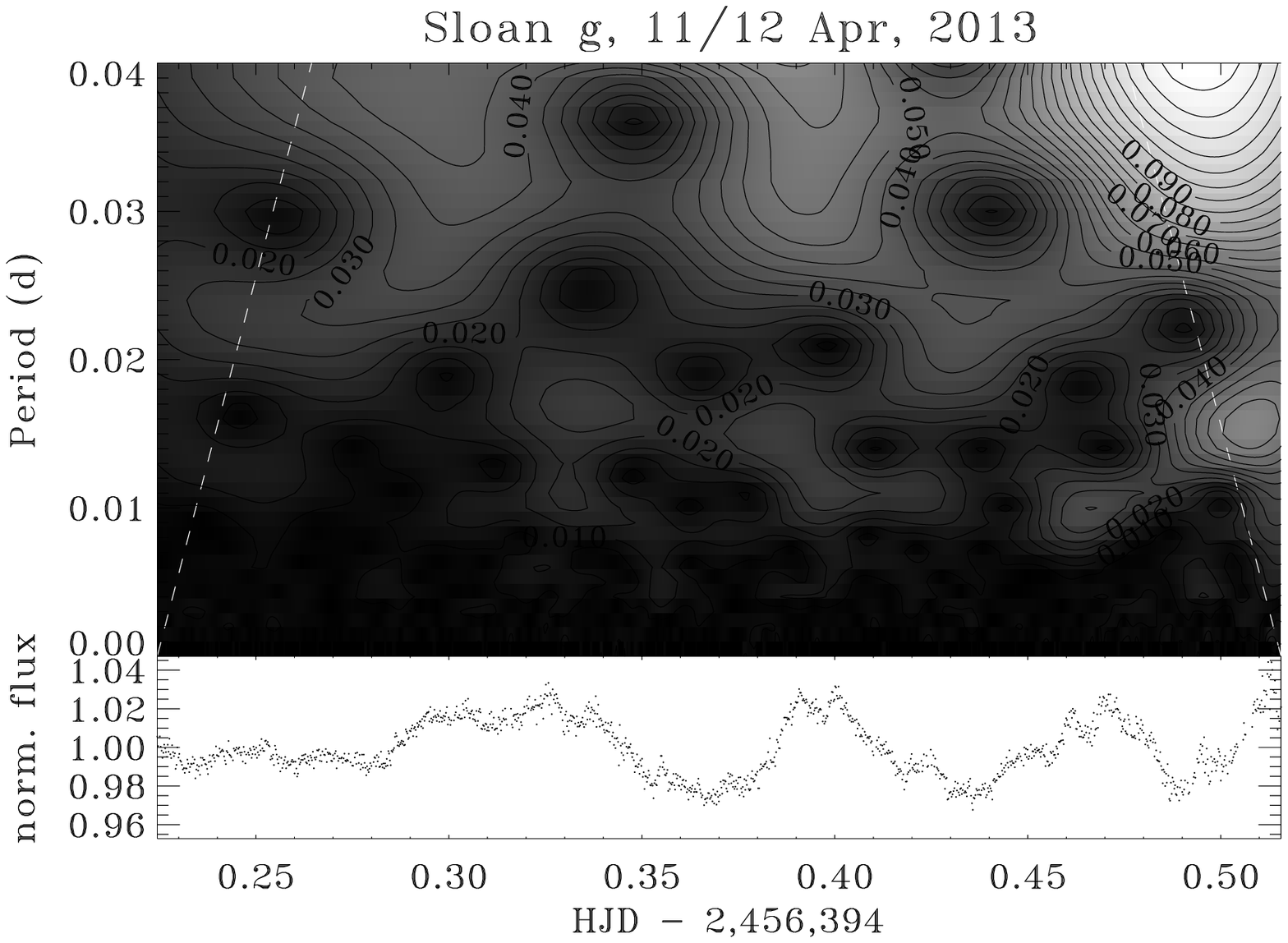}
\includegraphics[width=80mm]{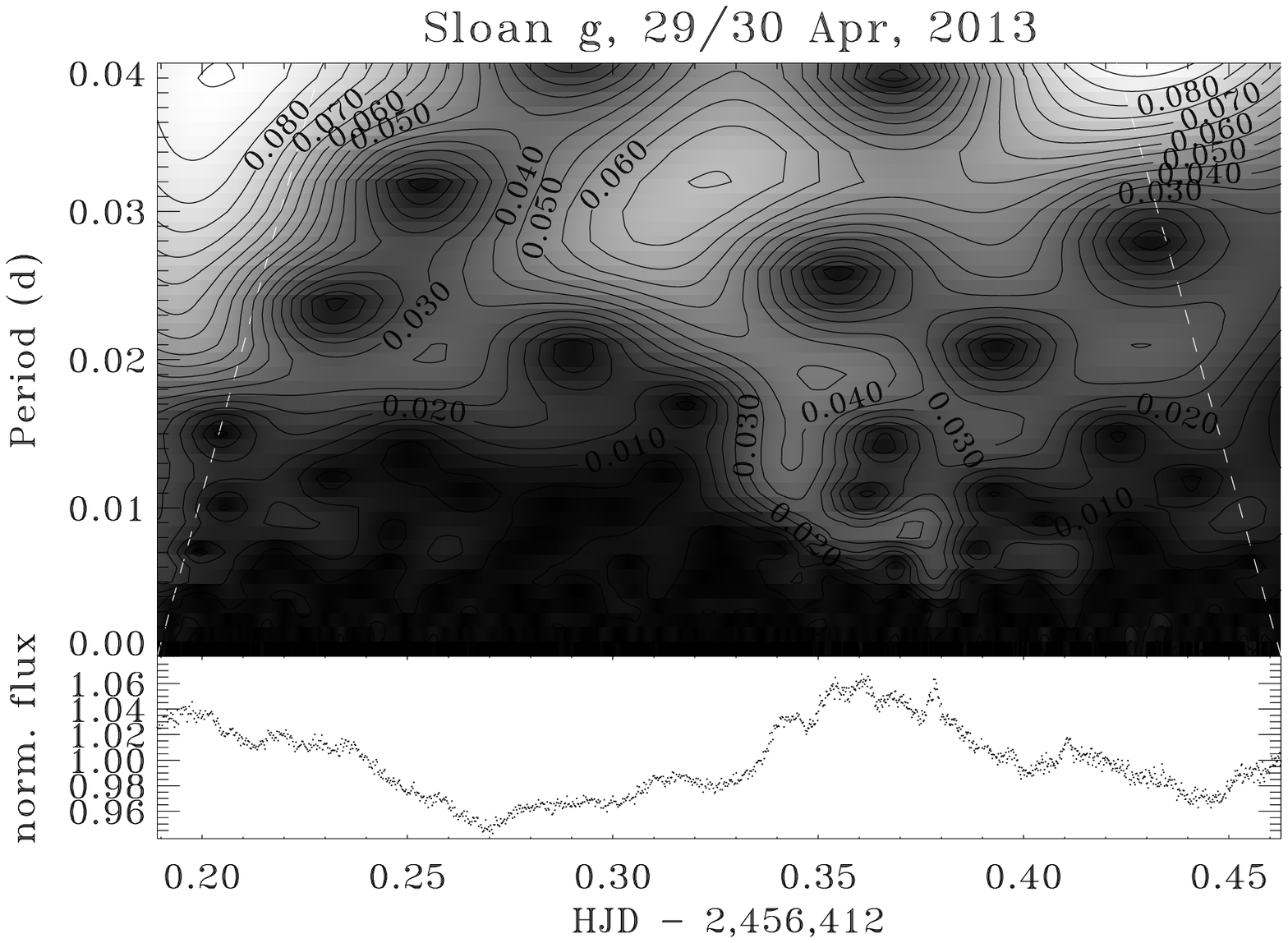}
\vskip 5pt
\includegraphics[width=80mm]{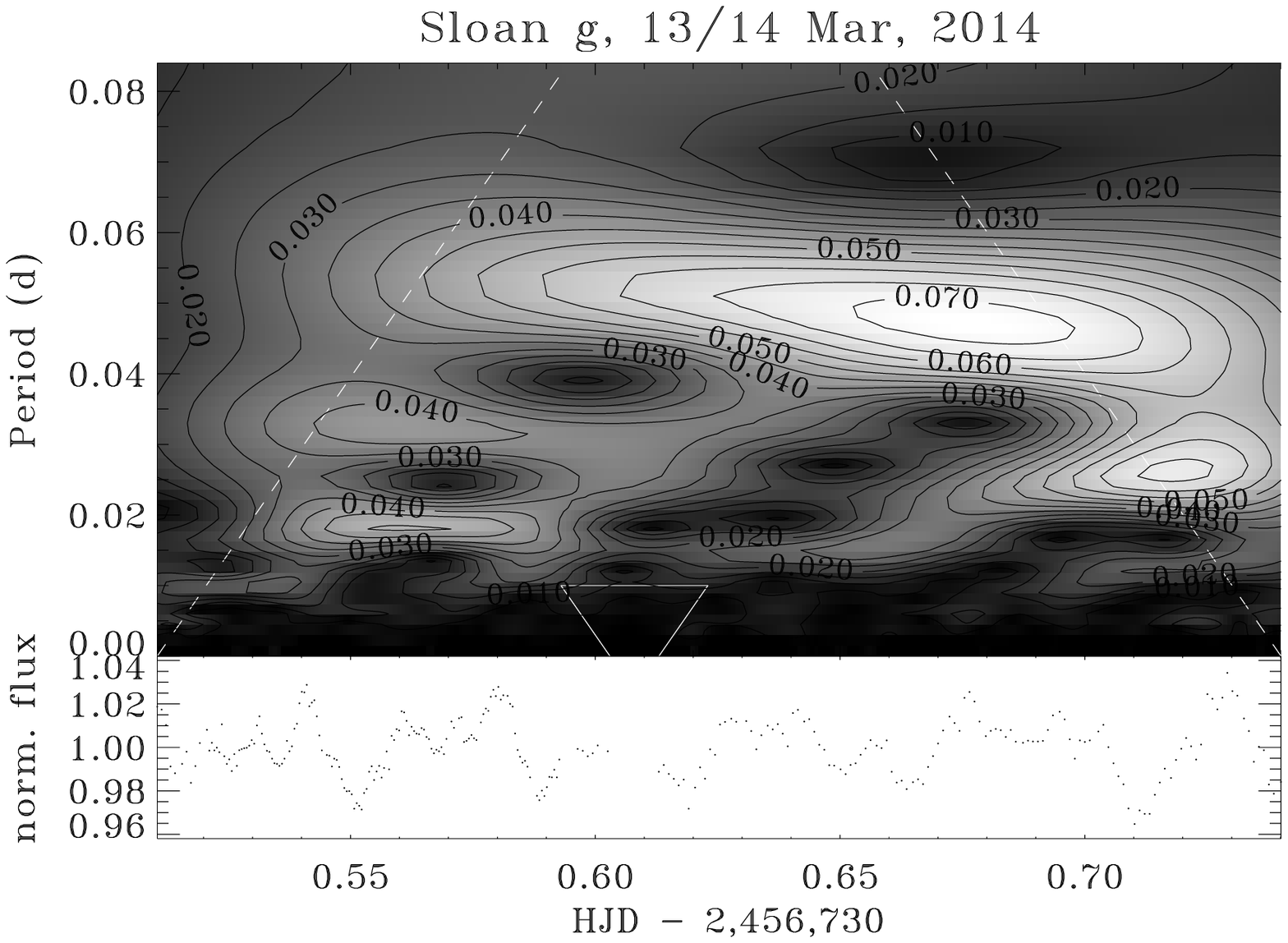}
\includegraphics[width=80mm]{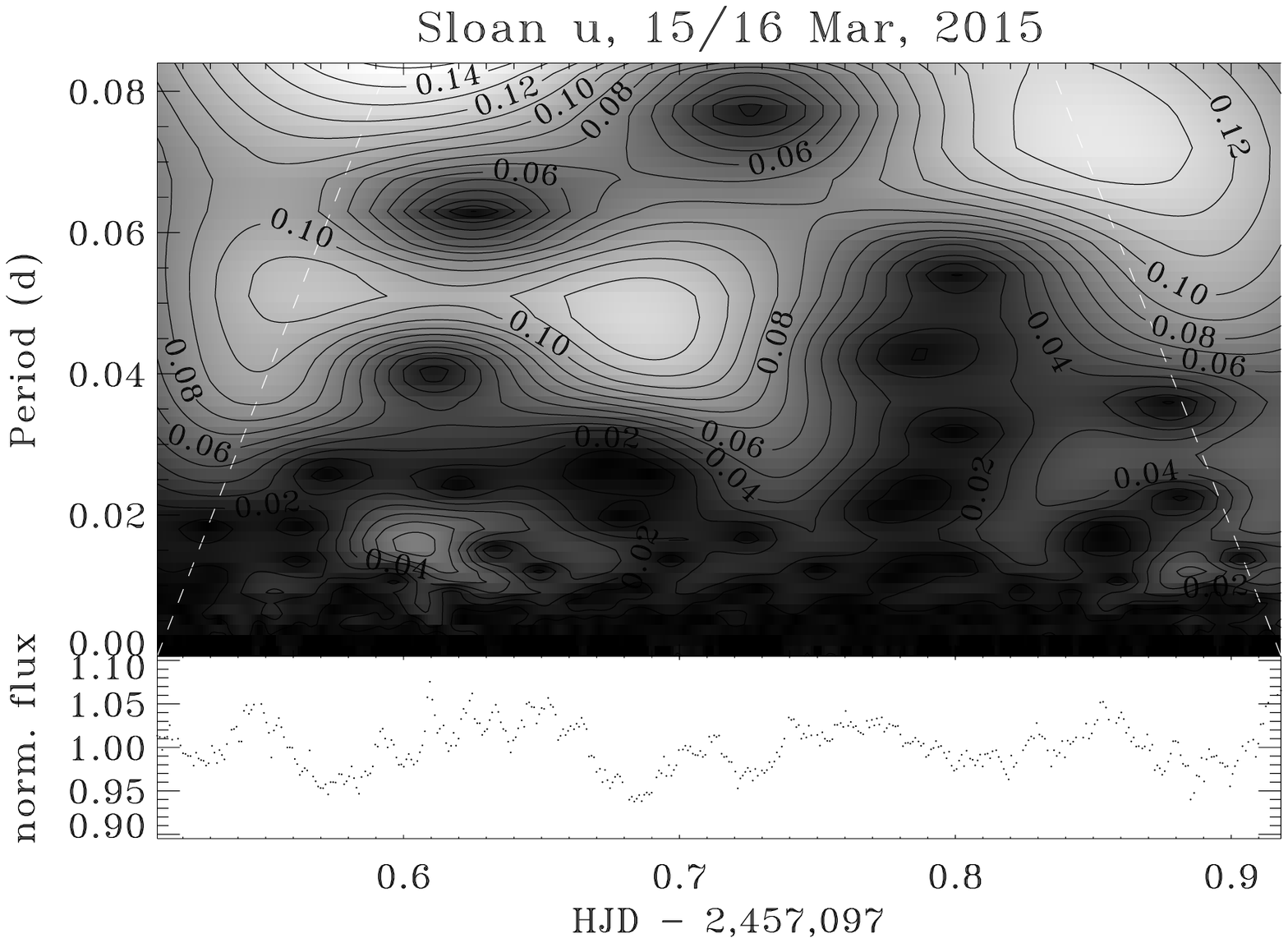}
\vskip 5pt
\includegraphics[width=80mm]{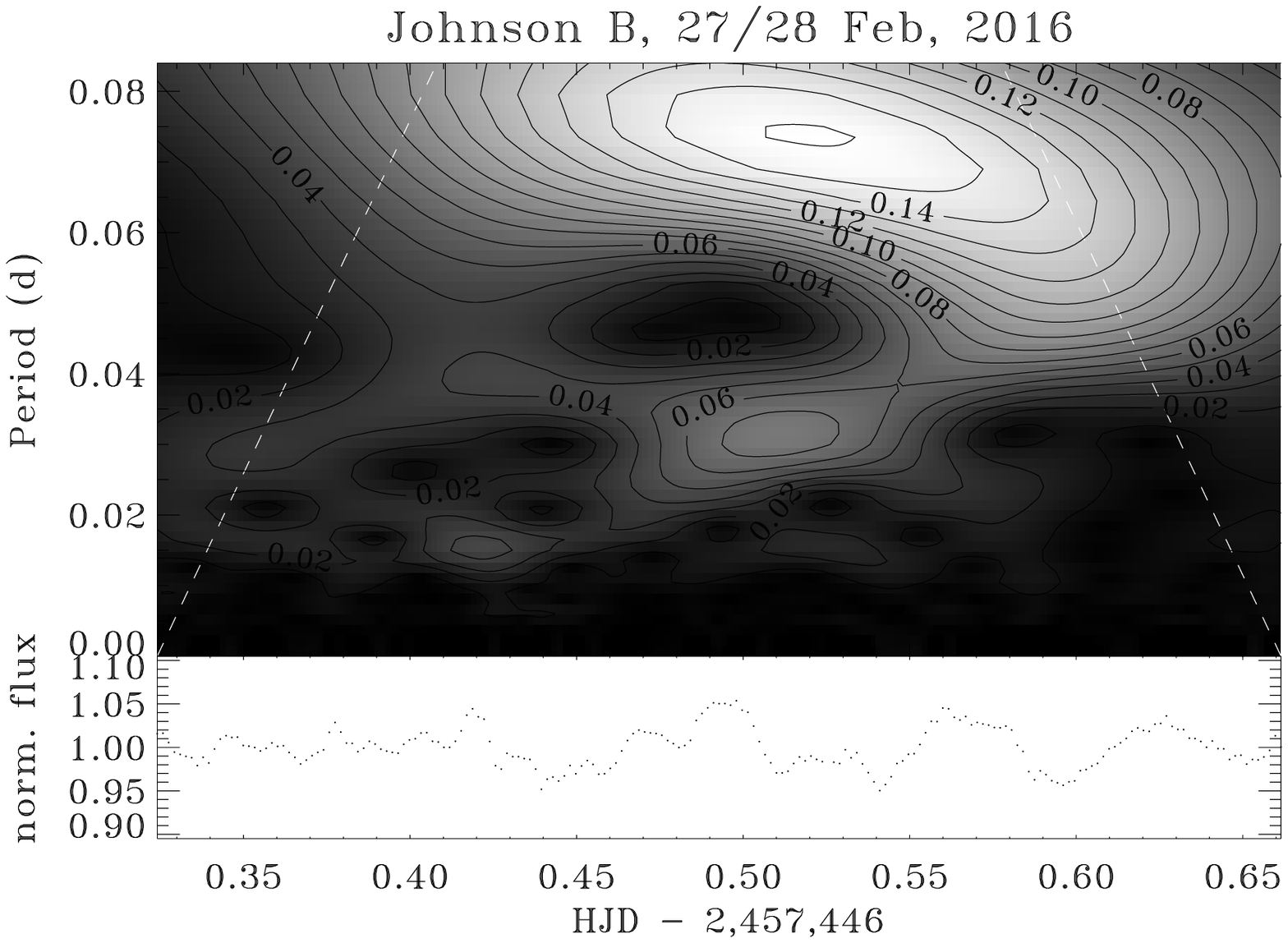}
\includegraphics[width=80mm]{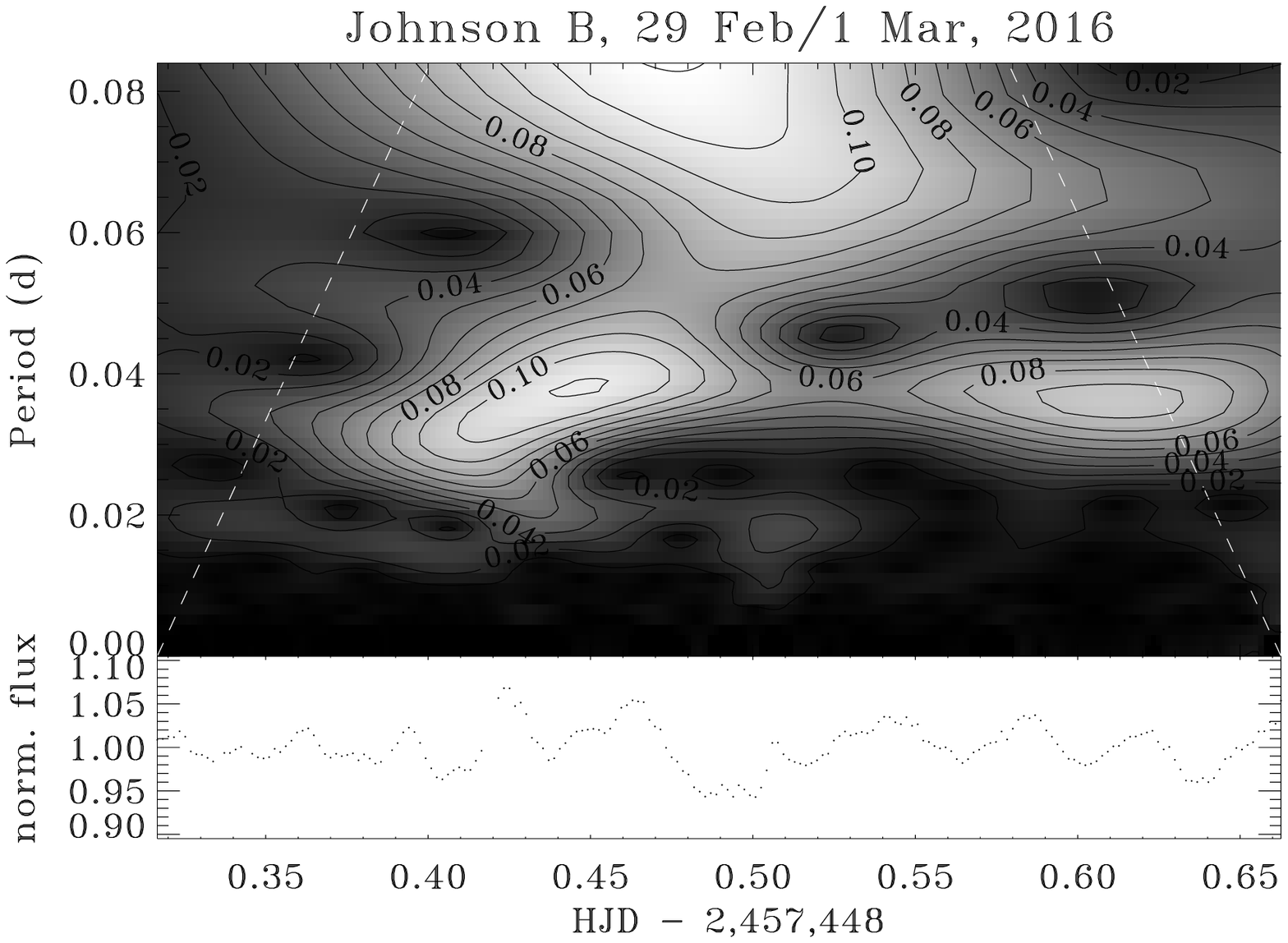}
\caption{The wavelet spectra for a few selected nights. 
Events in the local maximum observed on March 15, 2015, show best defined 
signs of an short-periodic oscillatory behaviour.}
\label{Fig.short_wav}
\end{figure*}
%----------------------------------------------------------------------

\subsection{Analysis of multi-colour 'accretion burst' dominated light curves.}
\label{accrburst}

\subsubsection{General description of the variability}

In the Appendix~\ref{app2} we present observations gathered at {\it SAAO} and {\it CTIO} 
during photometric or partly cloudy nights (which turned out to provide good-quality photometry anyway 
after careful reduction). 
Except for stellar flares, which can be well distinguished due to their characteristic fast 
rise time and slow exponential decay (Sec.~\ref{flares}), these data show a variety 
of light changes at almost all times. 
They appear to be irregular, but exceptions from this general rule are indicated on some nights. 
The highest amplitudes are observed in the u-band and they decrease with increasing effective wavelength 
of consecutive bands -- e.g. the $r'$ and $z'$-filter light curves show considerably less variation and lower amplitudes
than the accompanying $u'g'$-filter light curves.\newline 
The most pronounced variations, namely 'accretion bursts' as specified by \citet{gullbring96}, 
are of 'triangular' shape. 
Their rise and decay branches last for similar times and they are well isolated from underlying 
smooth light variations occurring on longer time-scales (e.g. on April 13, 2013). 
'Accretion bursts' may appear once, several times, or not at all during a night. 
They clearly avoid low brightness states when accretion activity is significantly 
reduced (e.g. on April 25, 2013). 
Sometimes they are observed in close pairs (e.g. on April 13\&14, 2013) and appear to be double-peaked 
with separation between consecutive maxima of a dozen minutes. 
Some of these close pairs are separated by a nearly flat light section, like on April 30, 2013. 
Interestingly, 'accretion bursts' and possibly QPOs of much shorter period (11-30~min) 
and small amplitude ($\sim0.01-0.05$~mag) are often superimposed on  primary ($\sim 0.1-0.5$~mag) 
bursts both in the rising and descending branches, or in the light maxima only. 
From time to time, we observe a significant accumulation of well-separated bursts occurring throughout 
the entire night: the variability may be either completely chaotic, like on April 22, 2013, or (nearly) periodic 
and possibly time-coherent, at least at first sight (e.g. March 9, 2014 and Feb 29, 2016). 
The lifetimes of such  single oscillatory packages are of merely 2-4 cycles. 
Amplitudes of these variations appear to scale as the whole light curve. They are highest 
in the ultraviolet, still well visible in blue bands, and usually only barely visible 
in the green and yellow parts of the spectrum. 
This is an important finding as it limits possible places of their formation to hot-spots and 
instabilities in the accretion process. 
%Within the framework of an inhomogenous accretion model, the range of observed QPOs could be 
%the result of impacts of many small density elements in an accretion flow, likely one after 
%the other in nearly regular intervals until resources are exhausted. 
 
All of the above phenomena observed in TW~Hya were already spotted and well characterised in BP~Tau (a CTTS) 
by \citet{gullbring96}. 
These authors explained these bursts as the result of inhomogeneous accretion, which is composed of numerous 
fragments that flow in a random way to the star (see Sec.~7 in their paper). 
Using realistic values of an accretion stream and the resulting hot-spots, they calculated cooling times 
after impacts of such fragments on the star, ranging from 30~min to several hours, what is in accord with their 
observations of the shortest bursts in BP~Tau lasting for 0.6~h.
They explained the range of rise times as dependent on the 'length' of the inhomogeneities in the accretion flow. 
Further 'shot-noise' simulations of light curves, binned appropriately for direct comparison with real 
observations, resulted in a range of peaks, either rounded (typical for variability observed through almost 
80 percent of the time), or sharp-like (typical for 'accretion bursts' of 'triangular' shapes), 
or even others, either having linear rises followed by exponential decays, or 'square' shapes, which are 
visible in the real photon-noised observations as an inverted 'U'.\newline
All the aforementioned structures can be easily identified in our light curves of TW~Hya.
This kind of variability occurring on longer time scales than investigated here was already noticed 
by \citet{ruc08}.
% although these authors were limited to time-scales longer than a single {\it MOST} orbit 
%(see above in Sec.~\ref{wavelet}).
Here in Figure~\ref{Fig.short} we present clear instances of such variations occurring 
in time-scales ranging from a dozen minutes to several hours.

\subsubsection{Fourier analysis}
To analyse the data from single nights with the Fourier technique we removed nightly trends 
using 2nd-3rd order polynomials. 
Thereafter magnitudes were transferred to fluxes which were normalised to unity at the nightly mean 
star brightness levels, as shown in the left panels of Figure~\ref{Fig.short} -- the right panels 
show corresponding periodograms obtained by a crude inversion 
of the amplitude-frequency spectra. 
Usually the blue-bands ($g'$-, $B$-filters) were analysed as they provide the best ratio 
of variability amplitude to photometric accuracy of a single data point. 

For illustrative purposes, in Figure~\ref{Fig.short} we selected data obtained during eight representative nights. 
For example, during the first night (April 11, 2013) long- and short-term superimposed oscillations 
coexisted at all times. 
The quasi-period of the longer variation was 1.6~h, while that of the shortest apparently 
was unstable, close to about 15~min. 
In addition, we observe a forest of peaks in the range of 0.6-1~h. 
Similar coexisting short- and long-term variability was also observed during other nights.

The peaks visible in the periodograms do appear randomly in the Fourier spectra. 
We arrived at this conclusion using average Fourier spectrum computed from all 16 individual 
Fourier spectra prepared from the u- and $g'$-band data, displayed in logarithmic scale; the result for 
the u-band is shown in Figure~\ref{Fig.flicker}. 
Instead however, in actual fact they clearly show a flicker-noise dependency $a(f)\propto\sqrt f$, 
indicated by the two straight lines. 
This result confirms the flicker-noise character of the short periodic variability in TW~Hya for time 
scales from minutes to hours, as was already proven for time-scales of days, weeks and years 
by \citet{ruc08}.\newline
Except for the above cases, we also made dedicated searches for the shortest quasi-periodicities 
possible for detection with the 2013 $g'$-filter data, which were sampled every 10-23 seconds. 
However, they do not show any significant oscillations in the range of 30~s--10~minutes.

\subsubsection{Wavelet analysis}

The lack of well-distinguished high-frequency peaks in the Fourier spectra appears to contrast 
with many clear instances of small-amplitude, possibly quasi-periodic variations in our light curves. 
Therefore we applied the wavelet technique for analysis of six representative nights. 
The original ground-based data were re-sampled with uniform steps (similar to the observing cadence 
during a given night) using the procedure previously applied to the {\it MOST} data (Sec.~\ref{wavelet}). 
%Appropriate steps were selected to match different observing cadences.\newline
As expected, the two-dimensional wavelet spectra shown in Figure~\ref{Fig.short_wav} are more 
informative than their one-dimensional counterparts: 
the spectra reveal a range of brief events, either limited to single or double brightening events, 
or showing oscillatory character extended over a significant fraction of a night. 
Period changes are noticeable in many cases, which may explain the lack of well-defined high-frequency peaks 
in the Fourier spectra.

The shortest significant peaks identified both in the Fourier spectra and directly in the light curves are 
at 0.008-0.02~d (11-30~min). 
The best defined oscillatory-like variations can be easily localised in the local maximum 
on March 15/16, 2015.  
They show a period change between 0.016-0.01~d (23-15~min).
Similar short-term 11-15~min light variations are observed during many (but not all) local maxima of 
'accretion bursts'. 
Interestingly, we did not identify similar 11-15~min oscillations anywhere else in the light curves.

\subsubsection{Discussion}
The findings obtained from the wavelet analyses prompted us to consider an accretion shock 
instability as the possible cause of the formation of the shortest QPOs next to the 'accretion burst' 
scenario, which well explains light variations lasting longer than 0.5~hr.
  
According to \citet{langer81} and \cite{chevalier82}, the fundamental mode of the thermal instability 
oscillation of an accretion shock has its characteristic period of the order of the cooling 
time of the post-shock gas and is mostly a function of electron density 
and different cooling functions (see in Sec.~4.1 of \citealt{drake09} for details). 
The oscillations were searched in CTTS (including TW~Hya) in the spectral range from X-rays 
to Johnson $U$-band but the results of \citet{drake09} and \citet{gunther10} were negative. 
The authors argued that the oscillations arising simultaneously in a few hot-spots can dampen each other 
and lead to negative detections.
The authors of the second paper were the first who searched for these oscillations in ultraviolet 
and blue bands, which pushed us to carry out similar searches in our data.

The shortest, directly visible oscillations in our light curves, wavelet spectra and in some 
periodograms are 11-20~min. 
This is much longer than typical periods 0.02-0.2~s predicted for CTTSs by \cite{koldoba08}. 
However, our values are of the same order as the period of $\sim400$~s predicted by \citet{sacco08}, 
$\sim500$~s by \citet{orlando13} and 80-610~s by \citet{costa17}.
Although \citet{matsakos13} presented a range of models, among them one with 
an oblique impacting surface leading to QPOs of 15~min -- very similar 
to those observed in TW~Hya -- the authors further argued that several perturbation types, 
such as clumps in the accretion stream or chromospheric fluctuations, may disrupt 
or even suppress expected oscillations.

We conclude that it is not clear whether the shortest light variations observed in TW~Hya can be assigned 
to post-shock plasma oscillations. 
Apparently new observational tests should be devised to check the advantage of this mechanism 
also over hypothetical, well-organized inhomogeneous accretion of small clumps. 
We speculate that for some reason the clumps could hit the star at quite regular intervals, 
which would lead in turn to light variations mimicking oscillatory variations.

%----------------------- Fig - CI ---------------------
\begin{figure}
\includegraphics[width=90mm]{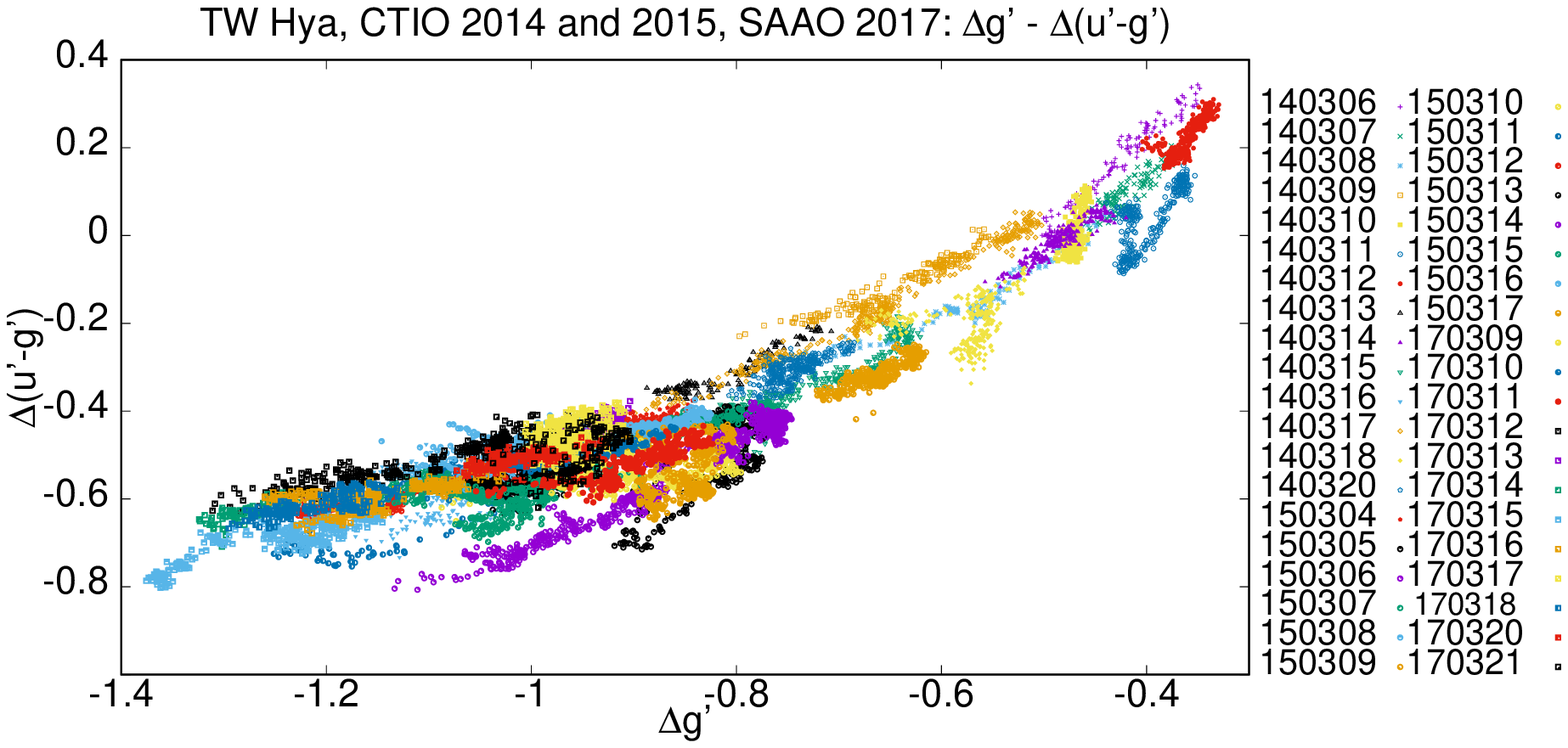}
\includegraphics[width=90mm]{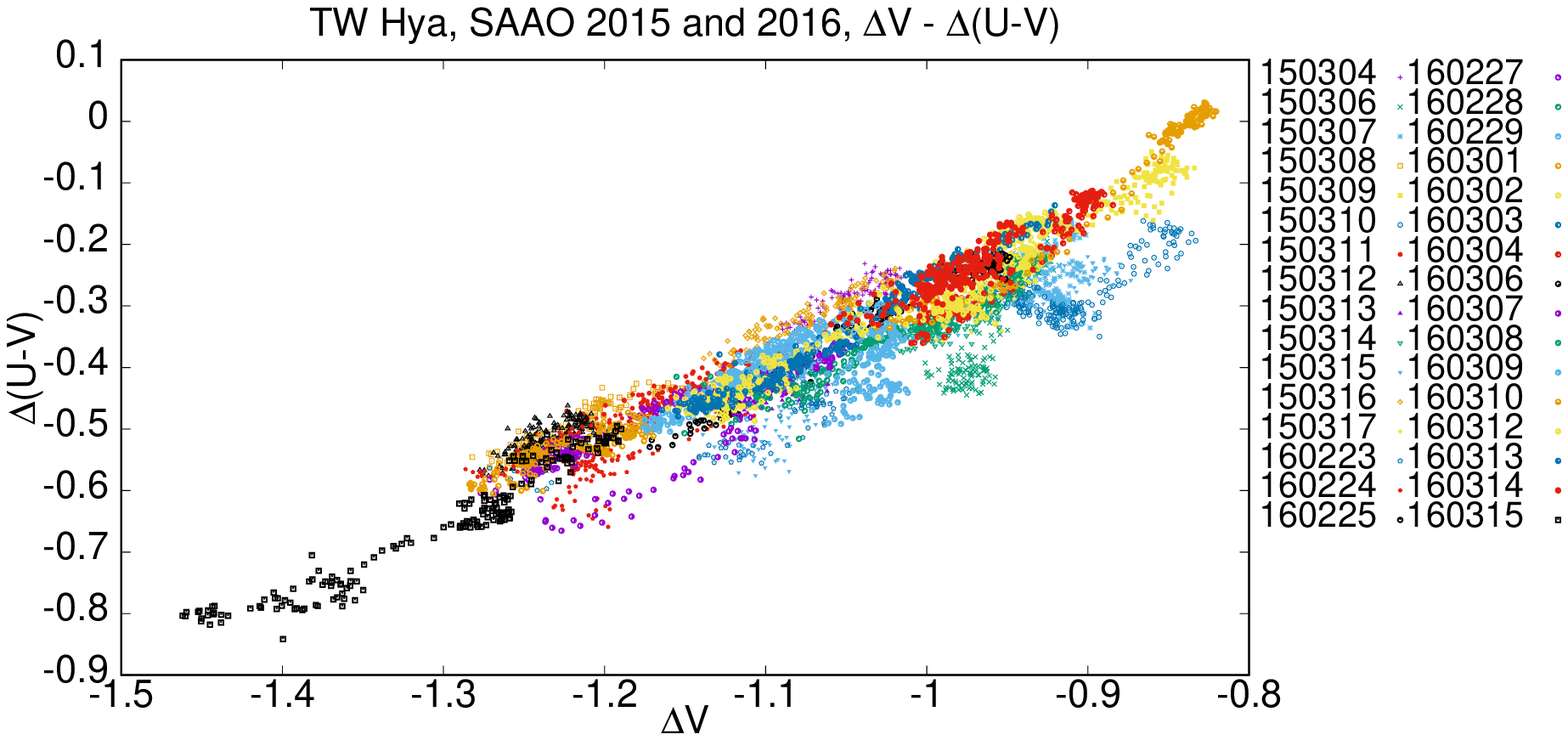}
\includegraphics[width=90mm]{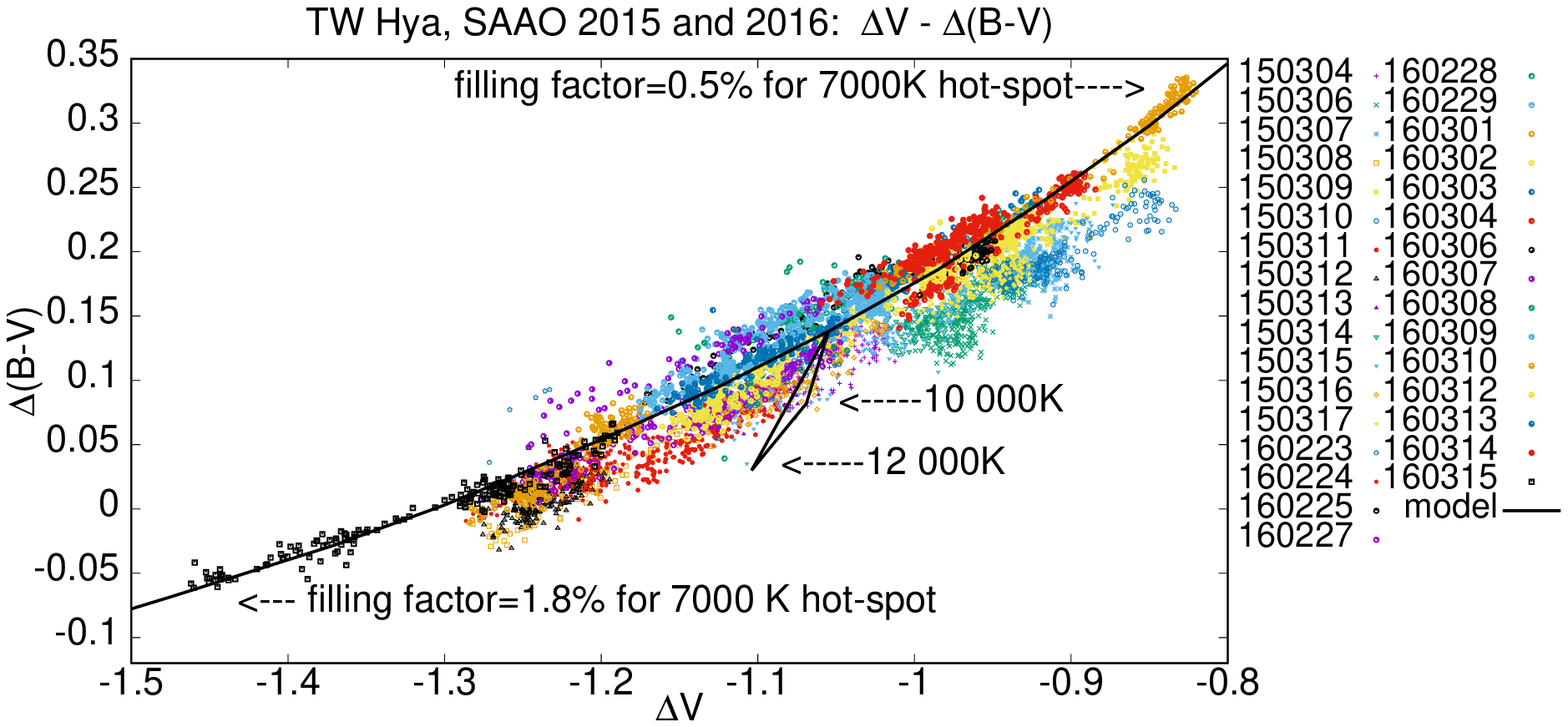}
\caption{Colour-magnitude diagrams in Sloan and Johnson filters, composed of observations obtained 
during nights indicated in legends ({\it yymmdd}). The differential magnitudes 
are computed with respect to the mean comparison star, computed from the first and third stars listed 
in Table~\ref{Tab.comp}. The theoretical diagram (namely 'model' in the legend) was manually 
matched only with the last diagram.} 
\label{Fig.colours-all}
\end{figure}
%----------------------------------------------------------------------

%----------------------- Fig - CI ---------------------
\begin{figure*}
\includegraphics[width=58mm]{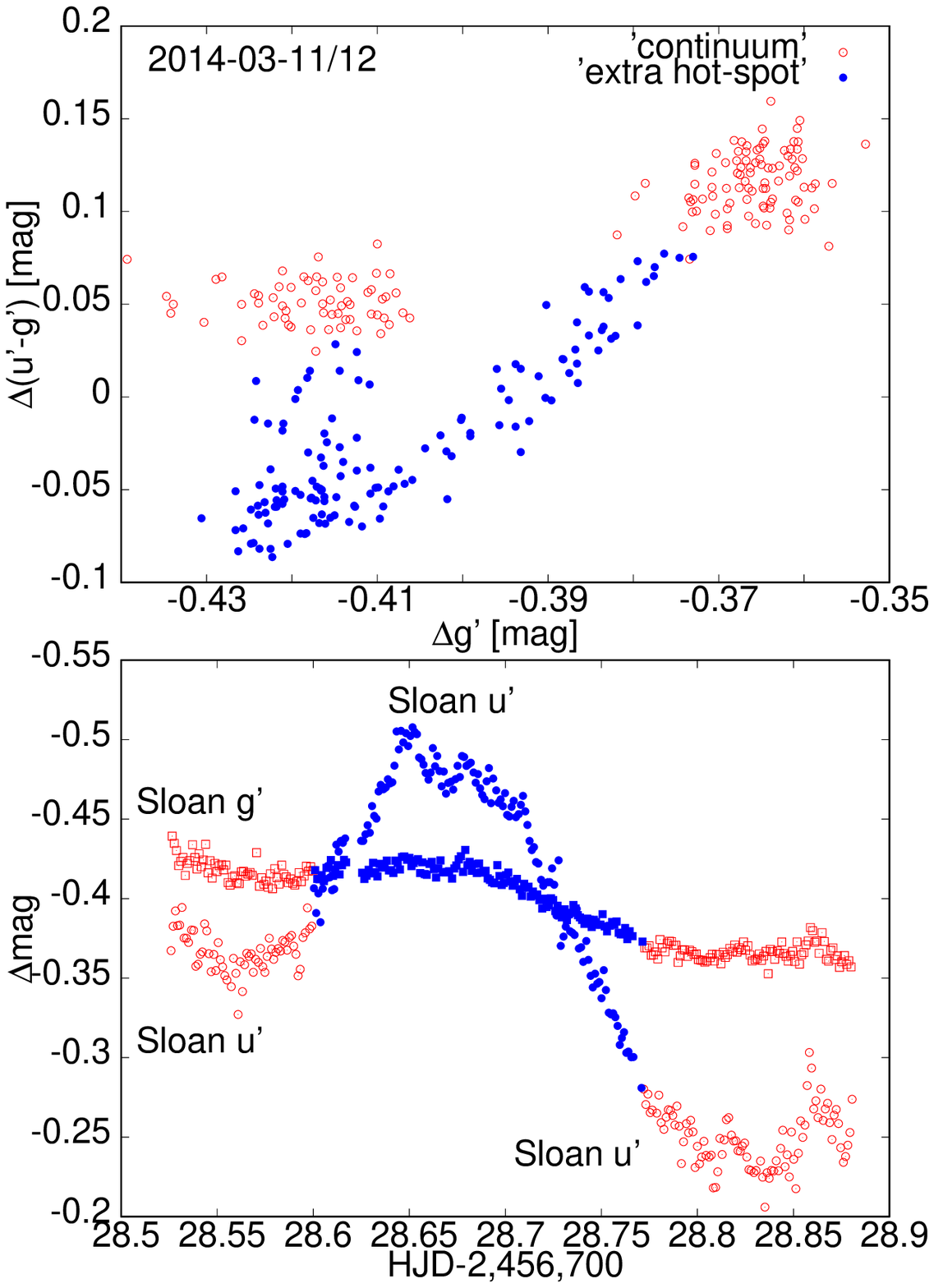}
\includegraphics[width=58mm]{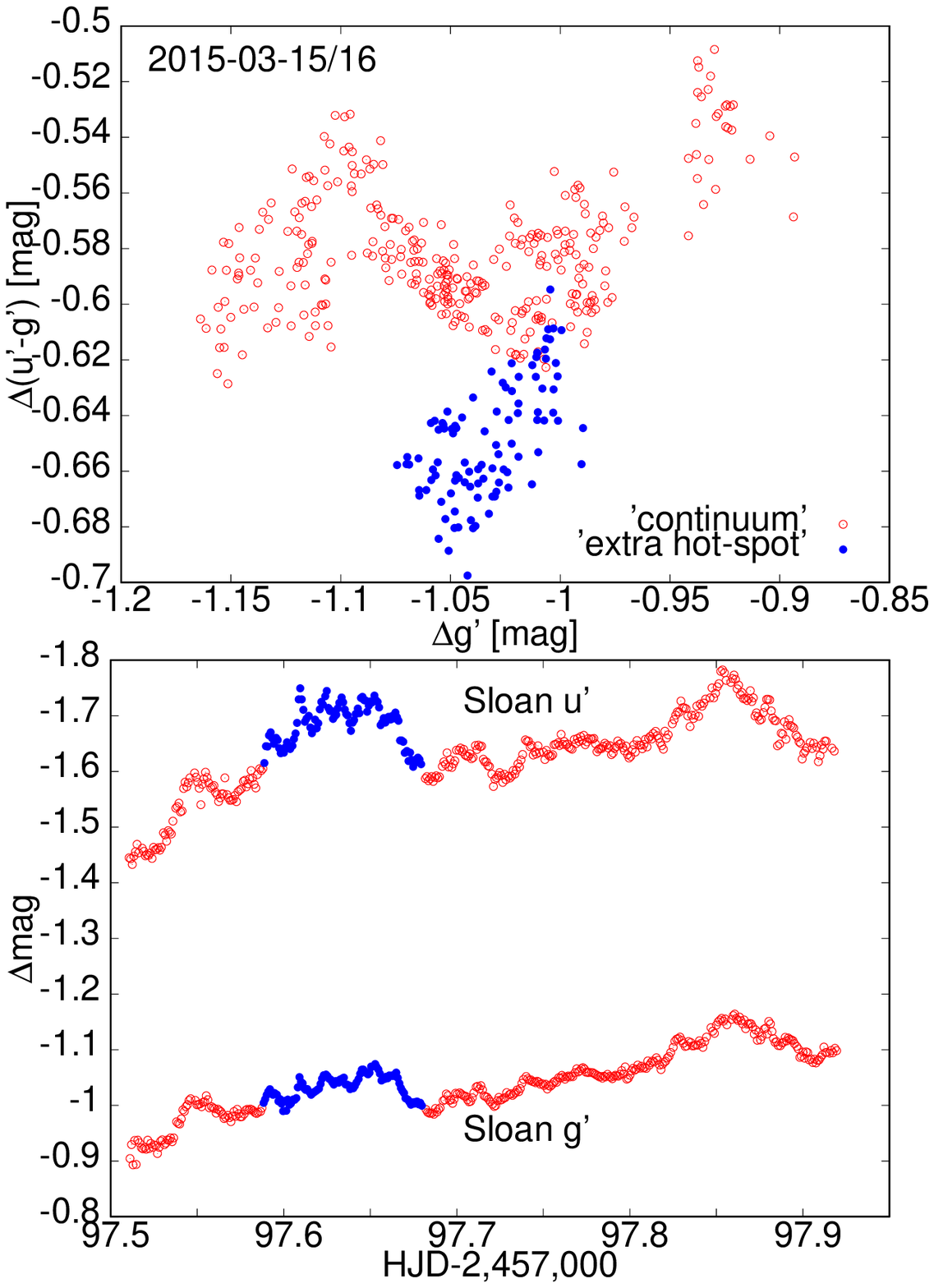}
\includegraphics[width=58mm]{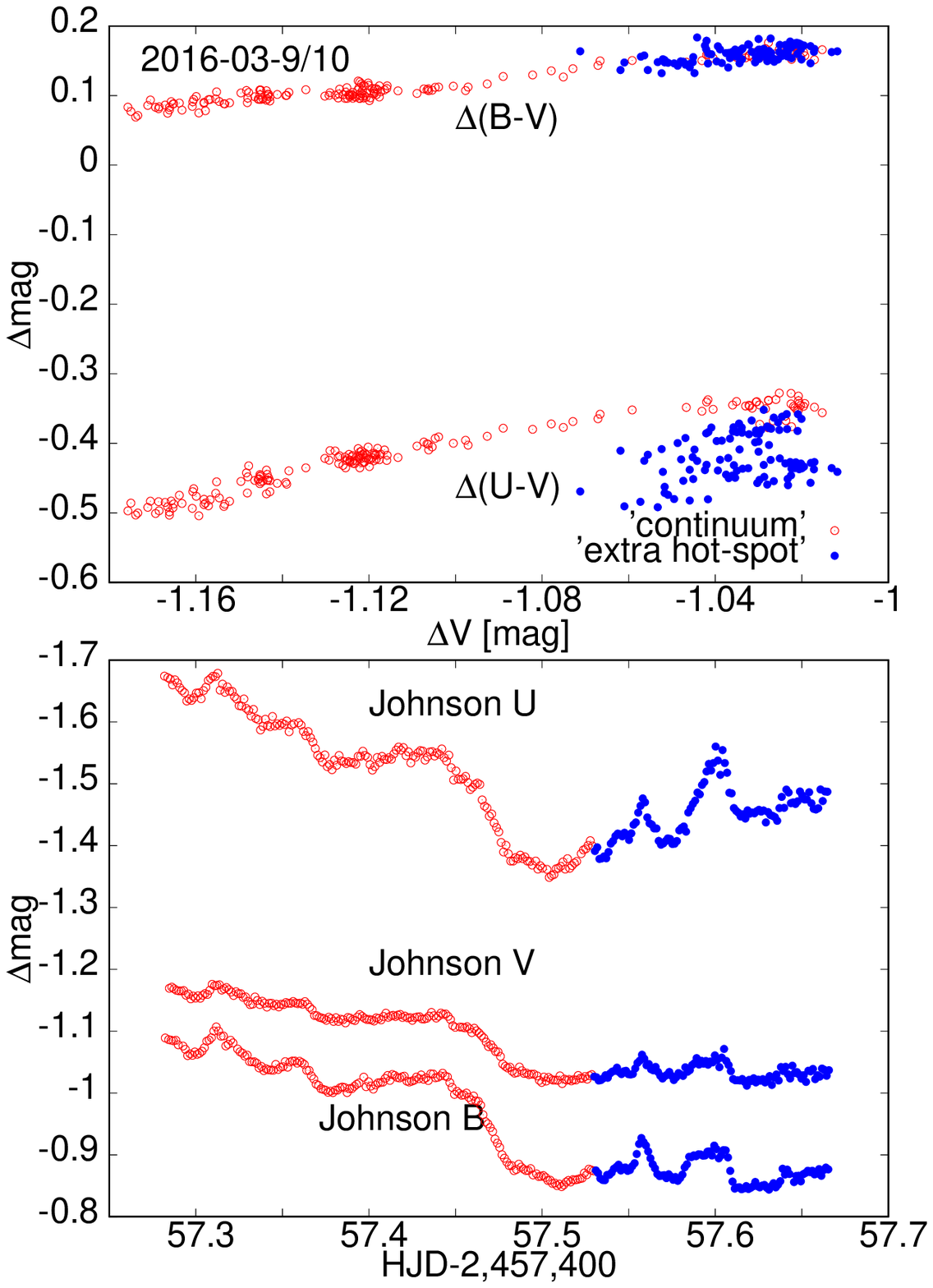}
\caption{Colour-magnitude diagrams for three particular nights (upper panels) and respective 
fragments of light curves (bottom panels). 
Filled symbols represent data and corresponding colour indicies obtained during additional hot-spot appearances.} 
\label{Fig.colours-singleN}
\end{figure*}
%---------------------------------------------------------------------

\subsection{Colour indexes vs. brightness variations}
\label{ci-diagrams}

Although our data suffer from a lack of photometric calibrations to the standard 
Johnson and Sloan systems, 
they contain rich information about colour index vs. brightness evolution of TW~Hya 
both during whole runs and during single nights. 
We prepared three colour-magnitude diagrams utilizing most of the 2014, 2015, 2016 and 2017 
multi-colour observations, corrected for atmospheric extinction effects (Figure~\ref{Fig.colours-all}). 
Relations for individual nights are not important on these plots, 
but the generall shape all these nights take when combined together. 

In general TW~Hya becomes bluer when brighter, as is typical for CTTSs where accretion 
is the rule. 
One can note considerable spread of colour indices for a given magnitude on each diagram. 
This amounts to 0.3~mag in the $\Delta g'-\Delta (u'-g')$ relation, gets slightly smaller (0.2~mag) 
in the $\Delta V-\Delta (U-V)$ diagram, and becomes a factor of two smaller (0.1~mag) 
in the last $\Delta V-\Delta (B-V)$ relation. 
These spreads are by far too large to be explained by any instrumental errors or imperfections 
in the atmospheric colour-extinction removal. 
Instead, they must be associated with the star -- or more specifically -- with the variety 
of hot-spot parameters.

The effects discussed above were observed in TW~Hya for the first time by \citet{ruc83}. 
Many other {\it Type~II} T~Tauri stars showed scatter in their colour indices based on 
$U$- and $B$-filter data by \citet{herbst94}.
\citet{fernandez96} explained this by the appearance of hot-spots of different temperature. 
These authors found more concise colour-magnitude relations in diagrams based on $VR$-filters, as they are less sensitive to variations in hot-spot temperatures.

In order to illustrate this effect in TW~Hya we present a theoretical colour-magnitude diagram. 
We utilized the results of \citet{calvet98}, who calculated the structure of the accretion 
column in CTTSs and obtained surface filling factors by all shocks to be smaller 
than 10, typically 0.1-1 per cent. 
%[<-- I don't understand. The filling factor, f, must normally satisfy 0 < f < 1 and you have f < 10 here! Perhaps you 
%mean 1/f?  TM]
As the authors obtained typical hot-spot temperatures of 6000-8000~K, we chose 
a mid-value of 7000~K during our computations. 
For comparison we also performed a second set of calculations assuming 10\,000~K.
We calculated synthetic magnitudes in $UBV$ Johnson filters using the {\it PHOENIX} library of spectral 
intensities \citep{husser13} and filter transmission curves from \citet{bessel90}. 
The effective temperature of the photosphere was assumed to be 4000~K and the respective intensity 
calculated for $\log g=4.0$ was used during the stellar light integration. 
Appropriate linear limb-darkening coefficients in Johnson filters for a stellar photosphere 
and considered hot-spot temperatures were taken from \citet{diaz-cordoves95} and \citet{claret95}.\newline 
We started to compute a grid of synthetic magnitudes from an unspotted star. 
In the next step we added a small $1\times 1$~deg square hot-spot on the observer's side of the star. 
Its size was increased by 1~deg in latitude and longitude in each consecutive step. 
As a result, we obtained synthetic $UBV$ magnitudes necessary for producing theoretical 
colour-magnitude diagrams.
As our {\it SAAO} observations lack absolute calibrations, the theoretical diagram was manually 
adjusted to match closely the observed colour-magnitude relations. 
We found that the 'theoretical' hot-spot responsible for the theoretical diagram which best matches 
our observations evolved in size either from $10\times10$ to $19\times19$~deg$^2$ 
for the effective temperature of 7000~K (as shown in the last panel in Figure~\ref{Fig.colours-all}), 
or between $5\times5$ to $10\times10$~deg$^2$ assuming 10\,000~K. 
Corresponding hot-spot surface filling factors would then change from 0.5 to 1.8 or from 0.1 to 0.5 per cent, 
which is in accord with typical values of 0.1-1 per cent \citep{calvet98}.\newline
In the last step we simulated effect caused by appearance of an additional 
hot-spot of temperature higher than a typical value of 7000~K. 
We considered single $2.5\times2.5$~deg$^2$ hot-spots of two temperatures: first of 10\,000~K 
and second of 12\,000~K. 
The effects triggered by both hot-spots are indicated on the last diagram by arrows. 
Once the additional hot-spot light was 'switched off', the colour-magnitude relation went to 
the original banana-shaped relation. 
This result is in accord with \citet{fernandez96} and may qualitatively explain the 'spread' 
observed in our diagrams.

The effect discussed above can also be well illustrated on diagrams specifically constructed for three particular 
nights, as shown in Figure~\ref{Fig.colours-singleN}. 
They show effects triggered by the appearance of hot-spots of apparently larger effective temperatures 
than that of an underlying hot continuum, leading to atypical behaviour in the colour-magnitude diagrams. 
To indicate the regions in the diagrams being disturbed by such hot-spots, we decided to split 
the data obtained before, during,  and after the appearance of the hot-spots and mark them with different colours and symbols:
\begin{itemize}
\item The first two left-hand panels in Fig.~\ref{Fig.colours-singleN} shows data obtained at {\it CTIO} during March 11, 2014. 
The night can be also quickly localised in the first panel of Figure~\ref{Fig.colours-all} as the star 
was then in a low state of its brightness. 
Yet, in the middle of the night its $u'$-filter brightness rose by 0.15~mag, while only 
0.02-0.03~mag rise was noted in the $g'$-filter at the same time. 
\item The second example (the two middle panels) is from March 15, 2015, when the short (possibly) QPO considered 
as presumably due to post-shock plasma oscillations (Sec.~\ref{accrburst}) were noticed by us so clearly for the first time. 
Apparently, the hot-spot associated with this QPO was of higher temperature than the mean temperature 
of the underlying hot continuum, since it added an extra 0.08~mag of ultraviolet light into 
the underlying relation.
\item The third example (the two right-hand panels) shows effects caused by two temporal hot-spots ('accretion bursts') 
of 0.5-0.8~h durations observed during March 9, 2016. 
Interestingly the $\Delta V-\Delta (B-V)$ relation remained stable, but the next one ($\Delta V-\Delta (U-V)$) 
which was more sensitive to the manifestation of additional hot-spots with higher effective temperatures, showed 
two separate relationships. 
\end{itemize}
We conclude that inhomogeneous accretion does not necessarily create hot-spots with equal temperatures 
as concluded by \citet{gullbring96}. 
Apparently inhomogeneous accretion may also create hot-spots by 1000-5000~K hotter from the assumed 'base level'. 
Our result is in qualitative accordance with \citet{ingleby13}, who analysed ultraviolet and optical spectra 
of low-mass T~Tauri stars assuming multiple accretion components.
Only columns carrying 'high-energy fluxes' ($3\times10^{11}$ and $10^{12}$~erg~s$^{-1}$~cm$^{-3}$) with equal 
filling factors of 0.0013 (0.13~per cent) were found for TW~Hya, as shown in Table~5 of their paper. 
According to the authors these 'high-energy flux' columns may produce hot-spots of effective temperatures 
up to 9000~K.

%The findings drawn in this and in the previous section clearly indicate that accretion 
%streams in TW~Hya are not homogeneous but clumpy. 
%Clumps approaching toward the star and channeled in a narrow tongue or a stable funnel flow 
%may potentially work as occulting screens of hotspots under favourable 
%geometry of visibility. 
%This will be considered below.

%----------------------- Fig - CI ---------------------
\begin{figure*}
\includegraphics[width=80mm]{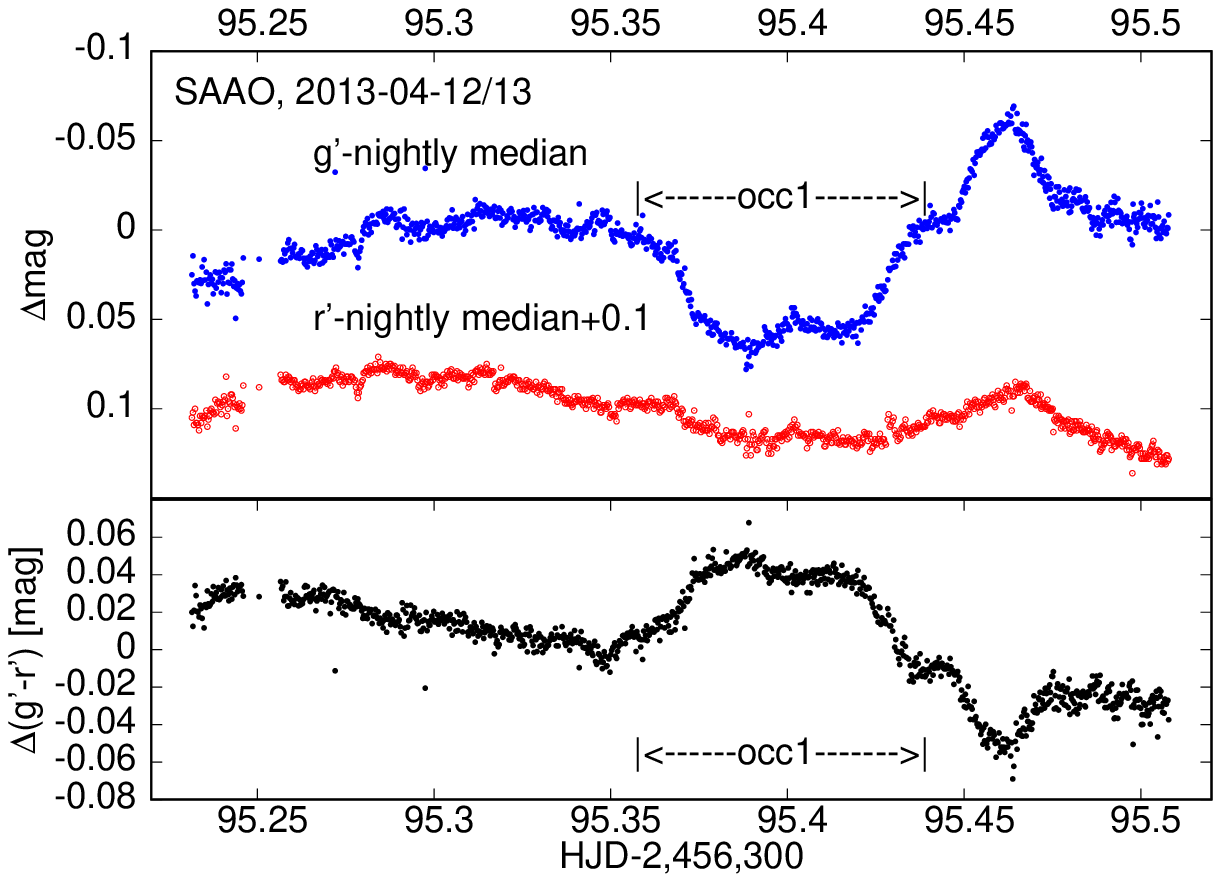}
\includegraphics[width=80mm]{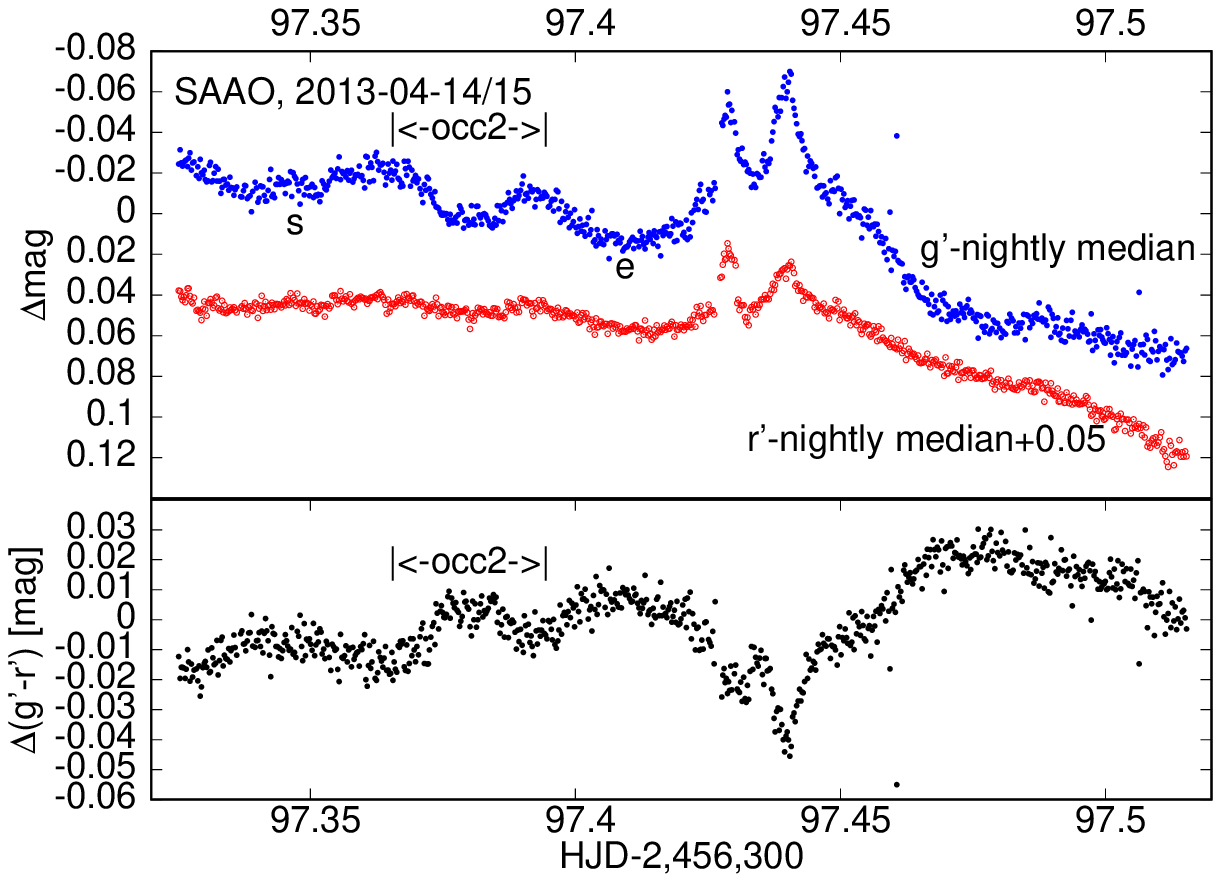}\\
\includegraphics[width=80mm]{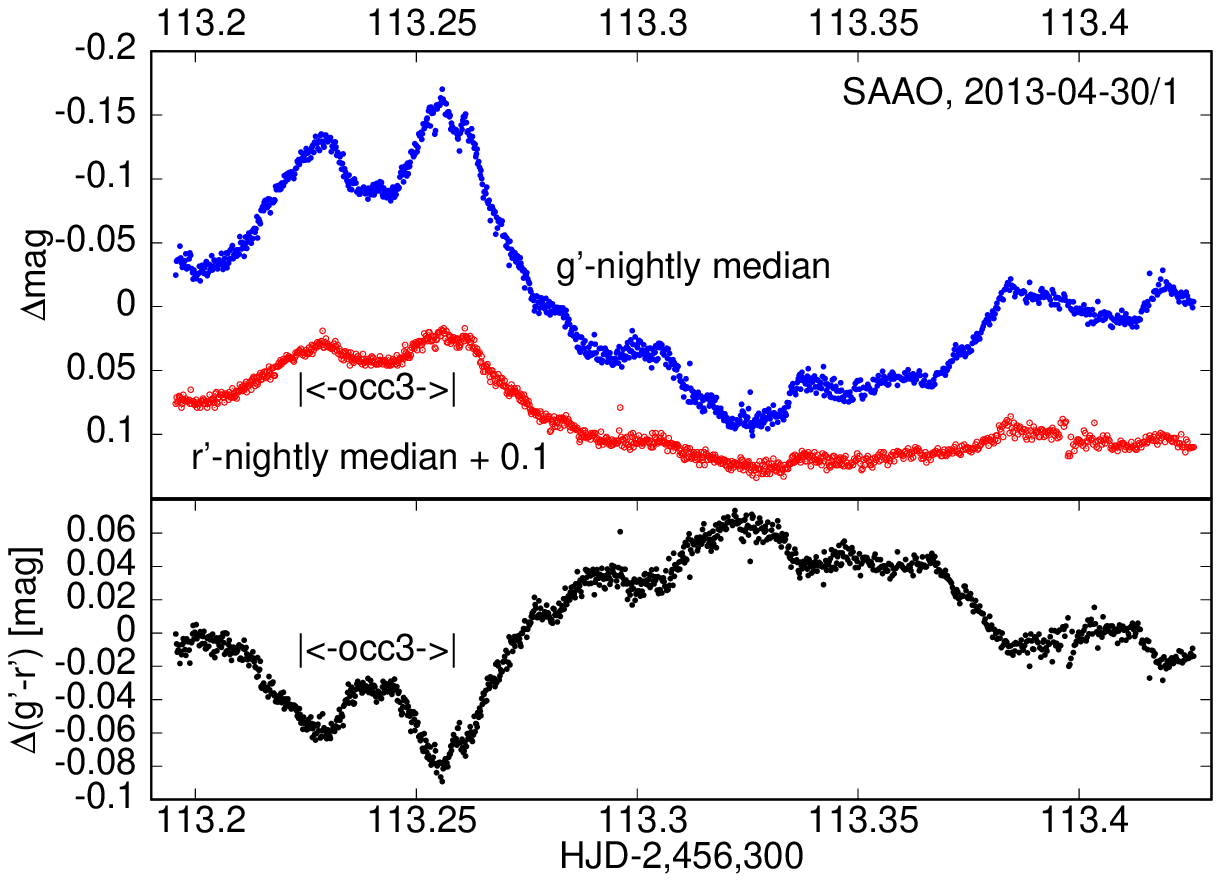}
\includegraphics[width=80mm]{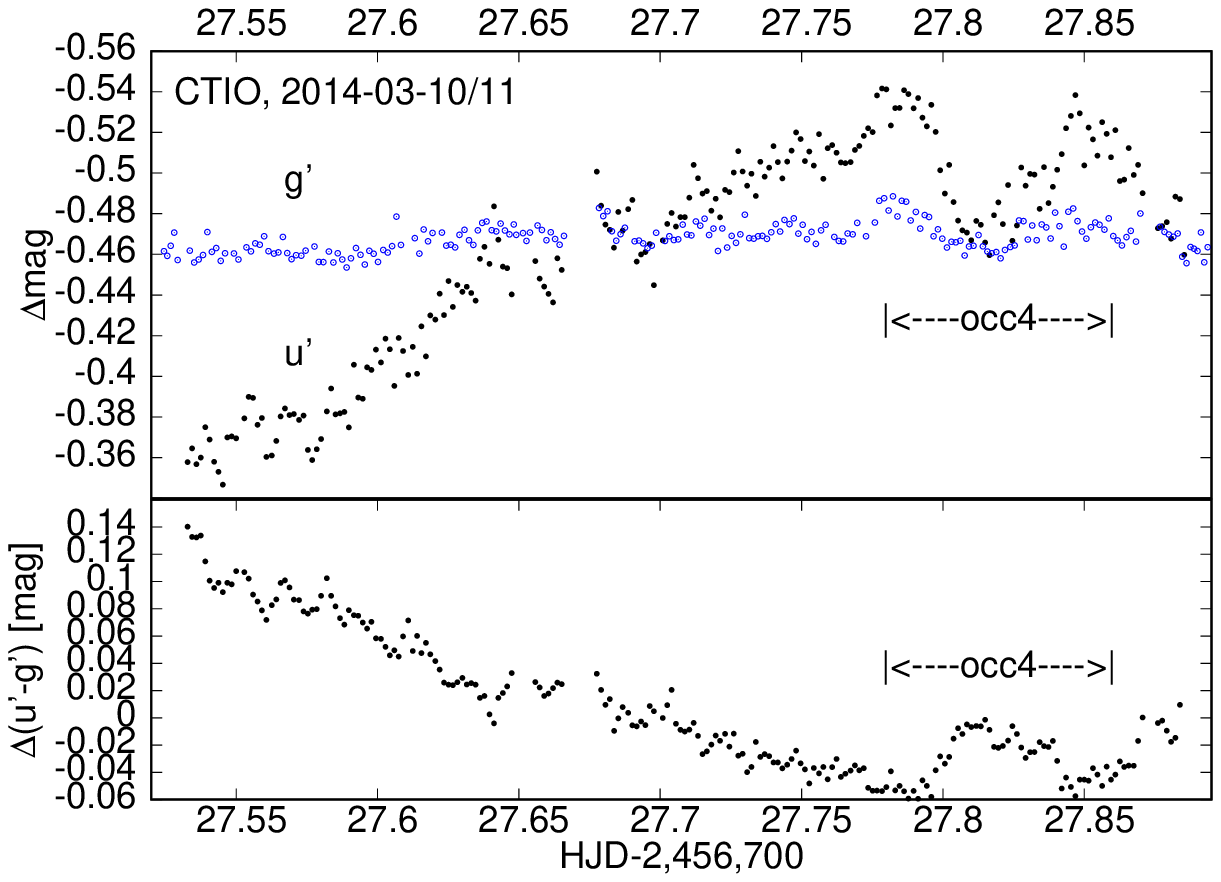}\\
\includegraphics[width=80mm]{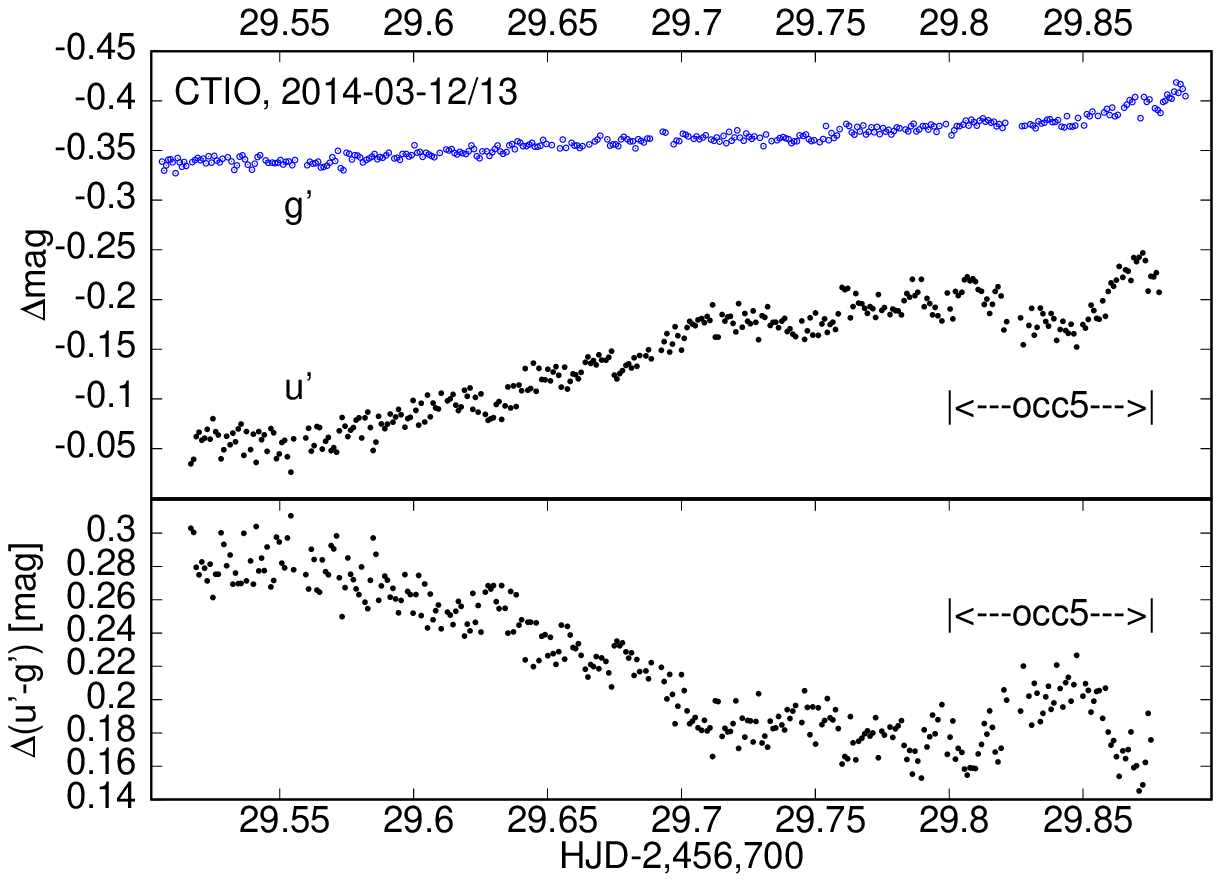}
\includegraphics[width=80mm]{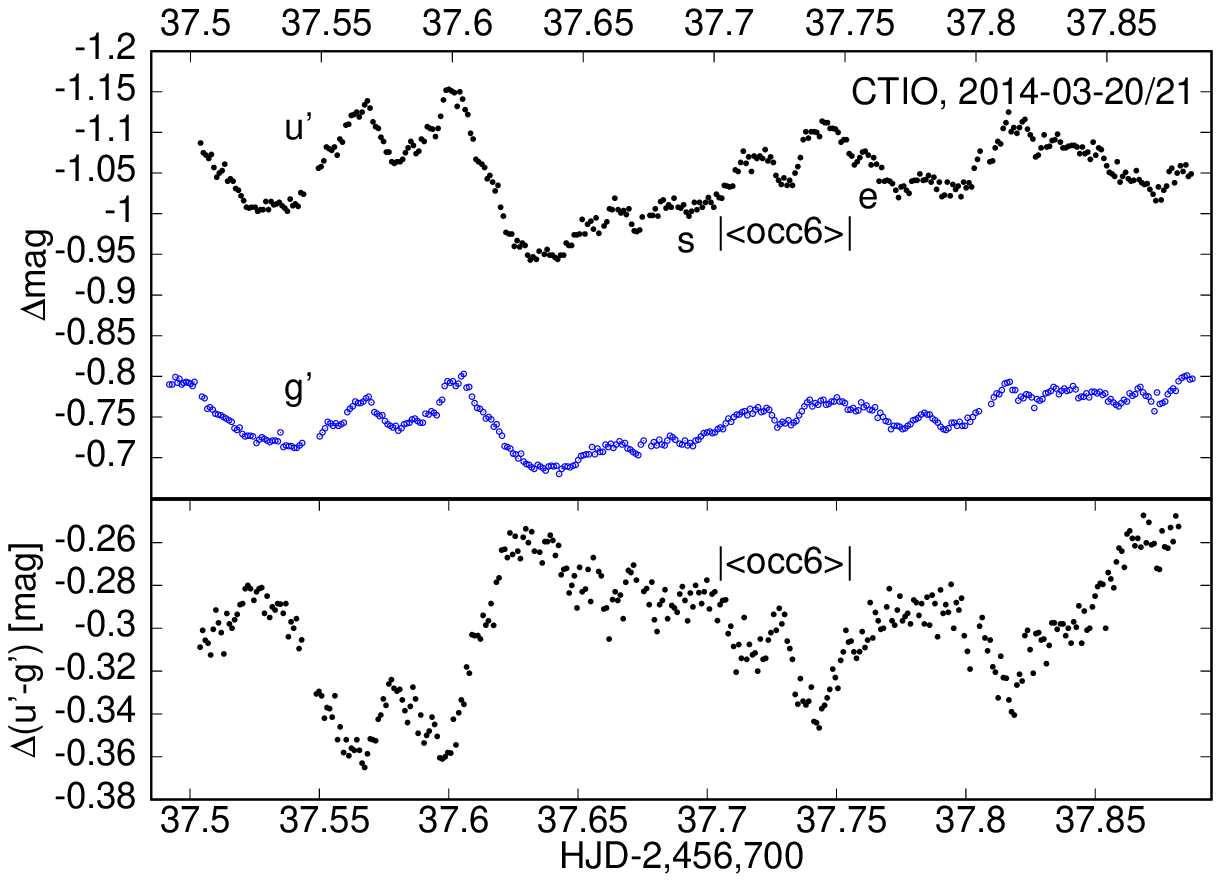}\\
\includegraphics[width=80mm]{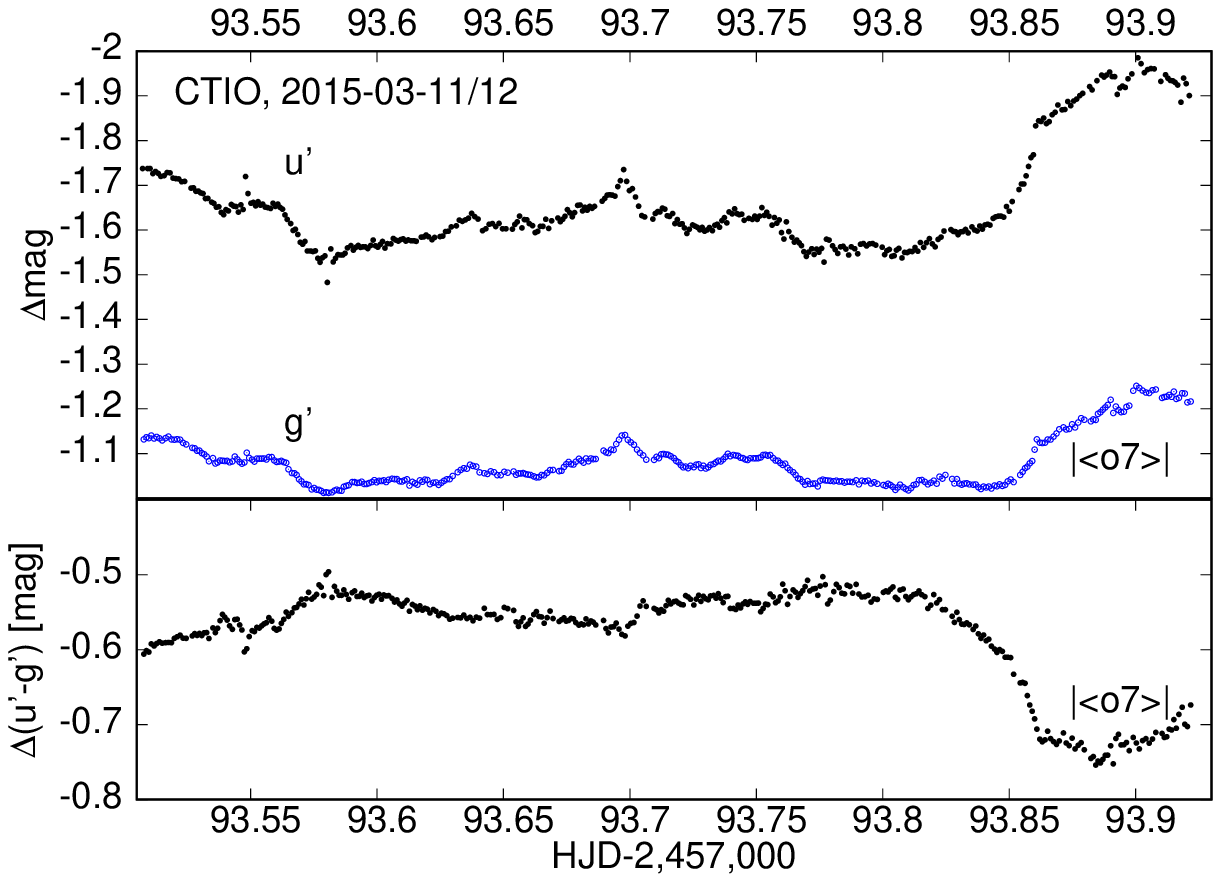}
\includegraphics[width=80mm]{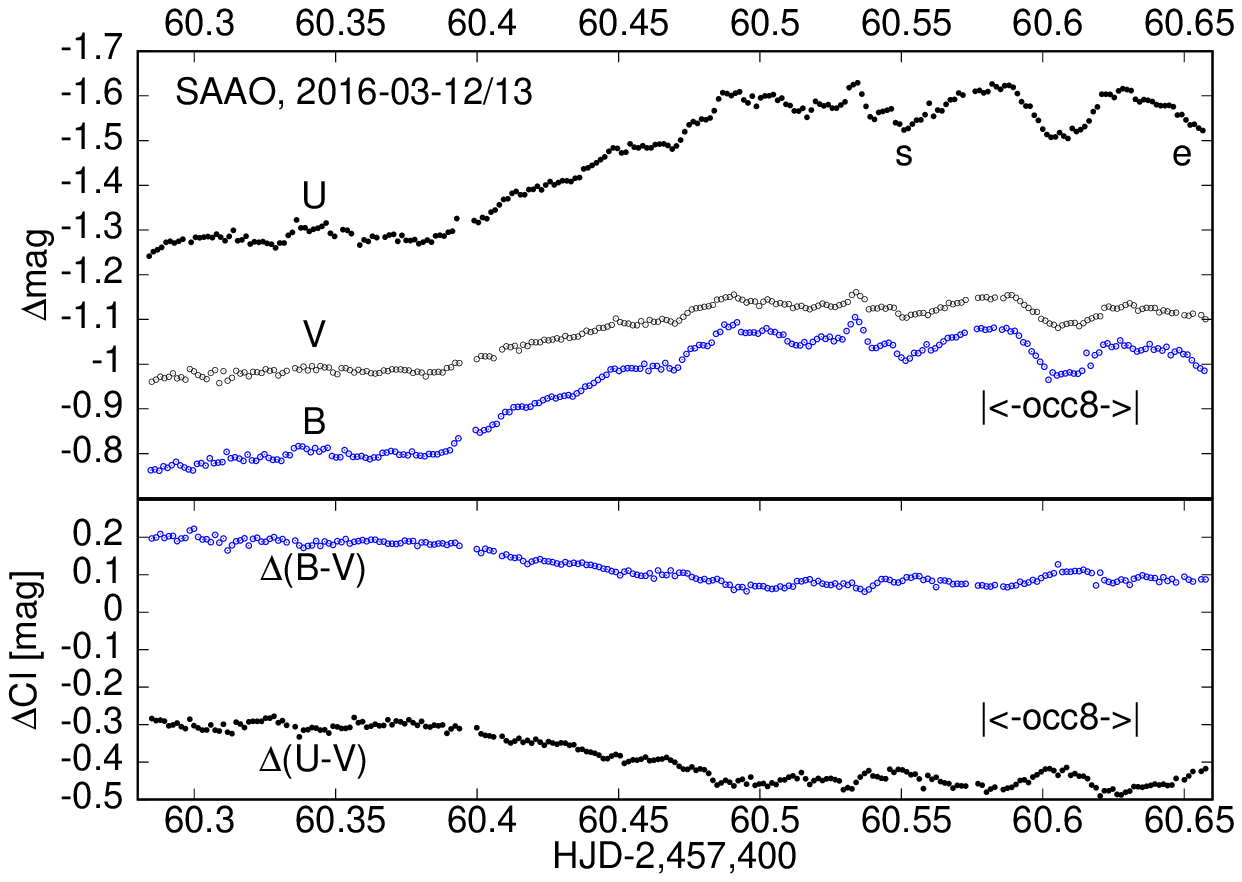}
\caption{Possible occultations of hot-spots in TW~Hya, marked by ranges and numbers.  
The upper panels show light curves corrected for colour extinction, while the bottom panel shows
colour index variations in time. Whenever possible, approximate moments when hot-spots associated 
with the occultations appeared and disappeared in the light curves are marked by 's' and 'e', respectively.} 
\label{Fig.occult}
\end{figure*}
%---------------------------------------------------------------------

%*************************************************TM

\subsection{The search for occultations}

The major motivation to conduct such an enormous observational effort from the ground was to investigate 
 the mysterious occultations discovered in the 2011 {\it MOST} data of TW~Hya. 
We were aware that this may be a risky venture, as previous space observations 
did not reveal any obvious similar events. 
Our first and minimal goal was to encounter occultations and check their spectral properties in at least 
two bands. 
In the best case, we expected to be able to determine their periodicity if they were sufficiently frequent 
in a particular season.

In \citet{siwak14} we temporarily concluded that these events could be due to hot-spot occultations 
by hypothetical 'dusty clumps'. 
Assuming this scenario to be true, it would be advisable to estimate the physical conditions 
which would lead to the 2-3 per cent brightness drops observed by {\it MOST} in 2011 and calculate 
their depths in consecutive bands in e.g. the Johnson system.
According to \citet{kulkarni08}, the total number of unstable tongues at any given time is of the order of a few. 
Let us assume that the tongues created five hot-spots with surface filling factors of about 0.7 or 0.1 per cent 
for the two already considered hot-spot temperatures of 7000~K and 10\,000~K, respectively. 
Assuming that one of the spots is completely obscured by a hypothetical optically-thick dusty clump, 
the light synthesis model (same as in Sec.~\ref{ci-diagrams}) predicts a brightness decrease 
by 0.03-0.04~mag in the wide-band {\it MOST}- and Johnson $R_c$-filters, 0.07-0.09~mag in the $V$-filter, 
0.14-0.20~mag in the $B$-filter, and 0.2-0.27~mag in the $U$-filter for both temperatures. 
Note that in all bands but $R_c$, the expected depths are significantly higher than in the {\it MOST} band.
Due to the fact that all the hypothetical hot-spots were placed centrally on the observer's side of the star during 
the computations, slightly smaller light drops should be observed in practice, closer to 3 per cent 
for the {\it MOST} filter and in accord with the 2011 data. 
We note that results of these calculations are sensitive functions of assumed hot-spot temperatures, filling-factors 
and occulted fraction, and are slightly affected by the lack of emission lines \citep{dodin18} 
in the {\it PHOENIX} library of intensities, used in our simple model.\newline 
The best way to verify the above predictions would be to catch at least a single occultation 
both by {\it MOST} and by the ground-based telescope. 
In spite of two simultaneous runs (2014, 2017) we were unfortunate that
not even a single occultation was seen in the {\it MOST} data themselves. 

%--------- Table 3 - Occultations and features ------------
\begin{table}
\caption{Basic properties of possible occultations: the central dip
mid-times $hjd_{min}=HJD-2\,456\,300$ are estimated from the inner contacts with an uncertainty of 0.001~d.
The depths are estimated in magnitudes with an uncertainty of 0.004~mag 
for appropriate filters. 
The outer $D$ and inner $d$ contact durations are in days; their typical 
uncertainty is 0.001~d (1.4~min).}
\begin{tabular}{c c c c c} 
\hline
no. &  $hjd_{min}$ [d] & depth (band) [mag]& $D$ [d] & $d$ [d] \\ \hline
 1  &  95.397          & $g'=0.056$ & 0.066   &  0.045  \\ 
    &                  & $r'=0.014$ &         &         \\  \hline
 2  &  97.380          & $g'=0.024$ & 0.018   &  0.010  \\
    &                  & $r'$-undef &         &         \\   \hline
 3  & 113.245          & $g'=0.062$ & 0.023   &  0.009  \\
    &                  & $r'=0.022$ &         &         \\  \hline
 4  & 427.822          & $u'=0.074$ & 0.051   &  0.033  \\
    &                  & $g'=0.014$ &         &         \\ \hline
 5  & 429.838          & $u'=0.059$ & 0.052   &  0.027  \\ 
    &                  & $g'$-undef &         &         \\   \hline
% 6  & 437.585          & $u'=0.073$ & 0.031   &  0.015  \\ 
%    &                  & $g'=0.045$ &         &         \\ \hline
 6  & 437.727          & $u'=0.059$ & 0.014   &  0.006  \\
    &                  & $g'=0.024$ &         &         \\   \hline
% 8  & 437.783          & $u'=0.090$ & 0.069   &  0.030  \\
%    &                  & $g'=0.040$ &         &         \\   \hline
 7  & 793.895          & $u'=0.044$ & 0.009   &  0.005  \\
    &                  & $g'=0.044$ &         &         \\   \hline 
 8  & 1160.609         & $U=0.106$  & 0.031   &  0.014  \\ 
    &                  & $B=0.076$  &         &         \\  
    &                  & $V=0.067$  &         &         \\   \hline \hline
\end{tabular}
\label{Tab.occult}
\end{table}
%---------------------------------------------- 

%----------------------- Fig 12 - branch-duration ---------------------
\begin{figure}
\includegraphics[width=82mm]{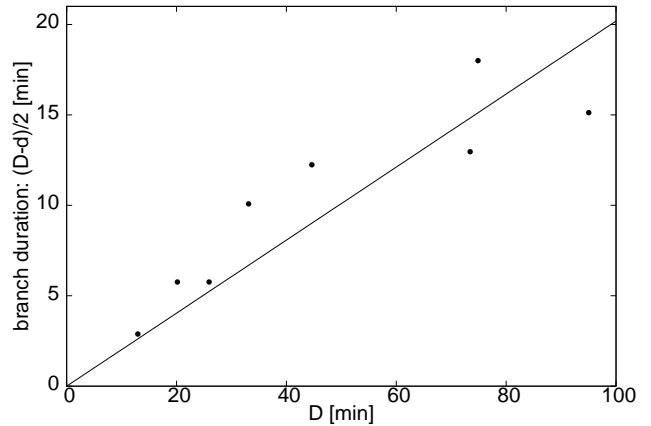}
\caption{The relation (a straight-line fit) between the overall durations $D$ and the branch durations $(D-d)/2$: 
$(D-d)/2=0.202(18)D$. The slope is very similar to that in 2011: 0.199(14).} 
\label{Fig.branch-dur}
\end{figure}
%---------------------------------------------------------------------

As described in Section~\ref{accrburst}, the ground-based data are dominated by 'accretion bursts' 
of various shape and by possible QPOs occurring on time-scales from 11 minutes to several hours. 
They are superimposed on each other and resemble the artificial light curves simulated by \cite{gullbring96}.
Therefore one would come to the conclusion that after sufficiently long monitoring it would be possible 
to identify a fake occultation-like event due to the superposition of a few accretion-induced light effects.\newline 
Prejudiced about this possibility we carefully studied all ground-based data. 
We checked light curves obtained differentially with respect to one or more comparison stars. 
We also separately examined the magnitudes of the variable and  comparison stars to make sure 
that weather conditions were stable and/or the telescope mirror was not obscured by the edge of the dome, 
which usually leads to creation of fake light changes. 
We identified eight events (at least), which -- according to the definition given in Sec.~4.1 
of \cite{siwak14} -- could be classified as 'occultations'. 
We mark them in the consecutive panels in Figure~\ref{Fig.occult}.

The new events are definitely not as securely defined among surrounding variations as in 2011.
The new events also have different characteristics from those seen in 2011: 
\begin{itemize}
\item Contrary to amazingly similar 
2-3 per cent depths observed in 2011 by {\it MOST}, the new events show quite different behaviour 
as illustrated in Table~\ref{Tab.occult}. 
They also exhibit much longer duration times between the outer ($D$) and inner ($d$) contacts. 
Nonetheless the relation {\it branch duration vs. total duration time} has amazingly similar slope ($0.202\pm0.018$) 
to that obtained from the 2011 {\it MOST} data ($0.199\pm0.014$), as shown 
in Figure~\ref{Fig.branch-dur}.\\
We note that only one event \#6 (March 20, 2014) was similar in duration, depth and shape 
to event \#3 observed by {\it MOST} in 2011.
\item Most of the 2011 dips were observed in random places -- only six events marked 
in the 2011 light curve as \#3, \#13,\#14,\#15 (possibly),\#17 and \#18 occurred at, 
or close to, the local brightness maxima.
In contrast, almost all new events appeared in the middle of the local brightness maxima. 
This is a very important finding, as it may indicate that they could be caused 
by total or partial hot-spot light dimming, rather than by chance.
\item Assuming the above scenario to apply, we can qualitatively estimate whether a given occultation 
was total or partial. 
Whenever possible, we identified and approximately marked the positions (by 's' and 'e') 
at which hot-spots associated with occultations appeared and disappeared among surrounding 
light variations. 
After that we drew a straight line between these positions. 
If the bottom of the occultation touched the line, we classified it as total. 
This is the case for events \#2, \#6, and \#8. 
Totality cannot be confirmed for events \#3, \#4, \#5, \#6, and \#7, 
as their bottoms fall much higher than base light levels from which associated hot-spots emerged. 
This indicates that the occulting bodies are not always fully optically-thick. 
\item Local brightness maxima observed in TW~Hya are similar to the 'square' and 'triangular' 
features found in artificial light curves by \citet{gullbring96}. 
Seven out of eight occultation-candidate events observed in TW~Hya during 2013-2016 
are localised in the middle of such light features. 
If these simulations included the possibility of hot-spot occultation by fully or partially 
optically thick clumps ordered in an accretion stream, a brightness drop similar to that observed 
by us would be produced under favourable geometry. 
Note that in 2017 no event was noticed.
\end{itemize}

The dips observed in TW~Hya are considerably shorter (15-90~min) and much shallower 
than those observed in AA~Tau, which last for about a day. 
They are due to a warp arising near the disc co-rotation radius \citep{bouvier03,bouvier07} in which 
optically-thick disc plasma is lifted high enough above the disc plane to cause semi-periodic 
occultations of the star. 
Most recently nine more similar semi-periodic dips were found in young stars in NGC~2264 by \citet{stauffer15}. 
At least three of the stars are locked with the stellar rotation, which suggests the same mechanism 
as for AA~Tau. 
Ten more similar stars were identified in {\it Kepler} spacecraft data for young members of the $\rho$ Oph 
and Upper Sco star-forming regions by \citet{ansdell16}.
The unified models of these mechanisms were presented by \citet{mcginnis15} and \citet{bodman17}. 
The first authors estimated that the maximum warp height amounts to about 20-30 per cent of the disc 
radius at which it originates and it may vary by 10-20 per cent on a time-scale of days, as inferred 
from analysis of consecutive occultations. 
The second authors added the possibility that the optical thickness of the warps decreases with their 
elevation over the disc mid-plane. 
These models do not foresee the possibility of occultations for CTTSs with low inclinations 
of visibility (as for TW~Hya), but we speculate that remnant dust can perhaps be lifted high 
above the star to act as an occulting screen for hot-spots revolving with an inner disc rotational 
frequency, as speculated in \citet{siwak14}.\newline
A similar hypothesis about toruses of gas or dust levitating in remnant funnel flows (or trapped 
by the stellar magnetosphere) over high stellar latitudes was considered for a dozen weak-lined 
T Tauri stars \citep{david17, stauffer17} and for so called 'scallop-shell' stars \citet{stauffer18}. 
Though these stars do not reveal signs of accretion in their light curves, they show 
semi-periodic persistent flux dips lasting for a few hours, or just smooth flux variations confined 
to certain phases only.
This hypothesis can also be considered for TW~Hya as absorption and scattering of 
u-band radiation by gas is much more efficient than on longer wavelengths, especially if emitted 
in the funnel stream direction. 
This scenario could especially well explain dips \#4 and \#5 well pronounced in u-band and separated 
by almost exactly 2 days, and only barely visible in $g'$-filter, as well as 
the full or partial light dimming of the hot-spots during remaining events. 

'Occulting screens' described above could also be carried in equatorial tongues instead of 
a stable funnel, what is especially important for TW Hya which does not necessarily always form a stable funnel flow. 
The plasma could first be transferred directly toward the star in a narrow tongue which 
is ultimately wound along the stellar magnetic field lines and forms miniature funnel flows. 
The funnel flows impact the star at moderate latitudes (35-65~deg) \citep{romanova08, kulkarni08, kulkarni09} 
or even lower at 10-30~deg, as strictly constrained in TW~Hya from X-ray observations 
by \citet{argiroffi17}. 
The scenario with gaseous 'blobs' carried in the tongues may be considered 
as most promising (dust inside a magnetosphere should quickly sublimate) given the results obtained for other young stars, and new indications 
for a clumpy accretion in TW~Hya, as discussed in Sections~\ref{accrburst} and \ref{ci-diagrams}. 
This scenario would also naturally explain the {\it occultation -- local light maximum} connection.

The last possibility is related to the fact that the 2011 events occurred during 
moderately stable accretion. 
According to \citet{cranmer08} inhomogeneous clumpy accretion may add energy to waves 
in the photosphere. 
The waves show enhanced dissipation in polar regions through open polar flux tubes which is sufficient 
to produce mass-loss rates of at least 0.01 times the accretion rate in the form of jet-like outflows. 
Whether or not the outflows could contain dusty clumps is also a matter of speculation. 
%Other possible scenario was proposed by \citet{scaringi16}, who investigated dips in EPIC~204278916. 
%They could be caused by transiting circumstellar clumps (the mass of  3.2 Halley Comets 
%for the whole clump was derived), likely on a highly eccentric orbit and considerably inclined 
%to the disc plane. 

\section{Summary}
\label{summary}

This is the last paper utilizing new {\it MOST} data for TW~Hya due to the constantly 
decreasing efficiency of the satellite solar panels. 
The data were analysed together with ground-based high-quality multi colour observations, 
which enabled studies of subtle light changes on time-scales ranging from 30 seconds to 5 years. 
We considered a few topics, each described and immediately discussed in dedicated subsections 
of Section~\ref{results}. 
The major results can be summarized as follows:
\begin{itemize}
\item We confirm primary findings from our last paper \citep{siwak14} that accretion in TW~Hya 
switches between an unstable and a moderately-stable state. 
In the moderately stable regime, quasi-periods caused by changing visibility of the major hot-spot are observed.
Although the new light curves were not as clearly dominated by equidistant peaks as in 2011, 
the values of 3.75 and 3.69~d observed in 2014 and 2017 are closer to the 3.57~d modulation, 
induced in spectral lines by a persistent high-latitude cold-spot \citep{donati11}.
%During all seasons, the most pronounced daily light variations in TW~Hya were seen in the form 
%of 'accretion burst' caused by inhomogeneous accretion. 
\item The new and important outcome from the ground-based observations is the characterisation of the 
'accretion burst' variability in TW~Hya occurring on time-scales unavailable for {\it MOST}. 
Usually smooth light variations are observed, but they are often disturbed by single, two or even 
a series of strong 'accretion bursts'. 
The oscillation spectrum of these bursts has a flicker-noise character and no periodicity stable 
over many nights is indicated.  
Some of the variations may be time coherent and quasi-periodic, as they appear at regular intervals 
not shorter than 11 minutes. 
The variations could be caused by inhomogeneous accretion, in which clumps ordered in the narrow accretion 
column hit the star in nearly regular intervals. 
Alternatively, the shortest 11-20~min light variations could be due to post-shock plasma oscillations. 
\item We also analysed colour-index versus stellar-magnitude variations occurring within 
single nights and during several years.
In particular, the comparison of colour-magnitude diagrams constructed from $UBV$ data 
with theoretical diagrams resulting from a simple star and a hot-spot light synthesis model indicates 
that these observations could be described by a hot-spot of an average temperature of about 7000~K 
and filling factor varying between 0.5-1.8 per cent during the 2015 and 2016 {\it SAAO} runs. 
A similar relation with assumed mean temperature of hot-spots at 10\,000~K results in a filling factor 
varying between 0.1-0.5 per cent. 
The complex colour-magnitude diagrams observed during a few particular nights can be assigned 
to short-term hot-spots of effective temperatures that are 1000-5000~K higher than that of the basic 
hot-spot continuum.
%Interestingly, many of the warmer hotspots are sporadic, sometimes they last for 0.5~h what could 
%be due to the small plasma clumps. 
\item We also found eleven flares -- a phenomenon common for many dwarf stars but only now observed 
in TW~Hya. 
They are all of different amplitudes and duration times. 
The occurrence rate estimated for 2013-2017 indicates the highest flaring activity 
at about one flare every two days during 2015-2016 but the significance of this result is low, 
as our monitoring lasted for barely 2-4 weeks each year. 
Yet, this may give a first clue for the existence of a magnetic cycle similar to that of the Sun. 
\item The ground-based observations obtained in 2013-2017 at {\it SAAO} and {\it CTIO} were 
obtained with the intention to catch and characterise short and shallow light dips discovered 
in the 2011 light curve of this star \citep{siwak14}. 
Although at least eight occultation-like events can be indicated in our new data, we stress that they 
are less securely defined among the surrounding light variations than in 2011, and may eventually 
turn out to be the result of an illusion caused by the superposition of 'accretion bursts'. 
However, new occultations are usually found in the local maxima of the stellar brightness. 
This finding prompted us to consider that these events may indeed be caused 
by occultations of hot-spots by gaseous and/or dusty clumps transferred toward the star 
in the associated magnetised accretion tongues. 
High-cadence spectroscopic observations may help to verify this picture.
\end{itemize}

\section*{Acknowledgments}
This study was based on data from the {\it MOST} satellite, a Canadian Space Agency 
mission jointly operated by Dynacon Inc., the University of Toronto Institute 
of Aerospace Studies, and the University of British Columbia, with the assistance 
of the University of Vienna.\newline 
This paper uses observations made at the South African Astronomical 
Observatory.
MS and WO acknowledge Dr. Hannah Worters and Dr. Francois van Wyk 
for introduction to the {\it SAAO} telescopes and the whole observatory 
staff for the very generous time allocations and their hospitality.\newline
This paper uses observations made at the Cerro Tololo Inter-American Observatory 
at the 0.9-m telescope operated by the {\it SMARTS} Consortium. 
MS is grateful to Dr. Jennifer G. Winters for help in efficient start 
of the {\it CTIO} run.\newline 
This paper made use of NASA's Astrophysics Data System (ADS) Bibliographic 
Services.\newline
This research has made use of the SIMBAD database,
operated at CDS, Strasbourg, France.\newline
MS and WO are grateful to the Polish National Science Centre for grant 2012/05/E/ST9/03915. 
The Natural Sciences and Engineering Research Council of Canada supports the research of DBG,
JMM, AFJM and SMR. 
Additional support for AFJM was provided by FQRNT (Qu{\'e}bec). 
CC was supported by the Canadian Space Agency. 
TK, RK and WWW are supported by the Austrian Science Funding Agency (P22691-N16).\newline
Special thanks are also due to an anonymous referee for very useful suggestions 
and comments on the previous version of the paper.

%\end{document}

\appendix 

\section{Extinction corrections of light curves} 
\label{app1}
Due to close proximity of TW~Hya and both comparison stars (4~arcmin), differential extinction effects 
are negligible and were unaccounted. 
However, in spite of very similar colour indexes of comparison stars and TW~Hya, the variations of airmass $X$ 
and colour indicies of TW~Hya itself do affect our nightly results 
($\Delta U_{obs}, \Delta B_{obs}, \Delta V_{obs}$, $\Delta u'_{obs}, \Delta g'_{obs}$) 
to an measurable extent -- correction on the colour extinction term can eliminate most 
of these harmful effects.

We used the mean colour extinction coefficients $\beta$ evaluated for {\it SAAO} and {\it CTIO} 
from \cite{fukugita96} and Mt.\ Palomar sites -- the values used in this work 
are as follows: 
-0.06, -0.052, -0.031, -0.02, -0.01, -0.008 for $U, u', B, g', V, r'$-filters, respectively. 
TW~Hya significantly changes its colour indicies within each night. 
To follow these changes we first interpolated $V$-filter differential magnitudes to get $\Delta V_{obs}^{int}$ 
at the moment of observations in $B$-filter (and vice versa). 
Then we obtained the corrected differential magnitudes using colour equation limited to the colour extinction 
term:\newline 
$\Delta B_{corr}= \Delta B_{obs}-\beta_B \times(\Delta B_{obs} - \Delta V_{obs}^{int})\times X_B$,\newline
$\Delta V_{corr}= \Delta V_{obs}-\beta_V \times(\Delta B_{obs}^{int} - \Delta V_{obs})\times X_V$.\newline 
Finally, we corrected observations in $U$-filter:\newline 
$\Delta U_{corr}= \Delta U_{obs}-\beta_U \times(\Delta U_{obs} - \Delta B_{corr}^{int})\times X_U$,\newline
where $\Delta B_{corr}^{int}$ are the corrected above $B$-filter magnitudes, interpolated 
for the moment of $U$-filter observations.
These equations automatically account for differencies between colour indicies of TW~Hya and comparison stars.
$X_U$, $X_B$, $X_V$ denote airmasses for indicated filters, calculated for the moments 
of mid-exposures.
Analogous procedure was applied to correct the data obtained in Sloan filters:\newline 
$\Delta u'_{corr}= \Delta u'_{obs}-\beta_{u'}\times(\Delta u'_{obs} - \Delta g_{obs}^{' int})\times X_{u'}$,\newline
$\Delta g'_{corr}= \Delta g'_{obs}-\beta_{g'}\times(\Delta u_{obs}^{' int} - \Delta g'_{obs})\times X_{g'}$.\newline
These results were used to construct reliable (though still left in the instrumental systems) 
colour-magnitude diagrams, i.e. $\Delta (B_{corr}-V_{corr})$, $\Delta (U_{corr}-V_{corr})$ 
and $\Delta (u'_{corr}-g'_{corr})$, as shown in Figures~\ref{Fig.colours-all} and \ref{Fig.colours-singleN}, and for investigation 
of long-term variability of TW~Hya in Sec~\ref{long-term} (Fig.~\ref{Fig.gb}~h,i).

\begin{figure*}
\includegraphics[width=80mm]{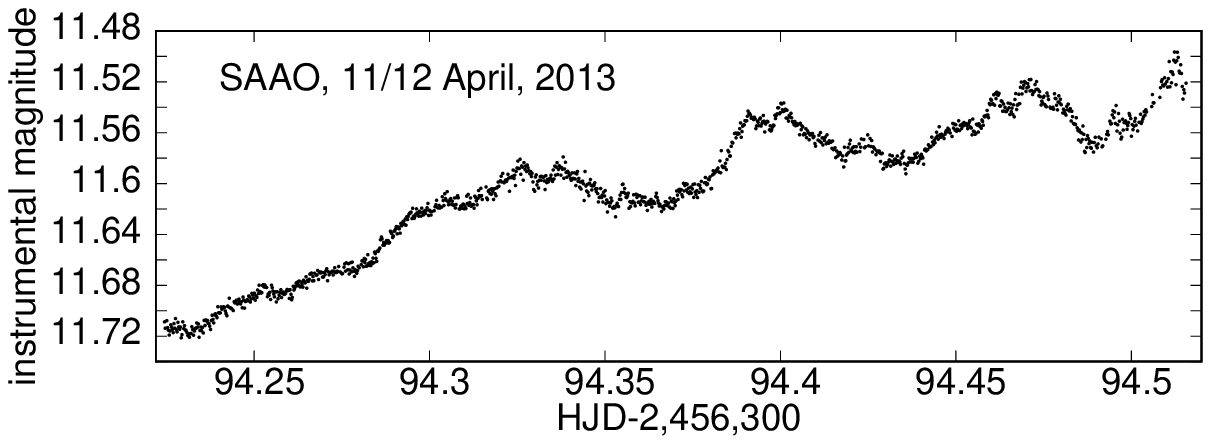}
\includegraphics[width=80mm]{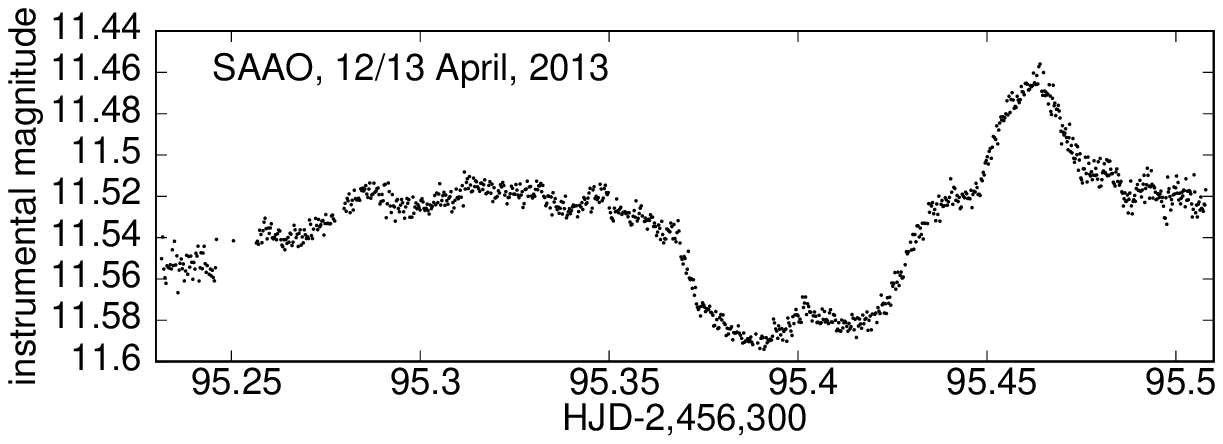}
\includegraphics[width=80mm]{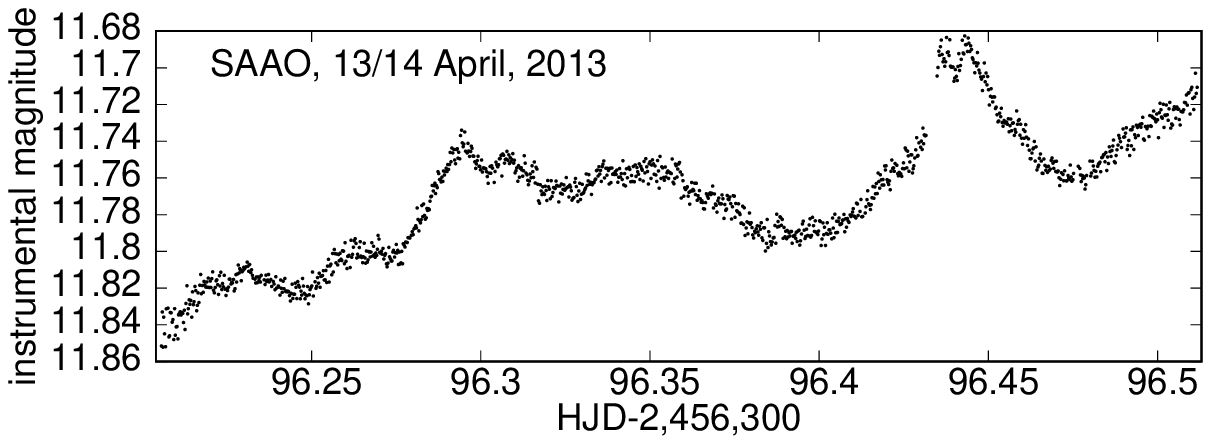}
\includegraphics[width=80mm]{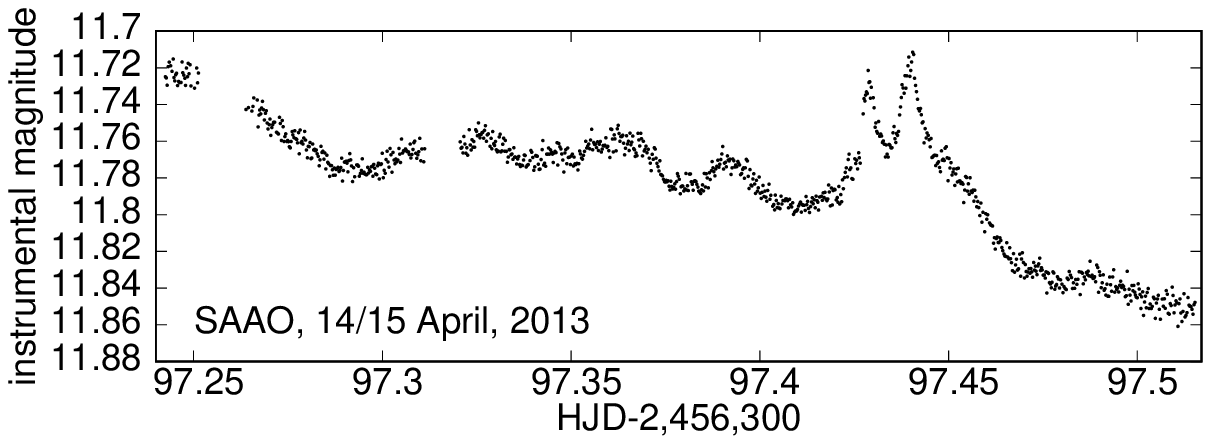}
\includegraphics[width=80mm]{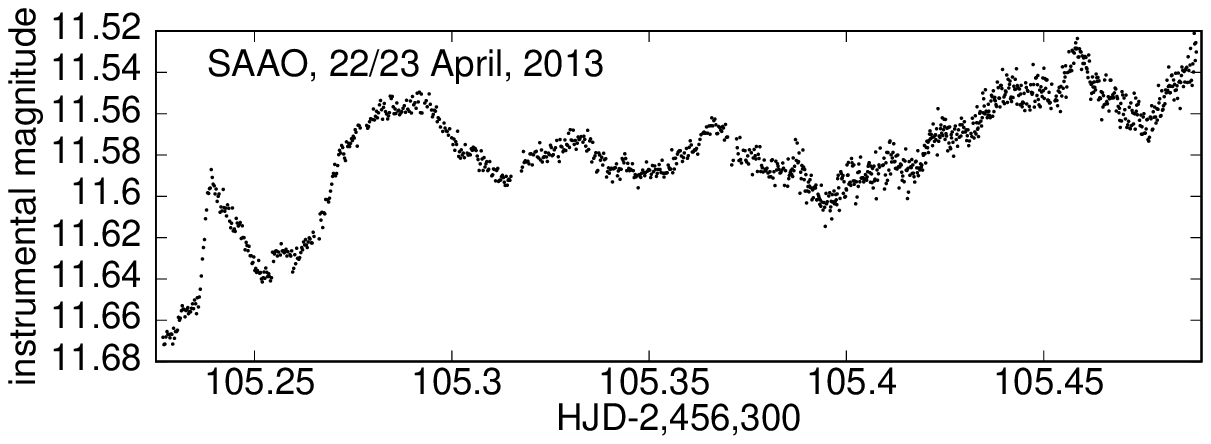}
\includegraphics[width=80mm]{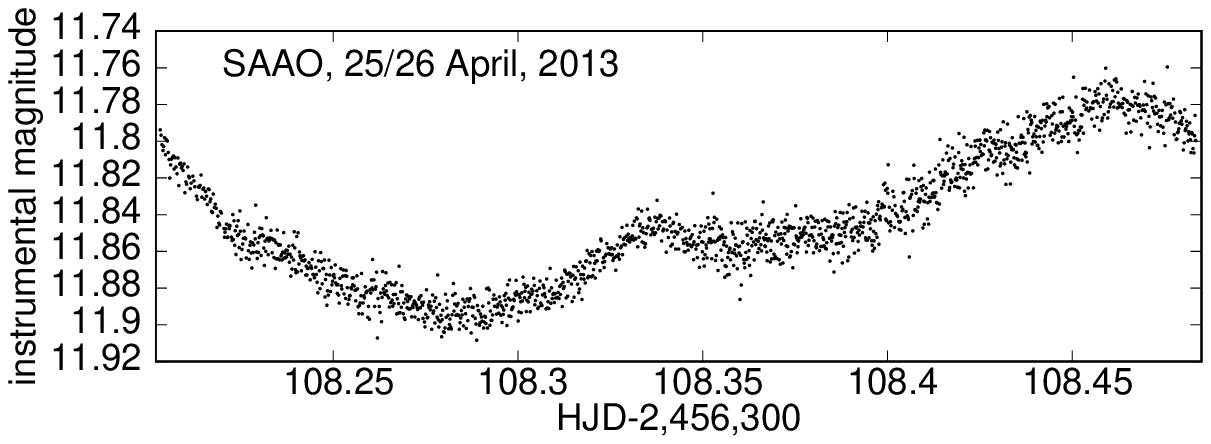}
\includegraphics[width=80mm]{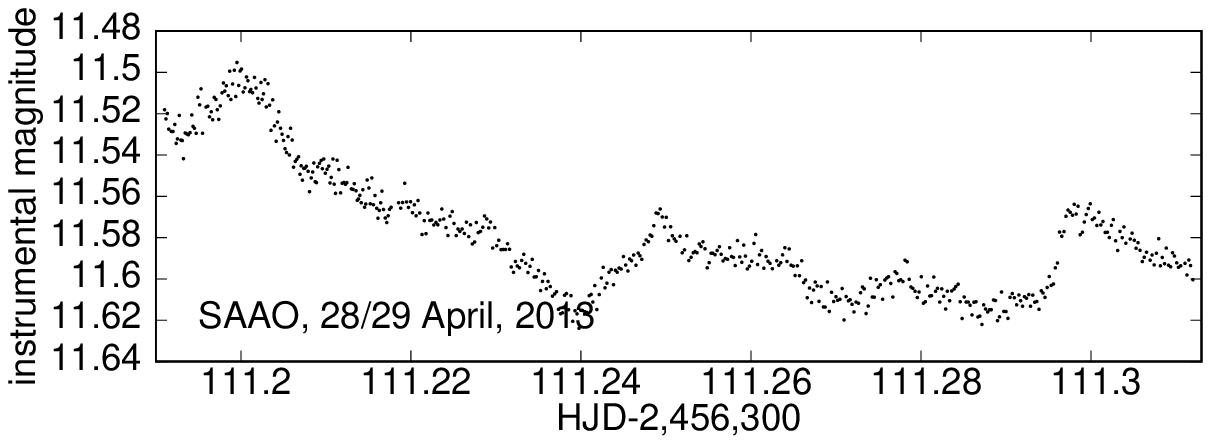}
\includegraphics[width=80mm]{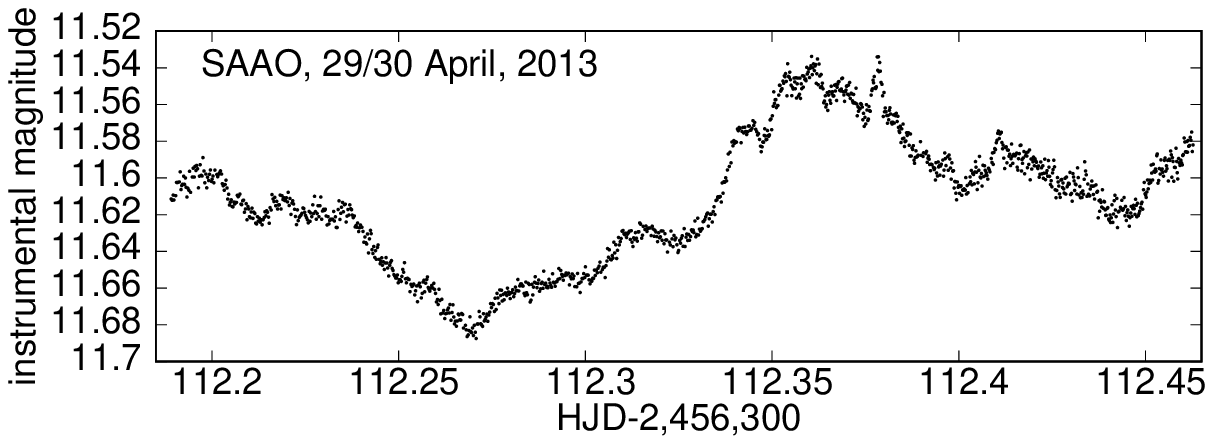}
\includegraphics[width=80mm]{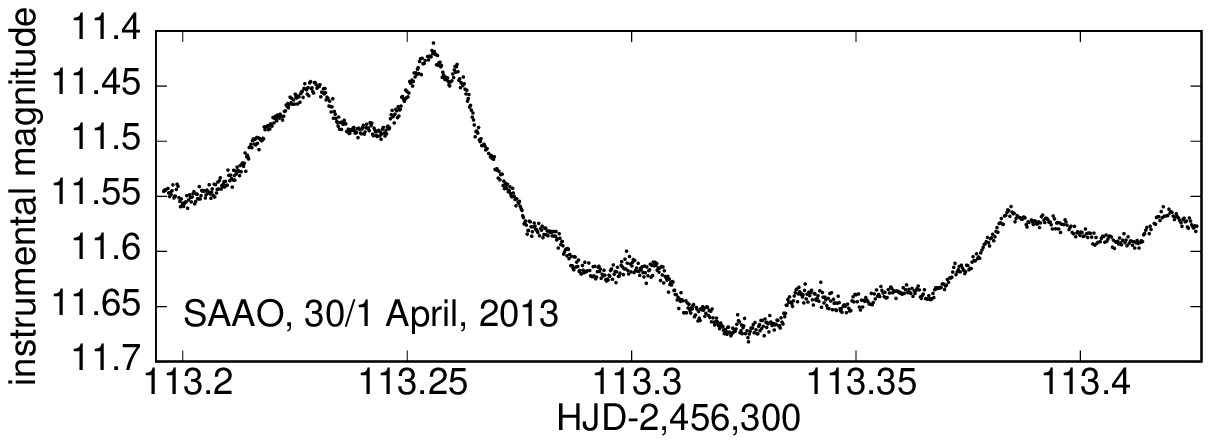}
\includegraphics[width=80mm]{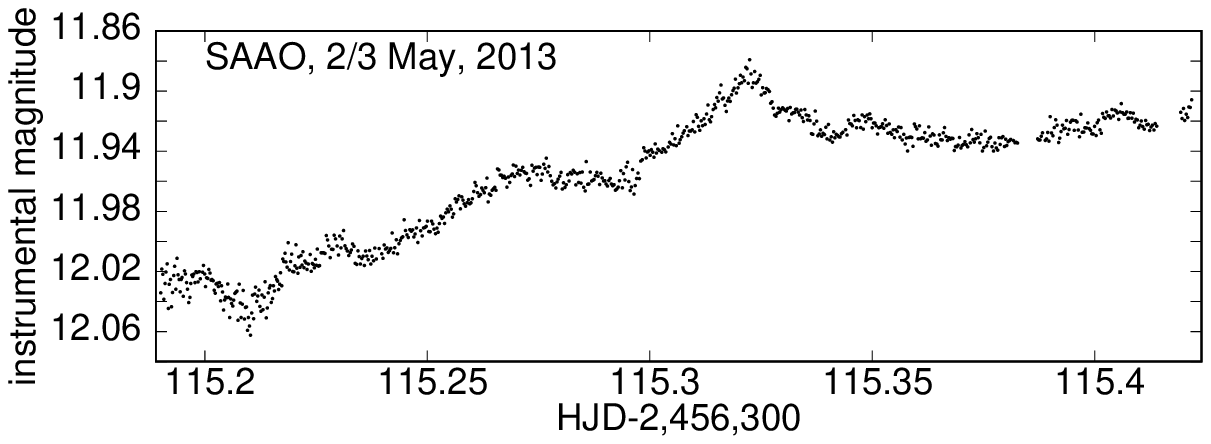}
\includegraphics[width=80mm]{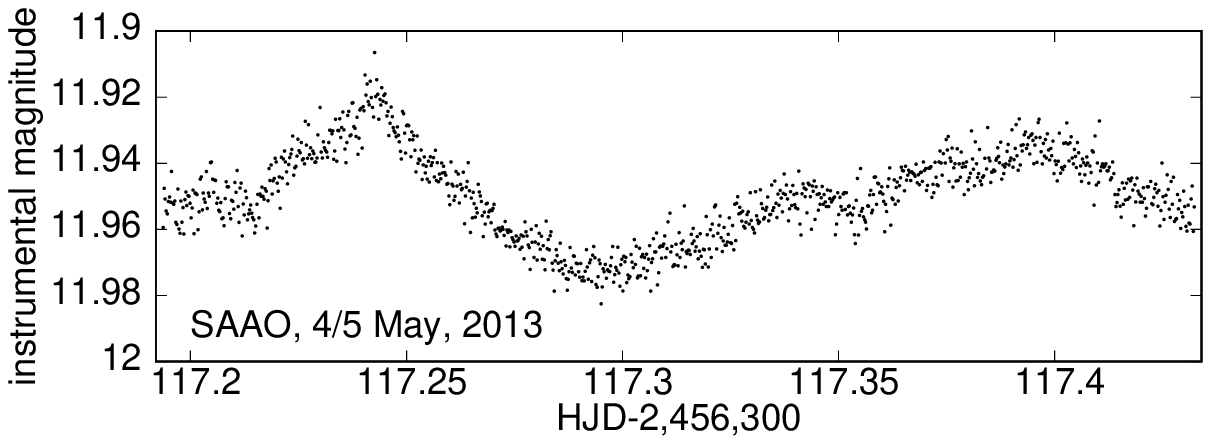}
\includegraphics[width=80mm]{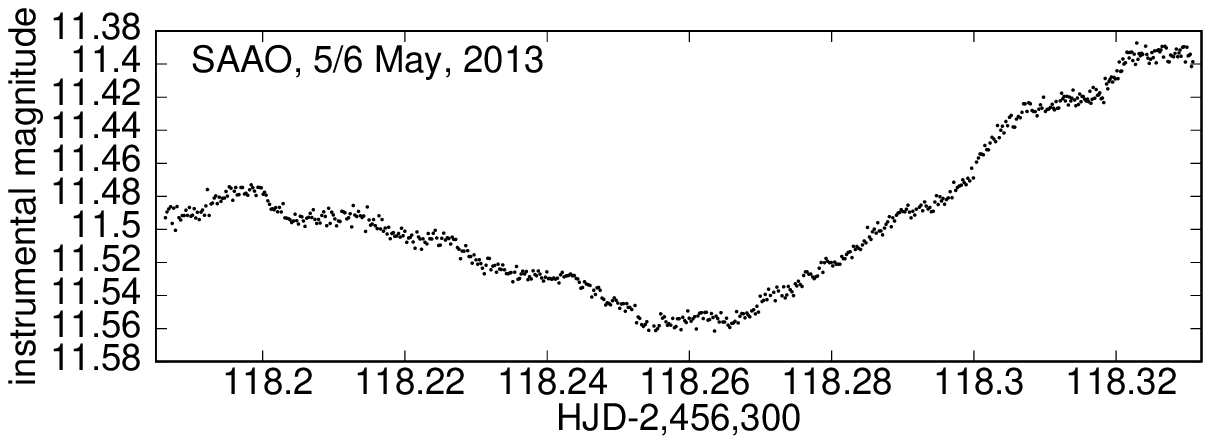}
\includegraphics[width=80mm]{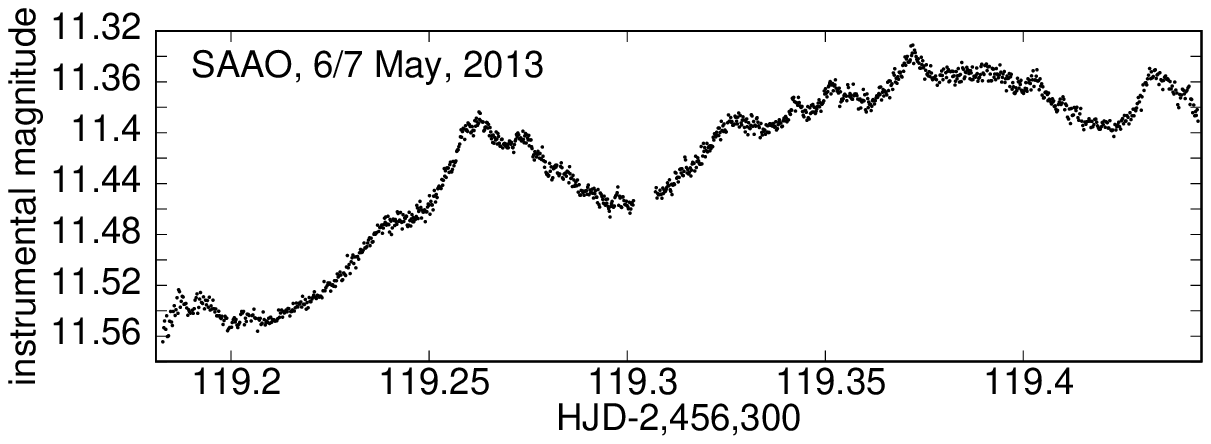}
\caption{The $g'$-filter data obtained in 2013 at {\it SAAO} during photometric nights. 
The data are expressed in an arbitrary magnitude level due to the lack of a good comparison star. 
The nightly trends were removed using mean atmospheric extinction coefficients (App~\ref{app1}). }
\label{Fig.saao13}
\end{figure*}

The instrumental raw data points for TW~Hya ($m_{obs}$) obtained at the {\it SAAO} with {\it TRIPOL} (Sec.~\ref{tripol}) 
during photometric nights in $g'r'$-filters were approximately corrected on nightly trends by means 
of the equation:\newline
$m_{corr}= m_{obs}-k \times X - \beta \times(g'-r')\times X$, \newline
where differential extinction coefficients $k$ were assumed as equal to +0.22 
and +0.1 for $g'$- and $r'$-filters, respectively, while $\beta$ coefficients 
were the same as used above. 
The constant value of $(g'-r')\approx1.22$ taken from SIMBAD database was used in this process. 
The results for $g'$-filter are shown in Fig.~\ref{Fig.saao13}, but were earlier used in transformed light curves 
in Figures~\ref{Fig.short}, \ref{Fig.short_wav} and ~\ref{Fig.occult}. 

\section{The {\it SAAO} and {\it CTIO} light curves}
\label{app2}
% ----------------------- Fig.1 appendix the SAAO 2013 g-light curves ---------------------
%----------------------------------------------------------------------

% ----------------------- Fig.2 appendix the CTIO 2014 light curves ---------------------
\begin{figure*}
\includegraphics[width=80mm]{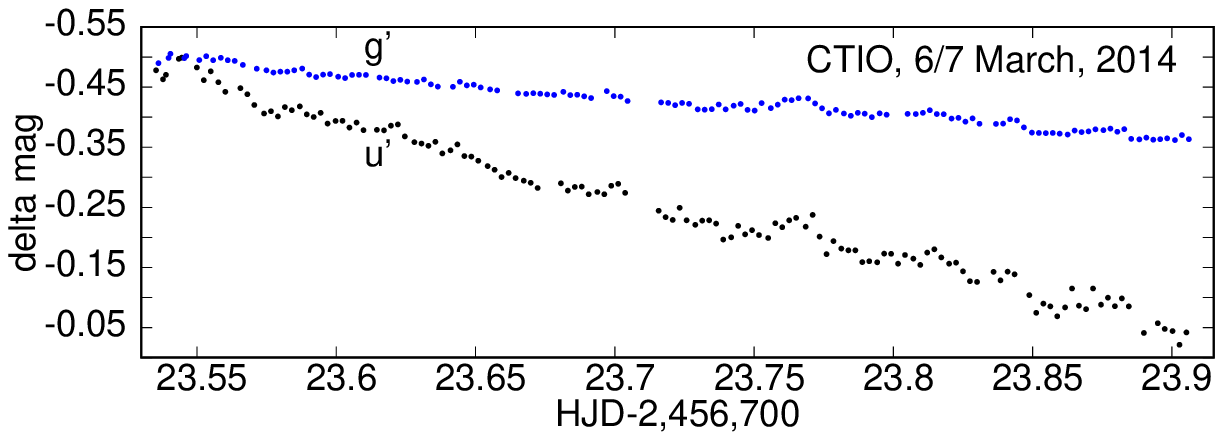}
\includegraphics[width=80mm]{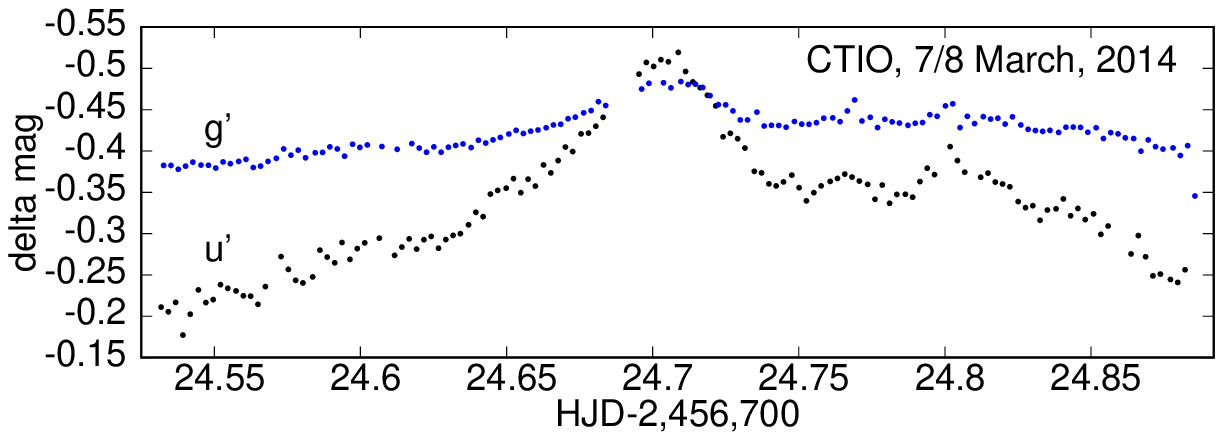}
\includegraphics[width=80mm]{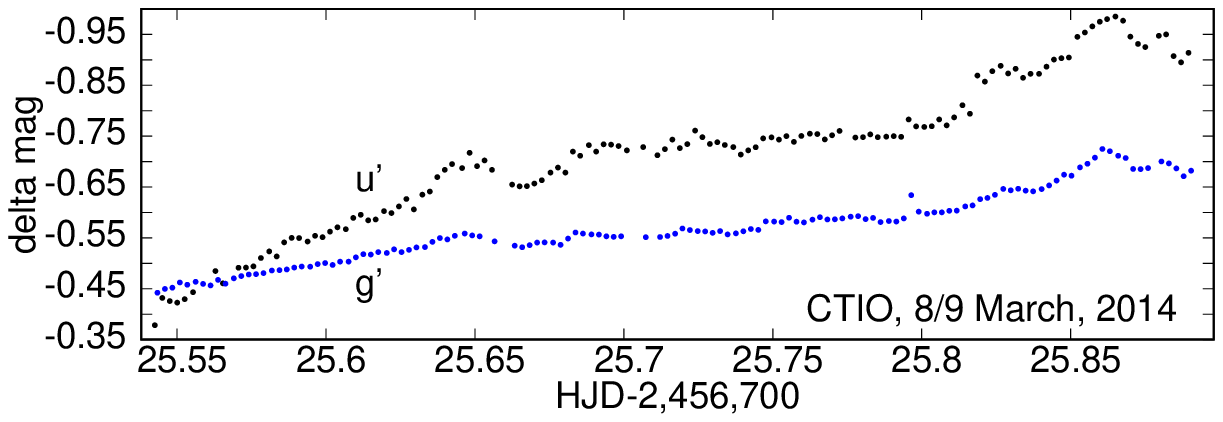}
\includegraphics[width=80mm]{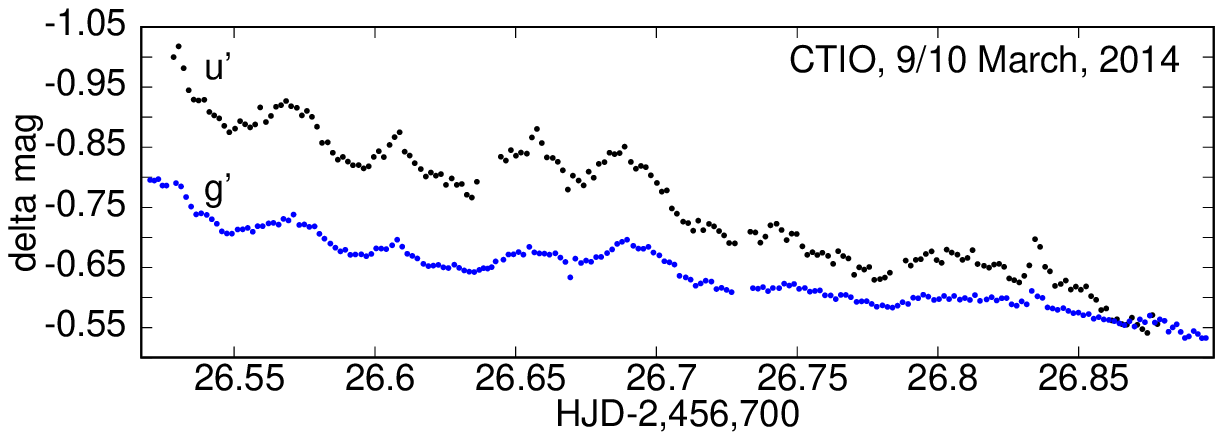}
\includegraphics[width=80mm]{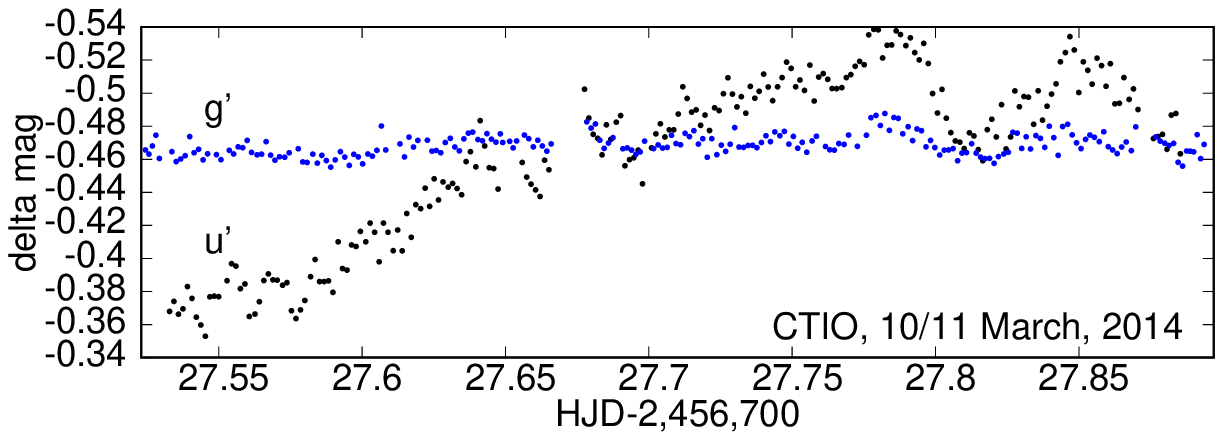}
\includegraphics[width=80mm]{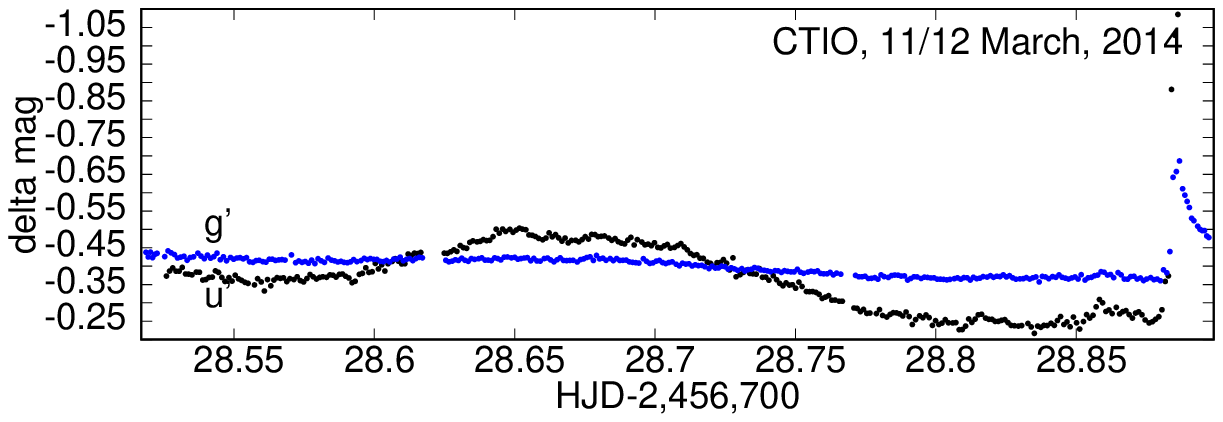}
\includegraphics[width=80mm]{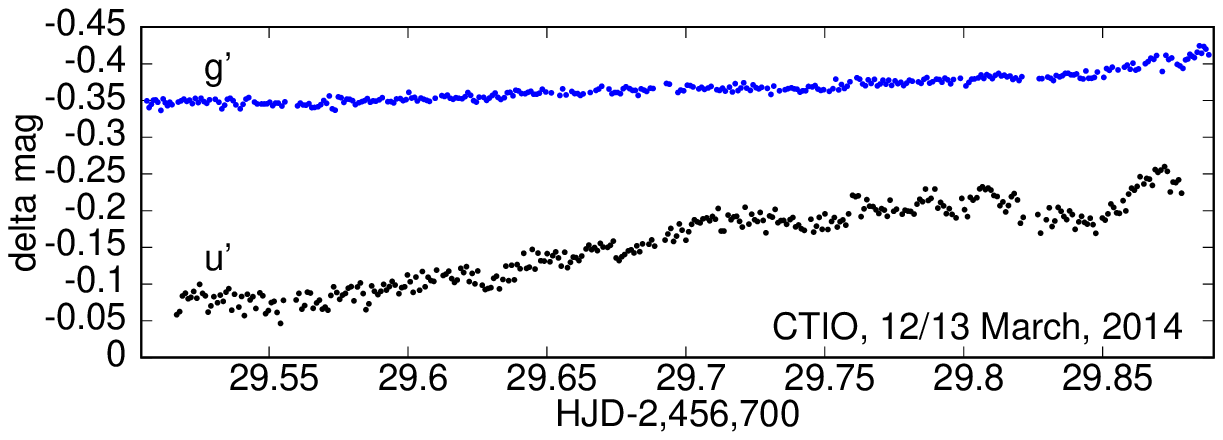}
\includegraphics[width=80mm]{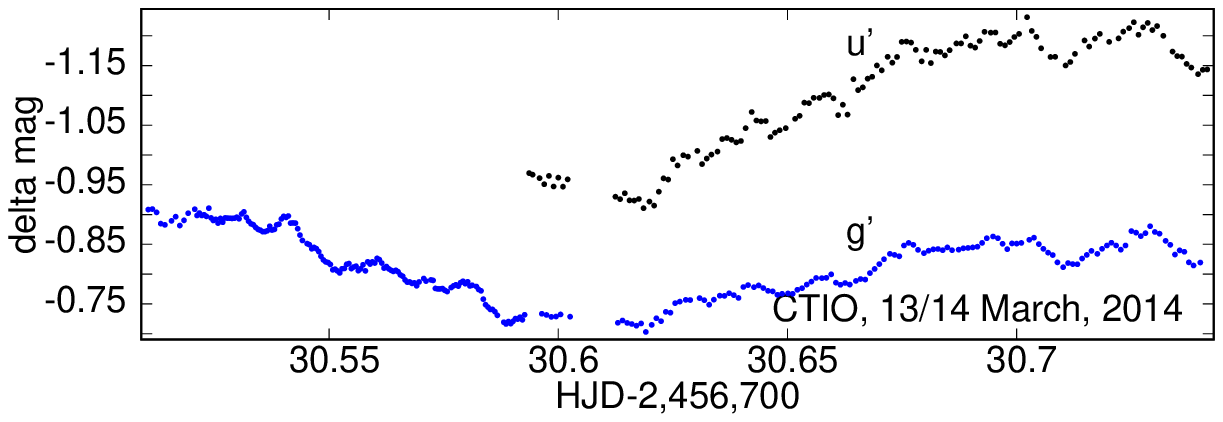}
\includegraphics[width=80mm]{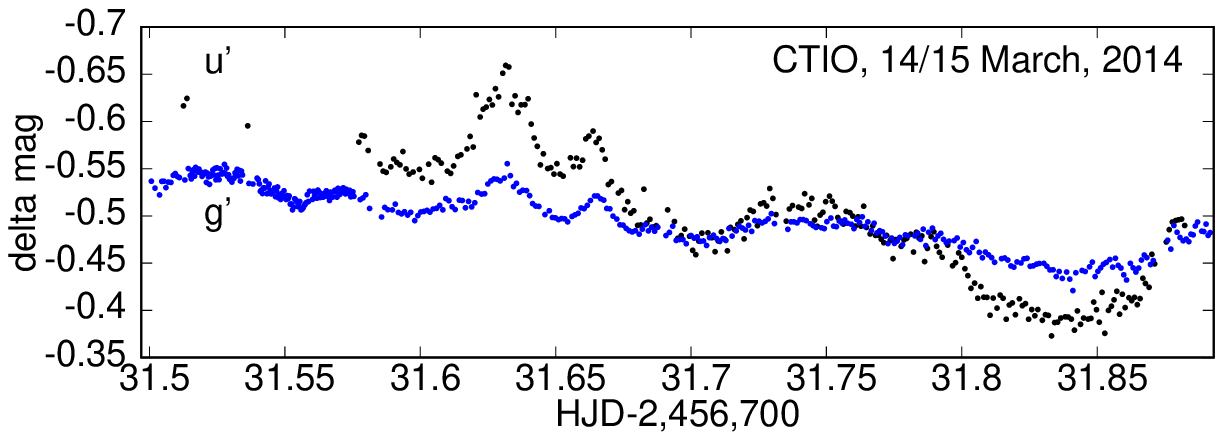}
\includegraphics[width=80mm]{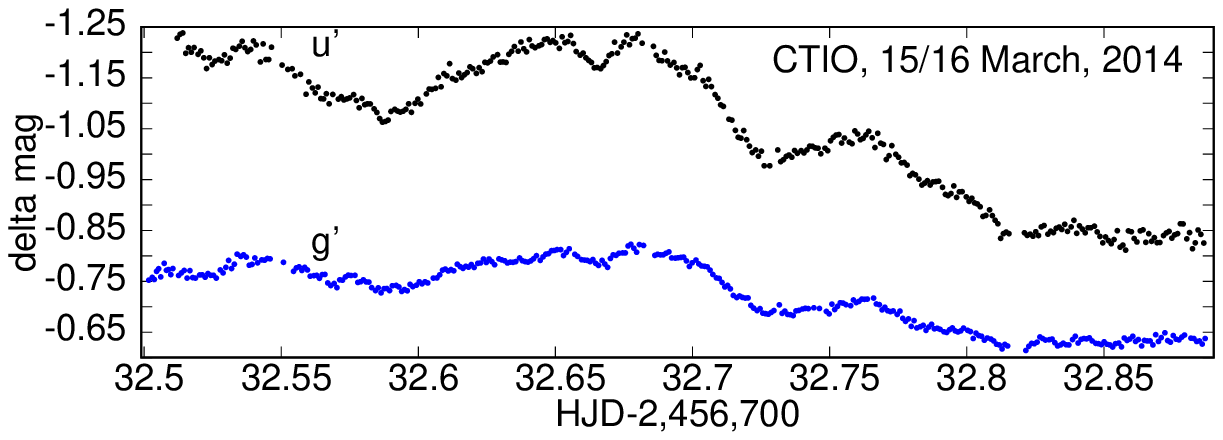}
\includegraphics[width=80mm]{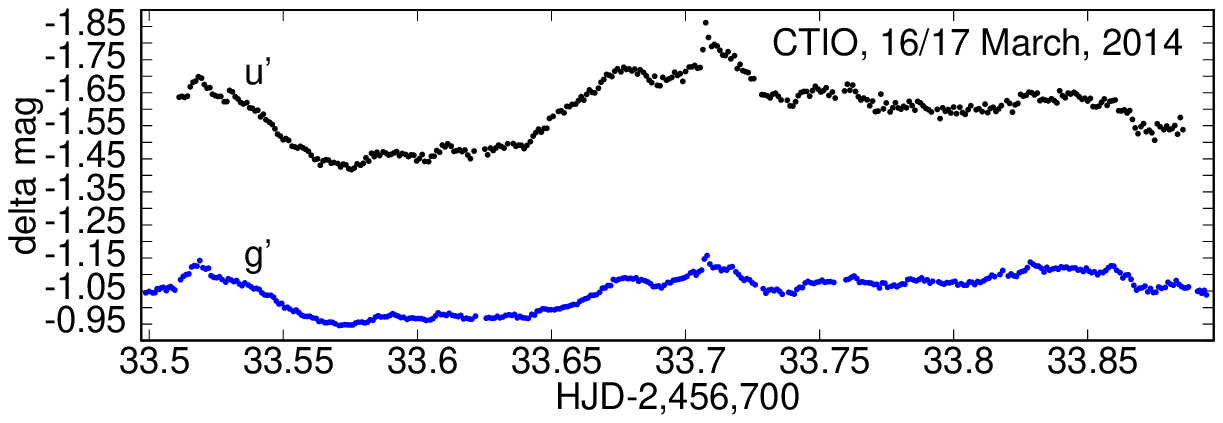}
\includegraphics[width=80mm]{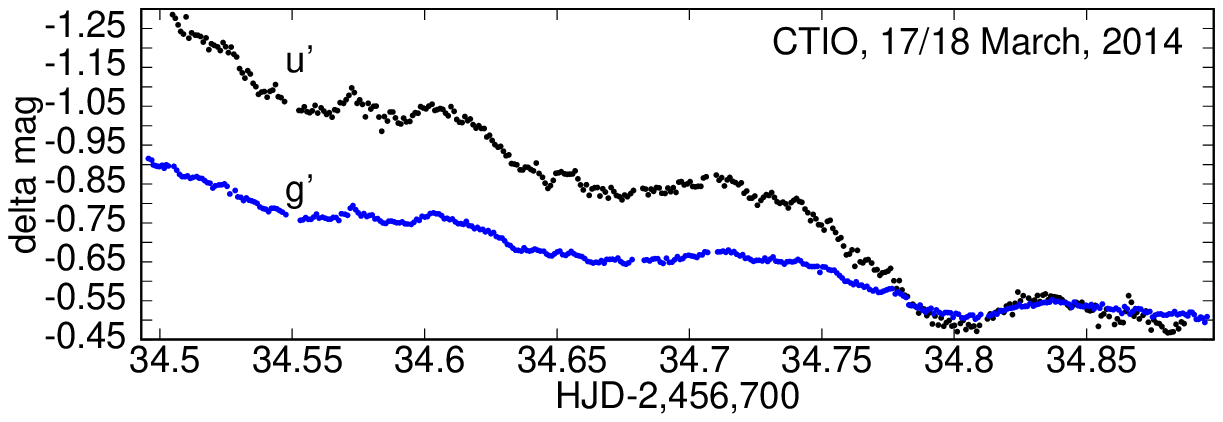}
\includegraphics[width=80mm]{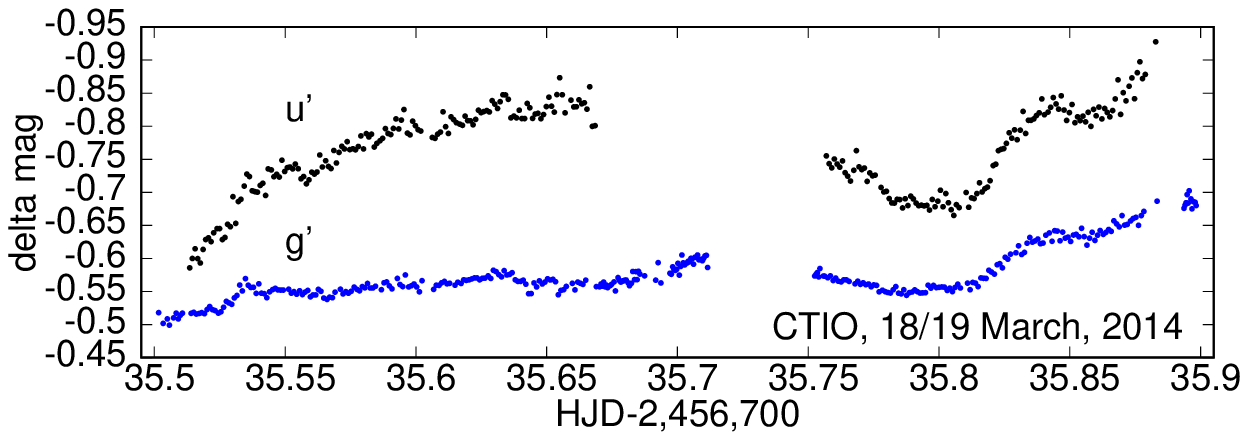}
\includegraphics[width=80mm]{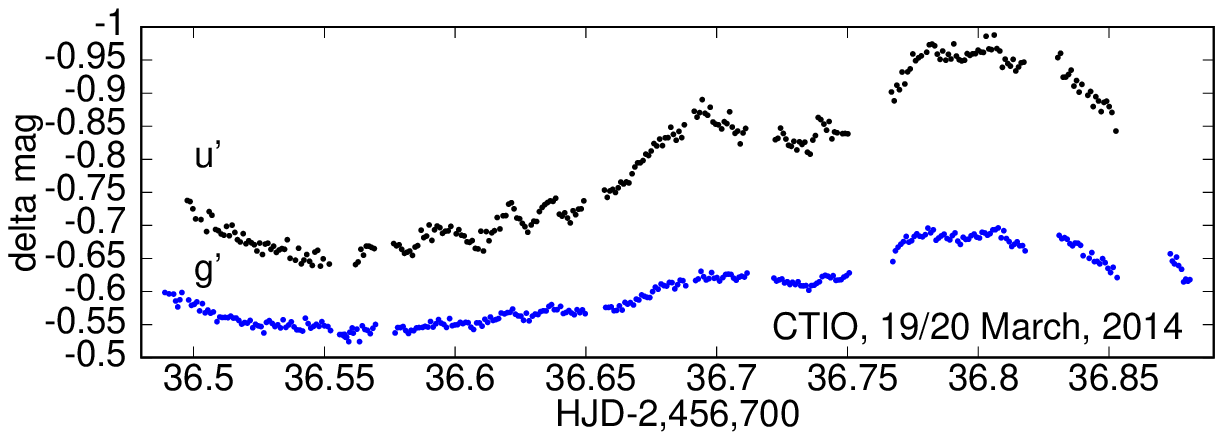}
\includegraphics[width=80mm]{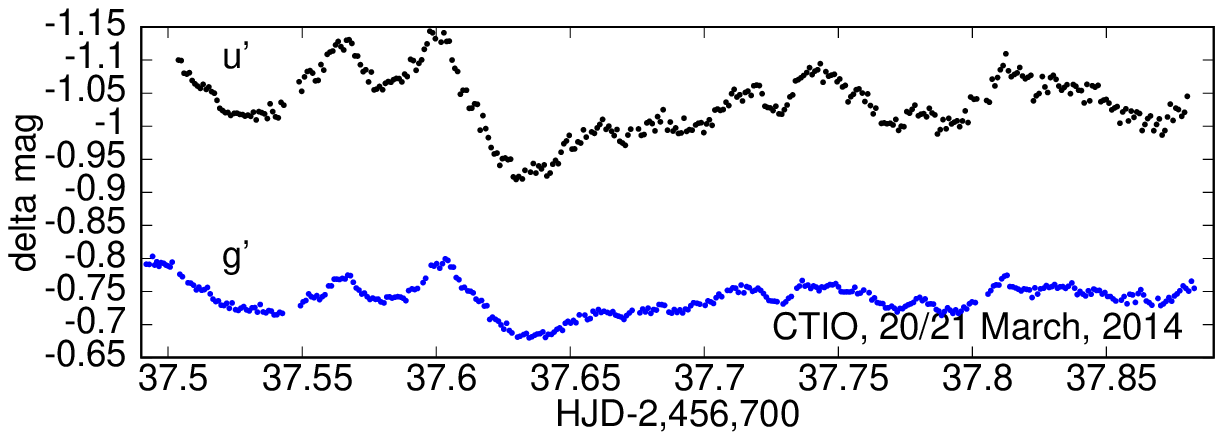}
\caption{The $u'$- and $g'$- filter data obtained at {\it CTIO} in 2014 with respect 
to the first and third comparison stars from Table~\ref{Tab.comp}.
The data were left in the instrumental system uncorrected for atmospheric extinction effects.}
\label{Fig.ctio14}
\end{figure*}
%----------------------------------------------------------------------

% -----------------------  Fig.2 appendix the CTIO 2015 light curves ---------------------
\begin{figure*}
\includegraphics[width=85mm]{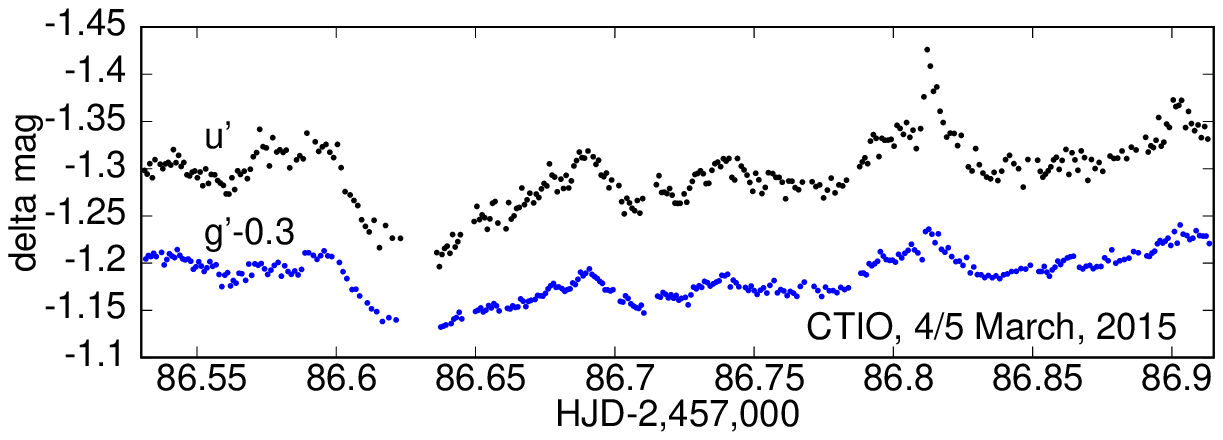}
\includegraphics[width=85mm]{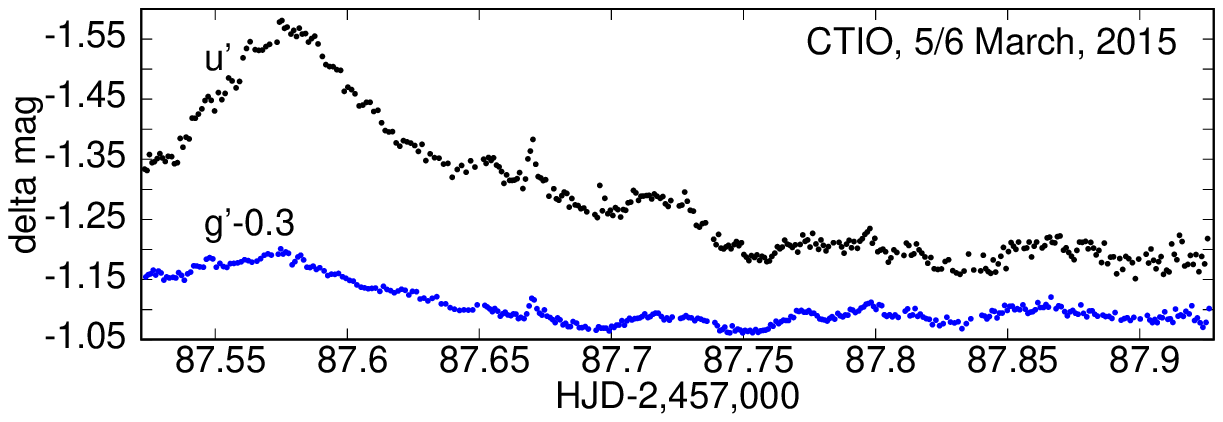}
\includegraphics[width=85mm]{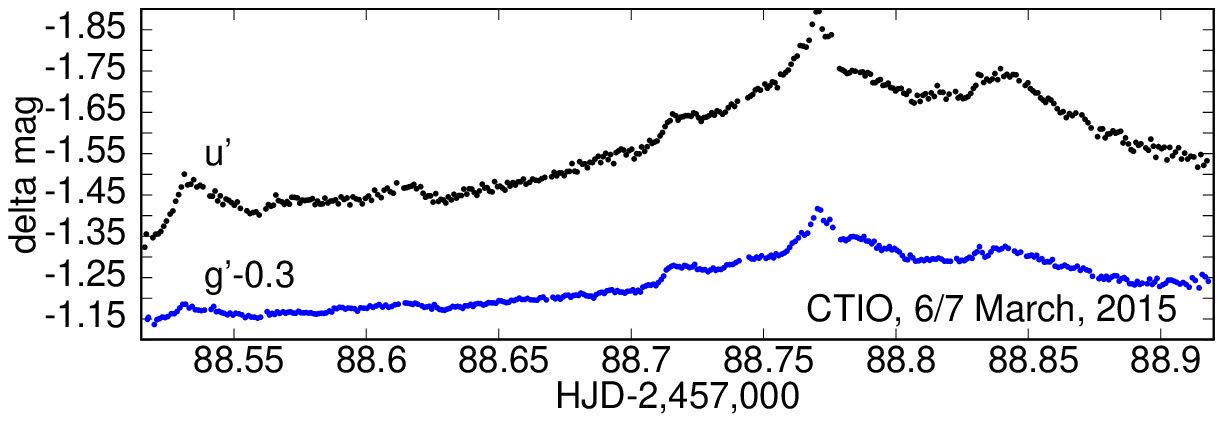}
\includegraphics[width=85mm]{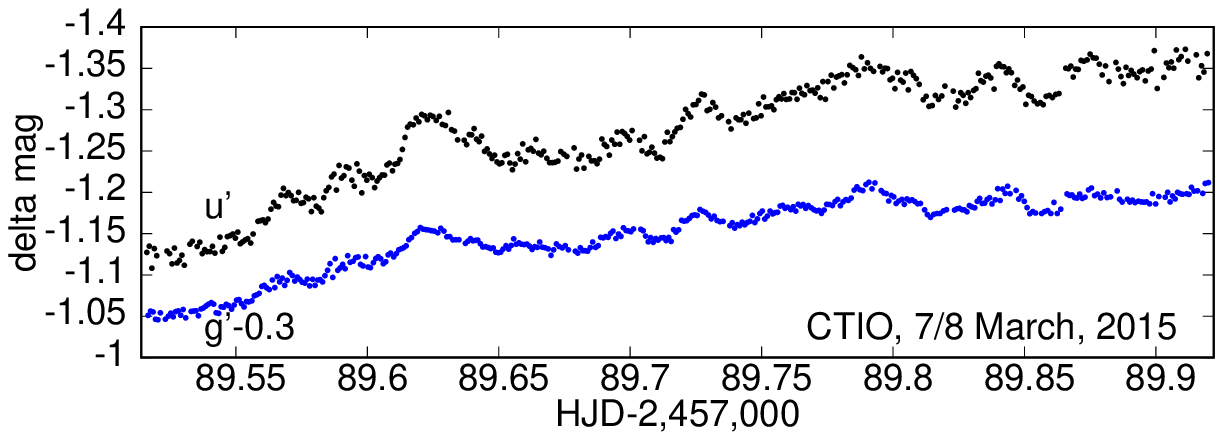}
\includegraphics[width=85mm]{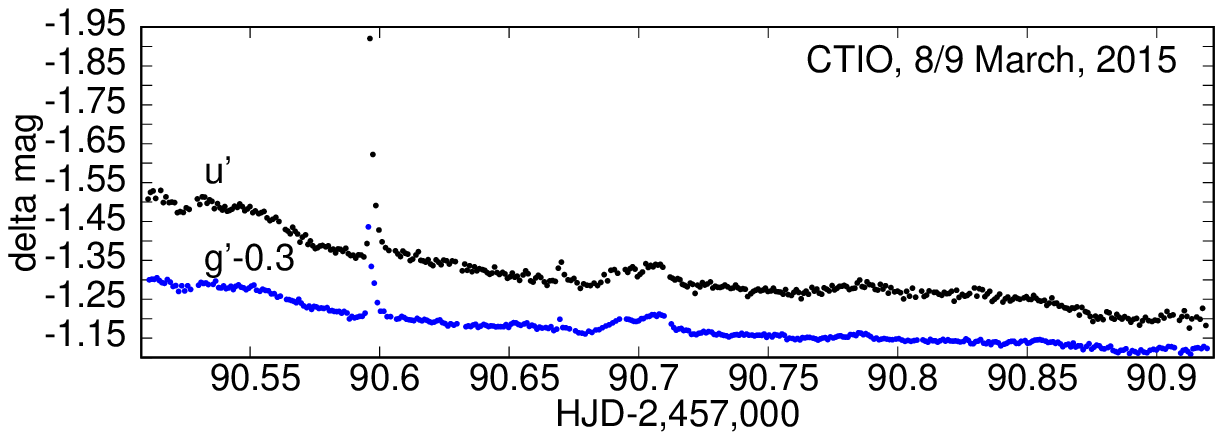}
\includegraphics[width=85mm]{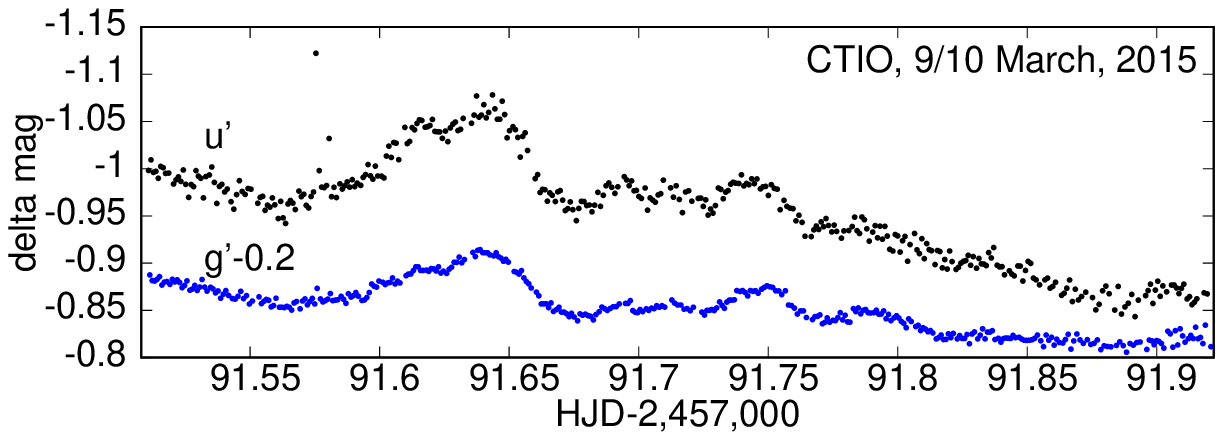}
\includegraphics[width=85mm]{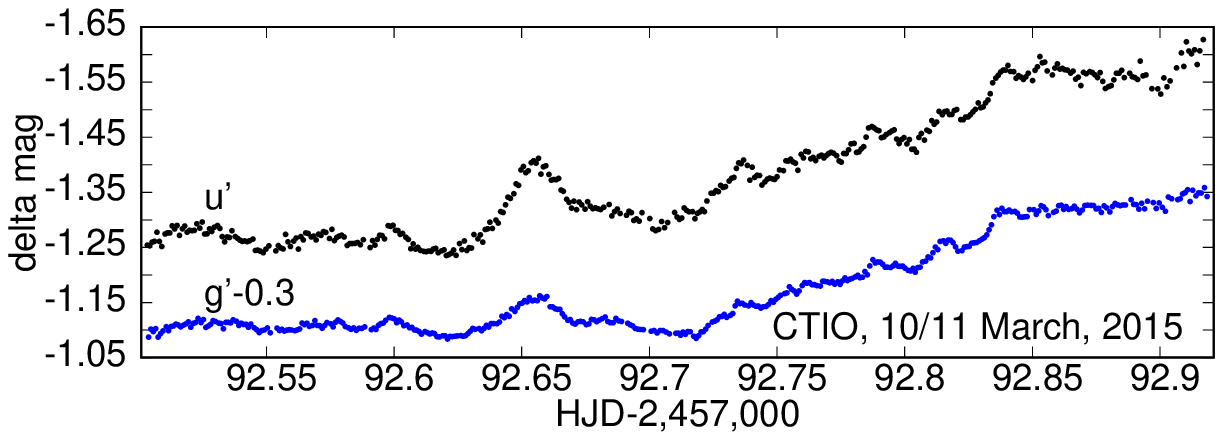}
\includegraphics[width=85mm]{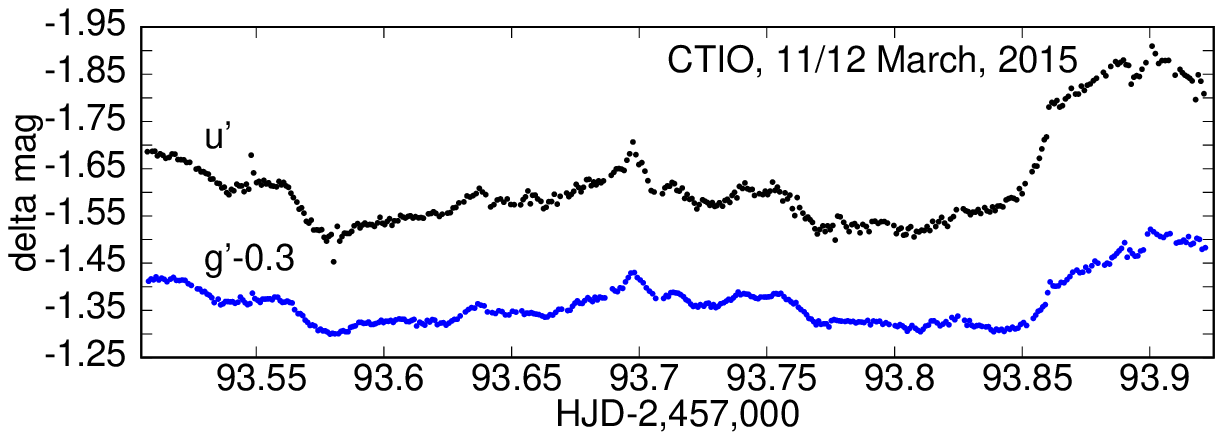}
\includegraphics[width=85mm]{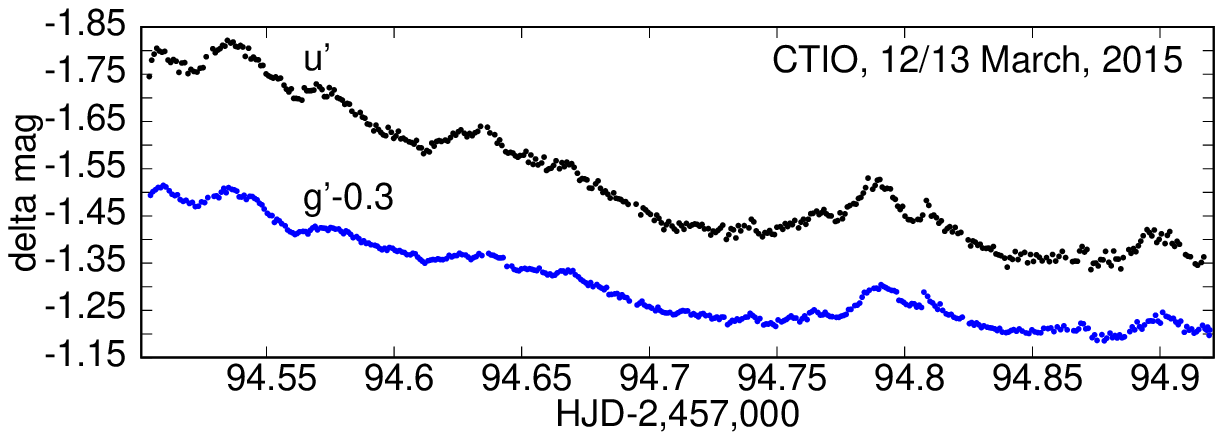}
\includegraphics[width=85mm]{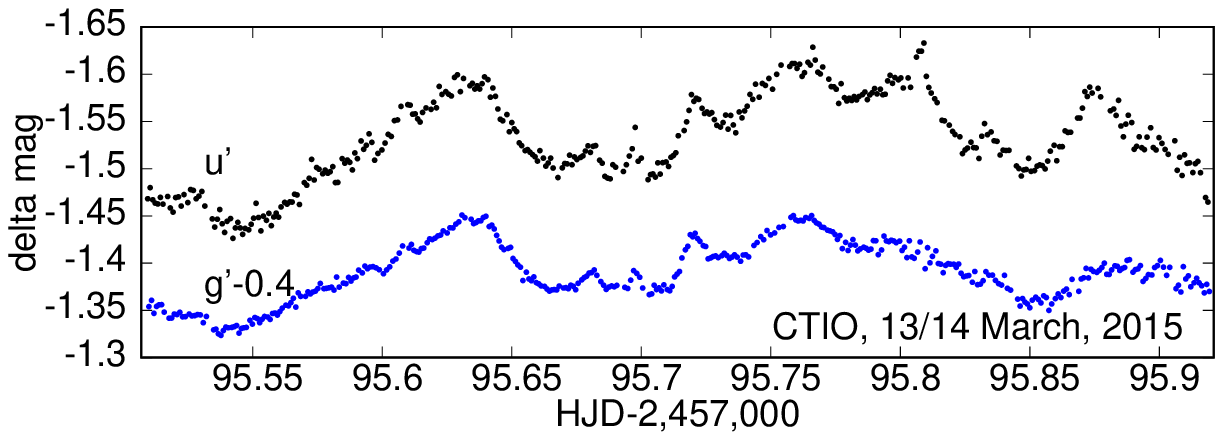}
\includegraphics[width=85mm]{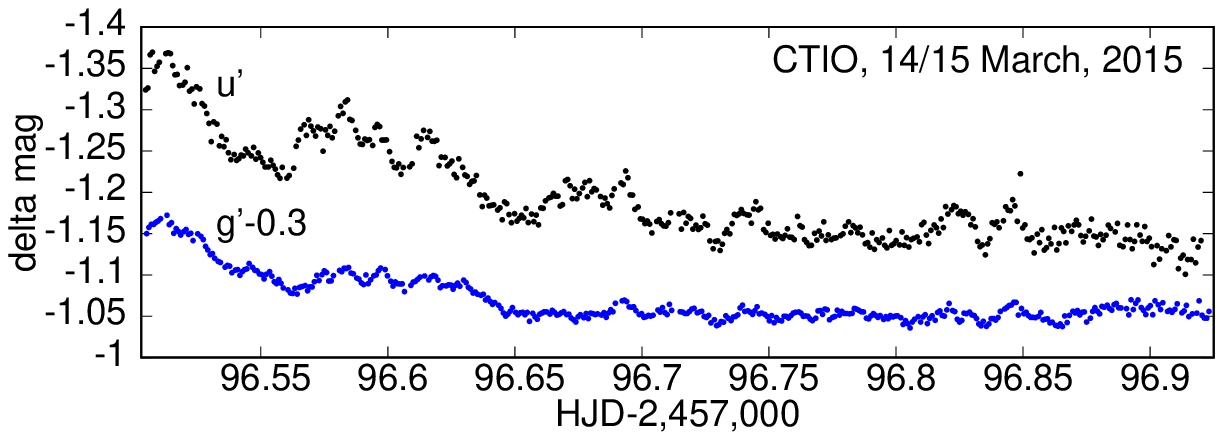}
\includegraphics[width=85mm]{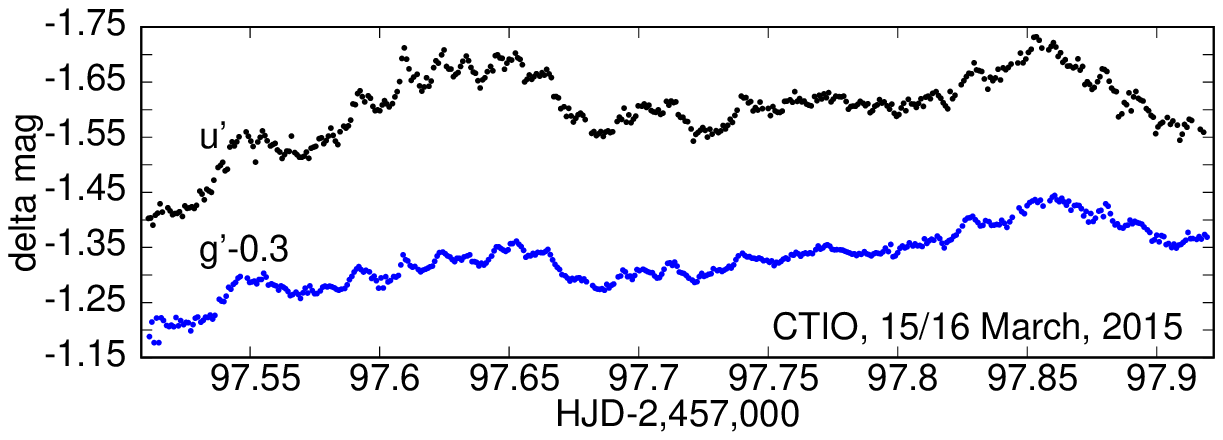}
\includegraphics[width=85mm]{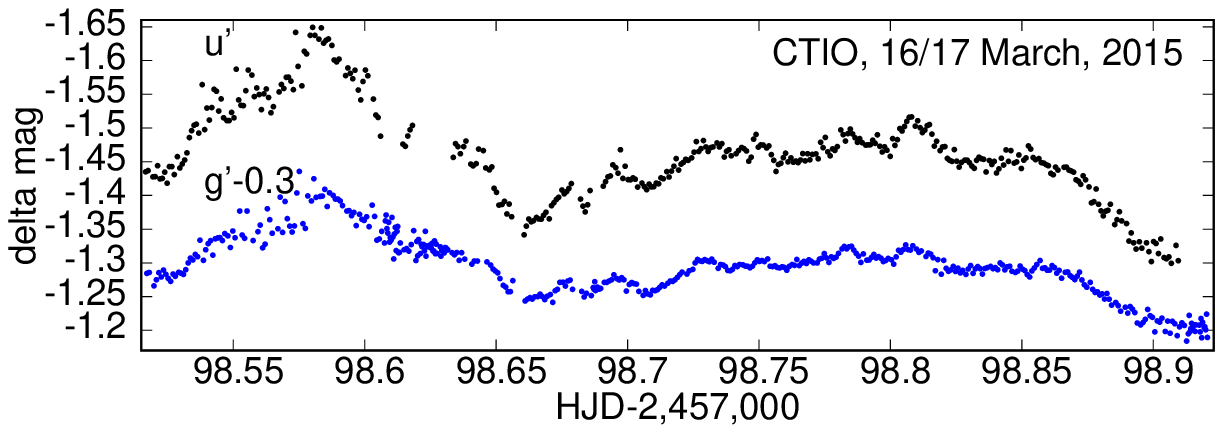}
\includegraphics[width=85mm]{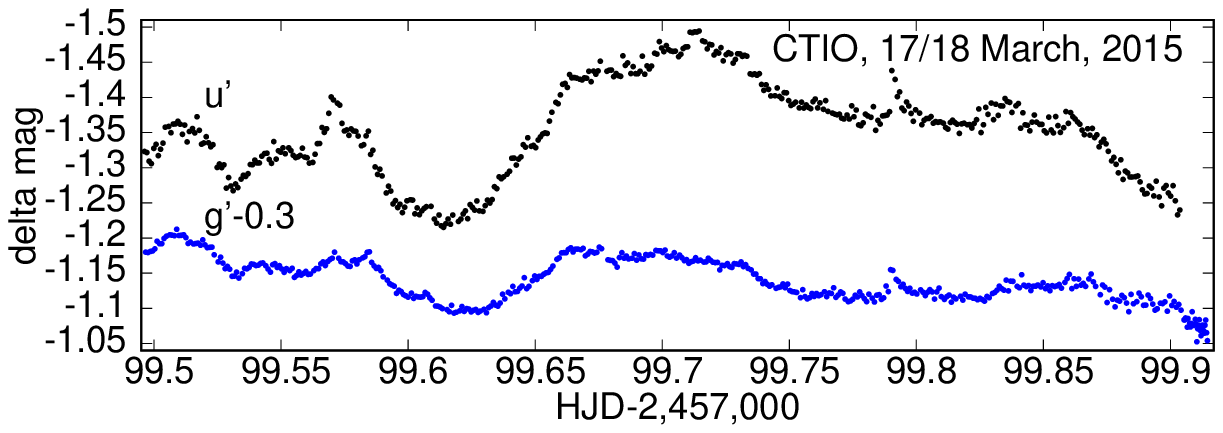}
\caption{The $u'$- and $g'$- filter data obtained at {\it CTIO} in 2015 with respect 
to the first and third comparison stars from Table~\ref{Tab.comp}.
The data were left in the instrumental system uncorrected for atmospheric extinction effects. 
In contrast to the previous run, the $g'$-filter light curves were shifted by the indicated 
value for clarity.}
\label{Fig.ctio15}
\end{figure*}
%----------------------------------------------------------------------

% -----------------------  Fig.3 appendix the SAAO 2015 light curves ---------------------
\begin{figure*}
\includegraphics[width=58mm]{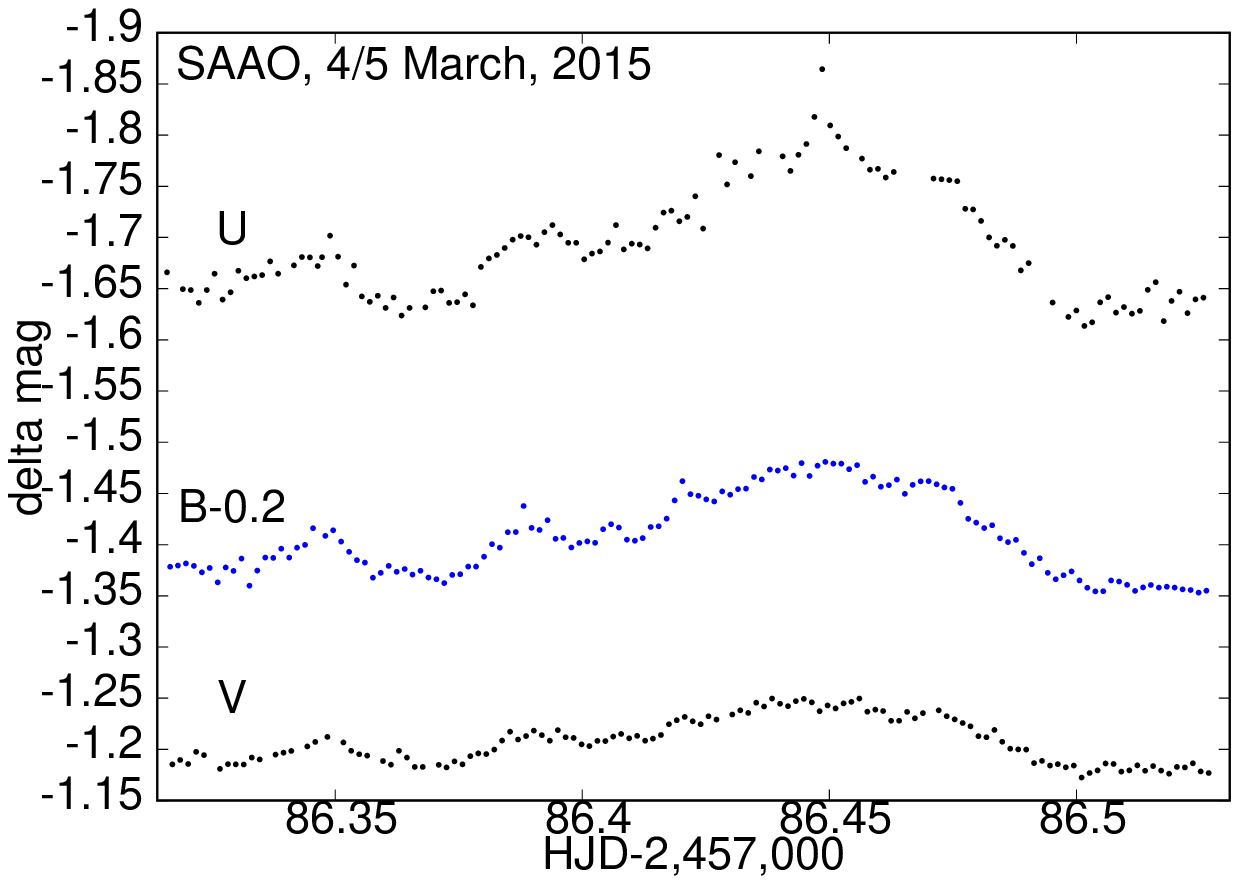}
\includegraphics[width=58mm]{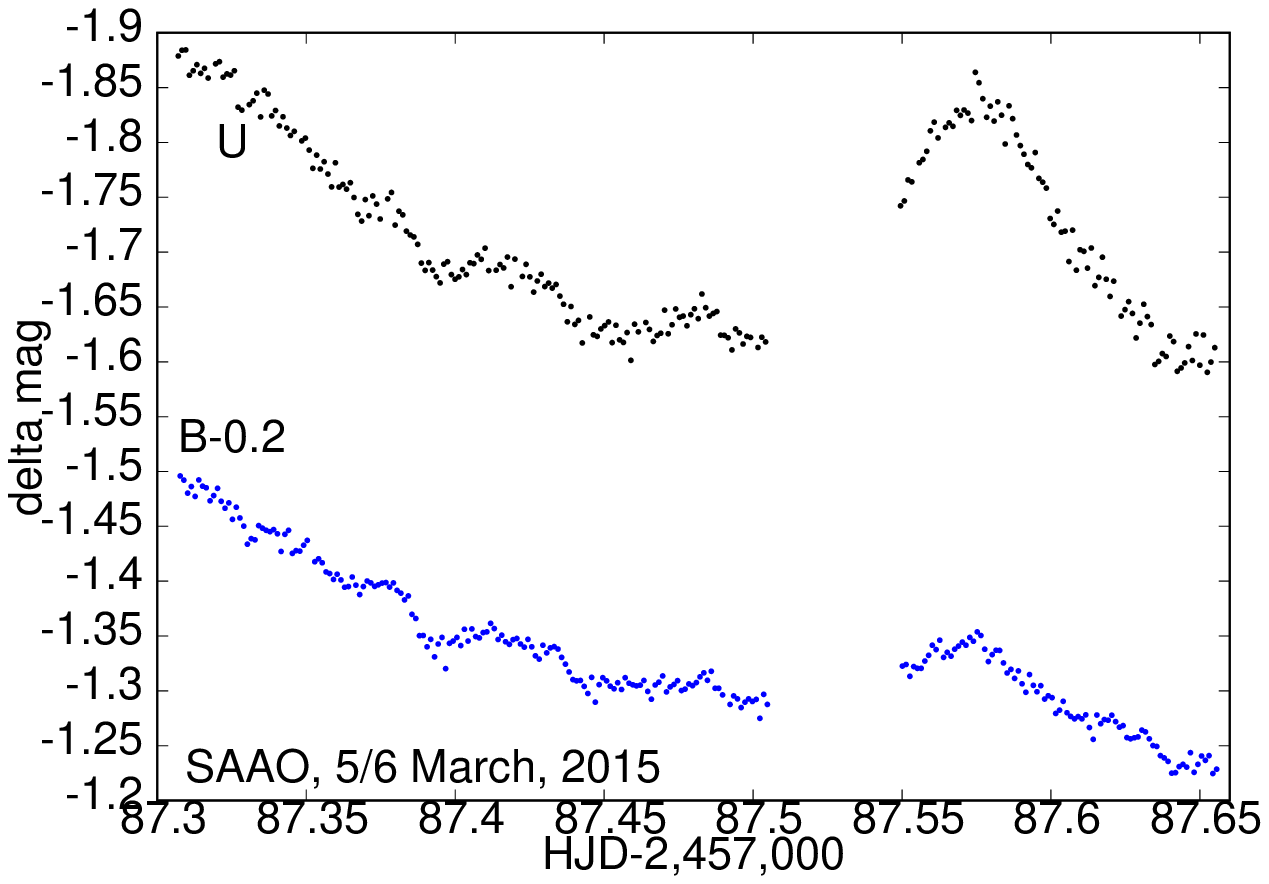}
\includegraphics[width=58mm]{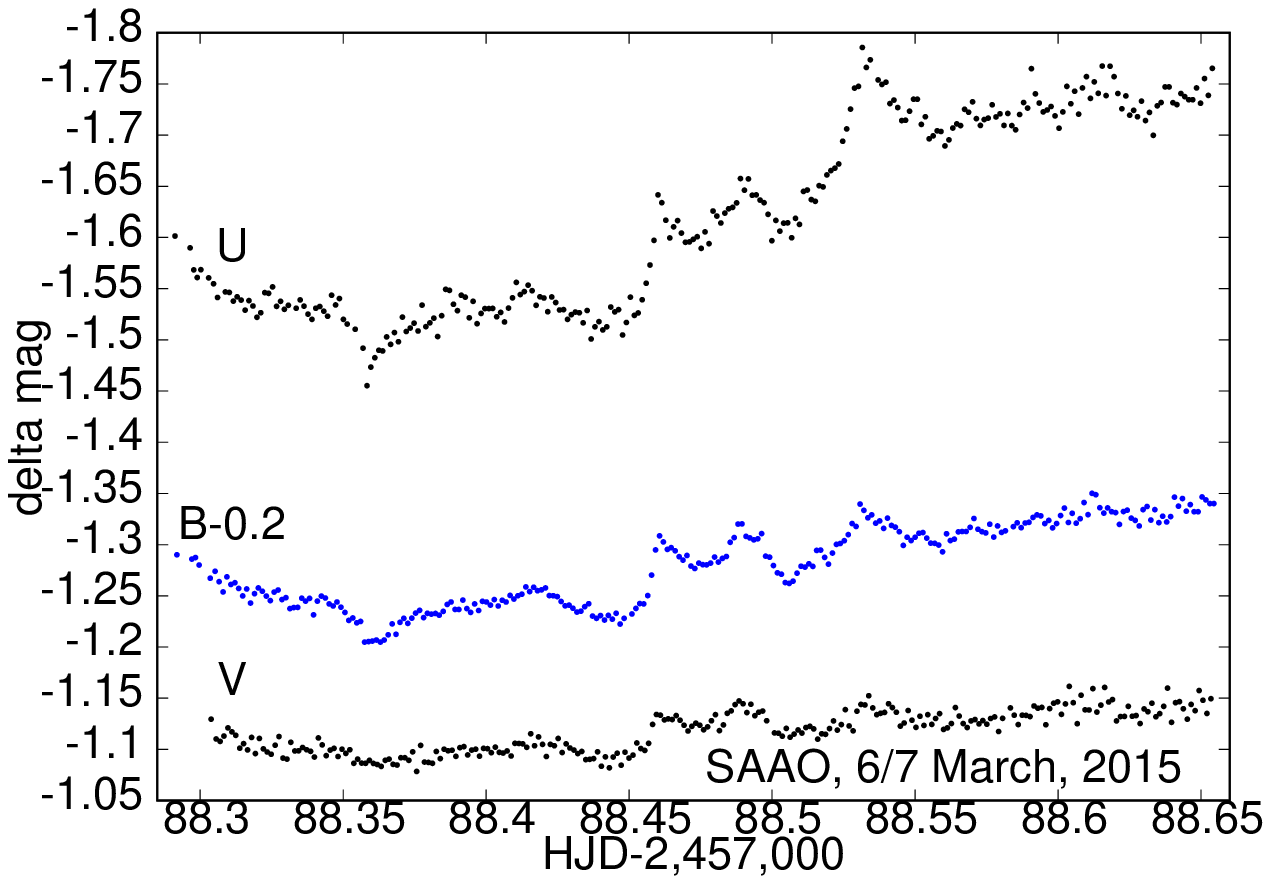}
\includegraphics[width=58mm]{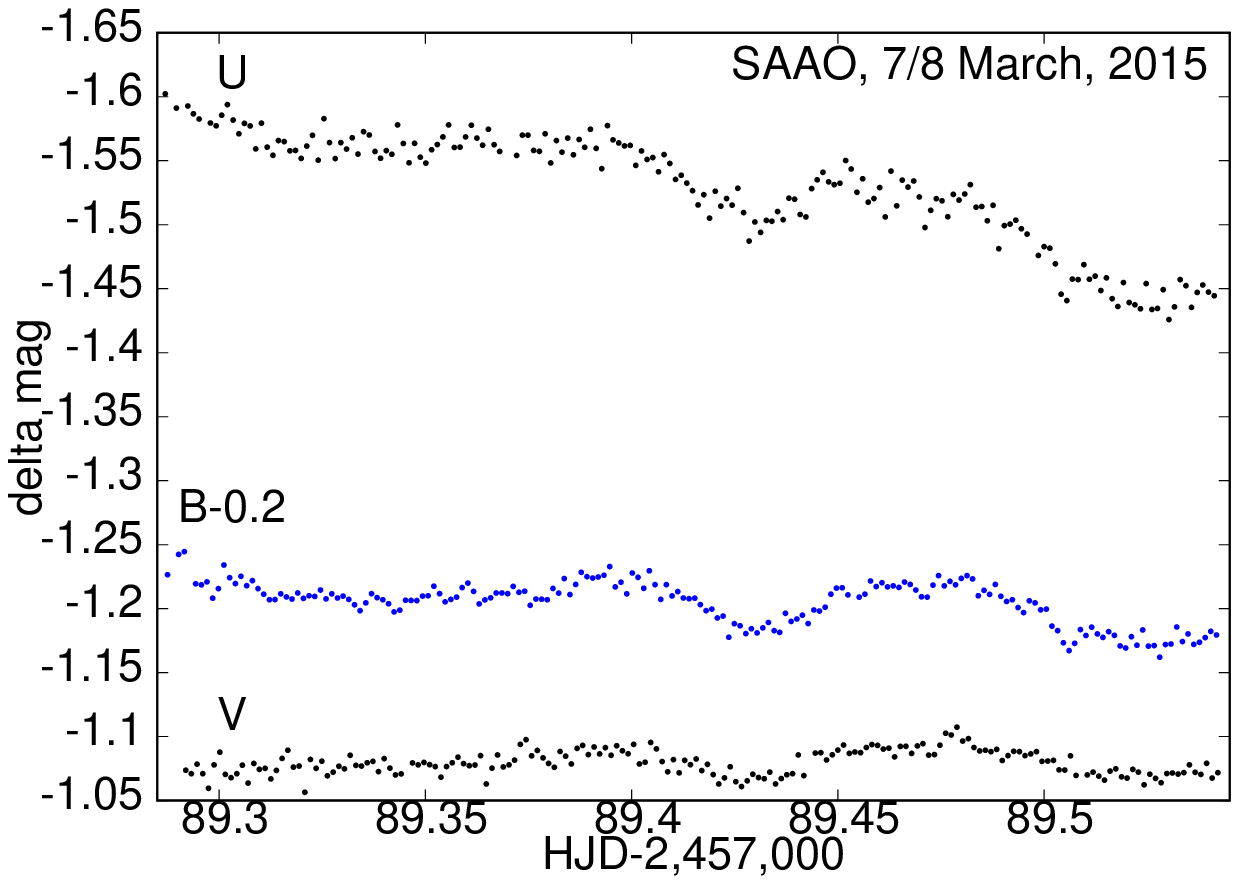}
\includegraphics[width=58mm]{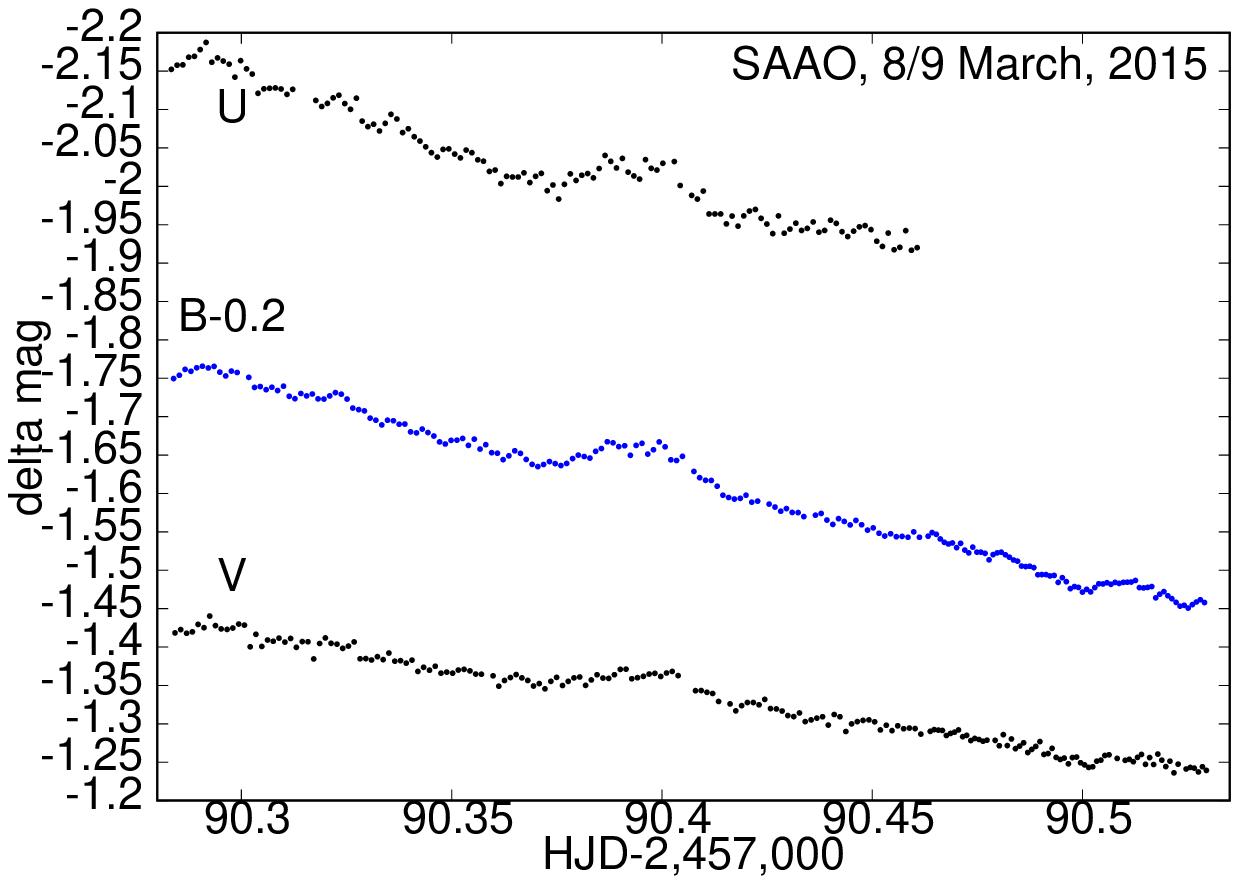}
\includegraphics[width=58mm]{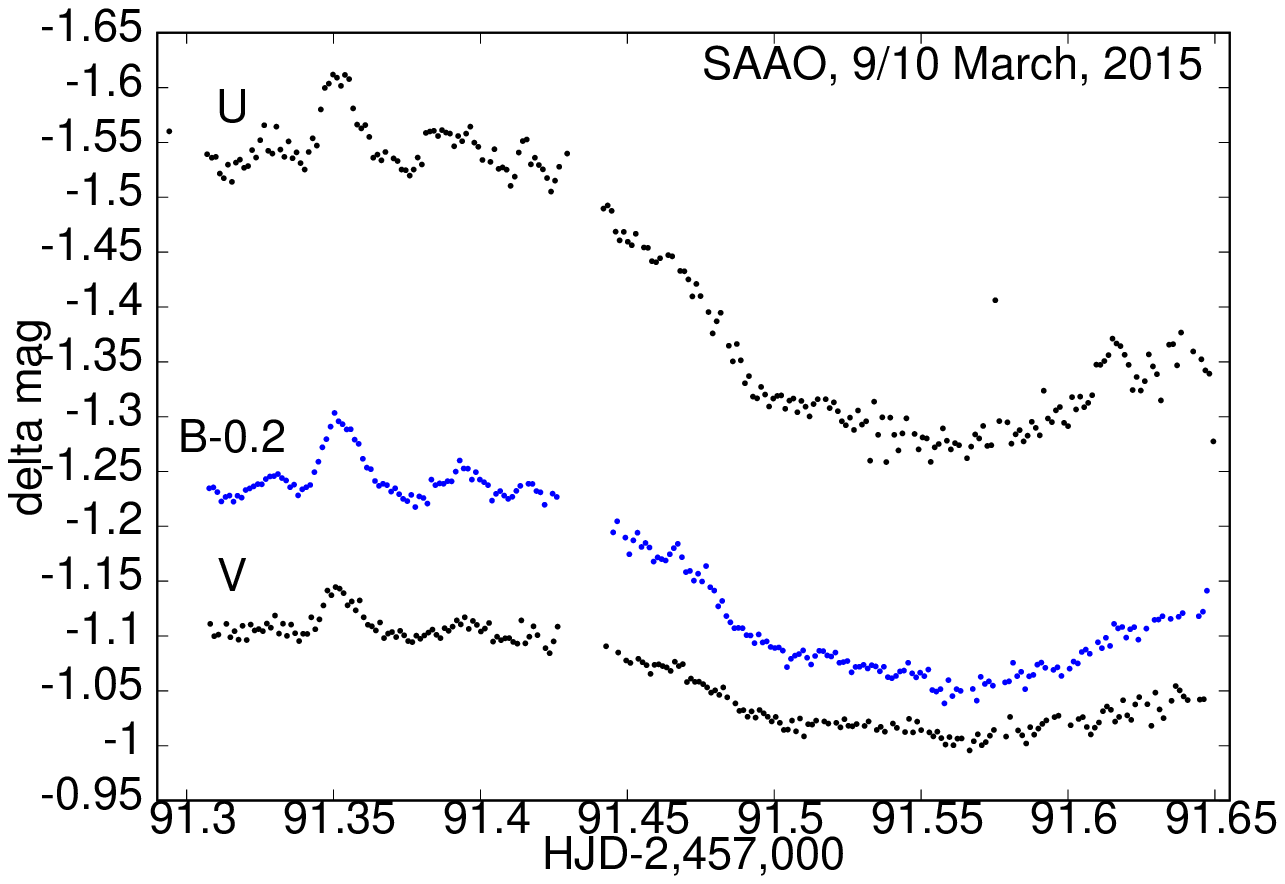}
\includegraphics[width=58mm]{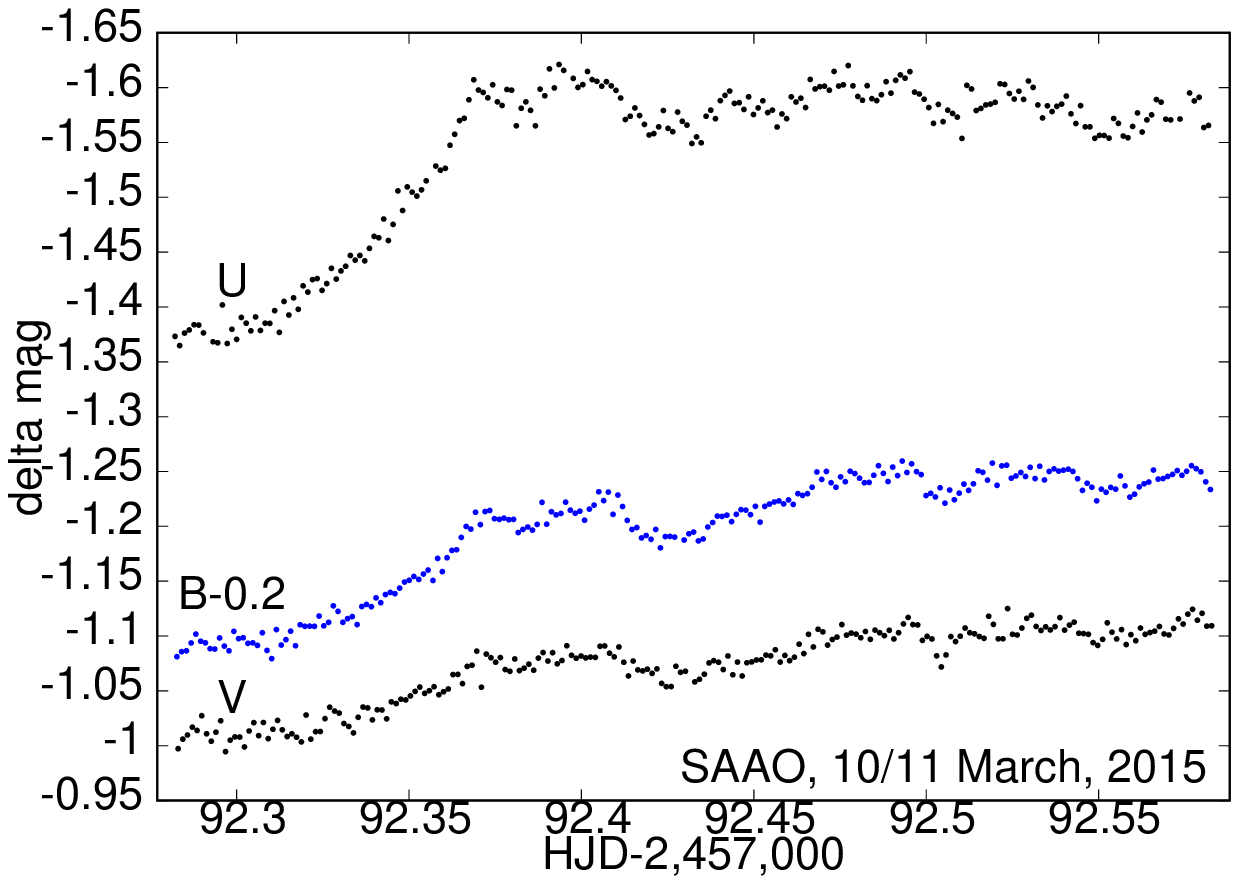}
\includegraphics[width=58mm]{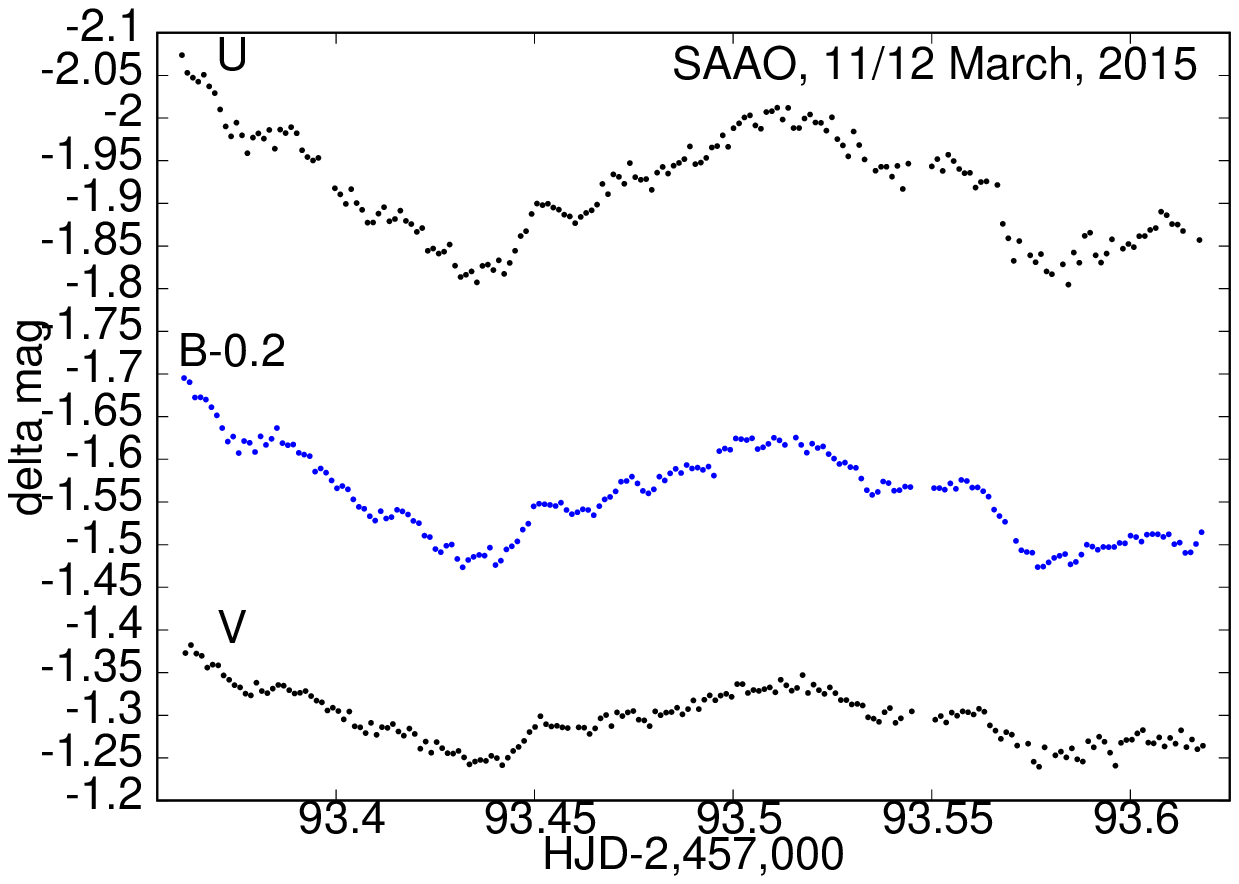}
\includegraphics[width=58mm]{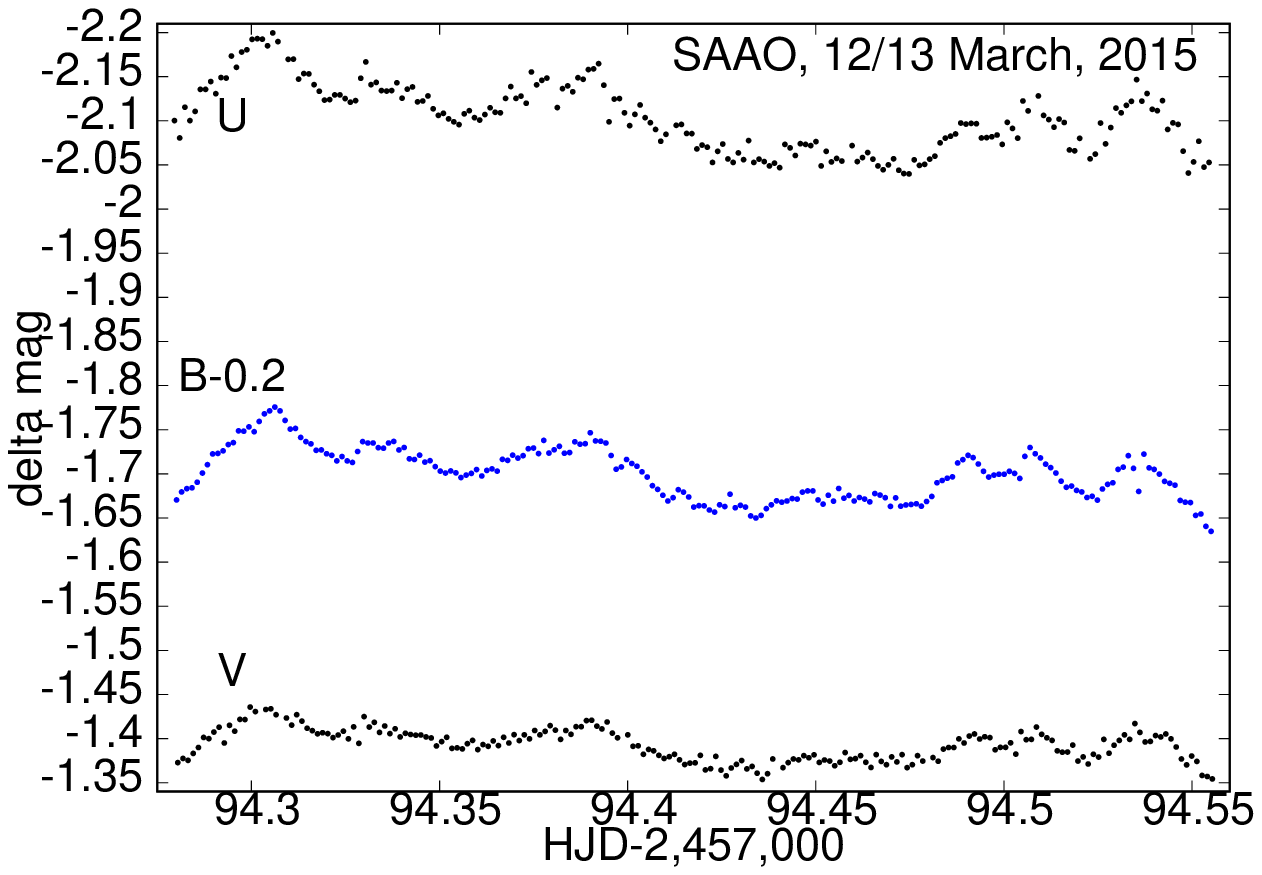}
\includegraphics[width=58mm]{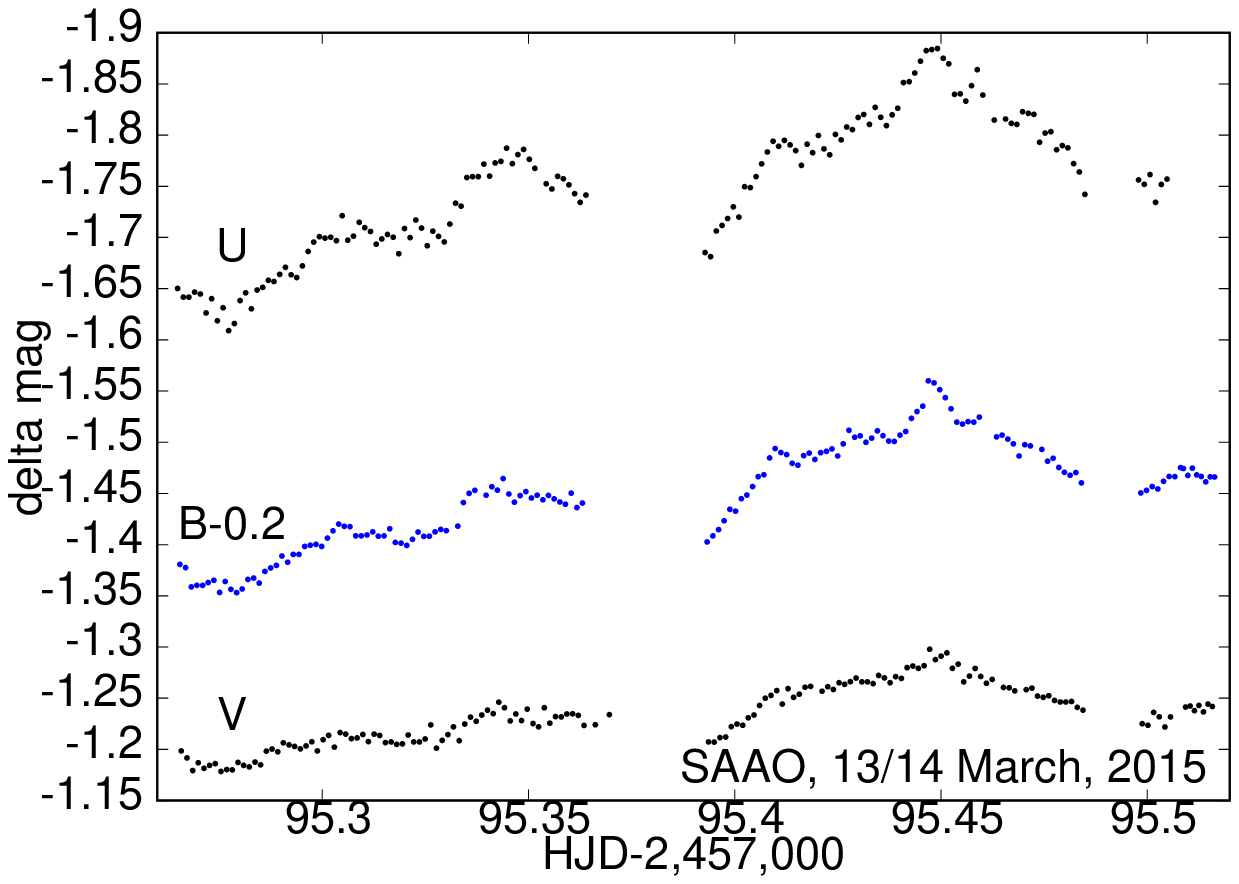}
\includegraphics[width=58mm]{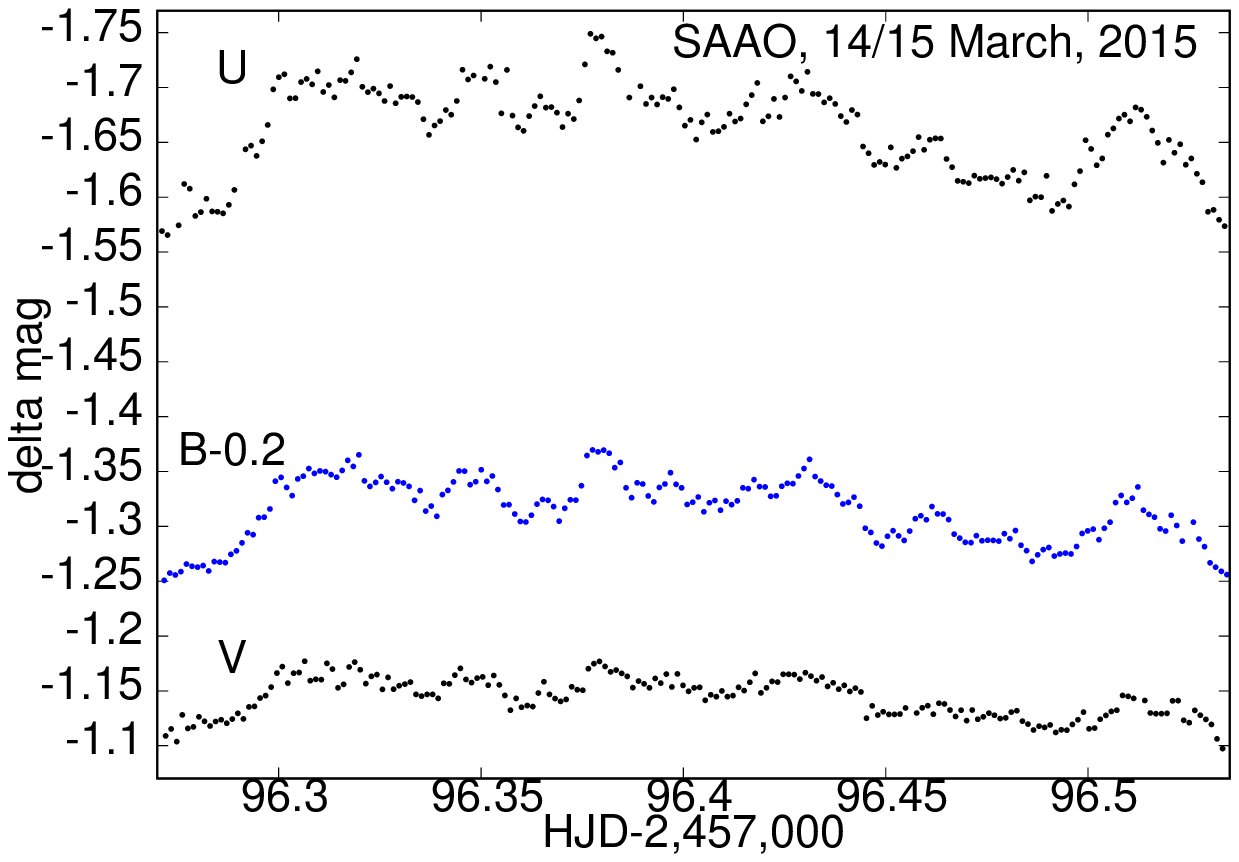}
\includegraphics[width=58mm]{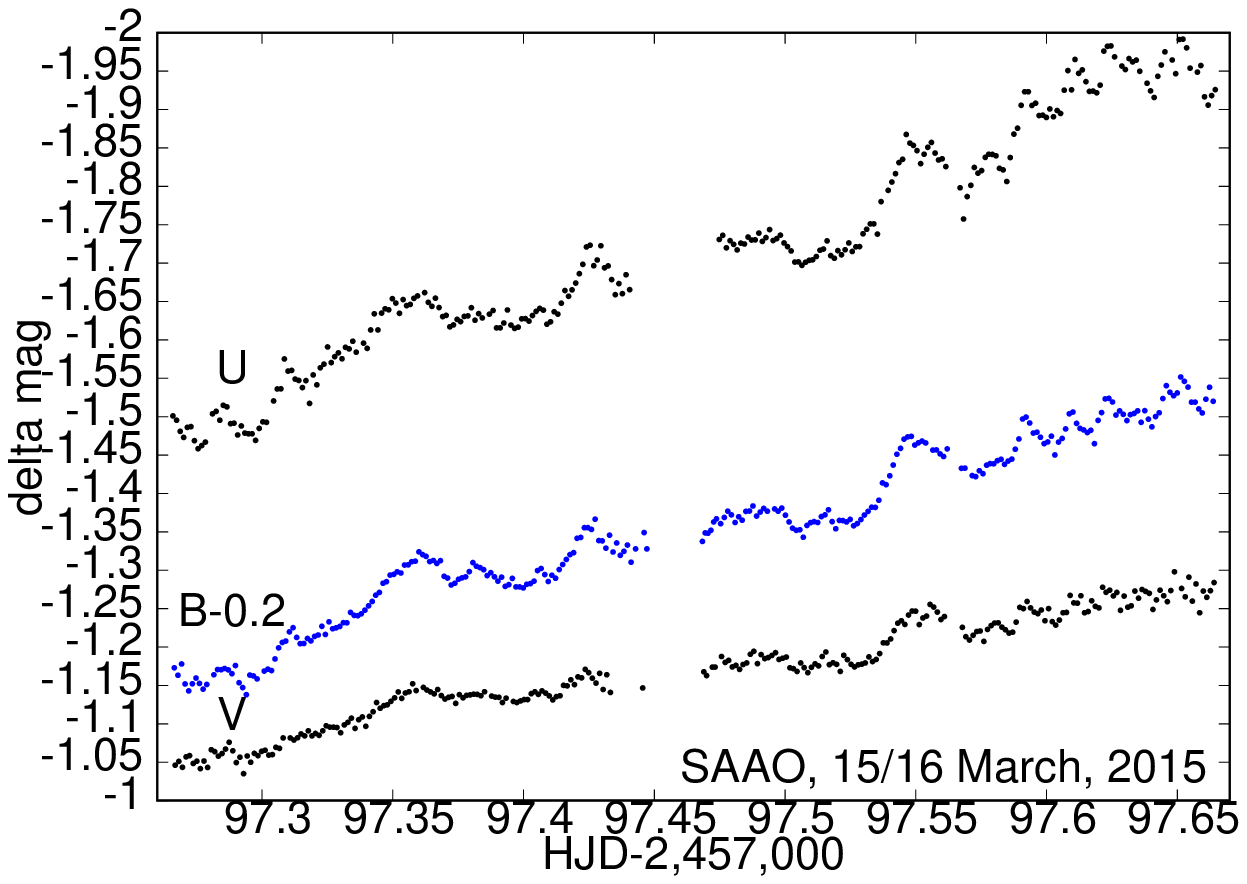}
\caption{The $UBV$ light curves obtained at {\it SAAO} in 2015. 
The light curves were obtained with respect to the first and second comparison stars 
from Table~\ref{Tab.comp} and were left in the instrumental system uncorrected 
for atmospheric extinction. 
Only runs lasting longer than 3 hours are shown.}
\label{Fig.saao15}
\end{figure*}
%----------------------------------------------------------------------

% -----------------------  Fig.4 appendix the SAAO 2016 light curves ---------------------
\begin{figure*}
\includegraphics[width=58mm]{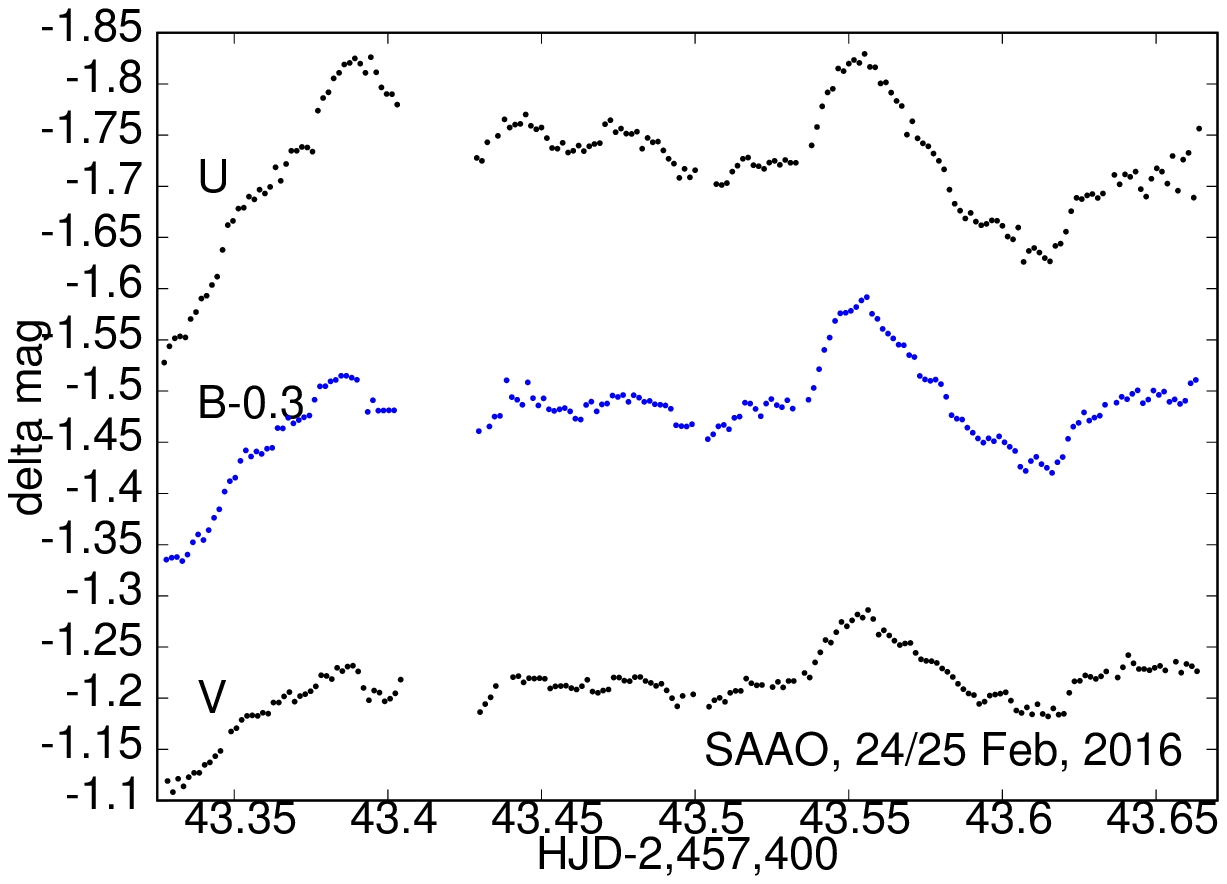}
\includegraphics[width=58mm]{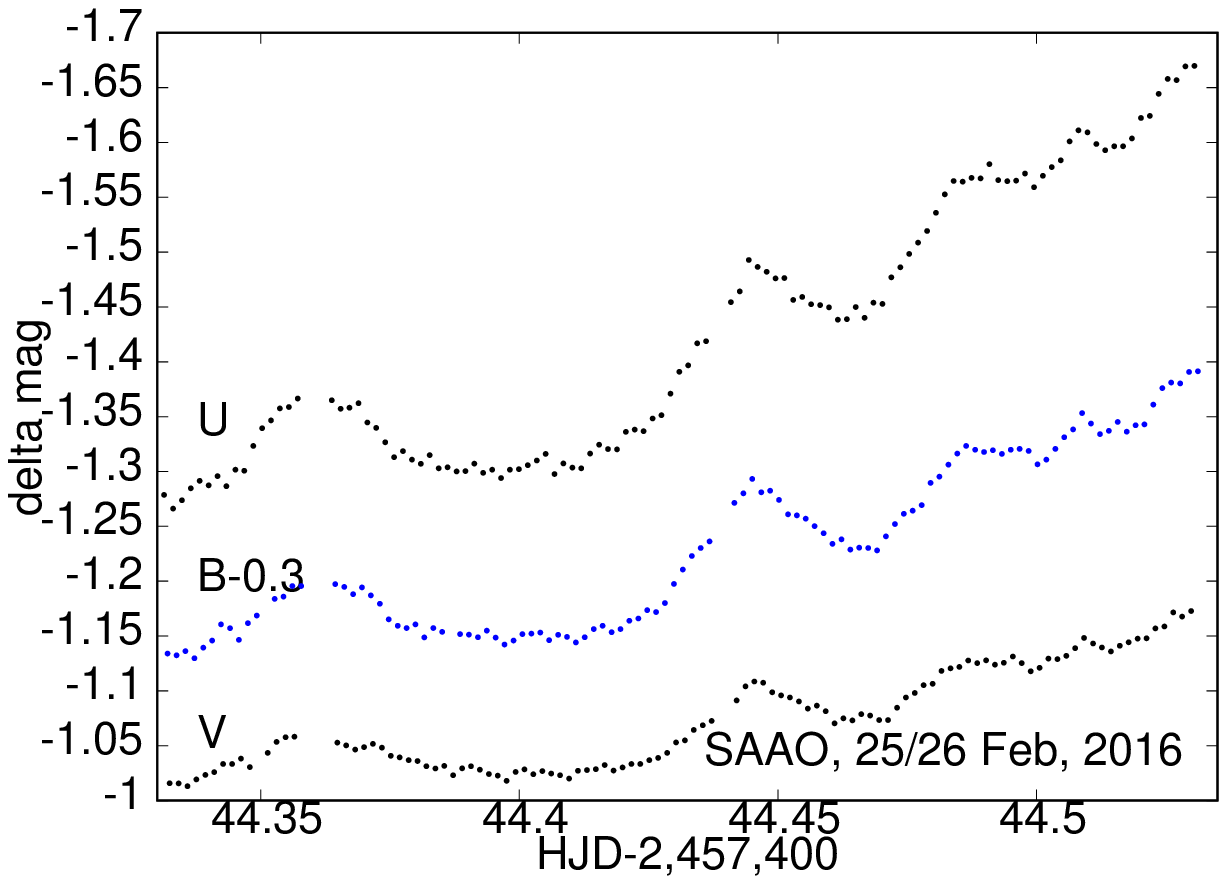}
\includegraphics[width=58mm]{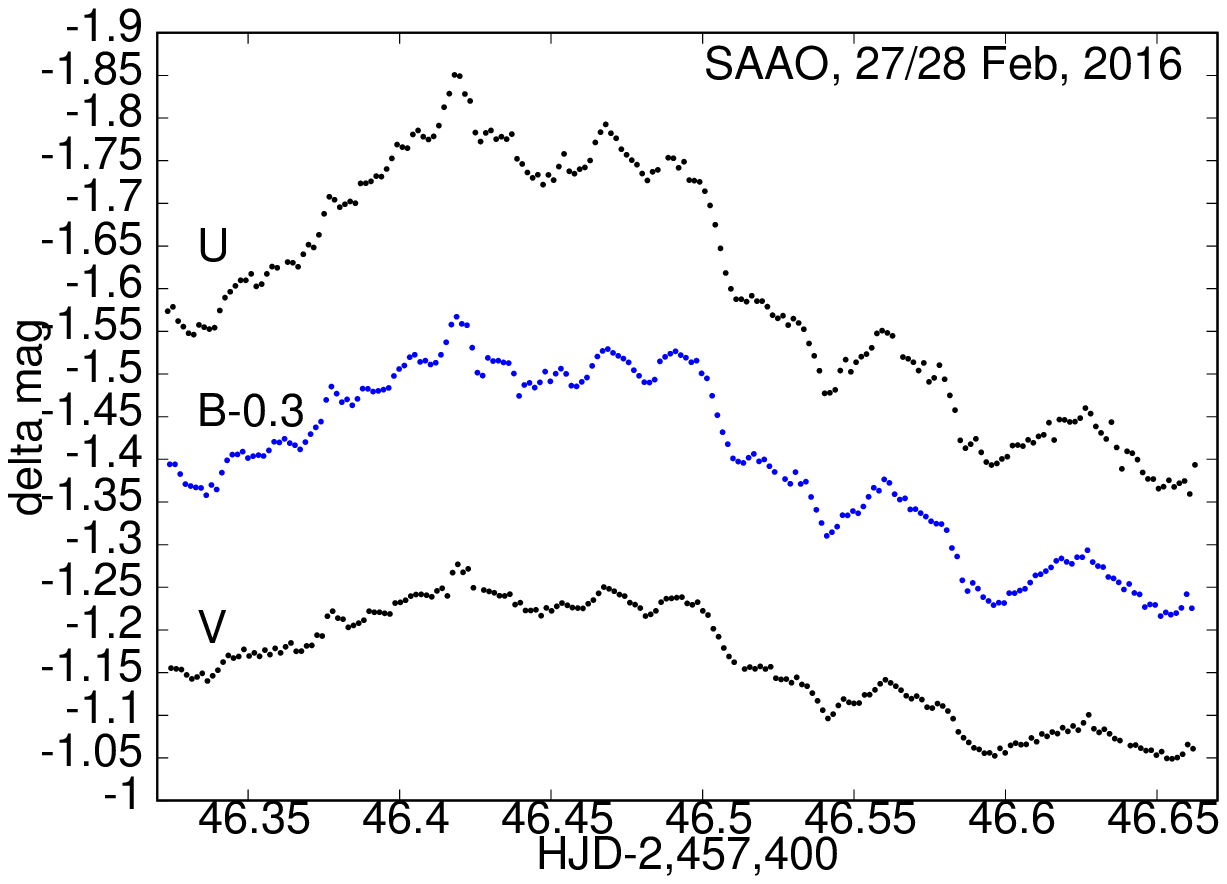}
\includegraphics[width=58mm]{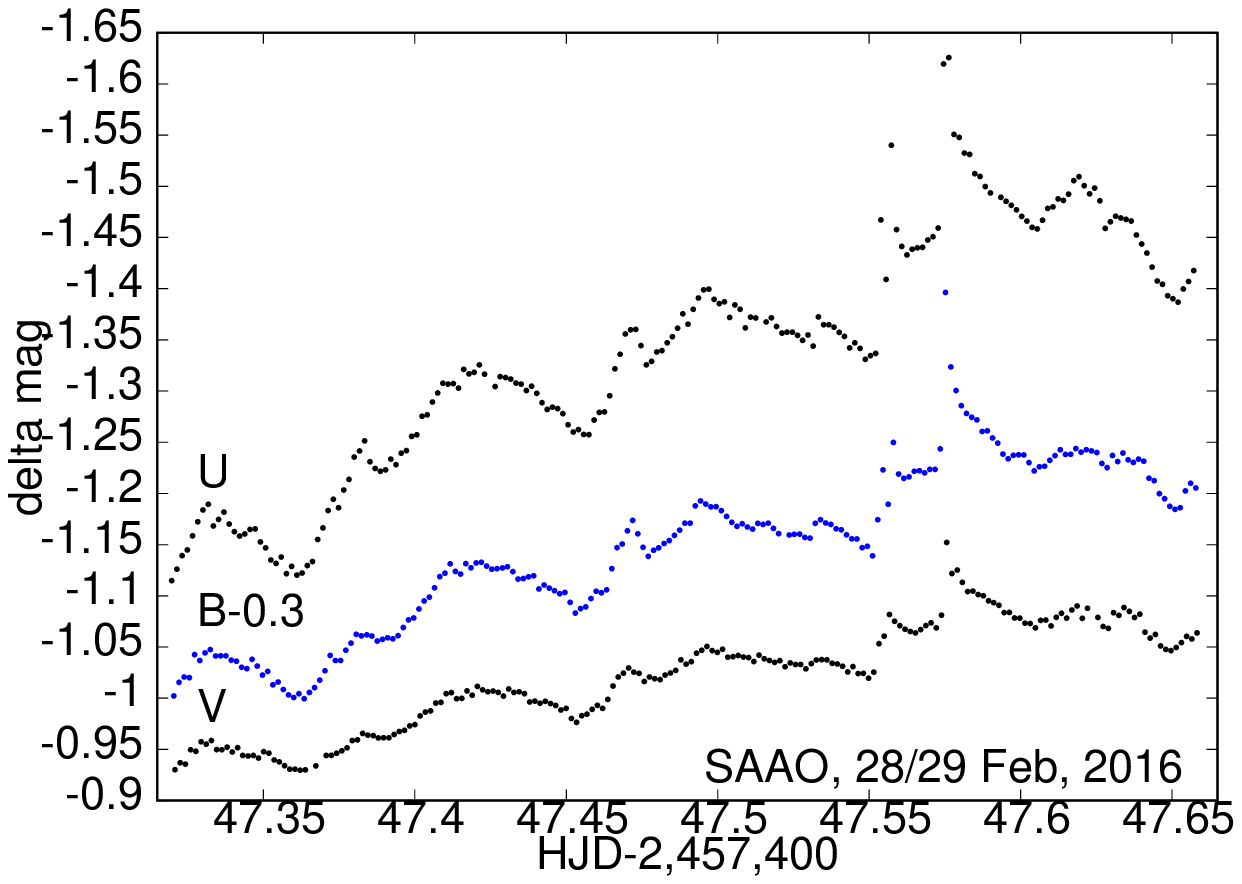}
\includegraphics[width=58mm]{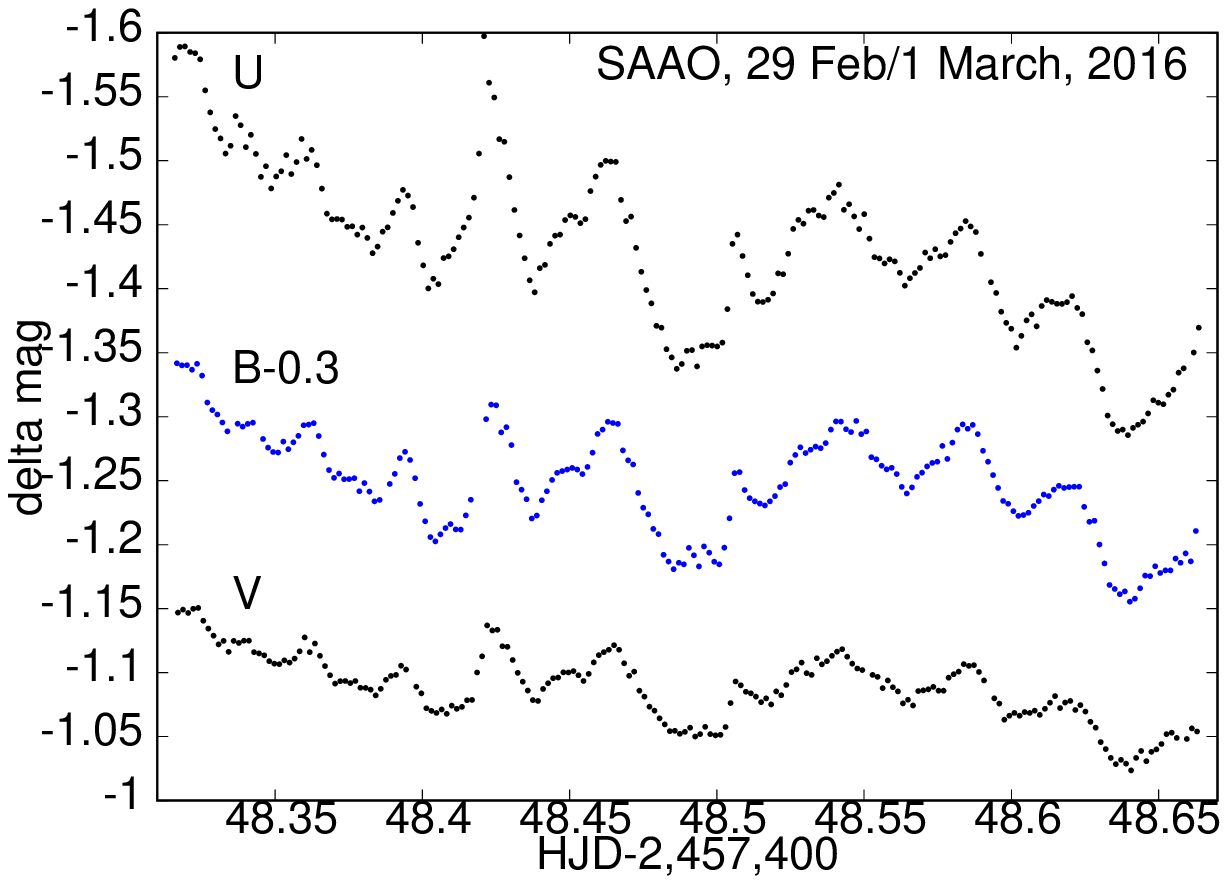}
\includegraphics[width=58mm]{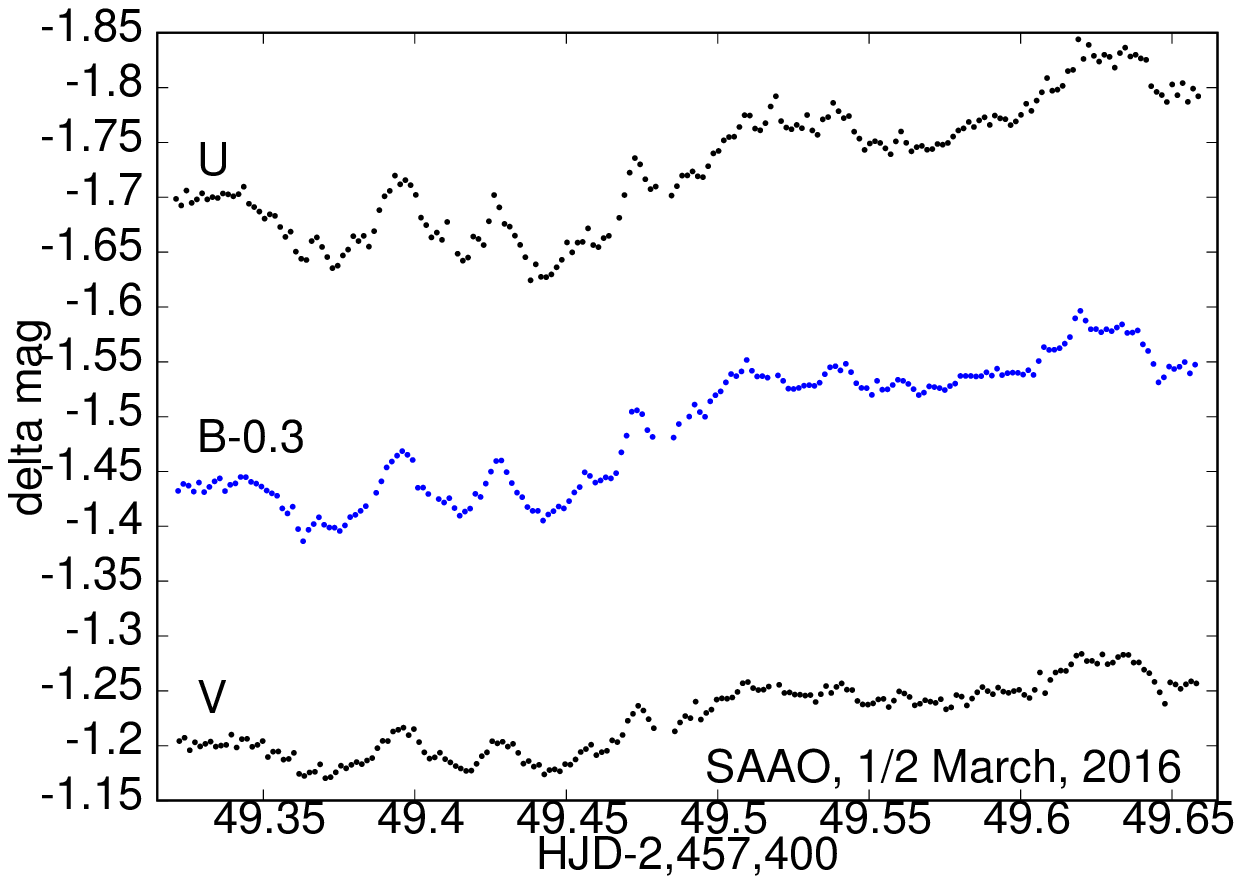}
\includegraphics[width=58mm]{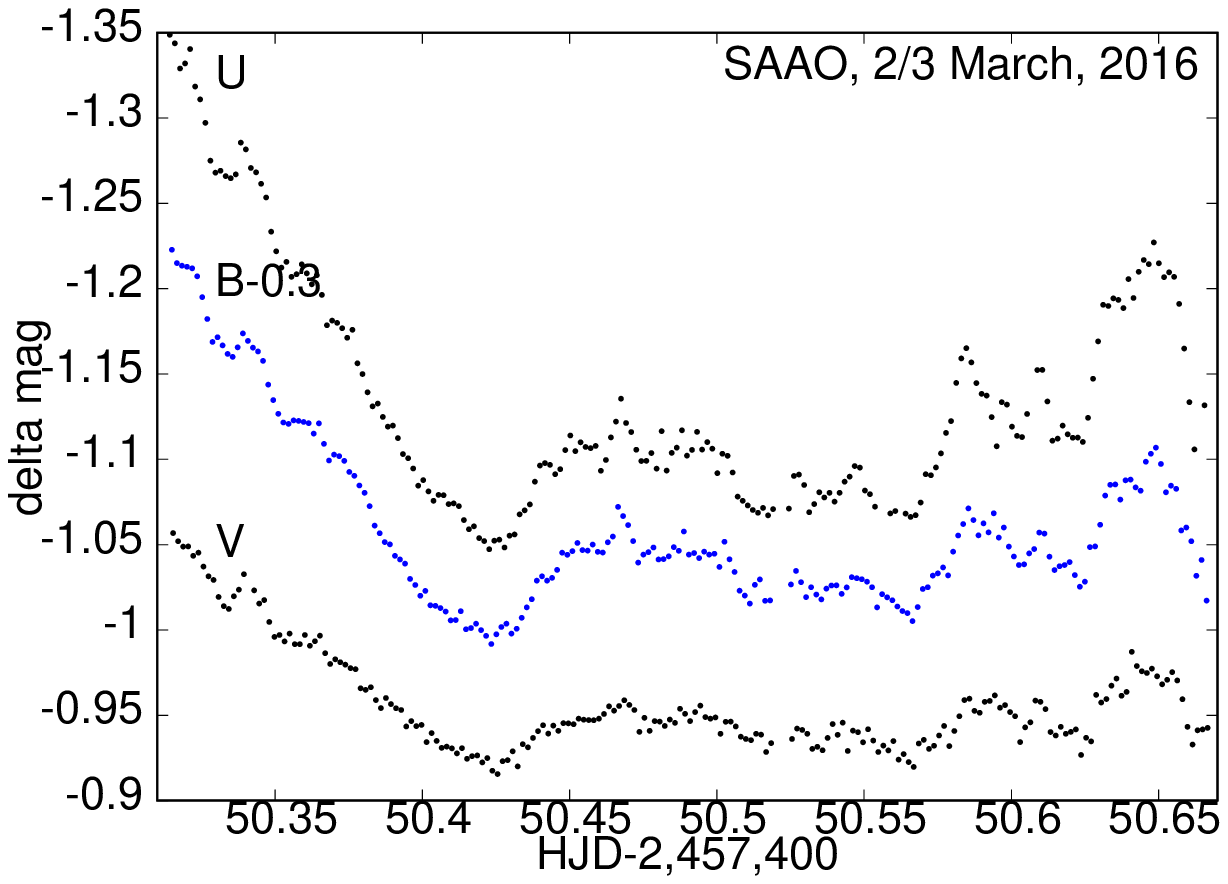}
\includegraphics[width=58mm]{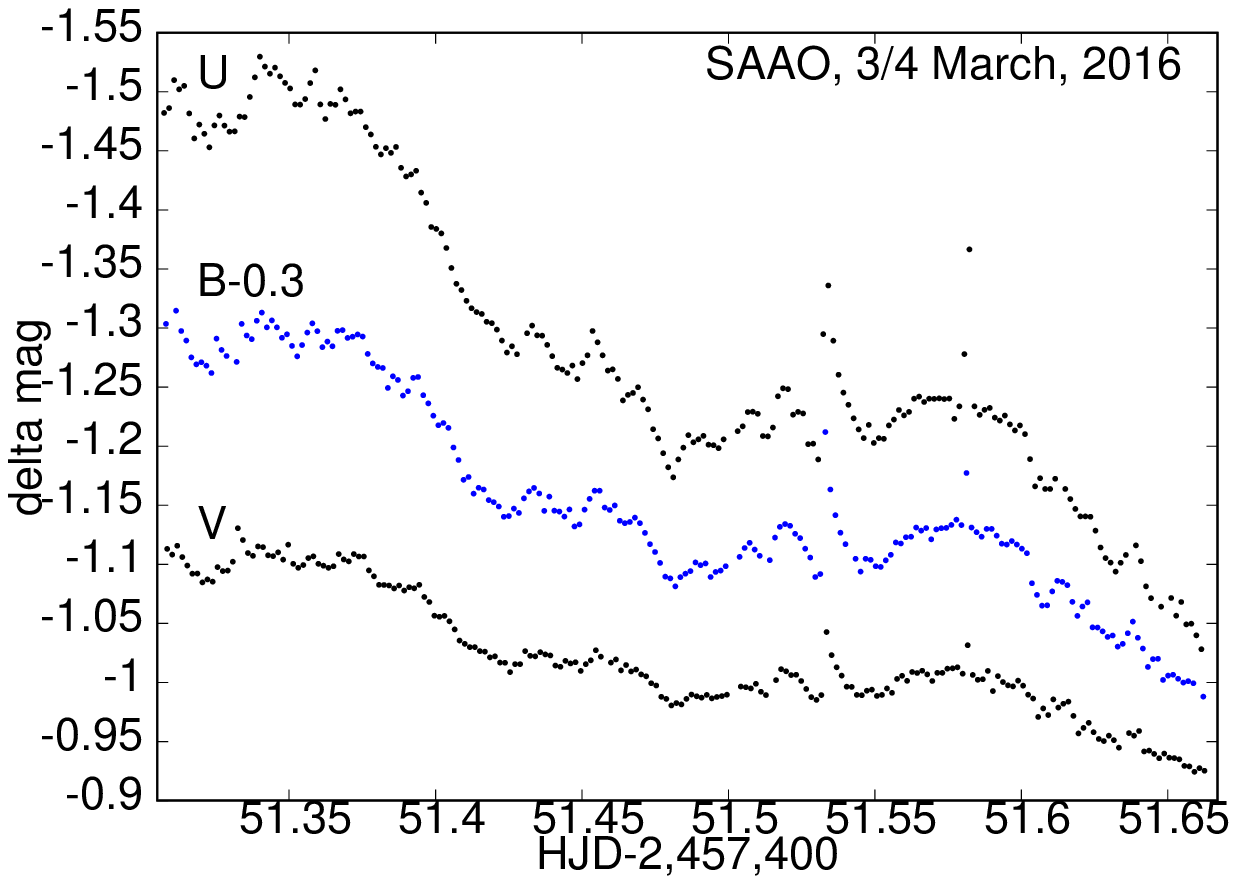}
\includegraphics[width=58mm]{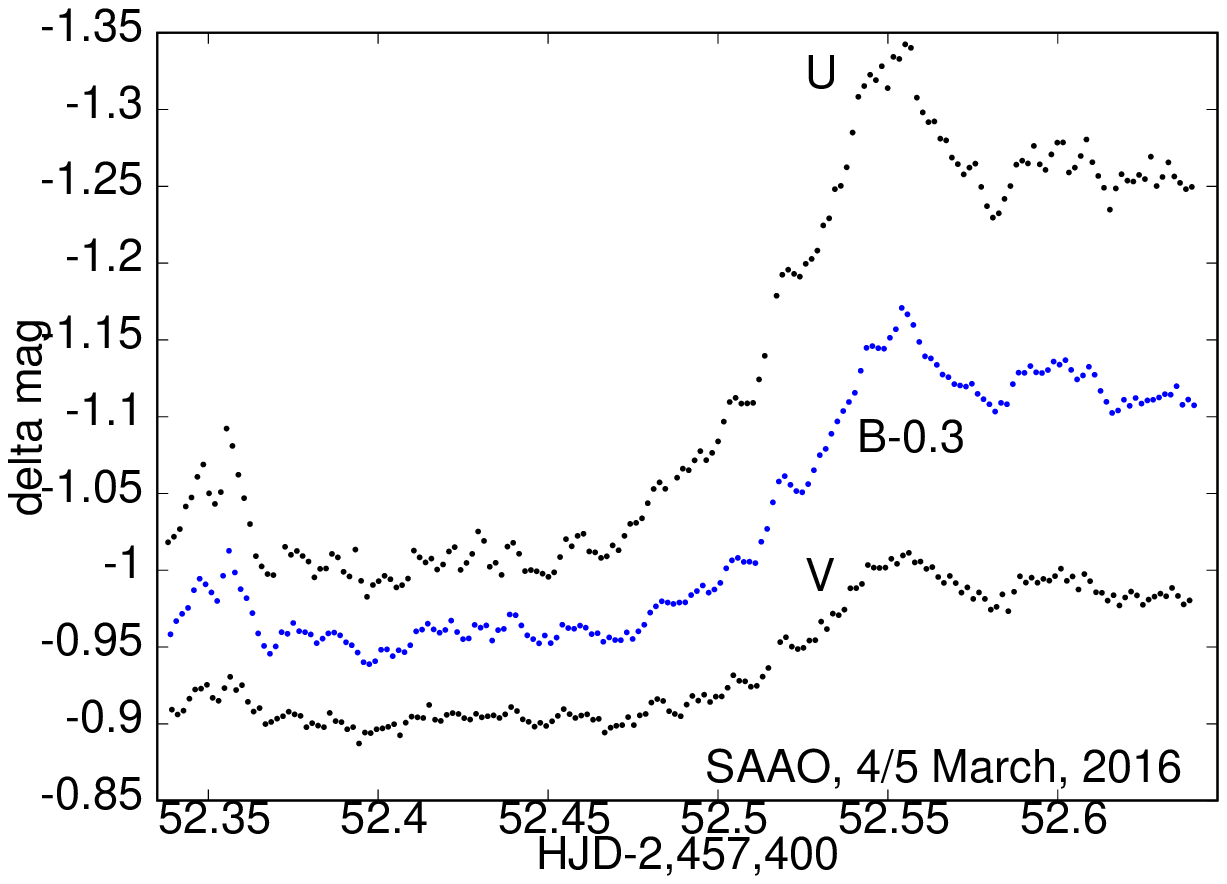}
\includegraphics[width=58mm]{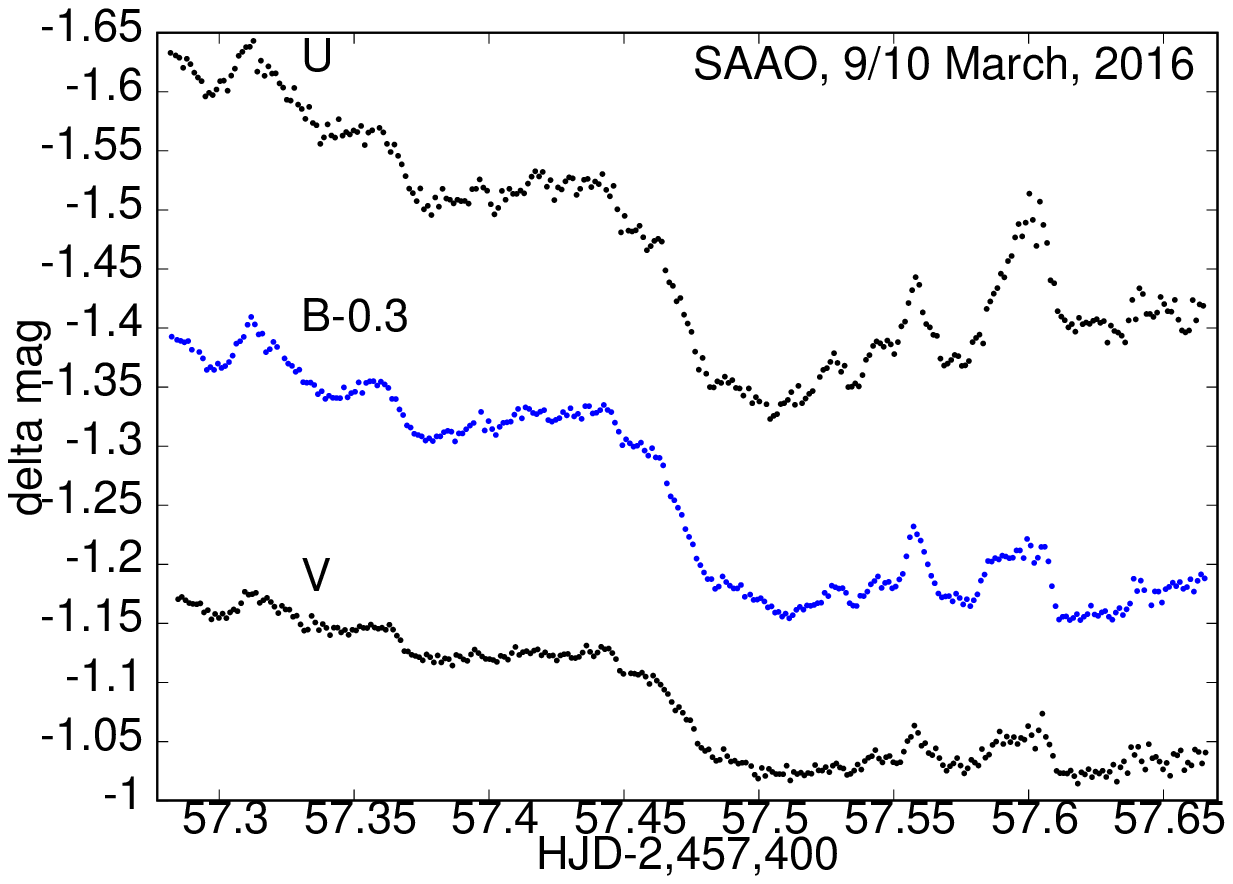}
\includegraphics[width=58mm]{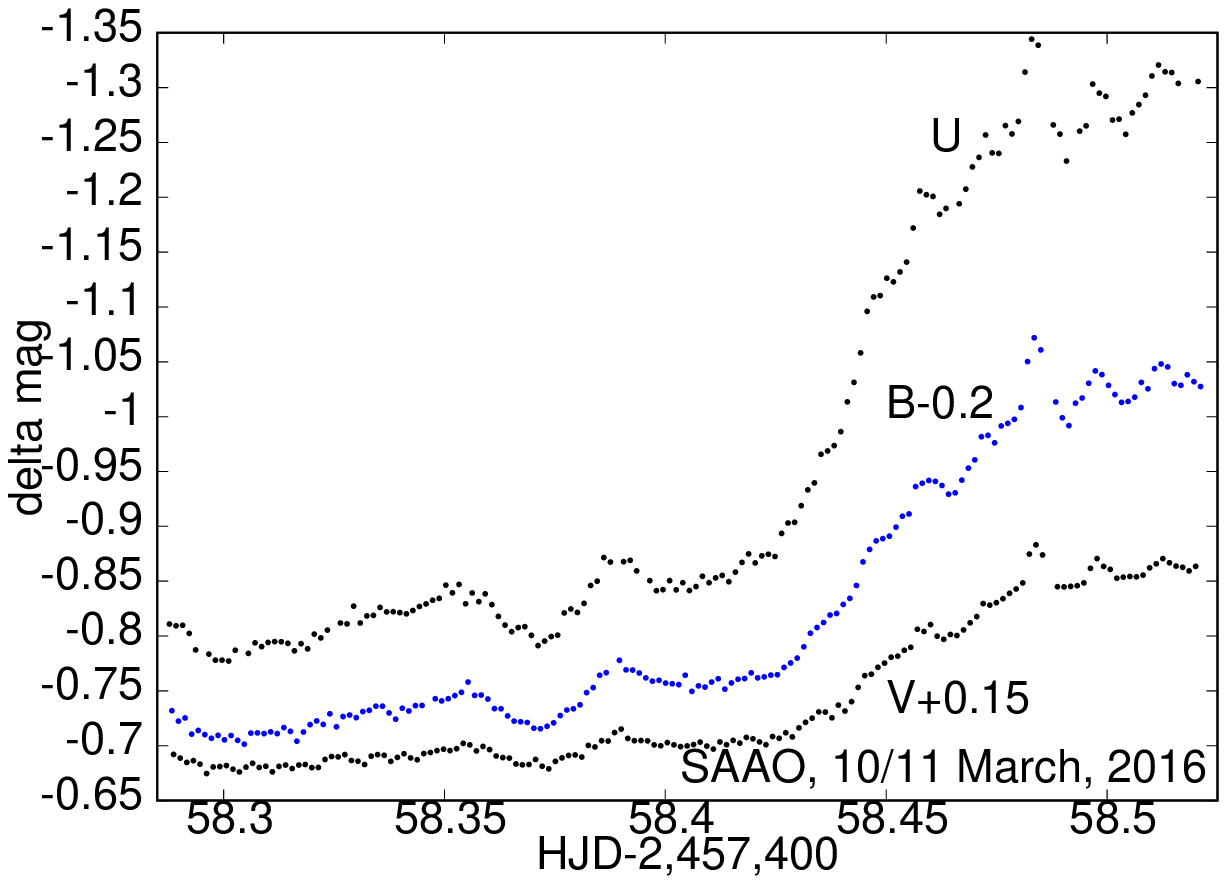}
\includegraphics[width=58mm]{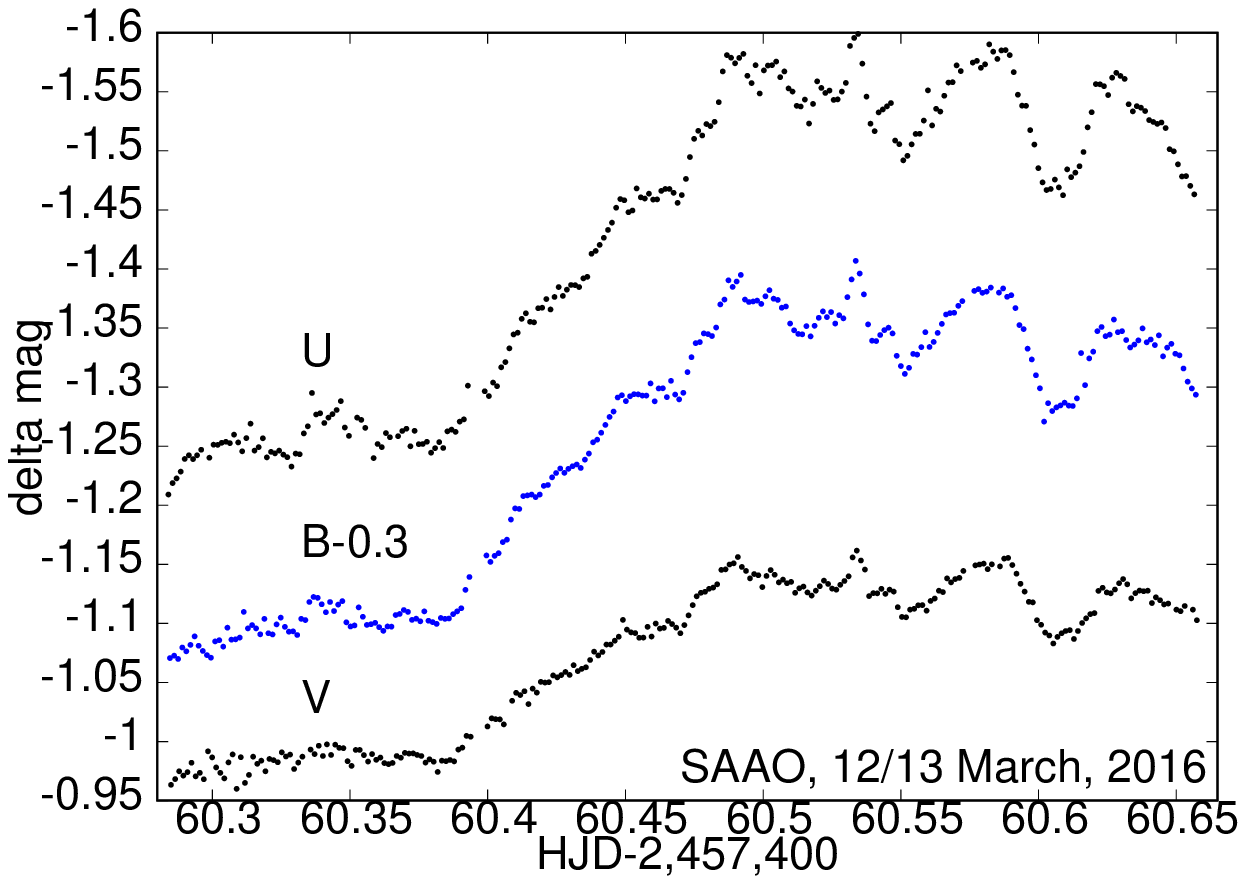}
\includegraphics[width=58mm]{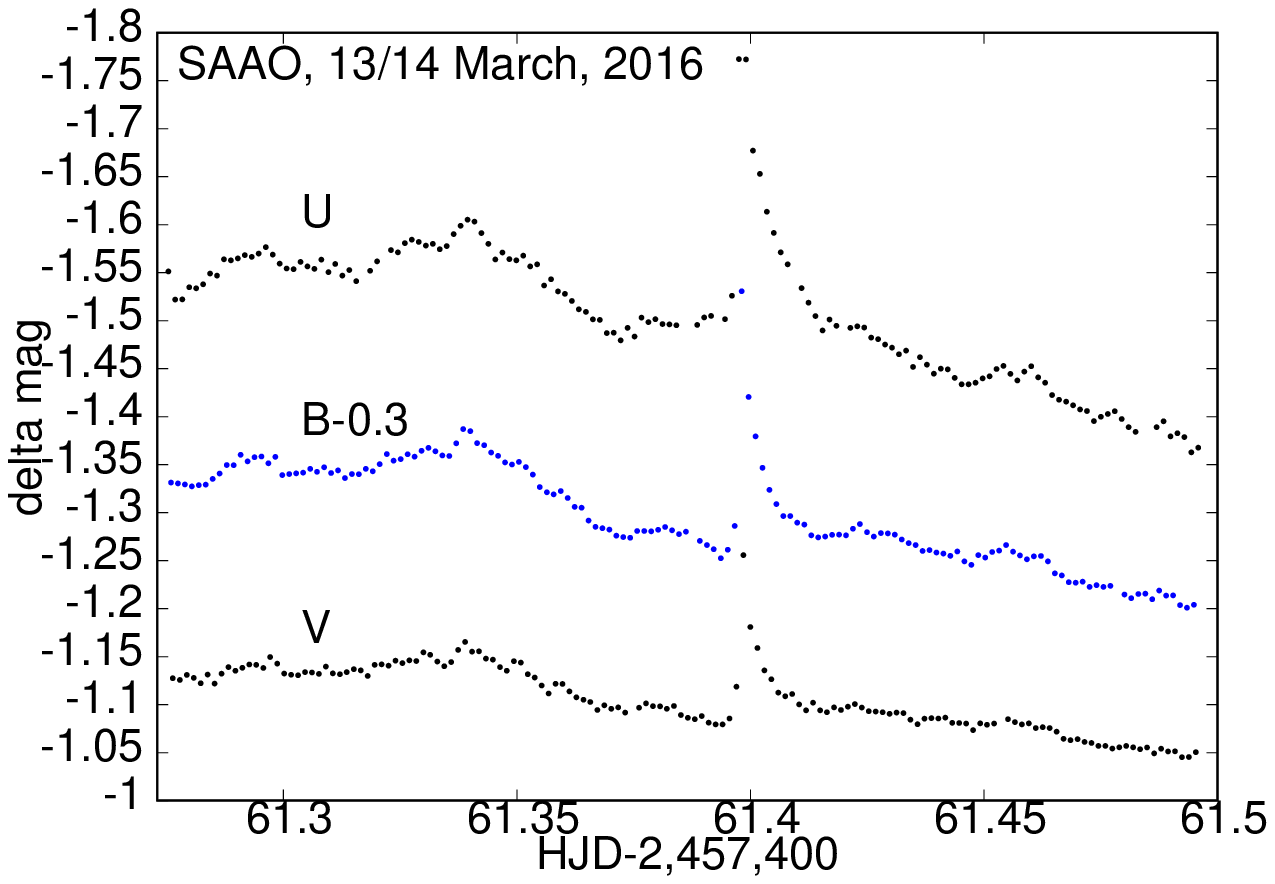}
\includegraphics[width=58mm]{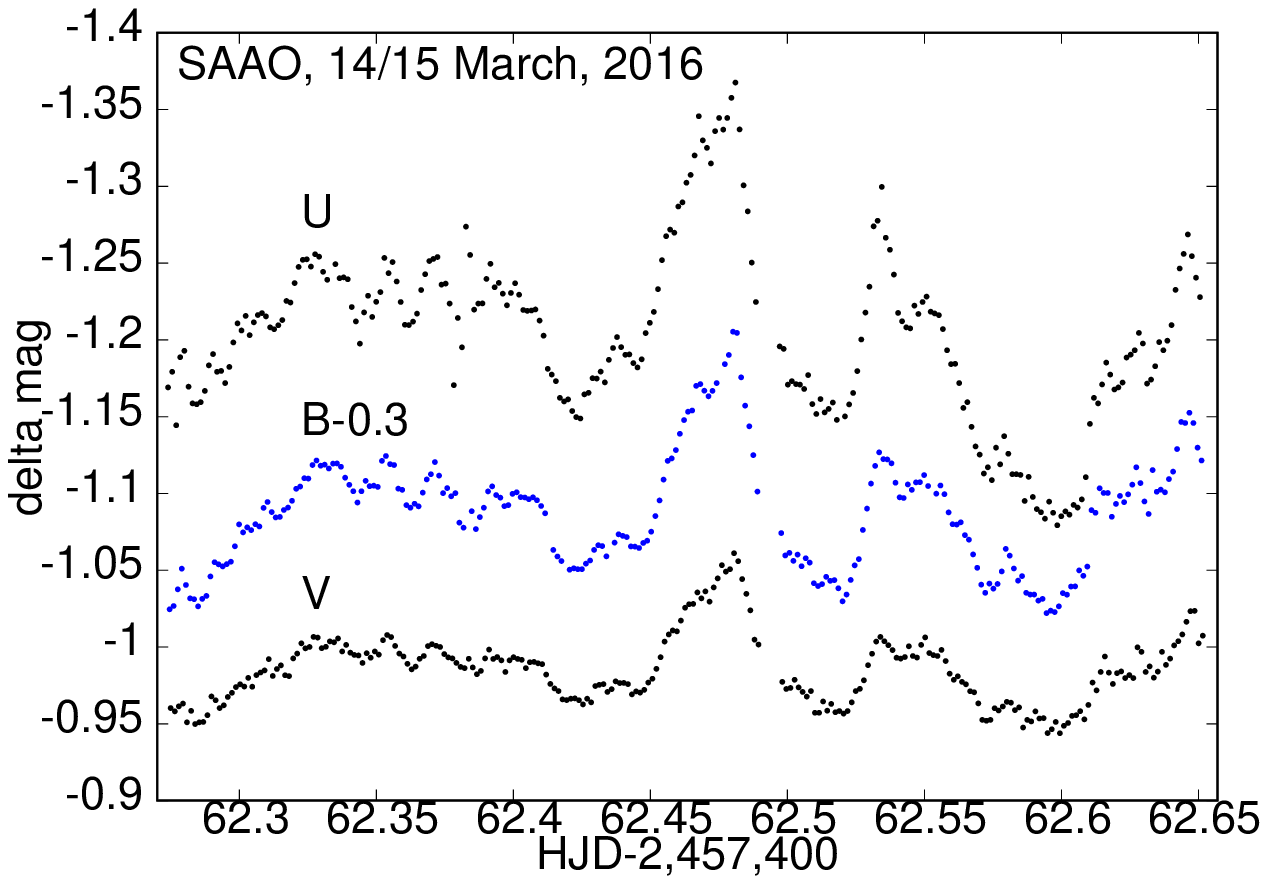}
\includegraphics[width=58mm]{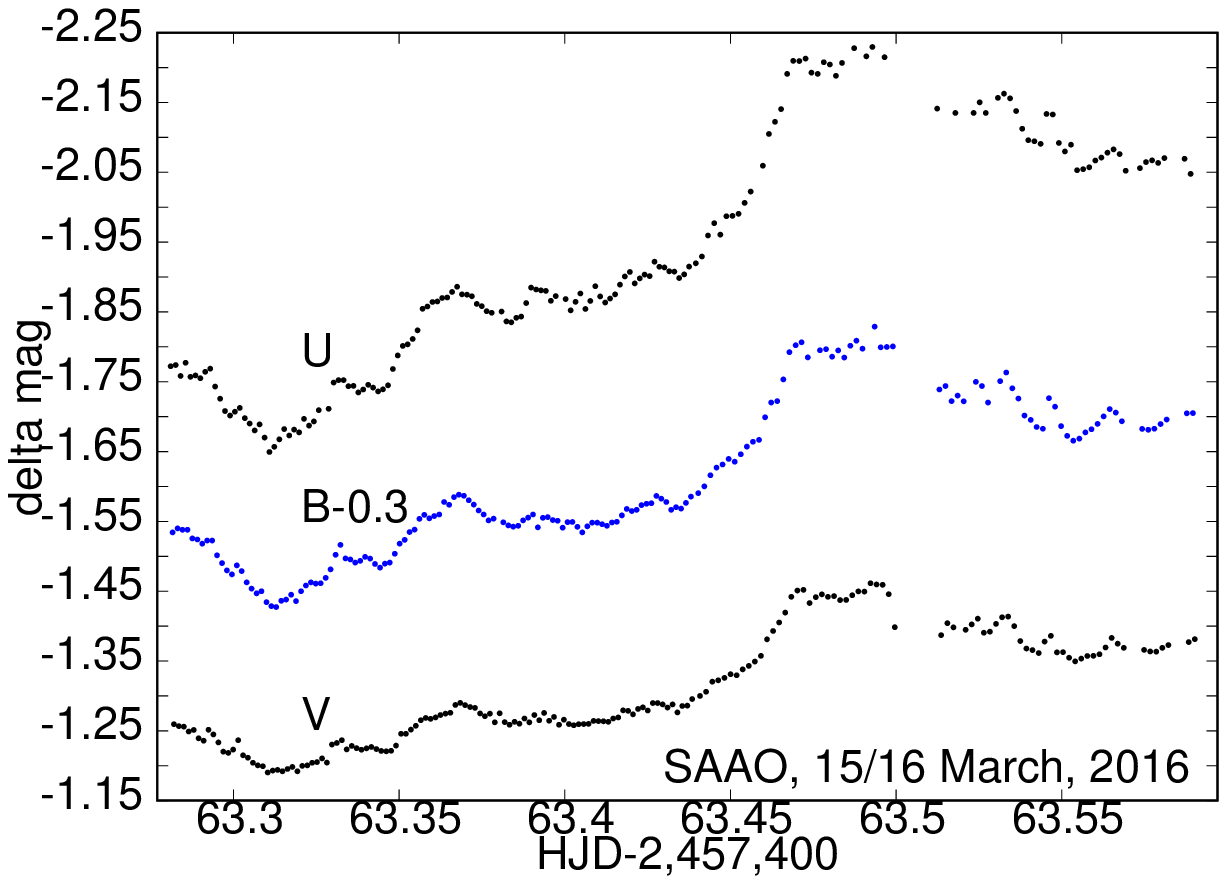}
\caption{The $UBV$ light curves obtained at {\it SAAO} in 2016. 
The light curves were obtained with respect to the first and third comparison stars 
from Table~\ref{Tab.comp}, and were left in the instrumental system  
uncorrected for atmospheric extinction. 
Only runs lasting longer than 3 hours are shown.}
\label{Fig.saao16}
\end{figure*}
%----------------------------------------------------------------------

% ----------------------- Fig.2 appendix the CTIO 2014 light curves ---------------------
\begin{figure*}
\includegraphics[width=58mm]{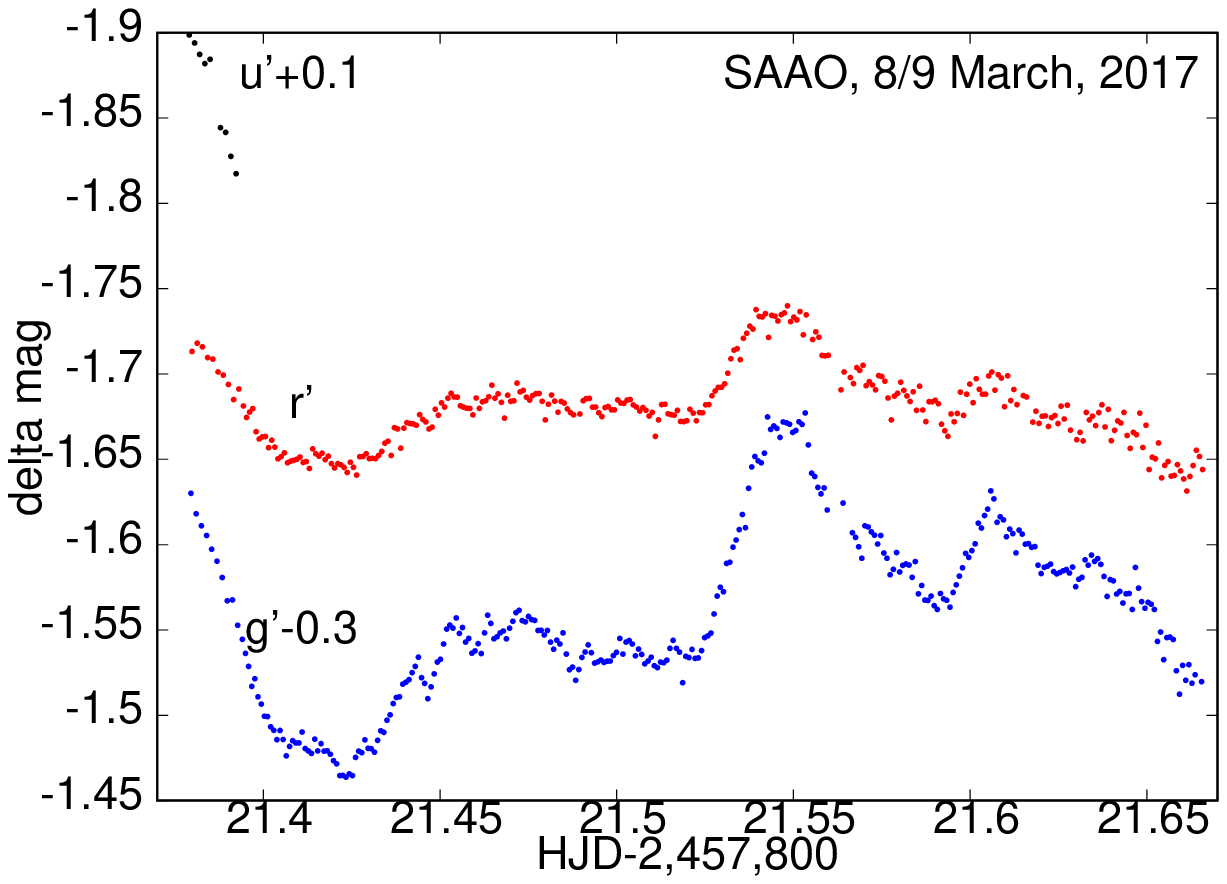}
\includegraphics[width=58mm]{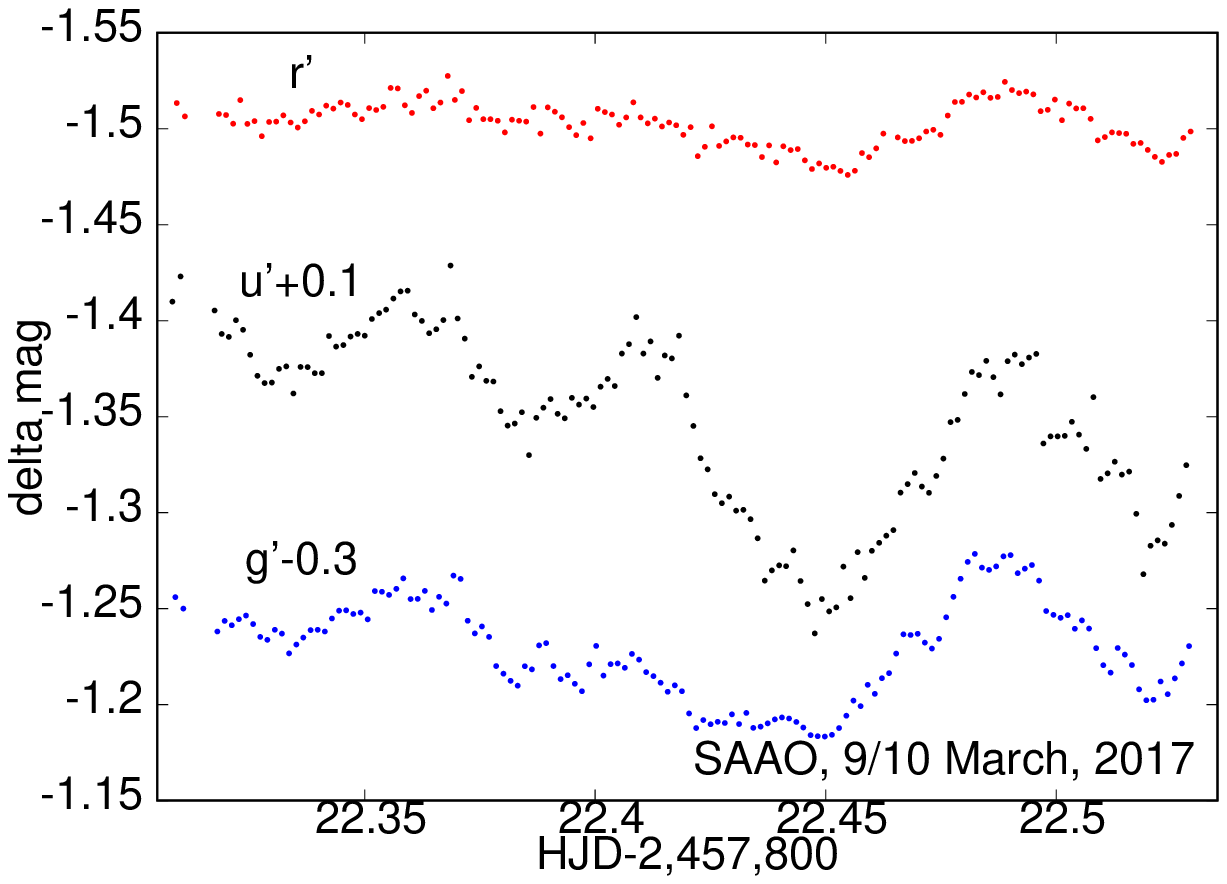}
\includegraphics[width=58mm]{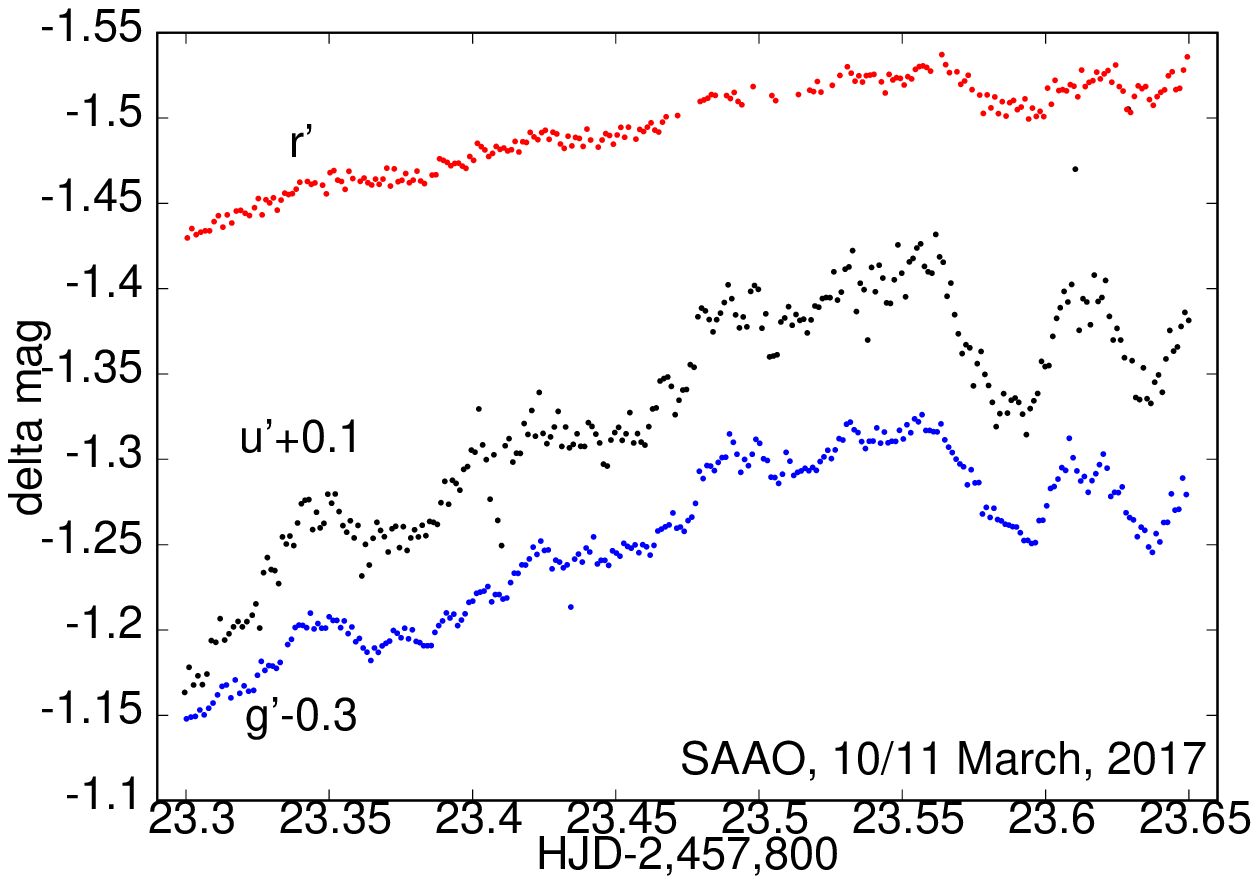}
\includegraphics[width=58mm]{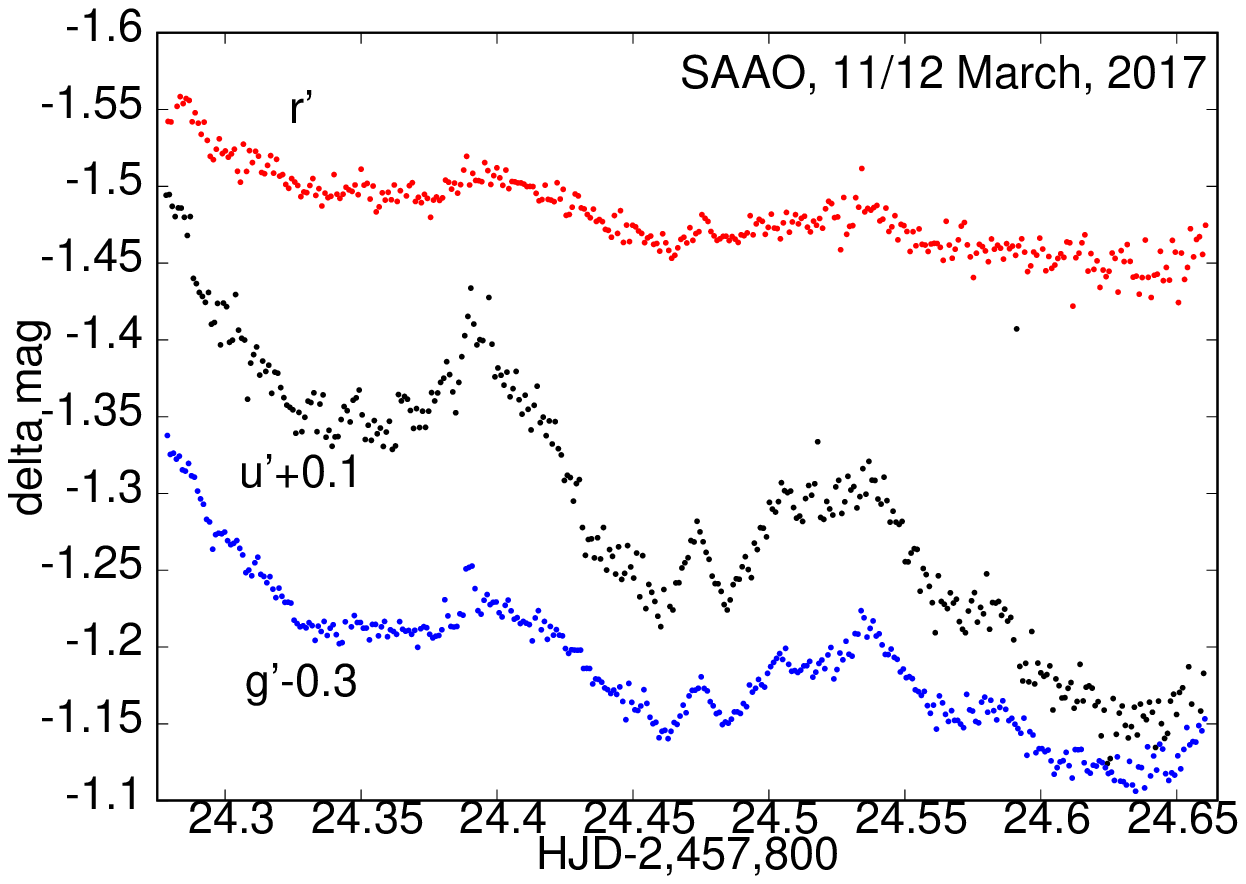}
\includegraphics[width=58mm]{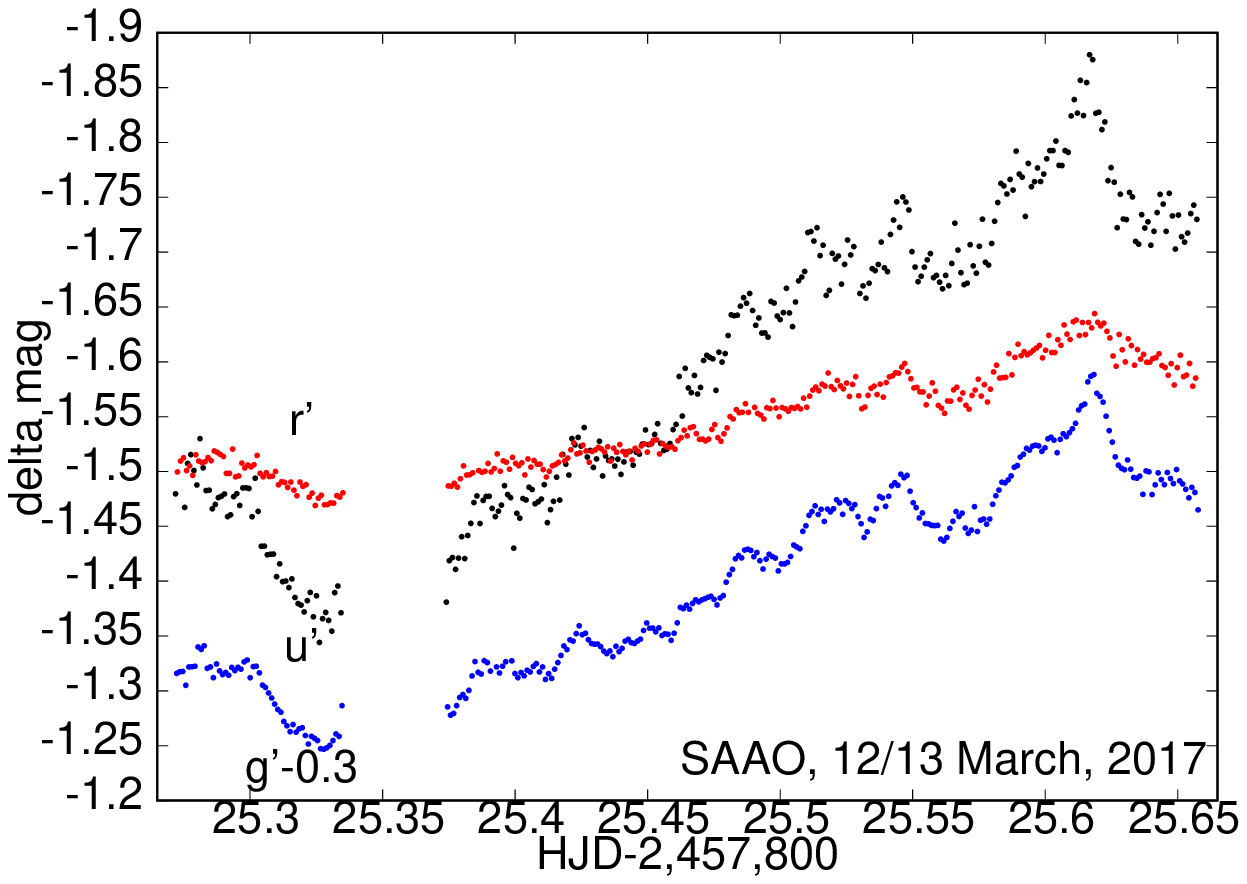}
\includegraphics[width=58mm]{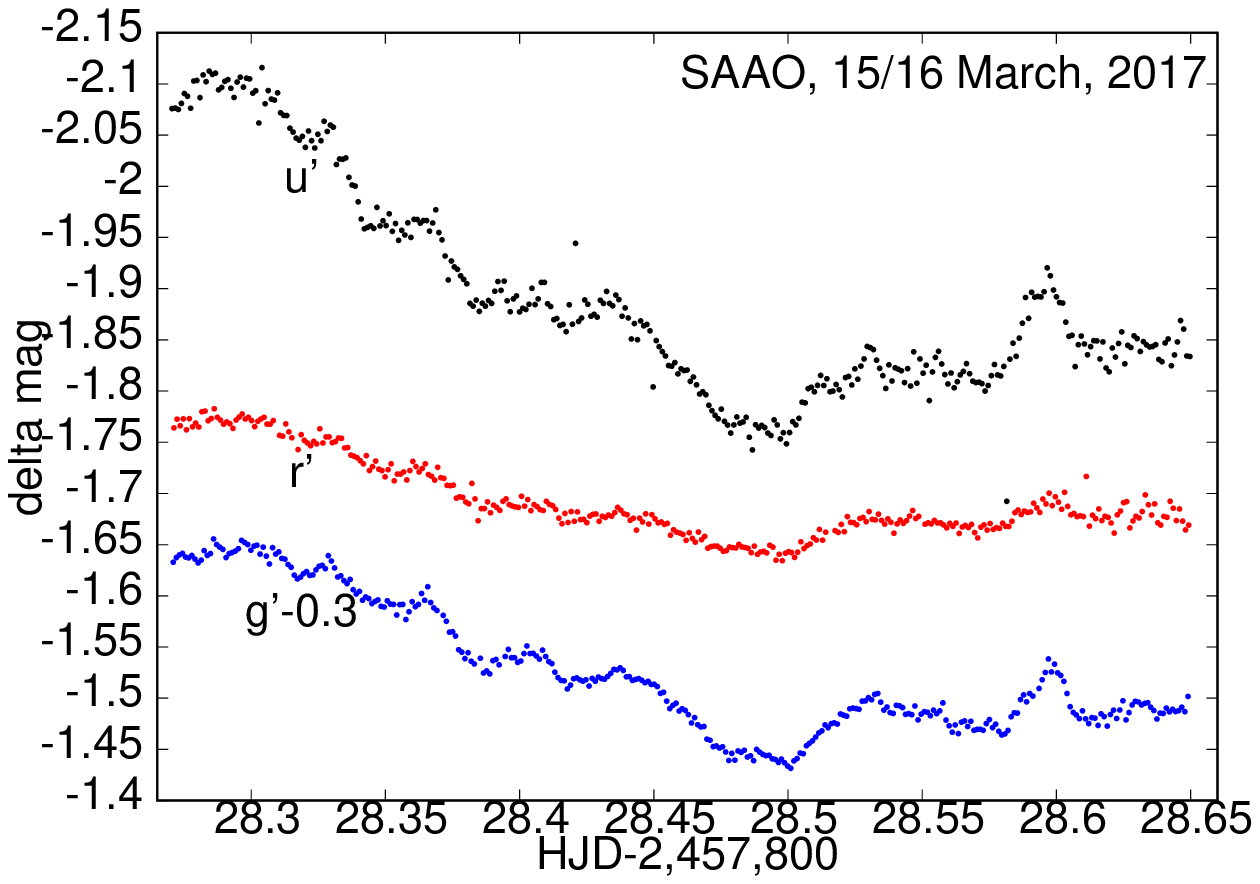}
\includegraphics[width=58mm]{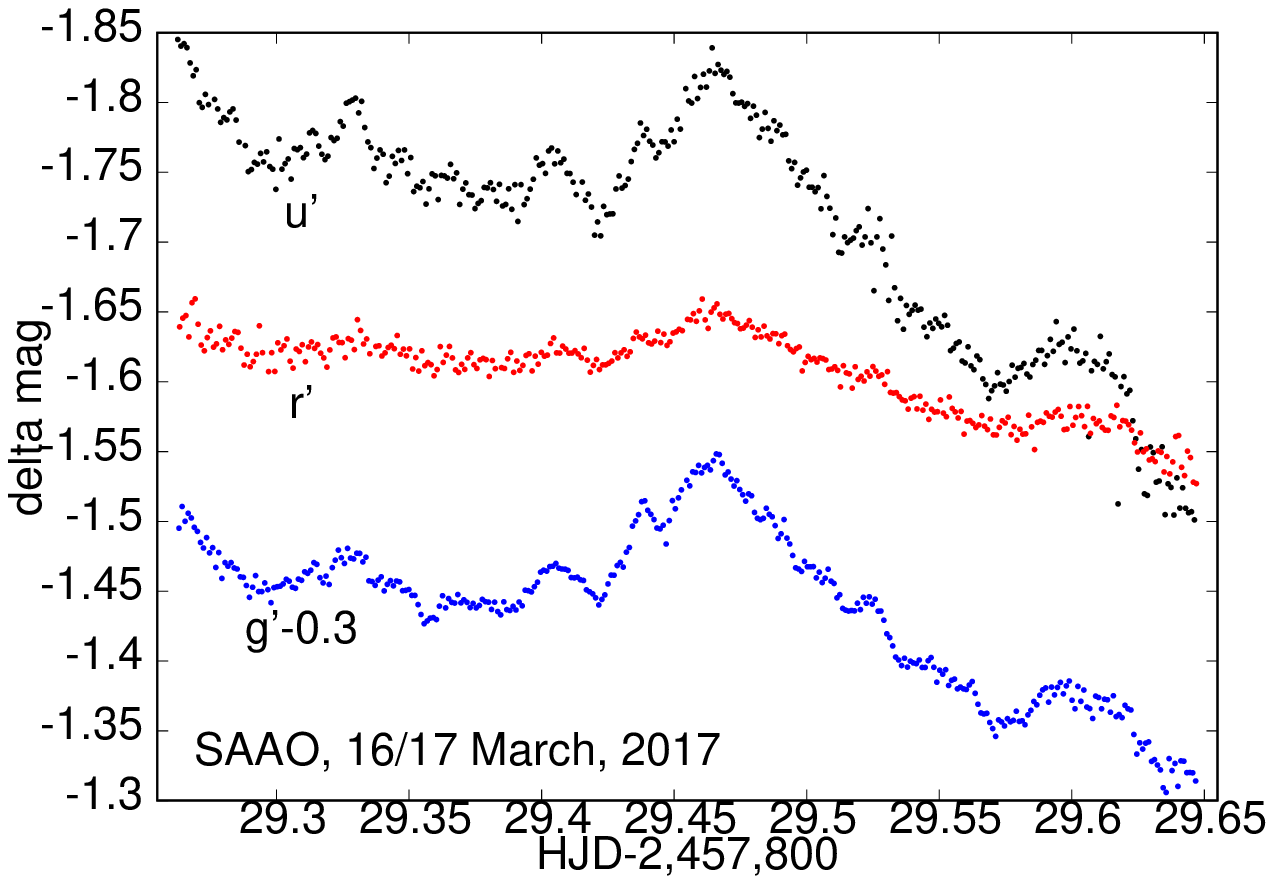}
\includegraphics[width=58mm]{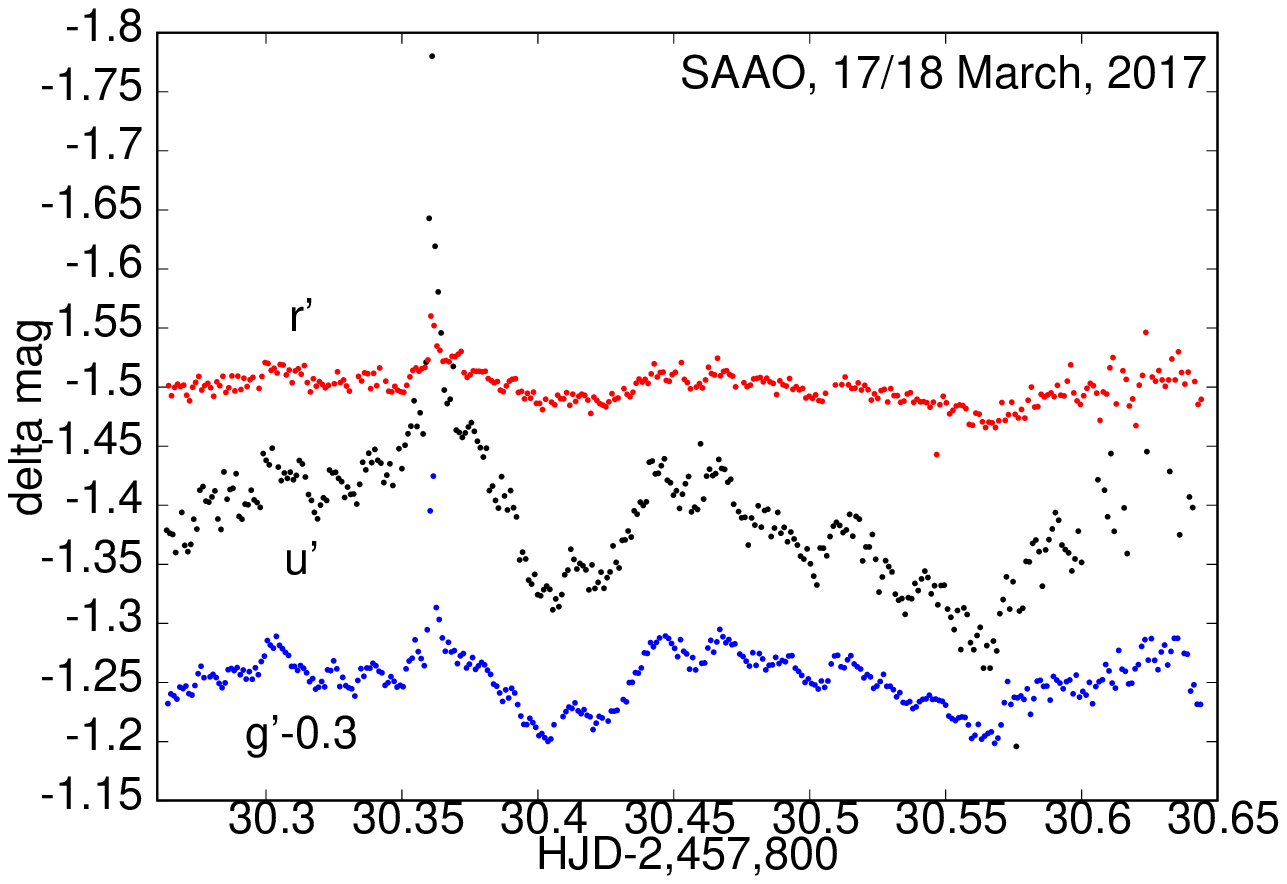}
\includegraphics[width=58mm]{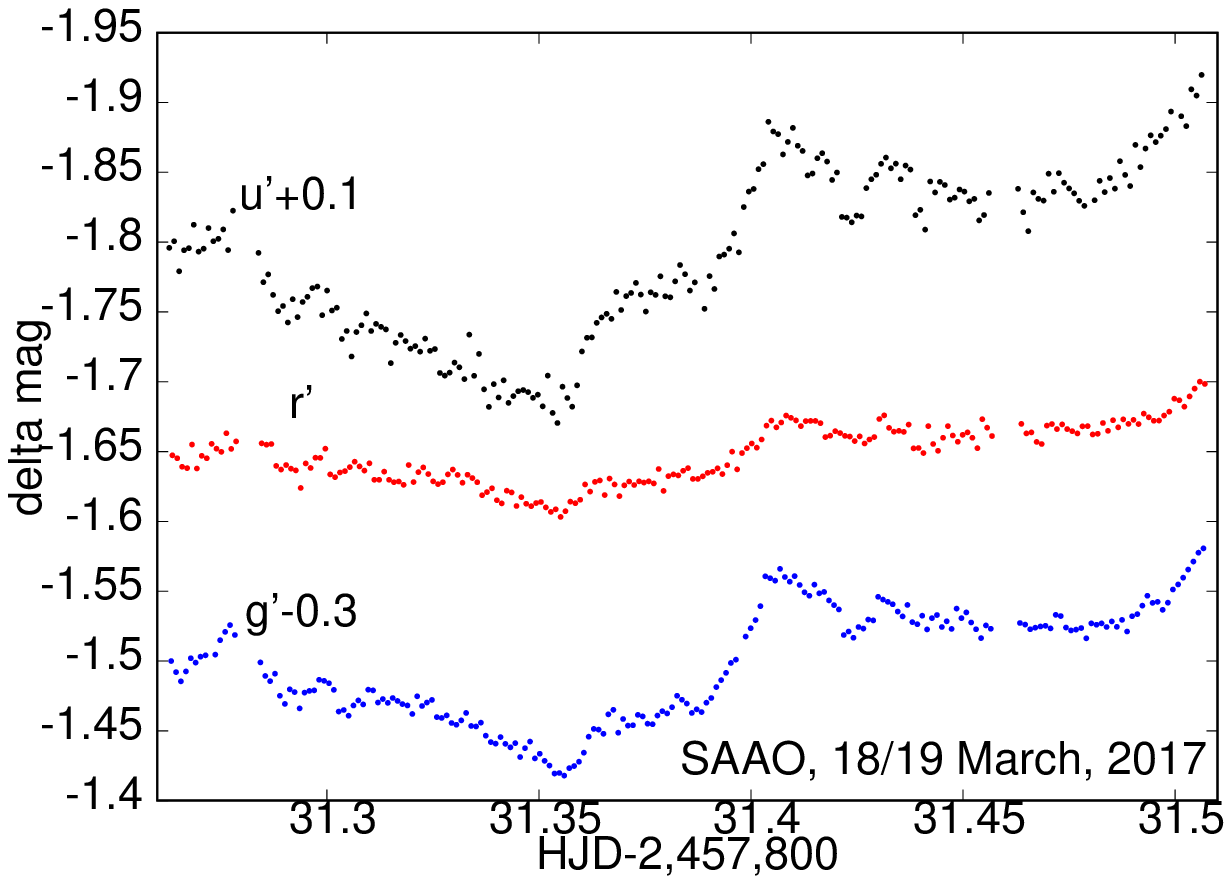}
\includegraphics[width=58mm]{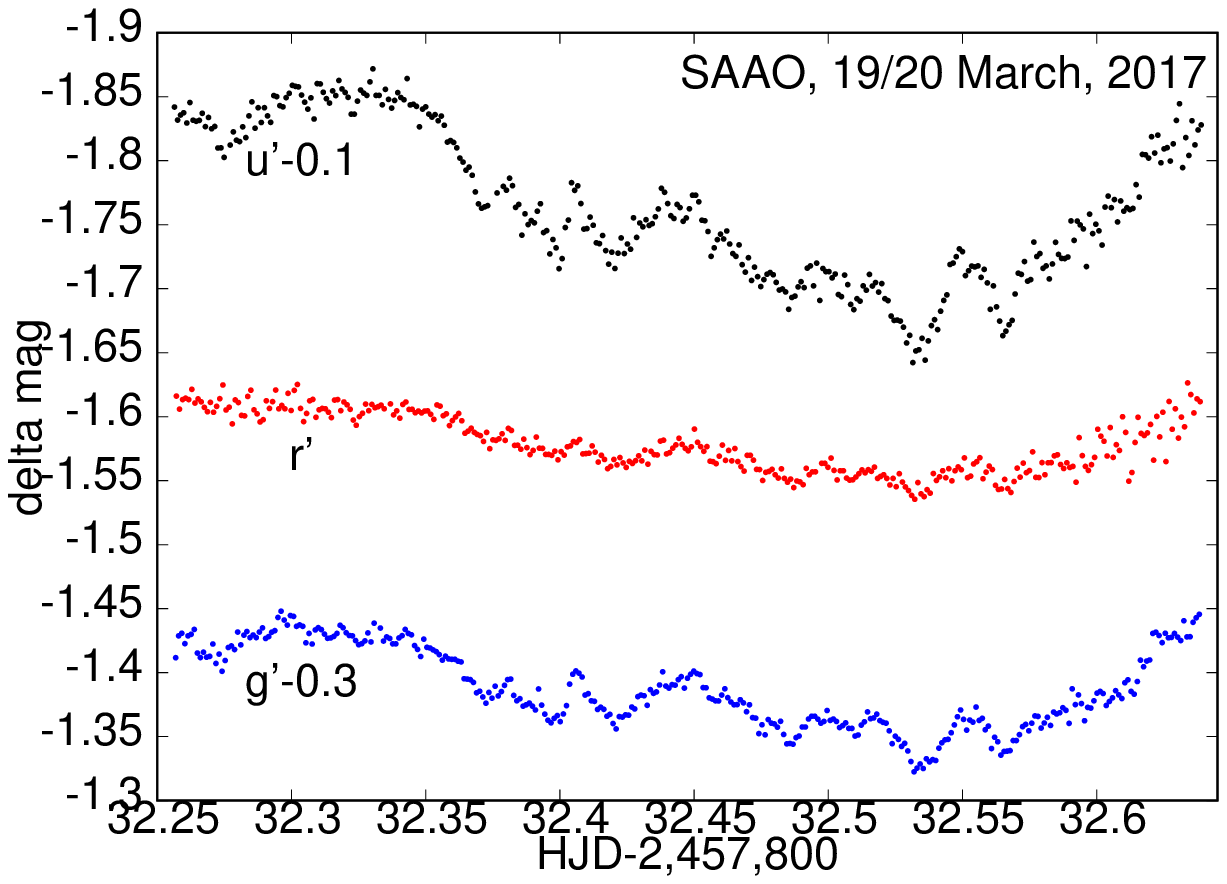}
\includegraphics[width=58mm]{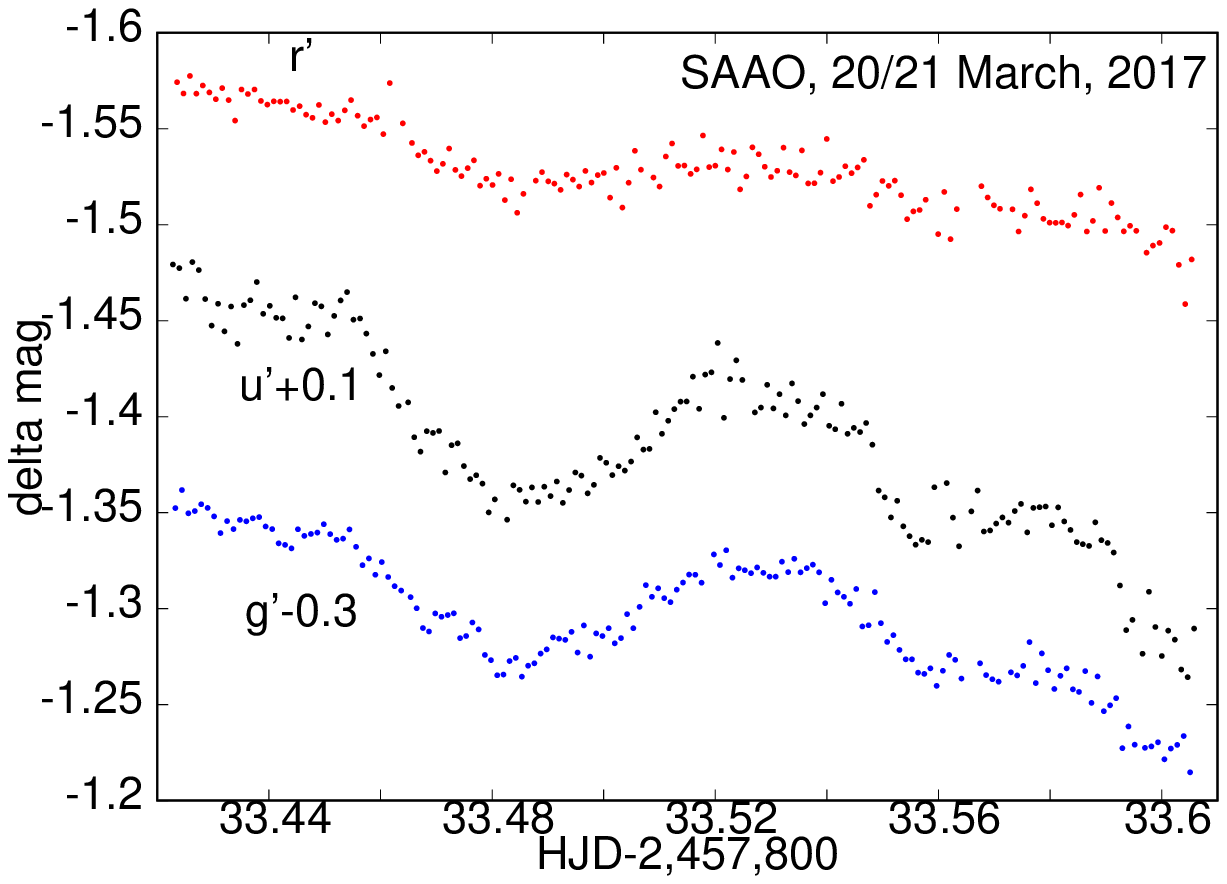}
\caption{The $u'g'r'$ data obtained at {\it SAAO} in 2017 with respect 
to the first and third comparison stars from Table~\ref{Tab.comp}. 
Data obtained both during photometric and partly-cloudy nights are shown.
The data were left in the instrumental system uncorrected for atmospheric extinction effects. A shift in the magnitude scale is sometimes added 
for clarity.}
\label{Fig.saao17}
\end{figure*}
%----------------------------------------------------------------------

\end{document}